\documentclass[final,reqno]{amsart}

\usepackage[colorlinks=true,allcolors=blue]{hyperref}
\hypersetup{hypertexnames=false}

\usepackage{amsaddr}
\usepackage{amssymb,amsfonts,amsmath}
\usepackage{amsthm}
\usepackage{thmtools}
\usepackage[capitalise]{cleveref}

\usepackage[square,numbers,sort]{natbib}
\usepackage{bm}
\usepackage{bbold}

\usepackage{tikz}

\crefname{equation}{}{}
\crefname{figure}{Figure}{Figures}
\Crefname{figure}{Figure}{Figures}

\newtheorem{theorem}{Theorem}[section]
\newtheorem{lemma}[theorem]{Lemma}
\crefname{lemma}{Lemma}{Lemmas}
\Crefname{lemma}{Lemma}{Lemmas}
\newtheorem{corollary}[theorem]{Corollary}
\crefname{corollary}{Corollary}{Corollaries}
\Crefname{corollary}{Corollary}{Corollaries}
\newtheorem{proposition}[theorem]{Proposition}
\crefname{proposition}{Proposition}{Propositions}
\Crefname{proposition}{Proposition}{Propositions}

\newtheorem*{claim}{Claim}
\theoremstyle{remark}
\newtheorem{remark}{Remark}
\crefname{remark}{Remark}{Remarks}
\Crefname{remark}{Remark}{Remarks}

\renewcommand{\le}{\leqslant}
\renewcommand{\ge}{\geqslant}
\let\phi=\varphi
\renewcommand{\epsilon}{\varepsilon}
\undef\ul
\newcommand{\ol}[1]{\overline{#1}}
\newcommand{\ul}[1]{\underline{#1}}
\newcommand{\wt}[1]{\widetilde{#1}}
\newcommand{\wh}[1]{\widehat{#1}}
\newcommand{\spu}{\uparrow}
\newcommand{\spd}{\downarrow}

\DeclareMathOperator{\diag}{diag}
\DeclareMathOperator{\ran}{ran}
\DeclareMathOperator{\supp}{supp}
\DeclareMathOperator{\Real}{Re}
\DeclareMathOperator{\diam}{diam}
\DeclareMathOperator{\Tr}{Tr}
\DeclareMathOperator{\Int}{int}
\newcommand{\Trs}{\Tr_{\CC^2}}

\newcommand{\RR}{\mathbb{R}}
\newcommand{\CC}{\mathbb{C}}
\newcommand{\ZZ}{\mathbb{Z}}
\newcommand{\NN}{\mathbb{N}}
\newcommand{\EE}{\mathbb{E}}

\newcommand{\SO}{\mathrm{SO}}
\newcommand{\SU}{\mathrm{SU}}
\newcommand{\LL}{\mathrm{LL}}
\newcommand{\xc}{\mathrm{xc}}
\newcommand{\TF}{\mathrm{TF}}
\newcommand{\LO}{\mathrm{LO}}
\newcommand{\UEG}{\mathrm{UEG}}
\newcommand{\SSW}{\mathrm{SSW}}

\newcommand{\pscal}[1]{\left\langle #1 \right\rangle}
\newcommand{\norm}[1]{\left\| #1 \right\|}
\newcommand{\iden}{\mathbb{1}}
\newcommand{\lapl}{\bm{\Delta}}
\newcommand{\grad}{\bm{\nabla}}
\newcommand{\tetra}{\mathbb{\Delta}}
\newcommand{\dd}{\mathrm{d}}

\newcommand{\TC}{\mathcal{T}}
\newcommand{\EC}{\mathcal{E}}
\newcommand{\Cc}{\mathcal{C}}
\newcommand{\LC}{\mathcal{L}}
\newcommand{\QC}{\mathcal{Q}}
\newcommand{\GC}{\mathcal{G}}

\newcommand{\HF}{\mathfrak{H}}

\def\Xint#1{\mathchoice
{\XXint\displaystyle\textstyle{#1}}
{\XXint\textstyle\scriptstyle{#1}}
{\XXint\scriptstyle\scriptscriptstyle{#1}}
{\XXint\scriptscriptstyle\scriptscriptstyle{#1}}
\!\int}
\def\XXint#1#2#3{{\setbox0=\hbox{$#1{#2#3}{\int}$ }
\vcenter{\hbox{$#2#3$ }}\kern-.6\wd0}}
\def\dashint{\Xint-}

\title[The TSGA in noncollinear SDFT]{The transverse spin-gradient approximation in noncollinear spin density functional theory}

\author[M.\,A. Csirik]{Mih\'aly~A.~Csirik}
\address{Department of Mathematical Sciences, Chalmers University of
Technology and the University of Gothenburg, G\"oteborg, Sweden}
\thanks{Email: \href{mailto:csirik@gmail.com}{\texttt{csirik@gmail.com}}}

\begin{document}

\begin{abstract}
We prove a quantitative error bound for a fixed-axis variant of the transverse spin-gradient approximation (TSGA) in noncollinear spin density functional theory. Inspired by Eich and Gross, we use a spin-spiral wave (SSW) as the local reference for the grand-canonical Levy--Lieb energy with the Hartree term subtracted. We construct the thermodynamic limit of the electron gas with constant total density and a prescribed SSW magnetization. Its energy per unit volume $e_\SSW$ is evaluated at the local wavevector $\grad\varphi$, where $\varphi$ is the spiral phase. When $\varphi$ is the azimuthal angle of the magnetization $\mathbf{m}=(m_1,m_2,m_3)$, the approximation reads
\[
 E(\rho,\mathbf{m})\approx\int_{\RR^3}e_\SSW\big(\rho(\mathbf{x}),\mathbf{m}(\mathbf{x}),\grad\varphi(\mathbf{x})\big)\,\dd \mathbf{x}.
\]
Here $\grad\varphi$ measures the local rotation of $\mathbf{m}$ about the third axis. Wherever $m_1^2+m_2^2>0$, it equals $(m_1\grad m_2-m_2\grad m_1)/(m_1^2+m_2^2)$. The approximation thus depends only on the total density $\rho$, the magnetization and its first derivatives.
When the rotating-frame density and the local wavevector vary on the length scale $N^{1/3}$, we obtain an $O(N^{5/6})$ error for an energy of order $N$ under the assumption that the $p$-th root of that density has an $L^p$ gradient. Here $N$ is the mean number of particles.
Setting the phase to zero gives the noncollinear local spin density approximation (LSDA) for arbitrary magnetization directions.
\end{abstract}

\maketitle

\clearpage
\tableofcontents

\begin{figure}[!b]
\centering
\begin{tikzpicture}[>=stealth]
\def\sswA{0.26}
\def\sswR{0.1}
\def\sswNx{25}
\def\sswNy{13}
\pgfmathsetmacro{\sswLx}{\sswNx*\sswA}
\pgfmathsetmacro{\sswLy}{\sswNy*\sswA}
\begin{scope}[shift={(-1.4,{\sswLy/2})}]
\draw[->] (0,0)--(0.5,0) node[right]{$\mathbf{e}_1$};
\draw[->] (0,0)--(0,0.5) node[above]{$\mathbf{e}_2$};
\draw (0,0) circle[radius=0.09]; \fill (0,0) circle[radius=0.025];
\node[below left] at (0,0) {$\mathbf{e}_3$};
\end{scope}
\foreach \i in {0,...,\sswNx}{\foreach \j in {0,...,\sswNy}{
  \pgfmathsetmacro{\a}{15*(2*\i+\j-27)}
  \draw[->,line width=0.45pt,blue!75!black] ({\i*\sswA-\sswR*cos(\a)},{\j*\sswA-\sswR*sin(\a)})--++(\a:2*\sswR);}}
\begin{scope}
\clip (-0.25,-0.25) rectangle (\sswLx+0.25,\sswLy+0.25);
\foreach \c in {27,51}{
  \draw[black!70,densely dashed,semithick] ({(\c*\sswA+0.25)/2},-0.25)--({(\c*\sswA-\sswLy-0.25)/2},\sswLy+0.25);}
\end{scope}
\foreach \c/\lab in {27/0,51/2\pi}{
  \node[above] at ({(\c*\sswA-\sswLy-0.25)/2},\sswLy+0.25) {$\mathbf{q}\cdot\mathbf{x}=\lab$};}
\draw[->,thick] (\sswLx+0.5,{\sswLy/2-0.5})--++({1.2*cos(atan(1/2))},{1.2*sin(atan(1/2))}) node[right]{$\mathbf{q}$};
\end{tikzpicture}
\caption{The SSW in a plane containing $\mathbf{q}$. The arrows show the transverse magnetization $(m_1,m_2)$, seen from the top along the third axis $\mathbf{e}_3$ of spin space.}\label{fig:SSW}
\end{figure}
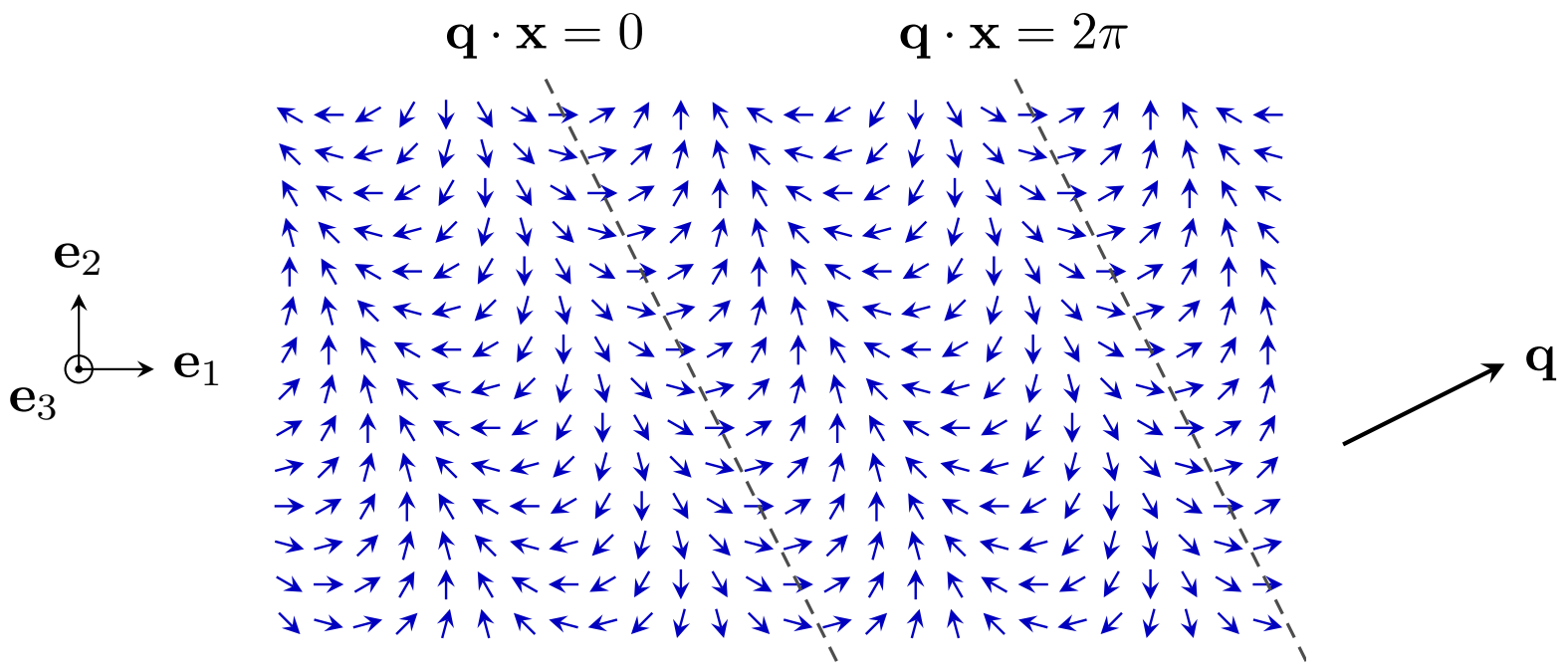

\clearpage
\section{Introduction}

One of the central objects of density functional theory (DFT) is the Levy--Lieb functional~\cite{levy1979universal,lieb1983density}, which is the lowest possible internal energy of all pure $N$-particle states having the prescribed density $\rho$, with $N=\int\rho$. As in~\cite{lewin2020local}, we use its grand-canonical version $F_\LL(\rho)$, in which the minimum is over mixed states on Fock space that commute with the particle number, so that $\int\rho$ need not be an integer. Throughout the paper we subtract the direct (Hartree) energy and, for simplicity, work with the indirect energy
\begin{equation}\label{eq:def_E_intro}
E(\rho):=F_\LL(\rho)-\frac{1}{2}\iint_{\RR^3\times\RR^3}\frac{\rho(\mathbf{x})\rho(\mathbf{y})}{|\mathbf{x}-\mathbf{y}|}\,\dd\mathbf{x}\,\dd\mathbf{y}.
\end{equation}
In fact, our estimates also hold for the physically more important exchange-correlation energy $E_\xc(\rho)=E(\rho)-T(\rho)$, where $T(\rho)$ is the lowest possible kinetic energy of mixed states at density $\rho$.

The \emph{local density approximation} (LDA) consists in replacing $E(\rho)$ by the integral of the energy per unit volume $e_\UEG$ of the uniform electron gas (UEG) taken at the local value of the density,
\[
 E(\rho)\approx\int_{\RR^3}e_\UEG\big(\rho(\mathbf{x})\big)\,\dd \mathbf{x}.
\]
In~\cite{lewin2020local}, Lewin, Lieb and Seiringer gave the first mathematically rigorous justification of this approximation. They proved~\cite[Theorem~3]{lewin2020local} that for $p>3$ and $0<\theta<1$ with $2\le p\theta\le1+p/2$ there exists a constant $C=C(p,\theta)$ such that
\begin{multline*}
 \left|E(\rho)-\int_{\RR^3}e_\UEG\big(\rho(\mathbf{x})\big)\,\dd\mathbf{x}\right|\\
 \le\epsilon\int_{\RR^3}(\rho+\rho^2)
 +\frac{C(1+\epsilon)}{\epsilon}\int_{\RR^3}|\grad\sqrt{\rho}|^2
 +\frac{C}{\epsilon^{4p-1}}\int_{\RR^3}|\grad\rho^\theta|^p
\end{multline*}
for every $\epsilon>0$ and every density $\rho\ge0$ such that $\rho\in L^1\cap L^2$, $\grad\sqrt{\rho}\in L^2$ and $\grad\rho^\theta\in L^p$.

This bound is interesting for slowly varying densities. Let $\rho_N(\mathbf{x})=\rho(N^{-1/3}\mathbf{x})$ with $\int_{\RR^3}\rho=1$, which describes $N$ electrons spread over a volume of order $N$ at a local density of order one. Taking $\epsilon=N^{-1/12}$ gives~\cite[Eq.~(9)]{lewin2020local}
\begin{multline}\label{eq:LLS_rate}
 \left|E(\rho_N)-N\int_{\RR^3}e_\UEG\big(\rho(\mathbf{x})\big)\,\dd\mathbf{x}\right|\\
 \le CN^{5/12}\int_{\RR^3}|\grad\sqrt{\rho}|^2+CN^{11/12}\int_{\RR^3}\big(\rho+\rho^2+|\grad\rho^\theta|^p\big).
\end{multline}
Since $E(\rho_N)$ is of order $N$, the LDA is exact to leading order for slowly varying densities, with an explicit rate. Moreover, the bound holds for every density, not only in the limit, and the gradient terms measure how far $\rho$ is from being locally constant. It has remained a challenge to extend this analysis to gradient approximations, whose energy per unit volume also depends on the derivatives of the density. In this paper we do so for a fixed-axis variant of the transverse spin-gradient approximation (TSGA) inspired by Eich and Gross~\cite{eich2013transverse}.

This approximation concerns magnetic systems, which are described by spin-polarized DFT (SDFT) as introduced by von Barth and Hedin~\cite{von1972local}. The basic variables in SDFT are the total density $\rho$ and the magnetization $\mathbf{m}=(m_1,m_2,m_3)$. They are combined into the \emph{spin-polarized density}
$$
\bm{\rho}(\mathbf{x})=\frac{1}{2}\begin{pmatrix}\rho(\mathbf{x})+m_3(\mathbf{x})&m_1(\mathbf{x})-im_2(\mathbf{x})\\ m_1(\mathbf{x})+im_2(\mathbf{x})&\rho(\mathbf{x})-m_3(\mathbf{x})\end{pmatrix}=\begin{pmatrix}\rho^{\spu\spu}(\mathbf{x})&\rho^{\spu\spd}(\mathbf{x})\\ \rho^{\spd\spu}(\mathbf{x})&\rho^{\spd\spd}(\mathbf{x})\end{pmatrix},
$$
which is a Hermitian $2\times 2$ matrix with $\Trs\bm{\rho}=\rho$. Most applications are restricted to the \emph{collinear} case where $\mathbf{m}$ points in a fixed direction, that is, $\bm{\rho}$ is diagonal in a fixed basis of $\CC^2$. In this case the quantitative LDA estimate above carries through, with $e_\UEG(\rho)$ replaced by the energy per unit volume $e_\UEG(\rho_\spu,\rho_\spd)$ of the spin-polarized uniform gas with prescribed densities of spin-up and spin-down electrons~\cite[Remark~6]{lewin2020local}.

The more general \emph{noncollinear} case, where the direction of $\mathbf{m}(\mathbf{x})$ depends on $\mathbf{x}$, is necessary to describe for instance frustrated magnets, spin-spiral waves or spin dynamics~\cite{kubler1988density}. The mathematical study of noncollinear SDFT was initiated by Gontier, who characterized in~\cite{gontier2013n} the set of representable spin-polarized densities and studied the corresponding Kohn--Sham models in~\cite{gontier2015existence}.

For a spin-polarized density we use the same notation for the functionals. The functional $F_\LL(\bm{\rho})$ is the lowest possible internal energy of all quantum states having the prescribed $\bm{\rho}$, and $E(\bm{\rho})$ is defined as in~\cref{eq:def_E_intro}, with $\rho=\Trs\bm{\rho}$ in the direct energy.

\medskip

\noindent\textbf{The local spin density approximation.} In the noncollinear setting, the \emph{local spin density approximation} (LSDA) reads~\cite{von1972local,kubler1988density}
\begin{equation}\label{eq:LSDA}
 E(\bm{\rho})\approx\int_{\RR^3}e_\UEG\big(\rho^+(\mathbf{x}),\rho^-(\mathbf{x})\big)\,\dd \mathbf{x}.
\end{equation}
Here the functions $\rho^\pm(\mathbf{x})=\big(\rho(\mathbf{x})\pm|\mathbf{m}(\mathbf{x})|\big)/2$ are the two eigenvalues of $\bm{\rho}(\mathbf{x})$. The rationale for~\cref{eq:LSDA} is that the Hamiltonian commutes with global rotations of the spin, so that the energy of a gas with constant spin-polarized density $\bm{\rho}_0$ can only depend on the eigenvalues of $\bm{\rho}_0$. The LSDA then consists in assuming that $\bm{\rho}$ is almost constant locally, and in rotating the local spin quantization axis to the direction of $\mathbf{m}(\mathbf{x})$. This rotation diagonalizes $\bm{\rho}(\mathbf{x})$, so that the local energy is that of the polarized uniform gas with spin-up and spin-down densities $\rho^+(\mathbf{x})$ and $\rho^-(\mathbf{x})$. Two matrices with the same eigenvalues have the same LSDA energy, however fast their magnetizations rotate.

\medskip

\noindent\textbf{The transverse spin-gradient approximation.} Eich and Gross~\cite{eich2013transverse} proposed using the \emph{spin-spiral wave} (SSW) state as a reference for the exchange-correlation energy. In this state the total density is constant and the magnetization rotates about a fixed axis with wavevector $\mathbf{q}$, see \cref{fig:SSW}. We call the electron gas in this state the \emph{SSW gas}. Eich and Gross determine the local spiral parameters from combinations of first and second derivatives of the magnetization that are invariant under global spin rotations. In our variant for the indirect energy $E(\bm{\rho})$, every reference spiral rotates about the third spin axis $\mathbf{e}_3$. Write the transverse magnetization as $m_1+im_2=a e^{i\varphi}$ with real amplitude $a$ and phase $\varphi$. With $\grad\varphi$ as the local wavevector, our approximation reads
\begin{equation}\label{eq:TSGA}
 E(\bm{\rho})\approx\int_{\RR^3}e_\SSW\big(\bm{\rho}(\mathbf{x}),\grad\varphi(\mathbf{x})\big)\,\dd \mathbf{x},\qquad\grad\varphi=\frac{m_1\grad m_2-m_2\grad m_1}{m_1^2+m_2^2}.
\end{equation}
The formula for $\grad\varphi$ holds where $m_1^2+m_2^2>0$. Here $e_\SSW(\bm{\rho}_0,\mathbf{q})$ is the energy per unit volume of the infinite electron gas whose spin-polarized density is obtained from the constant matrix $\bm{\rho}_0$ by a rotation of the spin of angle $\mathbf{q}\cdot\mathbf{x}$ about $\mathbf{e}_3$. For $\mathbf{q}=0$ we have $e_\SSW(\bm{\rho}_0,0)=e_\UEG(\rho_0^+,\rho_0^-)$, hence the TSGA reduces to the LSDA when the magnetization does not rotate about $\mathbf{e}_3$. 
The resulting functional depends on the choice of $\mathbf{e}_3$ and is invariant under spin rotations about this axis.  

The TSGA functional is a gradient-type semilocal functional, i.e. it is given by the integral of a function of $\bm{\rho}(\mathbf{x})$ and of its first derivatives. The SSW gas supplies this function, with the wavevector as an additional parameter. In contrast to the usual generalized gradient approximations of DFT~\cite{engel2011density}, whose dependence on the gradients is chosen by fitting or by imposing exact constraints, in our case this dependence is fixed by a reference system without adjustable parameters.

In this paper, we first establish in \cref{thm:spiral_limit} that the thermodynamic limit defining $e_\SSW(\bm{\rho}_0,\mathbf{q})$ exists. Our main result, \cref{thm:LSSWA}, gives a quantitative error bound for this fixed-axis approximation to the grand-canonical functional $E(\bm{\rho})$.
To state our main estimate, let $\bm{\rho}$ be a spin-polarized density and $\varphi$ a real function. Let
\[
 \mathbf{U}_\varphi(\mathbf{x})=\begin{pmatrix}e^{-\frac i2\varphi(\mathbf{x})}&0\\0&e^{\frac i2\varphi(\mathbf{x})}\end{pmatrix}
\]
be the rotation of the spin through the angle $\varphi(\mathbf{x})$ about $\mathbf{e}_3$, and write $\bm{\rho}=\mathbf{U}_\varphi\wt{\bm{\rho}}\,\mathbf{U}_\varphi^*$, where $\wt{\bm{\rho}}$ is the spin-polarized density in the rotating frame. Let $p>3$ and $2/p\le\theta<1$. Assume that $\rho\in L^1\cap L^2$, $\grad\sqrt{\wt{\bm{\rho}}}\in L^2$ and $\grad\wt{\bm{\rho}}^\theta\in L^p$. Assume also that $\varphi\in W^{2,p}_{\rm loc}(\RR^3)$ and that $\wt{\bm{\rho}}^{1/p}\in W^{1,p}_{\rm loc}$ on the open set $\{\grad\varphi\ne0\}$. If the three phase-dependent terms on the right are integrable, \cref{thm:LSSWA} gives
\begin{multline}\label{eq:TSGA_estim_intro}
 \left|E(\bm{\rho})-\int_{\RR^3}e_\SSW\big(\wt{\bm{\rho}}(\mathbf{x}),\grad\varphi(\mathbf{x})\big)\,\dd\mathbf{x}\right|\\
 \le \epsilon\int_{\RR^3}\big(\rho+\rho^2+\rho|\grad\varphi|^2\big)
 +\frac{C(1+\epsilon)}{\epsilon}\int_{\RR^3}|\grad\sqrt{\wt{\bm{\rho}}}|^2
 +\frac{C}{\epsilon^b}\int_{\RR^3}|\grad\wt{\bm{\rho}}^\theta|^p\\
 +\frac{C}{\epsilon^{2p-1}}\left(\int_{\RR^3}\rho\,|\mathbf{D}^2\varphi|^p
 +\int_{\{\grad\varphi\ne0\}}|\grad\varphi|^2\big|\grad\wt{\bm{\rho}}^{1/p}\big|^p\right)
\end{multline}
for every $\epsilon>0$, where $b=\max\{2p,(1+3\theta)p\}-1$.

Like the LDA bound above, this estimate is interesting when its right side is small compared with the energy scale $\int(\rho+\rho^2+\rho|\grad\varphi|^2)$. Since $-C\rho_0^{4/3}\le e_\SSW(\bm{\rho}_0,\mathbf{q})\le C\big(\rho_0^{5/3}+\rho_0|\mathbf{q}|^2\big)$ by \cref{prop:SSW_properties} and \cref{cor:lower_bound_E_rho}, this scale bounds the TSGA integral in absolute value, so the first term is a relative error of order $\epsilon$. The other terms vanish formally for a single spin-spiral wave, where $\wt{\bm{\rho}}$ is constant and $\varphi(\mathbf{x})=\mathbf{q}\cdot\mathbf{x}+c$ is affine, and they measure the deviation from it. The terms in $\grad\sqrt{\wt{\bm{\rho}}}$ and $\grad\wt{\bm{\rho}}^\theta$ are those of the LDA bound, now for the density in the rotating frame. The Hessian term measures the variation of the local wavevector $\grad\varphi$. The last term measures the variation of $\wt{\bm{\rho}}$ weighted by the squared wavevector, and it comes from the kinetic energy of order $\rho|\grad\varphi|^2$ carried by the spiral. Its weight vanishes where $\grad\varphi=0$, so it is integrated only over $\{\grad\varphi\ne0\}$, and no regularity of $\wt{\bm{\rho}}^{1/p}$ is needed where the phase is locally constant. The estimate is therefore useful when $\wt{\bm{\rho}}$ and $\grad\varphi$ vary slowly, so that the gradient terms are small compared with $\int(\rho+\rho^2+\rho|\grad\varphi|^2)$ and $\epsilon$ can be chosen to balance them. The wavevector itself need not be small, only its variation. The magnetization may rotate quickly about $\mathbf{e}_3$, provided that it locally looks like a spin-spiral wave whose density and wavevector change slowly.

When $\wt{\bm{\rho}}$ and $\grad\varphi$ are fixed profiles of $N^{-1/3}\mathbf{x}$ with finite error integrals and mean particle number $N$, \cref{eq:TSGA_estim_intro} gives an $O(N^{5/6})$ error when $\theta\le1/3$. If $\grad\wt{\bm{\rho}}^{1/p}\in L^p(\RR^3)$, a second estimate in \cref{thm:LSSWA} gives an $O(N^{5/6})$ error for all $p>3$. These rates are uniform up to full polarization, even for a wavevector of order one. The LSDA error can be proportional to $N$ for the same spin-spiral waves.

Taking $\varphi=0$ in \cref{thm:LSSWA} gives a quantitative bound for the noncollinear LSDA~\cref{eq:LSDA}. Then $\wt{\bm{\rho}}=\bm{\rho}$ with arbitrary magnetization direction, the SSW gas has energy $e_\UEG$, and the domain $\{\grad\varphi\ne0\}$ of the relative density gradient is empty. The resulting LSDA estimate of \cref{cor:LSDA} extends the LDA estimate of~\cite{lewin2020local} to noncollinear spin-polarized densities under the density assumptions $\rho\in L^1\cap L^2$, $\grad\sqrt{\bm{\rho}}\in L^2$ and $\grad\bm{\rho}^{\theta}\in L^p$, with the same exponent $b$ as in~\cref{eq:TSGA_estim_intro}. For $\bm{\rho}_N(\mathbf{x})=\bm{\rho}(N^{-1/3}\mathbf{x})$ with $\int\rho=1$, optimizing over $\epsilon$ gives an LSDA error of order at most $N^{\max\{5/6,\,1-1/(3+9\theta)\}}$. Since $\theta<1$, this improves the $O(N^{11/12})$ bound~\cref{eq:LLS_rate}, and the rate is $O(N^{5/6})$ when $\theta\le1/3$.

\medskip

\noindent\textbf{Organization of the paper.} We borrow the notation of~\cite{lewin2020local} as much as possible. \cref{sec:Prel} introduces the functionals of noncollinear spin-polarized density functional theory and \cref{sec:main} states our results. \cref{sec:proof} contains the proof of the TSGA bound of \cref{thm:LSSWA}. For the arguments which are identical to those of~\cite{lewin2020local} we only indicate the necessary changes. \cref{app:SSW_limit} constructs the thermodynamic limit of the SSW gas. The remaining appendices contain auxiliary results whose proofs would break the flow of the exposition.

\medskip

\noindent\textbf{Acknowledgments.} The author acknowledges funding from the European Research Council through ERC-2021-STG Grant Agreement No.~101041487 \mbox{(REGAL)}.

\medskip

\noindent\textbf{Lean formalization.} The Lean~4 formalization proves the main estimates and their supporting results under explicit hypotheses from analysis, operator theory and mathematical physics. The source files are in the accompanying \texttt{lean/} directory.\footnote{\url{https://github.com/xyzw/tsga_sdft}}\enlargethispage{2pt}

\medskip

\noindent\textbf{AI disclosure.} The large language models Claude Fable~5.2 and Claude Opus~5.5 (Anthropic), and GPT-6.2 Astra (OpenAI), were used throughout the development of this paper and its Lean formalization. The author has reviewed all content and takes full responsibility for the contents of this paper.

\clearpage
\section{Preliminary notions}\label{sec:Prel}

\subsection{Notation}\label{sec:notation}
For $2\times 2$ matrices $\mathbf{A}$ we use the Hilbert--Schmidt norm $|\mathbf{A}|^2:=\Trs\mathbf{A}^*\mathbf{A}$, which comes from the scalar product $\Trs\mathbf{A}^*\mathbf{B}$. Functions of a non-negative Hermitian matrix, such as $\sqrt{\bm{\rho}}$, $\bm{\rho}^\theta$ or $\wt{\bm{\rho}}^{1/p}$, are defined pointwise in $\mathbf{x}$ by the functional calculus. For a matrix-valued function $\mathbf{M}$ on $\RR^3$ we set
$$
|\grad\mathbf{M}(\mathbf{x})|^2:=\sum_{j=1}^3|\partial_{x_j}\mathbf{M}(\mathbf{x})|^2,
$$
which is equal to $\sum_j\Trs(\partial_{x_j}\mathbf{M})^2$ when $\mathbf{M}$ is Hermitian. This applies for instance to $|\grad\sqrt{\bm{\rho}}|^2$ and $|\grad\bm{\rho}^\theta|^p$. Similarly, $|\mathbf{D}^2\varphi|$ is the Hilbert--Schmidt norm of the Hessian matrix of a real function $\varphi$.

We denote by $(\mathbf{e}_1,\mathbf{e}_2,\mathbf{e}_3)$ the canonical basis of $\RR^3$, by $B_r$ the open ball of radius $r>0$ centered at the origin, and by $\bm{\sigma}=(\bm{\sigma}_1,\bm{\sigma}_2,\bm{\sigma}_3)$ the Pauli matrices
$$
\bm{\sigma}_1=\begin{pmatrix}0&1\\1&0\end{pmatrix},\qquad\bm{\sigma}_2=\begin{pmatrix}0&-i\\i&0\end{pmatrix},\qquad\bm{\sigma}_3=\begin{pmatrix}1&0\\0&-1\end{pmatrix}.
$$
Every Hermitian $2\times 2$ matrix can be written $\bm{\rho}=\frac12(\rho\,\iden_{\CC^2}+\mathbf{m}\cdot\bm{\sigma})$, with $\rho=\Trs\bm{\rho}$ and $m_k=\Trs\bm{\sigma}_k\bm{\rho}$. This is the decomposition into total density and magnetization used in the introduction. We use the same convention for decorated symbols, for instance $\wt{\rho}=\Trs\wt{\bm{\rho}}$.

\subsection{Noncollinear spin-polarized density functionals}\label{sec:functionals}
In this section, we introduce the density functionals of noncollinear spin-polarized density functional theory.

Let ${\HF^n:=L^2_a\big((\RR^3\times \{\spu,\spd\})^n,\CC\big)}$ be the $n$-particle space of antisymmetric square-integrable functions. The \emph{spin-polarized density} of a normalized wavefunction ${\Psi\in\HF^N}$ is the Hermitian $2\times 2$ matrix-valued function $\bm{\rho}_\Psi$ with entries~\cite{von1972local,gontier2013n}
$$
\rho_\Psi^{\sigma\sigma'}(\mathbf{x})=N\int\Psi(\mathbf{x},\sigma,Z)\,\overline{\Psi(\mathbf{x},\sigma',Z)}\,\dd Z,\qquad \sigma,\sigma'\in\{\spu,\spd\},
$$
where $Z=(\mathbf{x}_2,\sigma_2,\ldots,\mathbf{x}_N,\sigma_N)$ and $\int\dd Z$ denotes the integral over $\RR^{3(N-1)}$ together with the sum over $\{\spu,\spd\}^{N-1}$. It is non-negative and its trace $\rho_\Psi=\Trs\bm{\rho}_\Psi$ is the total density of $\Psi$, with $\int_{\RR^3}\rho_\Psi=N$. For a trace-class operator $\Gamma_n=\Gamma_n^*\ge0$ on $\HF^n$, the spin-polarized density $\bm{\rho}_{\Gamma_n}$ is defined by the same formula with $N=n$, where the integrand is replaced by the kernel $\Gamma_n(\mathbf{x},\sigma,Z;\mathbf{x},\sigma',Z)$ of $\Gamma_n$.

We denote by $\mathcal{D}$ the set of mixed states on the fermionic Fock space with finite kinetic energy which have the additional property that they commute with the particle number operator. Such a state is block diagonal, $\Gamma=\bigoplus_{n\ge0}\Gamma_n$, with $\Gamma_n=\Gamma_n^*\ge0$ on $\HF^n$, $\sum_{n\ge0}\Tr\Gamma_n=1$ and
$$
\sum_{n\ge1}\Tr_{\HF^n}\Bigg(-\sum_{j=1}^n\lapl_{\mathbf{x}_j}\Bigg)\Gamma_n<\infty.
$$
The spin-polarized density of $\Gamma\in\mathcal{D}$ is $\bm{\rho}_\Gamma:=\sum_{n\ge1}\bm{\rho}_{\Gamma_n}$. In other words, $\bm{\rho}_\Gamma(\mathbf{x})=\bm{\gamma}_\Gamma(\mathbf{x},\mathbf{x})$ where $\bm{\gamma}_\Gamma$ is the one-particle density matrix of $\Gamma$, seen as an operator on $L^2(\RR^3,\CC^2)$ with a $2\times 2$ matrix-valued kernel. The total density is $\rho_\Gamma=\Trs \bm{\rho}_\Gamma$.

More generally, we call \emph{spin-polarized density} a measurable function $\bm{\rho}$ on $\RR^3$ with values in the Hermitian $2\times 2$ matrices such that
\begin{equation}\label{eq:representable}
\boxed{\bm{\rho}(\mathbf{x})=\bm{\rho}(\mathbf{x})^*\ge0,\qquad \rho:=\Trs \bm{\rho}\in L^1(\RR^3),\qquad \sqrt{\bm{\rho}}\in H^1(\RR^3,\CC^{2\times 2})}
\end{equation}
Here $\sqrt{\bm{\rho}}(\mathbf{x})=\sqrt{\bm{\rho}(\mathbf{x})}$ is the square root of the non-negative matrix $\bm{\rho}(\mathbf{x})$. We denote by $\rho^+(\mathbf{x})\ge\rho^-(\mathbf{x})\ge0$ the two eigenvalues of $\bm{\rho}(\mathbf{x})$. We call~\cref{eq:representable} the \emph{$N$-representability condition}: by the following theorem, together with $\int_{\RR^3}\rho=N$ it characterizes the spin-polarized densities of mixed $N$-particle states with finite kinetic energy.

\begin{theorem}[Gontier~\cite{gontier2013n}]\label{thm:Gontier}
Let $N\ge1$ be an integer. A measurable function $\bm{\rho}$ on $\RR^3$ with values in the Hermitian $2\times 2$ matrices is the spin-polarized density of a mixed $N$-particle state with finite kinetic energy if and only if $\bm{\rho}\ge0$, $\sqrt{\bm{\rho}}\in H^1(\RR^3,\CC^{2\times2})$ and $\int_{\RR^3}\rho=N$.
\end{theorem}

The spin-polarized densities of pure states form in general a smaller set~\cite{gontier2013n}. For instance, for $N=1$ we have $\bm{\rho}_\Psi(\mathbf{x})=\bm{\Psi}(\mathbf{x})\bm{\Psi}(\mathbf{x})^*$ with $\bm{\Psi}(\mathbf{x})=\big(\Psi(\mathbf{x},\spu),\Psi(\mathbf{x},\spd)\big)\in\CC^2$, which has rank one, whereas \cref{thm:Gontier} allows densities of full rank.

The condition $\sqrt{\bm{\rho}}\in H^1$ in~\cref{eq:representable} controls both the spin populations and the direction of the magnetization, since the matrix square root separates the two. In terms of the eigenvalues $\rho^\pm$ of $\bm{\rho}$ we have
\begin{equation}\label{eq:formula_gradient_sqrt_R}
|\grad\sqrt{\bm{\rho}}|^2=\big|\grad\sqrt{\rho^+}\big|^2+\big|\grad\sqrt{\rho^-}\big|^2+\frac12\Big(\sqrt{\rho^+}-\sqrt{\rho^-}\Big)^2|\grad \mathbf{n}|^2,\qquad \mathbf{n}=\frac{\mathbf{m}}{|\mathbf{m}|},
\end{equation}
on the set where $\mathbf{m}\ne0$. Indeed,
\[
\sqrt{\bm{\rho}}=\tfrac12(\sqrt{\rho^+}+\sqrt{\rho^-})\iden+
\tfrac12(\sqrt{\rho^+}-\sqrt{\rho^-})\,\mathbf{n}\cdot\bm{\sigma}.
\]
Differentiating and using $\Trs(\bm{\sigma}_j\bm{\sigma}_k)=2\delta_{jk}$ and $\mathbf{n}\cdot\partial_j\mathbf{n}=0$ gives the identity.

For a spin-polarized density $\bm{\rho}$ satisfying~\cref{eq:representable}, the \emph{grand-canonical Levy--Lieb functional} is defined by
\begin{equation}\label{eq:Levy-Lieb-grand-canonical}
\boxed{F_\LL(\bm{\rho}):=\inf_{\substack{\Gamma\in\mathcal{D}\\ \bm{\rho}_\Gamma=\bm{\rho}}}\sum_{n\ge1}\Tr_{\HF^n}\Bigg(-\sum_{j=1}^n\lapl_{\mathbf{x}_j}+\sum_{1\le j<k\le n}\frac{1}{|\mathbf{x}_j-\mathbf{x}_k|}\Bigg)\Gamma_n}
\end{equation}
If $\bm{\rho}$ does not satisfy~\cref{eq:representable}, we set $F_\LL(\bm{\rho}):=+\infty$.

\begin{theorem}[Existence of a minimizing state]\label{thm:LL_attainment}
For every spin-polarized density $\bm{\rho}\ge0$ on $\RR^3$ with $\rho\in L^1(\RR^3)$ and $\sqrt{\bm{\rho}}\in H^1(\RR^3,\CC^{2\times2})$, a state $\Gamma\in\mathcal{D}$ with $\bm{\rho}_\Gamma=\bm{\rho}$ attains the infimum defining $F_\LL(\bm{\rho})$ in~\cref{eq:Levy-Lieb-grand-canonical}.
\end{theorem}

The proof is exactly the same as in the spinless case~\cite[Theorem~3.4]{lewin2023universal}. Trace-norm convergence of the one-particle density matrices preserves the prescribed spin density entry by entry.

The functional $F_\LL$ is convex. If $\bm{\rho}$ is diagonal, that is $\bm{\rho}=\diag(\rho_\spu,\rho_\spd)$, we recover the collinear functional of~\cite[Remark~6]{lewin2020local} where only the densities of spin-up and spin-down electrons are prescribed.

We denote by
$$
D(f,g):=\frac12\iint_{\RR^3\times\RR^3}\frac{f(\mathbf{x})g(\mathbf{y})}{|\mathbf{x}-\mathbf{y}|}\,\dd\mathbf{x}\,\dd\mathbf{y},\qquad D(f):=D(f,f),
$$
the direct (Hartree) energy. Recall from the introduction that $E(\bm{\rho})=F_\LL(\bm{\rho})-D(\rho)$ and that the (grand-canonical) exchange-correlation energy is $E_\xc(\bm{\rho})=E(\bm{\rho})-T(\bm{\rho})$, where
\begin{equation}\label{eq:def_T}
T(\bm{\rho}):=\inf_{\substack{0\le \bm{\gamma}=\bm{\gamma}^*\le 1\\ \bm{\rho}_{\bm{\gamma}}=\bm{\rho}}}\Tr(-\lapl)\bm{\gamma}
\end{equation}
is the lowest possible kinetic energy. Here $\bm{\gamma}$ is an operator on $L^2(\RR^3,\CC^2)$ and $\bm{\rho}_{\bm{\gamma}}(\mathbf{x})=\bm{\gamma}(\mathbf{x},\mathbf{x})$.

\begin{remark}[Collinear and unpolarized cases]\label{rmk:collinear}
Let $E(\rho_\spu,\rho_\spd)$ be the collinear functional of~\cite[Remark~6]{lewin2020local}, where only the diagonal entries of $\bm{\rho}_\Gamma$ are prescribed, and let $E(\rho)$ be the functional of~\cite{lewin2020local} with $q=2$, where only the total density is prescribed. Then
\begin{equation*}
E\begin{pmatrix}\rho_\spu&0\\0&\rho_\spd\end{pmatrix}=E(\rho_\spu,\rho_\spd),\qquad E\Big(\frac{\rho}{2}\iden_{\CC^2}\Big)=E(\rho).
\end{equation*}
The inequalities $\ge$ are obvious. For the converse, we average a state $\Gamma$ over the rotation $\mathbf{U}=\diag(i,-i)$ of the spin, that is, we take $\frac12(\Gamma+\GC_{\mathbf{U}}\Gamma \GC_{\mathbf{U}}^*)$ with $\GC_{\mathbf{U}}=\bigoplus_n\mathbf{U}^{\otimes n}$, respectively over all $\mathbf{U}\in\SU(2)$ with the Haar measure. This does not change the energy, and it replaces $\bm{\rho}_\Gamma$ by its diagonal part, respectively by $(\rho/2)\iden_{\CC^2}$.
\end{remark}

\section{Main results}\label{sec:main}

\subsection{The spin-spiral wave gas}\label{sec:spirals}

Recall that the LSDA~\cref{eq:LSDA} does not depend on the direction of the magnetization. The simplest spin-polarized densities for which this direction varies are the \emph{spin-spiral waves}, for which we have more precise information. For $\mathbf{q}\in\RR^3$ we set
\begin{equation}\label{eq:def_U_q}
\mathbf{U}_{\mathbf{q}}(\mathbf{x}):=e^{-\frac{i}2(\mathbf{q}\cdot\mathbf{x})\bm{\sigma}_3}\in\SU(2).
\end{equation}
If $\bm{\rho}$ has the total density $\rho$ and the magnetization $\mathbf{m}$, then $\mathbf{U}_{\mathbf{q}}\bm{\rho}\,\mathbf{U}_{\mathbf{q}}^*$ has the same total density and its magnetization is obtained by rotating $\mathbf{m}(\mathbf{x})$ by the angle $\mathbf{q}\cdot\mathbf{x}$ about $\mathbf{e}_3$. For a constant matrix $\bm{\rho}_0$ this is a spin-spiral wave of wavevector $\mathbf{q}$, see \cref{fig:SSW}. It is planar when the magnetization of $\bm{\rho}_0$ is orthogonal to $\mathbf{e}_3$ and conical otherwise. By \cref{lem:gauge}, $\mathbf{U}_{\mathbf{q}}\bm{\rho}\,\mathbf{U}_{\mathbf{q}}^*$ satisfies the $N$-representability condition~\cref{eq:representable} whenever $\bm{\rho}$ does.

We now consider the uniform electron gas whose magnetization is a spin-spiral wave. The matrix $\mathbf{U}_{\mathbf{q}}\bm{\rho}_0\mathbf{U}_{\mathbf{q}}^*$ is not invariant under translations, but a translation by $\bm{\tau}$ amounts to a rotation of the spin by the angle $\mathbf{q}\cdot\bm{\tau}$ about $\mathbf{e}_3$, which does not change the energy. This allows us to prove the existence of the thermodynamic limit. To regularize the boundary, we use a mollifier: let $\chi\in L^1(\RR^3)$ be a radial non-negative function of compact support such that $\int_{\RR^3}\chi=1$ and $\int_{\RR^3}|\grad\sqrt\chi|^2<\infty$, and denote $\chi_\delta(\mathbf{x})=\delta^{-3}\chi(\mathbf{x}/\delta)$.

\begin{theorem}[Thermodynamic limit of the SSW gas]\label{thm:spiral_limit}
Let $\bm{\rho}_0\ge0$ be a non-zero constant Hermitian $2\times 2$ matrix and fix $\mathbf{q}\in\RR^3$. Let $\{\Omega_N\}\subset\RR^3$ be a sequence of bounded connected domains with $|\Omega_N|\to\infty$, such that $\Omega_N$ has a uniformly regular boundary in the sense that
$$
|\partial\Omega_N+B_r|\le Cr|\Omega_N|^{2/3},\qquad\text{for all $r\le |\Omega_N|^{1/3}/C$,}
$$
for some constant $C>0$. Let $\delta_N>0$ be any sequence such that $\delta_N/|\Omega_N|^{1/3}\to0$ and $\delta_N|\Omega_N|^{1/3}\to\infty$. Then the limit
\begin{equation}\label{eq:spiral_limit}
\boxed{\lim_{N\to\infty}\frac{E\big(\mathbf{U}_{\mathbf{q}}\bm{\rho}_0\mathbf{U}_{\mathbf{q}}^*\,\iden_{\Omega_N}\ast\chi_{\delta_N}\big)}{|\Omega_N|}=:e_\SSW(\bm{\rho}_0,\mathbf{q})}
\end{equation}
exists and is independent of $\{\Omega_N\}$, $\delta_N$ and $\chi$. At $\mathbf{q}=0$ this limit defines the energy $e_\UEG(\bm{\rho}_0)$ of the spin-polarized uniform gas.
\end{theorem}

The energy per unit volume $e_\SSW$ has the following properties, which are also proved in \cref{sec:proof_spiral_limit}.

\begin{proposition}[Properties of the SSW gas energy]\label{prop:SSW_properties}
Let $\bm{\rho}_0\ge0$ be a non-zero constant Hermitian $2\times 2$ matrix.
\begin{itemize}
\item[(i)] \emph{(Isotropy)}. We have $e_\SSW(\bm{\rho}_0,\mathbf{R}\mathbf{q})=e_\SSW(\bm{\rho}_0,\mathbf{q})$ for every $\mathbf{R}\in\SO(3)$ and $\mathbf{q}\in\RR^3$.
\item[(ii)] \emph{(Concavity)}. The function $\RR^3\ni\mathbf{q}\mapsto e_\SSW(\bm{\rho}_0,\mathbf{q})-\frac14\rho_0|\mathbf{q}|^2$ is concave.
\item[(iii)] \emph{(Bounds)}. Let $\zeta\in[0,1]$ and $\vartheta\in[0,\pi]$ be such that the magnetization $\mathbf{m}_0$ of $\bm{\rho}_0$ satisfies $|\mathbf{m}_0|=\zeta\rho_0$ and $\mathbf{m}_0\cdot\mathbf{e}_3=\zeta\rho_0\cos\vartheta$. For every $\mathbf{q}\in\RR^3$,
\begin{equation}\label{eq:spiral_UEG}
e_\UEG\big(\tfrac{1+\zeta\cos\vartheta}2\rho_0,\tfrac{1-\zeta\cos\vartheta}2\rho_0\big)\le e_\SSW(\bm{\rho}_0,\mathbf{q})\le e_\UEG(\bm{\rho}_0)+\frac{\rho_0|\mathbf{q}|^2}{4}\big(1-\zeta^2\cos^2\vartheta\big),
\end{equation}
and the upper bound is an equality for $\zeta=1$.
\end{itemize}
\end{proposition}

At zero wavevector, global spin invariance gives
\begin{equation*}
e_\UEG(\mathbf{U}\bm{\rho}_0\mathbf{U}^*)=e_\UEG(\bm{\rho}_0)\qquad(\mathbf{U}\in\SU(2)).
\end{equation*}
Hence $e_\UEG(\bm{\rho}_0)$ only depends on the two eigenvalues $\rho_0^+\ge\rho_0^-\ge0$ of $\bm{\rho}_0$, and we write $e_\UEG(a,b)=e_\UEG(\diag(a,b))$ in terms of the two spin populations, in either order, so that $e_\UEG(\bm{\rho}_0)=e_\UEG(\rho_0^+,\rho_0^-)$. By \cref{rmk:collinear}, $e_\UEG(\rho^+,\rho^-)$ coincides with the spin-polarized UEG energy mentioned in~\cite[Remark~6]{lewin2020local}, and $e_\UEG(\rho_0/2,\rho_0/2)$ with the function $e_\UEG(\rho_0)$ of~\cite[Theorem 1]{lewin2020local} for $q=2$. Some properties of $e_\UEG$ are derived in \cref{app:Lipschitz_SSW}. In particular, we prove in \cref{prop:Lipschitz_SSW} that $e_\UEG$ is H\"older continuous and that
$$
\zeta\in[-1,1]\mapsto e_\UEG\left(\frac{1+\zeta}{2}\rho_0,\frac{1-\zeta}{2}\rho_0\right)
$$
is \emph{convex} and even for every fixed $\rho_0>0$. This is a consequence of the convexity of $F_\LL$, hence of the use of mixed states.\footnote{In particular, at fixed density the unpolarized gas always has the lowest energy and the fully polarized gas the highest. This does not contradict the possible occurrence of ferromagnetic phases at low density: with mixed states, a spin-polarized density with zero magnetization can be obtained by averaging a ferromagnetic state over all the directions of the spin. }

The function $e_\SSW$ is the energy per unit volume of the SSW gas, proposed in~\cite{eich2013transverse} as a reference system for functionals depending on the transverse gradients of the magnetization, to which the LSDA is insensitive.

\subsection{The transverse spin-gradient approximation}\label{sec:tsga}

To describe a spin rotation about $\mathbf{e}_3$ through a prescribed phase $\varphi(\mathbf{x})$, we set
\begin{equation}\label{eq:def_U_phi}
\boxed{\mathbf{U}_\varphi(\mathbf{x}):=e^{-\frac i2\varphi(\mathbf{x})\bm{\sigma}_3}}
\end{equation}
 For a spin-polarized density $\wt{\bm{\rho}}$, the magnetization of $\mathbf{U}_\varphi\wt{\bm{\rho}}\,\mathbf{U}_\varphi^*$ is that of $\wt{\bm{\rho}}$ rotated by the angle $\varphi(\mathbf{x})$ about $\mathbf{e}_3$, and $\grad\varphi(\mathbf{x})$ plays the role of a local wavevector. The TSGA announced in the introduction reads
\begin{equation}\label{eq:LSSWA}
E\big(\mathbf{U}_\varphi\wt{\bm{\rho}}\,\mathbf{U}_\varphi^*\big)\approx\int_{\RR^3}e_\SSW\big(\wt{\bm{\rho}}(\mathbf{x}),\grad\varphi(\mathbf{x})\big)\,\dd\mathbf{x}.
\end{equation}
Our main theorem shows that the TSGA is accurate when $\wt{\bm{\rho}}$ and $\grad\varphi$ vary slowly, without any smallness assumption on $\grad\varphi$ itself. 

\begin{theorem}[Transverse spin-gradient approximation]\label{thm:LSSWA}
Let $p>3$ and $0<\theta<1$ satisfy
\begin{equation*}
p\theta\ge2.
\end{equation*}
Let $\wt{\bm{\rho}}\ge0$ be a spin-polarized density such that $\rho\in L^1\cap L^2$, $\grad\sqrt{\wt{\bm{\rho}}}\in L^2$ and $\grad\wt{\bm{\rho}}^\theta\in L^p$ on $\RR^3$. Let $\varphi\in W^{2,p}_{\rm loc}(\RR^3)$ be real. Assume that $\wt{\bm{\rho}}^{1/p}\in W^{1,p}_{\rm loc}(\{\grad\varphi\ne0\})$ and
\begin{equation}\label{eq:hyp_phi}
\int_{\RR^3}\rho\big(|\grad\varphi|^2+|\mathbf{D}^2\varphi|^p\big)
+\int_{\{\grad\varphi\ne0\}}|\grad\varphi|^2\,|\grad\wt{\bm{\rho}}^{1/p}|^p<\infty .
\end{equation}
For every $\epsilon>0$,
\begin{multline}\label{eq:LSSWA_estim}
\left|E\big(\mathbf{U}_\varphi\wt{\bm{\rho}}\,\mathbf{U}_\varphi^*\big)-\int_{\RR^3}e_\SSW\big(\wt{\bm{\rho}}(\mathbf{x}),\grad\varphi(\mathbf{x})\big)\,\dd\mathbf{x}\right|\\
\le\epsilon\int_{\RR^3}\big(\rho+\rho^2+\rho|\grad\varphi|^2\big)
+\frac{C(1+\epsilon)}{\epsilon}\int_{\RR^3}|\grad\sqrt{\wt{\bm{\rho}}}|^2
+\frac C{\epsilon^{b}}\int_{\RR^3}|\grad\wt{\bm{\rho}}^\theta|^p\\
+\frac C{\epsilon^{2p-1}}\left(\int_{\RR^3}\rho\,|\mathbf{D}^2\varphi|^p+
\int_{\{\grad\varphi\ne0\}}|\grad\varphi|^2\,|\grad\wt{\bm{\rho}}^{1/p}|^p\right).
\end{multline}
Here $b=\max\{2p,(1+3\theta)p\}-1$ and $C=C(p,\theta)$.

If $\grad\wt{\bm{\rho}}^{1/p}\in L^p(\RR^3)$, the term $C\epsilon^{-b}\int|\grad\wt{\bm{\rho}}^\theta|^p$ may instead be replaced by
\begin{equation}\label{eq:LSSWA_sharp_bulk}
\frac{C}{\epsilon^{2p-1}}\int_{\RR^3}\Big(|\grad\wt{\bm{\rho}}^\theta|^p+|\grad\wt{\bm{\rho}}^{1/p}|^p\Big).
\end{equation}
In this bound all constants depend only on $p$.
\end{theorem}

Consider $\wt{\bm{\rho}}_N(\mathbf{x})=\wt{\bm{\rho}}(N^{-1/3}\mathbf{x})$ and $\varphi_N(\mathbf{x})=N^{1/3}\varphi(N^{-1/3}\mathbf{x})$ with $\int_{\RR^3}\rho=1$. The local wavevector $\grad\varphi_N(\mathbf{x})=(\grad\varphi)(N^{-1/3}\mathbf{x})$ is of order one. Using~\cref{eq:LSSWA_sharp_bulk} and taking $\epsilon=N^{-1/6}$ gives, for $N\ge1$,
\begin{multline*}
\left|E\big(\mathbf{U}_{\varphi_N}\wt{\bm{\rho}}_N\mathbf{U}_{\varphi_N}^*\big)-N\int_{\RR^3}e_\SSW\big(\wt{\bm{\rho}}(\mathbf{x}),\grad\varphi(\mathbf{x})\big)\,\dd\mathbf{x}\right|\\
\le CN^{5/6}\int_{\RR^3}\Big(\rho+\rho^2+\rho|\grad\varphi|^2+|\grad\sqrt{\wt{\bm{\rho}}}|^2+|\grad\wt{\bm{\rho}}^\theta|^p\\
+\rho\,|\mathbf{D}^2\varphi|^p+\big(1+|\grad\varphi|^2\big)\big|\grad\wt{\bm{\rho}}^{1/p}\big|^p\Big),
\end{multline*}
provided the integrals on the right are finite. This gives the same rate at full and partial polarization. When $\theta\le1/3$ we have $b=2p-1$ and the same rate follows from~\cref{eq:LSSWA_estim} itself, without the unweighted term $\int|\grad\wt{\bm{\rho}}^{1/p}|^p$. For the same matrix the LSDA makes an error equal to $N\int(e_\SSW(\wt{\bm{\rho}},\grad\varphi)-e_\UEG(\wt{\bm{\rho}}))$ at leading order, which is proportional to $N$ in general. Indeed, by~\cref{cor:lower_bound_E_rho} and~\cref{eq:formula_gradient_sqrt_R} we have
$$
e_\SSW(\bm{\rho}_0,\mathbf{q})\ge\frac{|\mathbf{q}|^2}{4}\Big(\sqrt{\rho_0^+}-\sqrt{\rho_0^-}\Big)^2\sin^2\vartheta-c_\LO\rho_0^{4/3},
$$
whereas $e_\UEG(\bm{\rho}_0)\le c_\TF\rho_0^{5/3}$.

\begin{remark}\label{rmk:hypotheses_TSGA}
We comment on the hypotheses of \cref{thm:LSSWA}.
\begin{itemize}
\item[(i)] The local wavevector is the gradient $\grad\varphi$ of the phase. If it is bounded, the assumption $\grad\wt{\bm{\rho}}^{1/p}\in L^p(\RR^3)$ made for~\cref{eq:LSSWA_sharp_bulk} ensures that the weighted root integral in~\cref{eq:hyp_phi} is finite.
\item[(ii)] The phase itself may be unbounded even for a constant wavevector, as in $\varphi(\mathbf{x})=\mathbf{q}\cdot\mathbf{x}$ with $\mathbf{q}\ne0$. When the wavevector is unbounded, the weighted root integral can diverge even if the unweighted one is finite. To see this, take $\varphi(\mathbf{x})=|\mathbf{x}|^2/2$ and $\wt{\bm{\rho}}=u^p\mathbf{P}$, where $\mathbf{P}$ is a fixed rank-one orthogonal projection and
\[
u(\mathbf{x})=\sum_{n=3}^{\infty}\frac{1}{n}\,\eta\big(n(\mathbf{x}-n\mathbf{e}_1)\big)
\]
for a nonzero, nonnegative $\eta\in C_c^\infty(B_1)$. The bumps have disjoint supports. Each contributes $n^{-3}\int|\grad\eta|^p$ to $\int|\grad u|^p$, while $|\grad\varphi|^2=|\mathbf{x}|^2$ is comparable to $n^2$ on its support. The weighted root integral therefore diverges like $\sum_{n\ge3}n^{-1}$. The $n$-th bump has mass $n^{-p-3}\int\eta^p$, so $\int\rho|\grad\varphi|^2$ is finite. The weighted Hessian integral is also finite since $\mathbf{D}^2\varphi$ is constant.
\item[(iii)] The assumption $\grad\wt{\bm{\rho}}^{1/p}\in L^p(\RR^3)$ made for~\cref{eq:LSSWA_sharp_bulk} implies the density hypotheses of \cref{thm:LSSWA} when $\rho\in L^1$. For an example distinguishing~\cref{eq:LSSWA_estim} from~\cref{eq:LSSWA_sharp_bulk}, take a scalar density vanishing quadratically across a smooth surface. Its $1/p$ power has a normal derivative whose $p$-th power behaves like $|s|^{2-p}$ at signed distance $s$ from the surface. For $p>3$ this is not integrable. If the phase is constant in a neighborhood of the surface, only the LSDA density conditions are needed there.
\end{itemize}
\end{remark}

The TSGA~\cref{eq:LSSWA} involves the pair $(\wt{\bm{\rho}},\varphi)$. For a given spin-polarized density $\bm{\rho}$ there is a natural choice, in which $\varphi$ is the angle of the magnetization in the plane orthogonal to $\mathbf{e}_3$. The right side of~\cref{eq:LSSWA} then becomes a functional of $\bm{\rho}$ of the generalized gradient type, that is, the integral of a function of $\bm{\rho}(\mathbf{x})$ and of its first derivatives.

\begin{corollary}[Gradient form of the TSGA]\label{cor:gradient}
Let $p$ and $\theta$ be as in \cref{thm:LSSWA}, and let $\bm{\rho}$ be a spin-polarized density, with total density $\rho$ and magnetization $\mathbf{m}=(m_1,m_2,m_3)$. Assume that
\begin{equation}\label{eq:canonical_phase}
m_1+im_2=a\,e^{i\varphi}
\end{equation}
for two real functions $a$ and $\varphi$ with $\varphi\in W^{2,p}_{\rm loc}(\RR^3)$, and let
\begin{equation*}
\wt{\bm{\rho}}:=\frac12\begin{pmatrix}\rho+m_3&a\\ a&\rho-m_3\end{pmatrix}.
\end{equation*}
Assume that $\wt{\bm{\rho}}$ and $\varphi$ satisfy the hypotheses of \cref{thm:LSSWA}. Then its reference energy can be written as
\begin{equation}\label{eq:gradient_approx}
\int_{\RR^3}e_\SSW(\wt{\bm{\rho}},\grad\varphi)
=\int_{\RR^3}e_\SSW(\bm{\rho},\grad\varphi).
\end{equation}
The error in approximating $E(\bm{\rho})$ by this integral satisfies the estimate~\cref{eq:LSSWA_estim} of \cref{thm:LSSWA} and admits~\cref{eq:LSSWA_sharp_bulk} under its additional hypothesis. Moreover,
\begin{equation}\label{eq:azimuthal_gradient}
\begin{aligned}
\grad\varphi&=\frac{m_1\grad m_2-m_2\grad m_1}{m_1^2+m_2^2},\\
|\grad\sqrt{\wt{\bm{\rho}}}|^2&=|\grad\sqrt{\bm{\rho}}|^2-\frac12\Big(\sqrt{\rho^+}-\sqrt{\rho^-}\Big)^2\sin^2\vartheta\,|\grad\varphi|^2,
\end{aligned}
\end{equation}
where $\vartheta$ is the angle between $\mathbf{m}$ and $\mathbf{e}_3$. The first formula holds on the set where $m_1+im_2\neq0$ and the second almost everywhere.
\end{corollary}

By~\cref{eq:azimuthal_gradient} and~\cref{eq:formula_gradient_sqrt_R}, the term $\int|\grad\sqrt{\wt{\bm{\rho}}}|^2$ only contains the derivatives of the two eigenvalues $\rho^\pm$ and of the angle $\vartheta$. It is the corresponding error term of \cref{cor:LSDA}, from which the contribution of the rotation of the magnetization about $\mathbf{e}_3$ has been removed. The remaining contributions of the azimuthal rotation to the error involve the weight $|\grad\varphi|^2$ and the Hessian $\mathbf{D}^2\varphi$.

In \cref{cor:gradient} the axis $\mathbf{e}_3$ plays a special role, and the functional depends on its choice. The approximation is accurate when the magnetization is locally a spin-spiral wave about this fixed axis, with a slowly varying angle $\vartheta$.

\begin{remark}[Global phase]\label{rmk:global_phase}
The assumption~\cref{eq:canonical_phase} in \cref{cor:gradient} requires a phase which is defined on the whole space. The function $a$ is allowed to change sign, which covers the case where $m_1+im_2$ vanishes on a surface. Where $m_1+im_2\ne0$, the transverse magnetization determines $\varphi$ locally up to a multiple of $\pi$, and hence determines $\grad\varphi$. At zeros, the choice of a regular extension can affect the error bound~\cref{eq:LSSWA_estim} of \cref{thm:LSSWA}. The reference energy there is always $e_\UEG(\bm{\rho})$, since the spin-polarized density is diagonal. A vortex line with nonzero winding precludes a global phase. Around a transverse circle of radius $r$ encircling such a line, the local phase gradient has circulation $2\pi k$ for a nonzero integer $k$. Cauchy--Schwarz then gives $\int_{\partial B_r}|\grad\varphi|^2\,\dd s\ge2\pi k^2/r$, so $\int\rho|\grad\varphi|^2$ diverges if $\rho$ stays bounded below by a positive constant near the line. We do not treat this case.
\end{remark}

\subsection{The noncollinear local spin density approximation}\label{sec:lsda}
Taking $\varphi=0$ in \cref{thm:LSSWA}, we obtain the noncollinear LSDA~\cref{eq:LSDA} for arbitrary magnetization directions, under the assumptions of~\cite{lewin2020local} imposed on the spin-polarized density.

\begin{corollary}[Noncollinear local spin density approximation]\label{cor:LSDA}
Let $p>3$ and $2/p\le\theta<1$. Let $\bm{\rho}\ge0$ be a spin-polarized density with $\rho\in L^1\cap L^2$, $\grad\sqrt{\bm{\rho}}\in L^2$ and $\grad\bm{\rho}^\theta\in L^p$ on $\RR^3$. For every $\epsilon>0$,
\begin{multline}\label{eq:LDA_main_estim}
\left|E(\bm{\rho})-\int_{\RR^3}e_\UEG\big(\bm{\rho}(\mathbf{x})\big)\,\dd\mathbf{x}\right|
\le\epsilon\int_{\RR^3}(\rho+\rho^2)\\
+\frac{C(1+\epsilon)}{\epsilon}\int_{\RR^3}|\grad\sqrt{\bm{\rho}}|^2
+\frac{C}{\epsilon^b}\int_{\RR^3}|\grad\bm{\rho}^\theta|^p.
\end{multline}
Here $b=\max\{2p,(1+3\theta)p\}-1$ and $C=C(p,\theta)$.
\end{corollary}

\begin{proof}
Take $\wt{\bm{\rho}}=\bm{\rho}$ and $\varphi=0$ in \cref{thm:LSSWA}. Then $\mathbf{U}_0=\iden$ and $e_\SSW(\bm{\rho},0)=e_\UEG(\bm{\rho})$ by \cref{thm:spiral_limit}. The Hessian vanishes and $\{\grad\varphi\ne0\}$ is empty, so the local Sobolev assumption on $\bm{\rho}^{1/p}$ is vacuous and the last two integrals in~\cref{eq:LSSWA_estim} of \cref{thm:LSSWA} vanish.
\end{proof}

\section{Proofs}\label{sec:proof}
In the proofs $C$ denotes a constant that may change from line to line and may depend on $p$ and $\theta$. We write $\pscal{\mathbf{A},\mathbf{B}}:=\Trs\mathbf{A}^*\mathbf{B}$ for the Hilbert--Schmidt scalar product of $2\times2$ matrices and $\norm{\mathbf{A}}$ for the operator norm. We use the mean value notation
$$
\dashint_A u:=\frac{1}{|A|}\int_A u
$$
for any measurable set $A\subset\RR^3$ with $0<|A|<\infty$ and any function $u$ integrable on $A$, possibly matrix- or operator-valued. The Fourier transform of $f\in L^1(\RR^3)$ is $\wh{f}(\mathbf{p}):=(2\pi)^{-3/2}\int_{\RR^3}f(\mathbf{x})e^{-i\mathbf{p}\cdot\mathbf{x}}\,\dd\mathbf{x}$.

\subsection{Strategy of the proof}\label{sec:strategy}
The proof of \cref{thm:LSSWA} is a modification of the proof of the local density approximation in~\cite[Theorem~3]{lewin2020local}. It combines a localization in tetrahedral tiles with a comparison, on each tile, with a spin-spiral wave of constant parameters.

\medskip

\noindent\textbf{Localization.} As in~\cite[Section 4]{lewin2020local}, we approximate the energy by a sum over tiles,
\begin{equation*}
E(\bm{\rho})\approx \sum_k  E\big(\bm{\rho}\chi_k\big),
\end{equation*}
where $\{\chi_k\}$ is a smooth partition of unity made of \emph{scalar} functions. Localization acts identically on both spin components, so its errors only involve the total density $\rho$. The tiles are smeared tetrahedra, and the bounds are averaged over the size, the orientation and the position of the tiling. In the proof of \cref{thm:LSSWA} the tiles have size $\ell=\epsilon^{-1}$ and are smeared on the scale $\delta=1$.

\medskip

\noindent\textbf{Reduction to spin-spiral waves.} A local spin rotation preserves the Coulomb terms. \Cref{lem:gauge} bounds the change in kinetic energy in terms of the initial kinetic energy and $\int\rho\norm{\grad\mathbf{U}}^2$. On a ball the relative spin rotation $\mathbf{U}_{\varphi-\mathbf{q}\cdot\mathbf{x}}$ compares the phase $\varphi$ with the linear phase $\mathbf{q}\cdot\mathbf{x}$, as in \cref{lem:constant_wavevector}. In the rotating frame the spin-polarized density $\wt{\bm{\rho}}$ varies slowly, and \cref{prop:replace_spiral} replaces it by a constant matrix. Each tile then carries a spin-spiral wave with constant parameters. In strongly polarized regions this comparison also requires a local spin rotation, whose cost is controlled by the gradient of the matrix square root.

\medskip

\noindent\textbf{Averaging over orientations.} A translation of a spin-spiral wave amounts to a rotation of the spin, and a rotation of space changes the direction of the wavevector, see~\cref{eq:covariance_spiral}. A tetrahedron is not invariant under rotations, so the energy of a spin-spiral wave on a tile depends on the direction of the wavevector relative to the tile. The localization bounds, however, already contain an average over the orientations of the tiles. We choose the wavevector and the comparison densities on a ball around the center of mass of the tile, independently of its orientation and size. The average over orientations then gives the energy averaged over the directions of the wavevector. These averaged energies converge to the SSW gas energy $e_\SSW$, and \cref{prop:rate_SSW} bounds the rate of convergence.

\medskip

\noindent\textbf{Freezing the wavevector.} On each ball we take for $\mathbf{q}$ the value of $\grad\varphi$ at a point where $|\grad\varphi|$ is minimal, see \cref{lem:weighted_wavevector}. Then $|\mathbf{q}|\le|\grad\varphi|$ on the ball, so the kinetic energy at the frozen wavevector is bounded by $\int\rho|\grad\varphi|^2$. The cost of freezing is controlled by Morrey's inequality, applied to a function weighted by the density which vanishes at the minimum point. This only requires the integrability conditions~\cref{eq:hyp_phi} of \cref{thm:LSSWA}, in which the derivatives of the phase are weighted by the density.

\medskip

\noindent\textbf{Kinetic energy of the spin-spiral wave.} The kinetic energy of the spin-spiral wave is of order $\rho|\grad\varphi|^2$, which can remain comparable to $\rho$ at low density. Every error which depends on the wavevector is therefore a small multiple of $\rho|\mathbf{q}|^2$. On balls where $\wt{\bm{\rho}}$ oscillates strongly, no comparison with a constant matrix is available, and the whole kinetic term must be controlled by the gradient of $\wt{\bm{\rho}}^{1/p}$. This accounts for the wavevector weight on the relative density gradient in~\cref{eq:TSGA_estim_intro}.

\medskip

\noindent\textbf{Regularity of the SSW gas energy.} Concavity gives regularity in the wavevector. To control the dependence on the matrix up to full polarization, we remove electrons according to their spin in the rotating frame of a finite tetrahedron. The kinetic energy and the density removed control the change in spin current. Optimizing the size of the tetrahedron gives a H\"older estimate of exponent $1/3$. Convexity at fixed density then yields a Lipschitz estimate in the cube root of the minority spin fraction.

\medskip

\noindent\textbf{Assembly.} In \cref{sec:proof_LSSWA} we combine these ingredients, separately for the upper bound of \cref{prop:upper_bound} and the lower bound of \cref{prop:lower_bound}. Near each tile we freeze the wavevector, which produces the error terms containing the derivatives of the phase. Where $\wt{\bm{\rho}}$ varies little compared with its size, we replace it by a constant, average over the orientations of the tiles and compare with the SSW gas energy. For the lower bound we also average over the sizes of the tiles. The regularity of $e_\SSW$ then brings us back to the local values $\wt{\bm{\rho}}(\mathbf{x})$ and $\grad\varphi(\mathbf{x})$. Where $\wt{\bm{\rho}}$ oscillates strongly, the \emph{a priori} bounds suffice. Summing over the tiles gives the estimate~\cref{eq:LSSWA_estim} of \cref{thm:LSSWA}. The tiles have size $\epsilon^{-1}$, so the localization errors are of order $\epsilon\int(\rho+\rho^2)$, while the gradient terms carry negative powers of $\epsilon$.

\subsection{Transformation properties}\label{sec:transformations}
The internal Hamiltonian in~\cref{eq:Levy-Lieb-grand-canonical} does not depend on the spin. For a constant unitary matrix $\mathbf{U}\in \SU(2)$, the unitary $\bigoplus_{n\ge0}\mathbf{U}^{\otimes n}$ on Fock space therefore commutes with the Hamiltonian and transforms the spin-polarized density $\bm{\rho}_\Gamma$ into $\mathbf{U}\bm{\rho}_\Gamma \mathbf{U}^*$. Similarly, the rotations and translations of $\RR^3$ act on the space variable only. We deduce that
\begin{equation}\label{eq:invariance}
E(\mathbf{U}\bm{\rho}\,\mathbf{U}^*)=E(\bm{\rho}),\qquad E\big(\bm{\rho}(\mathbf{R}\,\cdot\,-\bm{\tau})\big)=E(\bm{\rho})
\end{equation}
for every $\mathbf{U}\in \SU(2)$, $\mathbf{R}\in \SO(3)$ and $\bm{\tau}\in\RR^3$. We emphasize that the rotations of the spin and the rotations of space are independent, because we do not include spin-orbit coupling.

For a constant Hermitian matrix $\bm{\rho}_0\ge0$ and $\mathbf{q}\in\RR^3$, consider the spin-spiral wave
$$
\bm{\rho}_{\mathbf{q}}:=\mathbf{U}_{\mathbf{q}}\bm{\rho}_0\mathbf{U}_{\mathbf{q}}^*,
$$
with $\mathbf{U}_{\mathbf{q}}(\mathbf{x})=e^{-\frac{i}2(\mathbf{q}\cdot\mathbf{x})\bm{\sigma}_3}$, and its behavior under translations and rotations of space. Let $f\ge0$ be a scalar function such that $\sqrt f\in H^1(\RR^3)$. We claim that
\begin{equation}\label{eq:covariance_spiral}
E\big(\bm{\rho}_{\mathbf{q}}\,f(\cdot-\bm{\tau})\big)=E\big(\bm{\rho}_{\mathbf{q}}\,f\big),\qquad E\big(\bm{\rho}_{\mathbf{q}}\,f(\mathbf{R}\,\cdot)\big)=E\big(\bm{\rho}_{\mathbf{R}\mathbf{q}}\,f\big)
\end{equation}
for all $\bm{\tau}\in\RR^3$ and all $\mathbf{R}\in\SO(3)$. Indeed, we have $\mathbf{U}_{\mathbf{q}}(\mathbf{x}-\bm{\tau})=\mathbf{U}_{\mathbf{q}}(\bm{\tau})^*\mathbf{U}_{\mathbf{q}}(\mathbf{x})$, hence
$$
\bm{\rho}_{\mathbf{q}}\,f(\cdot-\bm{\tau})=\mathbf{U}_{\mathbf{q}}(\bm{\tau})\,\big(\bm{\rho}_{\mathbf{q}}f\big)(\cdot-\bm{\tau})\,\mathbf{U}_{\mathbf{q}}(\bm{\tau})^*.
$$
A translation of the spin-spiral wave is a constant rotation of the spin. Similarly, $(\mathbf{R}\mathbf{q})\cdot(\mathbf{R}\mathbf{x})=\mathbf{q}\cdot\mathbf{x}$ gives $\bm{\rho}_{\mathbf{q}}\,f(\mathbf{R}\,\cdot)=(\bm{\rho}_{\mathbf{R}\mathbf{q}}f)(\mathbf{R}\,\cdot)$, that is, a rotation of space amounts to rotating $\mathbf{q}$. The two equalities in~\cref{eq:covariance_spiral} then follow from~\cref{eq:invariance}.

\subsection{Known bounds}\label{sec:known_bounds}
We collect the inequalities from the literature which are used in the proofs.

The following estimate on the indirect Coulomb energy will be used repeatedly. For a state $\Gamma=\bigoplus_{n\ge0}\Gamma_n$ on the fermionic Fock space, we denote by
$$
\Cc(\Gamma):=\sum_{n\ge2}\Tr_{\HF^n}\Bigg(\sum_{1\le j<k\le n}\frac1{|\mathbf{x}_j-\mathbf{x}_k|}\Bigg)\Gamma_n
$$
its Coulomb energy.

\begin{theorem}[Lieb--Oxford inequality]\label{thm:LO}
For every state $\Gamma=\bigoplus_{n\ge0}\Gamma_n$ on the fermionic Fock space, the bound
\begin{equation*}
\Cc(\Gamma)-D(\rho_\Gamma)\ge-c_\LO\int_{\RR^3}\rho_\Gamma^{4/3}
\end{equation*}
holds true with a universal constant $c_\LO>0$.
\end{theorem}

The inequality is due to~\cite{lieb1979lower,lieb1981improved}, see also~\cite{lieb2010stability}. The constant $c_\LO\le1.58$ is the best known at present~\cite{lewin2022improved}. The proof of~\cite{lewin2022improved} works all the same in the grand-canonical setting.

We use the tiling of~\cite[Section 4.1]{lewin2020local}. Up to their faces, the unit cube $C_1=(-1/2,1/2)^3$ is the disjoint union of 24 congruent tetrahedra $\tetra_1,\ldots,\tetra_{24}$. To be definite, we take the tetrahedra which have as vertices the center of the cube, the center of one of its faces and two adjacent vertices of this face. We have $\tetra_j=T_j\tetra$ where $\tetra$ is a reference tetrahedron with $0$ as its center of mass and $T_j=(\mathbf{z}_j,\mathbf{R}_j)\in C_1\times \SO(3)$ acts as $T_j \mathbf{x}=\mathbf{R}_j\mathbf{x}-\mathbf{z}_j$. For $0<\lambda\le1$ the tetrahedron $T_j(\lambda\tetra)$ is obtained from $\tetra_j$ by a dilation of ratio $\lambda$ about its center of mass $-\mathbf{z}_j$. The distances $d_1,\ldots,d_4$ from the center of mass of $\tetra$ to the planes containing its four faces are $1/8$, $1/(8\sqrt2)$, $1/(4\sqrt2)$ and $1/(4\sqrt2)$. We call
$$
d_{\min}:=\frac{1}{8\sqrt2}
$$
the smallest of them. For $\ell>0$ we write $C_\ell:=\ell C_1=(-\ell/2,\ell/2)^3$.

Following~\cite{csirik2026non}, we let
\begin{equation}\label{eq:def_eta_delta}
\eta_\delta(\mathbf{x}):=\frac{1}{(b\delta)^{3}}\,\eta_\star\left(\frac{\mathbf{x}}{b\delta}\right),\qquad b:=\frac1{100},
\end{equation}
where $\eta_\star$ is a fixed $C^\infty_c$ non-negative radial function with support in the unit ball and such that $\int_{\RR^3}\eta_\star=1$. The function $\eta_\delta$ is supported in the ball of radius $b\delta$. In~\cite{lewin2020local} the constant was $1/10$ instead of $b$. The construction of the upper bound in~\cref{sec:upper_bound} requires that
\begin{equation}\label{eq:condition_b}
b\le\frac{d_{\min}}3=\frac{1}{24\sqrt2}\simeq0.029.
\end{equation}
This condition ensures that the skeleton function used in~\cref{sec:upper_bound} vanishes on the supports of the smeared tetrahedra, and it is the reason for our choice of $b$.

We also use the smooth partition of unity without holes
\begin{equation}\label{eq:def_xi_ell_tiling}
 \boxed{\sum_{\mathbf{z}\in\ZZ^3}\sum_{j=1}^{24} (\iden_{\ell\tetra_j}\ast\eta_\delta)(\mathbf{x}-\ell \mathbf{z})=1}
\end{equation}

The indirect Coulomb energy can be localized with the smeared Graf--Schenker inequality. For a function $0\le f\le1$, we denote by $\Gamma_{|f}$ the geometrically $f$-localized state on Fock space~\cite{lewin2011geometric}, that is, the unique state which has the $k$-particle reduced density matrices $f^{\otimes k}\Gamma^{(k)}f^{\otimes k}$.

\begin{theorem}[Smeared Graf--Schenker inequality]\label{thm:GS}
There are universal constants $\kappa,C>0$ such that for every state $\Gamma=\bigoplus_{n\ge0}\Gamma_n$ on the fermionic Fock space and all $\ell>\kappa\delta>0$,
\begin{multline*}
\Cc(\Gamma)-D(\rho_\Gamma)\ge\Big(1-\frac{\kappa\delta}{\ell}\Big)\sum_{\mathbf{z}\in\ZZ^3}\sum_{j=1}^{24}\int_{\SO(3)}\dd\mathbf{R}\,\dashint_{C_\ell}\dd\bm{\tau}\,\Big(\Cc\big(\Gamma_{|\sqrt{(\iden_{\ell\tetra_j}\ast\eta_\delta)(\mathbf{R}\,\cdot\,-\ell \mathbf{z}-\bm{\tau})}}\big)\\
-D\big((\iden_{\ell\tetra_j}\ast\eta_\delta)(\mathbf{R}\,\cdot\,-\ell \mathbf{z}-\bm{\tau})\rho_\Gamma\big)\Big)-\frac{C}{\ell}\int_{\RR^3}\rho_\Gamma-\frac{C\delta^3}{\ell}\int_{\RR^3}\rho_\Gamma^2.
\end{multline*}
\end{theorem}

This is~\cite[Eq.~(53)]{lewin2020local}, which relies on the smeared Graf--Schenker inequality of~\cite[Lemma~6]{graf1995molecular}. Our normalization of $\eta_\delta$ in~\cref{eq:def_eta_delta} only changes the constants. The inequality only involves the positions of the particles, hence it holds independently of the spin.

The kinetic energy is localized with the IMS formula~\cite{cycon2009schrodinger}. For a state $\Gamma=\bigoplus_{n\ge0}\Gamma_n$ on the fermionic Fock space, we denote by
$$
\TC(\Gamma):=\sum_{n\ge1}\Tr_{\HF^n}\Bigg(-\sum_{j=1}^n\lapl_{\mathbf{x}_j}\Bigg)\Gamma_n=\Tr(-\lapl)\bm{\gamma}_\Gamma
$$
its kinetic energy. For a real Lipschitz function $\chi$ the operator identity $\chi(-\lapl)\chi=\frac12\big(\chi^2(-\lapl)+(-\lapl)\chi^2\big)+|\grad\chi|^2$ holds in the sense of quadratic forms, and the one-particle density matrix of $\Gamma_{|\chi}$ is $\chi\bm{\gamma}_\Gamma\chi$. Averaging over a family of localization functions gives the following, see also~\cite[proof of Lemma~5.6]{lewin2018statistical}. We need it for locally Lipschitz functions, whose gradients may be unbounded, and therefore give the short proof.

\begin{theorem}[IMS localization formula]\label{thm:IMS}
Let $(A,\mu)$ be a measure space and let $(\chi_a)_{a\in A}$ be a measurable family of locally Lipschitz functions on $\RR^3$ with $0\le\chi_a\le1$, such that $\int_A\chi_a(\mathbf{x})^2\,\dd\mu(a)=1$ for every $\mathbf{x}$ and $\int_A|\grad\chi_a(\mathbf{x})|^2\,\dd\mu(a)$ is bounded. For every state $\Gamma$ on the fermionic Fock space with $\TC(\Gamma)<\infty$ and $\rho_\Gamma\in L^1(\RR^3)$,
\begin{equation*}
\int_A\dd\mu(a)\,\TC\big(\Gamma_{|\chi_a}\big)=\TC(\Gamma)+\int_{\RR^3}\bigg(\int_A\dd\mu(a)\,|\grad\chi_a|^2\bigg)\rho_\Gamma.
\end{equation*}
\end{theorem}

\begin{proof}
Since $\rho_\Gamma\in L^1(\RR^3)$ and $\TC(\Gamma)<\infty$, we can write $\bm{\gamma}_\Gamma=\sum_k\lambda_k|\bm{\phi}_k\rangle\langle\bm{\phi}_k|$ with $\lambda_k\ge0$, $\sum_k\lambda_k<\infty$ and $\bm{\phi}_k\in H^1(\RR^3,\CC^2)$. Then $\rho_\Gamma=\sum_k\lambda_k|\bm{\phi}_k|^2$ and $\TC(\Gamma)=\int_{\RR^3}\tau_\Gamma$ with $\tau_\Gamma:=\sum_k\lambda_k|\grad\bm{\phi}_k|^2$, and the Cauchy--Schwarz inequality gives $|\grad\rho_\Gamma|\le2\sqrt{\rho_\Gamma\tau_\Gamma}\in L^1(\RR^3)$. By Tonelli's theorem, $\int_{\RR^3}|\grad\chi_a|^2\rho_\Gamma<\infty$ for $\mu$-almost every $a$. For such $a$ we have $\chi_a\bm{\phi}_k\in H^1(\RR^3,\CC^2)$ whenever $\lambda_k>0$, and expanding the square gives
\begin{align*}
\TC\big(\Gamma_{|\chi_a}\big)&=\Tr(-\lapl)\chi_a\bm{\gamma}_\Gamma\chi_a=\sum_k\lambda_k\int_{\RR^3}|\grad(\chi_a\bm{\phi}_k)|^2\\
&=\int_{\RR^3}\Big(\chi_a^2\tau_\Gamma+|\grad\chi_a|^2\rho_\Gamma+\chi_a\grad\chi_a\cdot\grad\rho_\Gamma\Big).
\end{align*}
By the Cauchy--Schwarz inequality, $\int_A\chi_a|\grad\chi_a|\,\dd\mu(a)$ is bounded, so that we can integrate over $a$ and exchange the integrals. The bounded function $\int_A\chi_a\grad\chi_a\,\dd\mu(a)$ equals $\frac12\grad\int_A\chi_a^2\,\dd\mu(a)=0$ in the sense of distributions, hence the last term vanishes, and $\int_A\chi_a^2\,\dd\mu(a)=1$ gives the formula.
\end{proof}

The localization functions are square roots of smooth functions, whose gradients we control by the following inequality of Glaeser~\cite{glaeser1963racine}.

\begin{theorem}[Glaeser's inequality]\label{thm:Glaeser}
Let $g\in C^2(\RR^3)$ be non-negative with $M:=\|\mathbf{D}^2g\|_{L^\infty}<\infty$. Then
\begin{equation}\label{eq:Glaeser}
|\grad g|^2\le2Mg\qquad\text{on $\RR^3$.}
\end{equation}
In particular, $\sqrt{g}$ is Lipschitz and $|\grad\sqrt{g}|^2\le M/2$ almost everywhere.
\end{theorem}

\begin{proof}
Fix $\mathbf{x}$ and let $\mathbf{v}=\grad g(\mathbf{x})$. By Taylor's formula and since $g\ge0$, we have $0\le g(\mathbf{x}-t\mathbf{v})\le g(\mathbf{x})-t|\mathbf{v}|^2+Mt^2|\mathbf{v}|^2/2$ for all $t\in\RR$. If $M=0$, letting $t\to\infty$ gives $\mathbf{v}=0$. Otherwise, taking $t=1/M$ gives~\cref{eq:Glaeser}. Where $g>0$ we obtain $|\grad\sqrt{g}|^2=|\grad g|^2/(4g)\le M/2$. Since $\sqrt{g}$ is continuous and vanishes wherever $g$ does, restricting to segments shows that $\sqrt{g}$ is Lipschitz with constant $\sqrt{M/2}$, and the gradient bound holds almost everywhere.
\end{proof}

Finally, we use the following form of Morrey's inequality on convex sets. To obtain the stated constant from the segment argument in~\cite[Lemma~7.16]{gilbarg2001elliptic}, let $r=|\mathbf{x}-\mathbf{y}|>0$ and average over $\mathbf{x}+(r/\diam K)(K-\mathbf{x})$. The segments from $\mathbf{x}$ and $\mathbf{y}$ to this set have lengths at most $r$ and $2r$, respectively. Applying H\"older's inequality in the spatial average leaves the factor $\int_0^1 t^{-3/p}\,\dd t=p/(p-3)$.

\begin{theorem}[Morrey's inequality]\label{thm:Morrey}
Let $3<p<\infty$ and let $K\subset\RR^3$ be a compact convex set with nonempty interior. For any finite-dimensional Hilbert space $V$, every $u\in W^{1,p}(\Int K;V)$ has a continuous representative on $K$ satisfying
\begin{equation}\label{eq:Morrey}
|u(\mathbf{x})-u(\mathbf{y})|\le c_{\mathrm{Mo}}\,|\mathbf{x}-\mathbf{y}|^{1-3/p}\,\|\grad u\|_{L^p(K)}\qquad\text{for all $\mathbf{x},\mathbf{y}\in K$,}
\end{equation}
with $c_{\mathrm{Mo}}=\dfrac{3p}{p-3}\big((\diam K)^3/|K|\big)^{1/p}$.
\end{theorem}

\subsection{\emph{A priori} bounds for spin-polarized density functionals}\label{sec:a_priori}
We now summarize the \emph{a priori} bounds on $T(\bm{\rho})$ and $E(\bm{\rho})$ which are used throughout the paper. They generalize those of~\cite[Section 3]{lewin2020local}. The proofs of \cref{lem:HO} and \cref{thm:upper_bound_T} are given in \cref{app:HO,app:upper_bound_T}. As in~\cite{lewin2020local} we work here in any dimension $d\ge1$, with spin-polarized densities defined on $\RR^d$ in the same way as in~\cref{eq:representable}, except for the bounds on $E(\bm{\rho})$, which is only defined here for $d=3$. No assumption on the polarization is needed.

In the spinless case, the Hoffmann-Ostenhof inequality~\cite{hoffmann1977schrodinger} states that the kinetic energy is bounded from below by $\int|\grad\sqrt{\rho_{\bm{\gamma}}}|^2$. The next result is its equivalent for spin-polarized densities. That $\sqrt{\bm{\rho}_{\bm{\gamma}}}\in H^1$ for a one-particle density matrix $\bm{\gamma}$ of finite kinetic energy was proved in~\cite{gontier2013n} for $d=3$, and our proof gives it in every dimension.

\begin{lemma}[Hoffmann-Ostenhof-type inequality]\label{lem:HO}
Let $\bm{\gamma}\ge0$ be a trace-class operator on $L^2(\RR^d,\CC^2)$ with finite kinetic energy. Its spin-polarized density $\bm{\rho}_{\bm{\gamma}}$ satisfies $\sqrt{\bm{\rho}_{\bm{\gamma}}}\in H^1(\RR^d,\CC^{2\times2})$ and
\begin{equation}\label{eq:Hoffmann-Ostenhof}
\Tr(-\lapl)\bm{\gamma}\ge \frac12\int_{\RR^d} \left|\grad\sqrt{\bm{\rho}_{\bm{\gamma}}}(\mathbf{x})\right|^2\,\dd \mathbf{x}.
\end{equation}
The constant $1/2$ is optimal.
\end{lemma}

The proof of \cref{lem:HO} relies on the convexity of $\bm{\rho}\mapsto\int_{\RR^d}|\grad\sqrt{\bm{\rho}}|^2$, see \cref{lem:convexity}, and is given in \cref{app:HO}. Using~\cref{eq:Hoffmann-Ostenhof} and~\cref{thm:LO}, we obtain the following.

\begin{corollary}[Lower bound on $E(\bm{\rho})$]\label{cor:lower_bound_E_rho}
For every spin-polarized density $\bm{\rho}\ge0$ on $\RR^3$ with $\rho\in L^1(\RR^3)$ and $\sqrt{\bm{\rho}}\in H^1(\RR^3,\CC^{2\times2})$, we have
\begin{equation*}
E(\bm{\rho})\ge \frac12\int_{\RR^3}|\grad\sqrt{\bm{\rho}}(\mathbf{x})|^2\,\dd \mathbf{x}-c_\LO\int_{\RR^3}\rho(\mathbf{x})^{4/3}\,\dd \mathbf{x}.
\end{equation*}
\end{corollary}

We turn to upper bounds. The following is the generalization of~\cite[Theorem~8]{lewin2020local} to the lowest kinetic energy $T(\bm{\rho})$ in~\cref{eq:def_T}. The main difficulty is again to construct a state having the exact given spin-polarized density $\bm{\rho}$. The proof is given in \cref{app:upper_bound_T}. We denote by
\begin{equation}\label{eq:c_TF}
c_\TF=\frac{ 4\pi^2 d}{(d+2)} \left( \dfrac{d}{| \mathbb{S}^{d-1} |} \right)^{2/d}
\end{equation}
the semi-classical constant.

\begin{theorem}[Upper bound on the kinetic energy functional]\label{thm:upper_bound_T}
There are two constants $\kappa_1,\kappa_2>0$ depending only on the space dimension $d$ such that
\begin{equation}\label{eq:upper_bound_T_kappa}
 T(\bm{\rho})\le c_\TF\left(1+\kappa_1\epsilon\right)\int_{\RR^d}\Trs \bm{\rho}(\mathbf{x})^{1+2/d}\dd \mathbf{x}
 +\frac{\kappa_2(1+\sqrt\epsilon)^2}{\epsilon}\int_{\RR^d}|\grad\sqrt{\bm{\rho}}(\mathbf{x})|^2\,\dd \mathbf{x},
\end{equation}
for every spin-polarized density $\bm{\rho}\ge0$ on $\RR^d$ with $\rho\in L^1(\RR^d)$ and $\sqrt{\bm{\rho}}\in H^1(\RR^d,\CC^{2\times2})$, and every $\epsilon>0$, where $c_\TF$ is the Thomas--Fermi constant~\cref{eq:c_TF}.
\end{theorem}

For $\bm{\rho}=(\rho/2)\iden_{\CC^2}$ this is exactly~\cite[Theorem~8]{lewin2020local} with $q=2$ (with a different $\kappa_2$). In general the kinetic energy per unit volume of the free gas with spin densities $\rho^\pm$ is $c_\TF\Trs \bm{\rho}^{1+2/d}=c_\TF\big((\rho^+)^{1+2/d}+(\rho^-)^{1+2/d}\big)$. We have
\begin{equation}\label{eq:compare_trace_power}
2^{-2/d}\rho^{1+2/d}\le \Trs \bm{\rho}^{1+2/d}\le \rho^{1+2/d}.
\end{equation}

Any fermionic one-particle density matrix $\bm{\gamma}$ on $L^2(\RR^d,\CC^2)$ is representable by a quasi-free state $\Gamma_{\bm{\gamma}}$ in Fock space~\cite{bach1994generalized}, the interaction energy of which is
$$
\frac12 \iint_{\RR^d\times\RR^d}\frac{\rho_{\bm{\gamma}}(\mathbf{x})\rho_{\bm{\gamma}}(\mathbf{y})-\Trs|\bm{\gamma}(\mathbf{x},\mathbf{y})|^2}{|\mathbf{x}-\mathbf{y}|}\,\dd\mathbf{x}\,\dd\mathbf{y}.
$$
Since the exchange term is non-positive, we immediately obtain the following in dimension $d=3$ from \cref{thm:upper_bound_T} and~\cref{eq:compare_trace_power}.

\begin{corollary}[Upper bound on $E(\bm{\rho})$]\label{cor:upper_bound_E_rho}
We have
\begin{equation*}
E(\bm{\rho})\le c_\TF(1+\kappa_1\epsilon)\int_{\RR^3} \Trs \bm{\rho}(\mathbf{x})^{5/3}\,\dd \mathbf{x}+\frac{\kappa_2(1+\sqrt\epsilon)^2}{\epsilon}\int_{\RR^3}|\grad\sqrt{\bm{\rho}}(\mathbf{x})|^2\,\dd \mathbf{x}
\end{equation*}
for every spin-polarized density $\bm{\rho}\ge0$ on $\RR^3$ with $\rho\in L^1(\RR^3)$ and $\sqrt{\bm{\rho}}\in H^1(\RR^3,\CC^{2\times2})$, and every $\epsilon>0$.
\end{corollary}

As in~\cite{lewin2020local}, the same bound holds in the mixed canonical case when $\int_{\RR^3}\rho\in\NN$, by~\cite{lieb1981variational}. In particular, \cref{cor:upper_bound_E_rho} gives a quantitative version of the representability theorem of~\cite{gontier2013n}.

\subsection{An improved spatial decoupling upper bound}\label{sec:upper_bound}

In~\cite[Proposition~13]{lewin2020local} every tetrahedron of the tiling was replaced by a slightly smaller one, in order to create holes between the tiles. The states in different tiles then have disjoint supports, so that one can take their tensor product. The density then needed to be multiplied by a normalization factor slightly larger than one in the whole tile. The exact density was then recovered after averaging over the translations of the tiling. In this procedure the charge which is missing in the holes is compensated by a charge spread over the whole tile. This costs an electrostatic energy of the order of $\delta^2\log(\alpha^{-1})\int\rho^2$, where $\delta$ is the size of the holes and the tiling is also averaged over dilations by factors in $(1-\alpha,1+\alpha)$.

Already for the non-uniform electron gas this construction is not sufficient, since the rescaling changes the inhomogeneity, and a different one was proposed in~\cite{csirik2026non}. There the density is not rescaled in the tiles. Instead, the holes are filled by an additional state, whose density is given by a ``skeleton function'' supported in a neighborhood of the faces of the tiles. A rough bound on the energy of this state is sufficient, since the skeleton occupies only a fraction $\delta/\ell$ of the volume, where $\ell$ is the size of the tiles. The upper bound of \cref{cor:upper_bound_E_rho} on this energy gives the second and third terms in the bound of \cref{prop:upper_bound} below. In~\cite{csirik2026non} the electrostatic error was estimated by $\delta^2\log(\alpha^{-1})\int\rho^2$, exactly as in~\cite{lewin2020local}. Here we use the same skeleton function and we show that, because the missing charge is now compensated \emph{locally}, the error is in fact of the order of $(\delta^3/\ell)\log(\alpha^{-1})\int\rho^2$. This bound gives tetrahedra of size $1/\epsilon$ in \cref{sec:proof_LSSWA} and controls the regularity of the gas energy up to full polarization in \cref{app:Lipschitz_SSW}.

\begin{figure}[t]
\centering
\begin{tikzpicture}[x=5cm,y=5cm]
\begin{scope}
\clip (-0.6,-0.6) rectangle (0.6,0.6);
\fill[green!50!black!40] (-1,-1) rectangle (1,1);
\foreach \cx in {-1,0,1}{\foreach \cy in {-1,0,1}{\foreach \rot in {0,90,180,270}{\foreach \refl in {1,-1}{
\begin{scope}[shift={(\cx,\cy)},rotate=\rot,yscale=\refl]
\fill[white,rounded corners=1.25mm] (0.075,0.05)--(0.475,0.05)--(0.475,0.45)--cycle;
\fill[blue!35,rounded corners=2.5mm] (0.15,0.1)--(0.45,0.1)--(0.45,0.4)--cycle;
\draw[gray,densely dotted] (0,0)--(0.5,0)--(0.5,0.5)--cycle;
\end{scope}}}}}
\foreach \t in {-0.5,0.5}{\draw[thick] (\t,-0.6)--(\t,0.6); \draw[thick] (-0.6,\t)--(0.6,\t);}
\end{scope}
\draw[|<->|] (-0.5,-0.66)--(0.5,-0.66) node[midway,below]{$\ell$};
\draw[|<->|] (0.66,0)--(0.66,0.5) node[midway,right]{$\ell/2$};
\path (-0.66,0.25) node[left]{\phantom{$\ell/2$}};
\end{tikzpicture}
\caption{Two-dimensional representation of the smeared tetrahedra $(\iden_{\ell T_j(1-\delta/\ell)\tetra}\ast\eta_\delta)(\cdot-\ell\mathbf{z})$ (blue) and of the skeleton function $\Upsilon_{\ell,\delta}$ (green). Here $\delta/\ell=0.4$ and the rounding of the corners due to the mollification is exaggerated.}\label{fig:skeleton}
\end{figure}
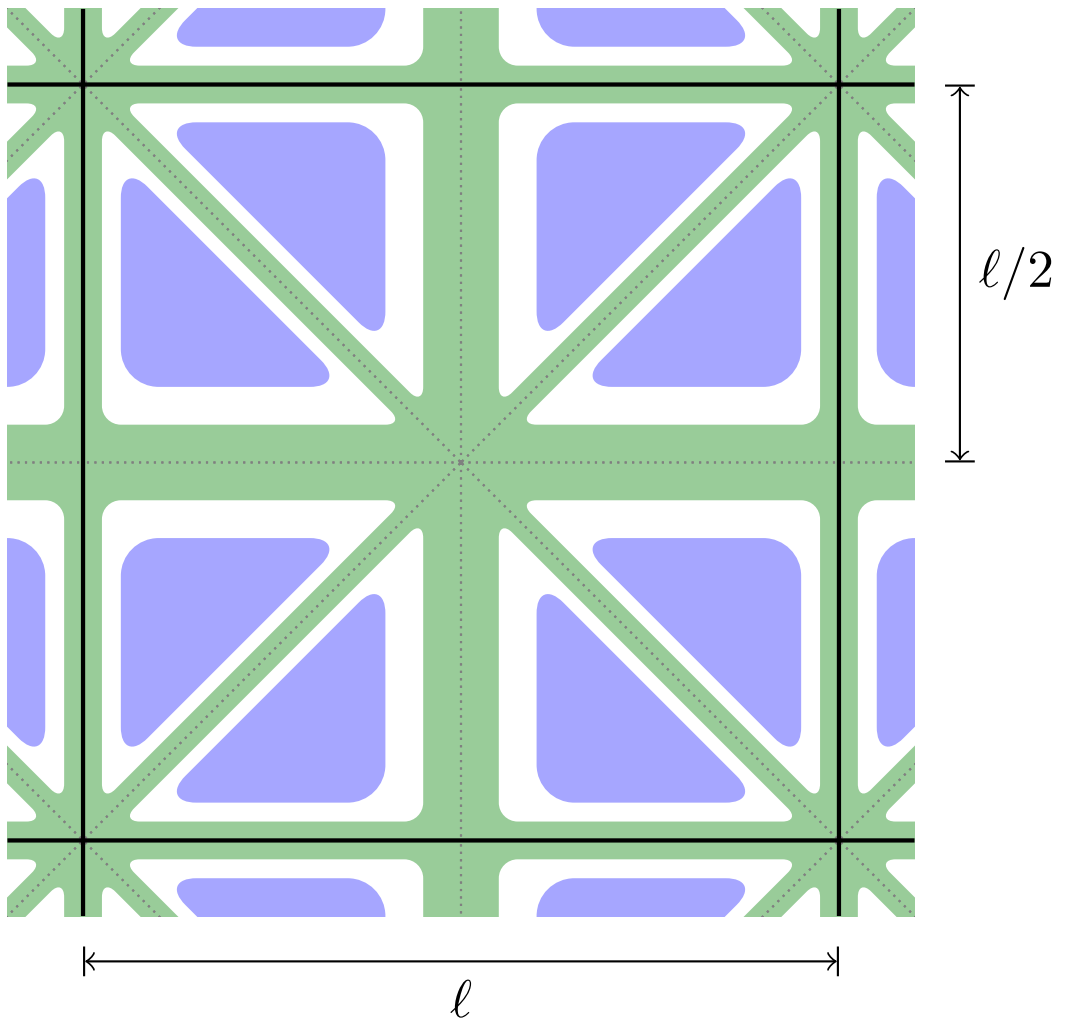

For $0<\delta\le\ell/2$ we use the smeared tetrahedra $\iden_{\ell T_j(1-\delta/\ell)\tetra}\ast\eta_\delta$, which are shrunk by the factor $1-\delta/\ell$, and the \emph{skeleton function} of~\cite{csirik2026non}
\begin{equation*}
 \boxed{\Upsilon_{\ell,\delta}:=\Lambda_{\delta/\ell}\Bigg(1-\sum_{\mathbf{z}\in\ZZ^3}\sum_{j=1}^{24}\iden_{\ell T_j(1-\delta/(2\ell))\tetra}\ast \eta_{\delta/2}(\cdot-\ell \mathbf{z})\Bigg)}
\end{equation*}
with the constant
$$
\Lambda_s:=\frac{1-(1-s)^3}{1-(1-s/2)^3}\le2.
$$
The function $\Upsilon_{\ell,\delta}$ is non-negative, smooth and $(\ell\ZZ^3)$-periodic. It is supported at a distance of order $\delta$ from the faces of the tiles, see \cref{fig:skeleton}.

Let us explain why the supports are disjoint. For $j=1,\ldots,24$, let $\bm{\nu}_1,\ldots,\bm{\nu}_4$ be the outward normals to the faces of $\tetra_j$ and $d_1,\ldots,d_4\ge d_{\min}$ be the distances from its center of mass $\mathbf{c}_j=-\mathbf{z}_j$ to these faces, so that
$$
\ell T_j(\lambda\tetra)=\big\{\mathbf{x}\ :\ \bm{\nu}_i\cdot(\mathbf{x}-\ell \mathbf{c}_j)<\lambda\ell d_i,\ i=1,\ldots,4\big\}.
$$
Since $\eta_\delta$ is supported in the ball of radius $b\delta$, the function $\iden_{\ell T_j(1-\delta/\ell)\tetra}\ast\eta_\delta$ is supported in the set
$$
\big\{\mathbf{x}\ :\ \bm{\nu}_i\cdot(\mathbf{x}-\ell \mathbf{c}_j)\le(\ell-\delta)d_i+b\delta,\ i=1,\ldots,4\big\}.
$$
This set is at a distance at least $(d_{\min}-b)\delta$ from the boundary of $\ell\tetra_j$, hence the functions $(\iden_{\ell T_j(1-\delta/\ell)\tetra}\ast\eta_\delta)(\cdot-\ell \mathbf{z})$ have disjoint supports. On the other hand, the function $\iden_{\ell T_j(1-\delta/(2\ell))\tetra}\ast \eta_{\delta/2}$ is equal to one at all the points which are at a distance at least $b\delta/2$ from the complement of $\ell T_j(1-\delta/(2\ell))\tetra$, that is, on
$$
\big\{\mathbf{x}\ :\ \bm{\nu}_i\cdot(\mathbf{x}-\ell \mathbf{c}_j)\le(\ell-\delta/2)d_i-b\delta/2,\ i=1,\ldots,4\big\}.
$$
Since $(\ell-\delta/2)d_i-b\delta/2-(\ell-\delta)d_i-b\delta=(d_i-3b)\delta/2\ge(d_{\min}-3b)\delta/2$, our condition~\cref{eq:condition_b} ensures that $\Upsilon_{\ell,\delta}$ vanishes on the support of all the functions $(\iden_{\ell T_j(1-\delta/\ell)\tetra}\ast\eta_\delta)(\cdot-\ell \mathbf{z})$. For $b=1/100$ the supports are at a distance at least $(d_{\min}-3b)\delta/2\ge\delta/40$ from each other. 

Finally, the constant $\Lambda_{\delta/\ell}$ has been chosen to ensure that
$$
\dashint_{C_\ell}\Upsilon_{\ell,\delta}=\Lambda_{\delta/\ell}\Big(1-\Big(1-\frac{\delta}{2\ell}\Big)^3\Big)=1-\Big(1-\frac\delta\ell\Big)^3=1-\frac{1}{\ell^3}\sum_{j=1}^{24}\int_{\RR^3}\iden_{\ell T_j(1-\delta/\ell)\tetra}\ast\eta_\delta.
$$
We therefore obtain a partition of unity after averaging over the translations of the tiling:
\begin{equation}\label{eq:resolution_identity}
\boxed{\dashint_{C_\ell}\Bigg(\sum_{\mathbf{z}\in\ZZ^3}\sum_{j=1}^{24} (\iden_{\ell T_j(1-\delta/\ell)\tetra}\ast\eta_\delta)(\mathbf{x}-\ell \mathbf{z}-\bm{\tau})+\Upsilon_{\ell,\delta}(\mathbf{x}-\bm{\tau})\Bigg)\,\dd\bm{\tau}=1}
\end{equation}
for almost every $\mathbf{x}\in\RR^3$. The skeleton function has a bounded hessian of order $\delta^{-2}$. Indeed, the tetrahedra $\ell T_j(1-\delta/(2\ell))\tetra+\ell\mathbf{z}$ are disjoint, hence $\Upsilon_{\ell,\delta}=\Lambda_{\delta/\ell}(1-\iden_S\ast\eta_{\delta/2})$ with $S$ their union, and $\|\mathbf{D}^2\Upsilon_{\ell,\delta}\|_{L^\infty}\le\Lambda_{\delta/\ell}\|\mathbf{D}^2\eta_{\delta/2}\|_{L^1}\le C\delta^{-2}$ by~\cref{eq:def_eta_delta}. Since $|\grad\sqrt{\Upsilon_{\ell,\delta}}|^2\le C\delta^{-2}$ by~\cref{thm:Glaeser} and since $\grad\Upsilon_{\ell,\delta}$ is supported in a set of volume $C\ell^2\delta$ per unit cell, we also have
\begin{equation}\label{eq:skeleton_bounds}
\dashint_{C_\ell}\Upsilon_{\ell,\delta}\le\frac{C\delta}{\ell},\qquad \dashint_{C_\ell}|\grad\sqrt{\Upsilon_{\ell,\delta}}|^2\le\frac{C}{\ell\delta}.
\end{equation}

We build a trial state on the partition of unity~\cref{eq:resolution_identity} and also average over the rotations and dilations of the tiling, as in~\cite{lewin2020local}. For the average over the dilation factor $t$ we use, here and henceforth, the weighted mean
$$
\dashint_a^bf(t)\,\frac{\dd t}{t^4}:=\left(\int_a^b\frac{\dd t}{t^4}\right)^{-1}\int_a^bf(t)\,\frac{\dd t}{t^4}
$$
for $0<a<b$. The construction gives the following decoupling upper bound, whose last term is the electrostatic error discussed above.

\begin{theorem}[Improved decoupling upper bound]\label{prop:upper_bound}
There exists a universal constant $C$ such that for any spin-polarized density $\bm{\rho}\ge0$ on $\RR^3$ with $\rho\in L^1(\RR^3)$ and $\sqrt{\bm{\rho}}\in H^1(\RR^3,\CC^{2\times2})$, any $0<\delta\le\ell/2$ and any $0<\alpha<1/2$,
\begin{multline}\label{eq:upper_bound_average}
 E(\bm{\rho})\le
  \dashint_{1-\alpha}^{1+\alpha}\frac{\dd t}{t^4}\int_{\SO(3)}\dd \mathbf{R}\,\dashint_{C_{t\ell}}\dd\bm{\tau}\sum_{\mathbf{z}\in\ZZ^3}\sum_{j=1}^{24} E\Big((\iden_{t\ell T_j(1-\delta/\ell)\tetra}\ast\eta_{t\delta})(\mathbf{R}\,\cdot\,-t\ell \mathbf{z}-\bm{\tau})\bm{\rho}\Big)\\
 +\frac{C\delta}{\ell}\int_{\RR^3}\Big(\rho^{5/3}+|\grad\sqrt{\bm{\rho}}|^2\Big)+\frac{C}{\ell\delta}\int_{\RR^3}\rho+C\frac{\delta^3}{\ell}\log(\alpha^{-1})\int_{\RR^3}\rho^2.
\end{multline}
\end{theorem}

The main new ingredient of the proof is an estimate on the Fourier coefficients of the periodic function which measures the defect of the partition of unity before the average over translations. For $0<s\le1/2$, let $f_s$ be the $\ZZ^3$-periodic function which is equal to $\sum_{j=1}^{24}\iden_{T_j(1-s)\tetra}\ast\eta_s+\Upsilon_{1,s}-1$ on the unit cube $C_1$.

\begin{lemma}[Fourier estimate for the skeleton]\label{lem:Fourier_skeleton}
There exists a universal constant $C$ such that for all $0<s\le1/2$
\begin{equation}\label{eq:Fourier_skeleton}
\sum_{\substack{\mathbf{k}\in 2\pi\ZZ^3\\ \mathbf{k}\neq0}}\frac{1}{|\mathbf{k}|^2}\left|\int_{C_1}f_{s}(\mathbf{z})e^{-i\mathbf{k}\cdot \mathbf{z}}\,\dd \mathbf{z}\right|^2\le Cs^3.
\end{equation}
\end{lemma}

\begin{proof}
Let $\mathbf{k}\in2\pi\ZZ^3\setminus\{0\}$. By~\cref{eq:def_eta_delta} the convolution with $\eta_{s}$ multiplies the Fourier coefficients by $(2\pi)^{3/2}\wh{\eta}_\star(bs\mathbf{k})$, whereas the convolution with $\eta_{s/2}$ multiplies them by $(2\pi)^{3/2}\wh{\eta}_\star(bs\mathbf{k}/2)$. Let $\lambda_j(\mathbf{x})$ be the infimum of the scales $\lambda>0$ for which $\mathbf{x}\in T_j(\lambda\tetra)$ and, for a bounded function $\psi$ on $[0,1]$,
$$
\GC_\psi(\mathbf{k}):=(2\pi)^{3/2}\sum_{j=1}^{24}\int_{\tetra_j}\psi\big(\lambda_j(\mathbf{x})\big)e^{-i\mathbf{k}\cdot \mathbf{x}}\,\dd \mathbf{x}.
$$
Since $\sum_j\iden_{\tetra_j}=1$ almost everywhere on $C_1$, the constants do not contribute to the Fourier coefficients and we obtain
\begin{equation}\label{eq:Fourier_two_terms}
\int_{C_1}f_{s}(\mathbf{z})e^{-i\mathbf{k}\cdot \mathbf{z}}\,\dd \mathbf{z}=\wh{\eta}_\star(bs\mathbf{k})\,\GC_\phi(\mathbf{k})+\Lambda_s\left(\wh{\eta}_\star\Big(\frac{bs\mathbf{k}}{2}\Big)-\wh{\eta}_\star(bs\mathbf{k})\right)\GC_{\phi_1}(\mathbf{k})
\end{equation}
with
$$
\phi(\lambda):=\Lambda_s\iden(\lambda\ge1-s/2)-\iden(\lambda\ge1-s),\qquad \phi_1(\lambda):=\iden(\lambda\ge1-s/2).
$$
The first term in~\cref{eq:Fourier_two_terms} would be the only one if the same regularizing function was used for the tiles and for the skeleton. The function $\phi$ is bounded, supported in the interval $[1-s,1]\subset[1/2,1]$ and it satisfies
\begin{equation}\label{eq:neutral_skeleton}
\int_0^1\phi(\lambda)\lambda^2\,\dd\lambda=0
\end{equation}
by definition of $\Lambda_s$. This is the local neutrality of the skeleton.

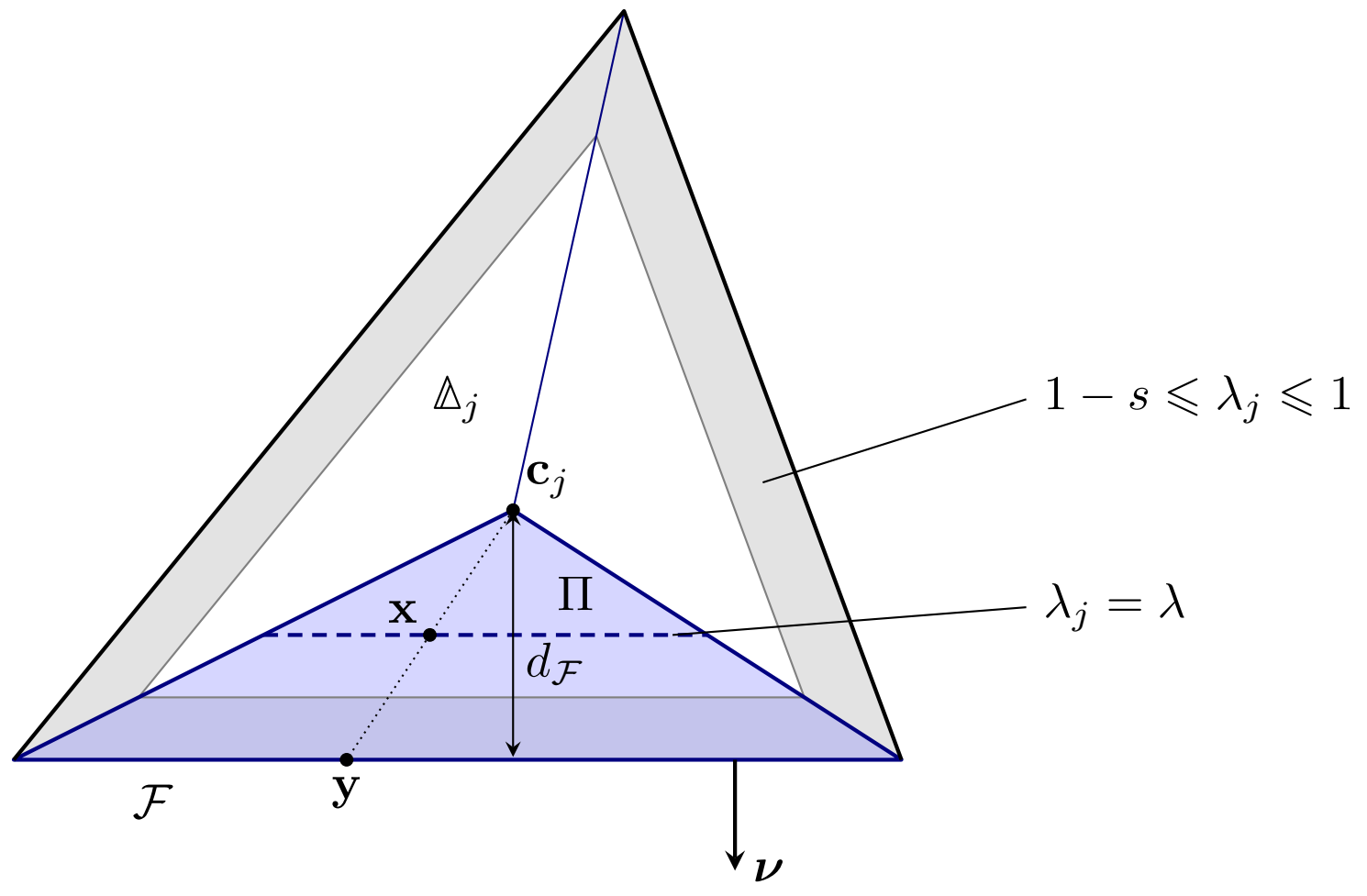
\begin{figure}[t]
\centering
\begin{tikzpicture}[>=stealth,line join=round]
\coordinate (P1) at (0,0); \coordinate (P2) at (6.4,0); \coordinate (P3) at (4.4,5.4);
\coordinate (c) at (3.6,1.8);
\coordinate (Q1) at (0.9,0.45); \coordinate (Q2) at (5.7,0.45); \coordinate (Q3) at (4.2,4.5);
\fill[gray!22,even odd rule] (P1)--(P2)--(P3)--cycle (Q1)--(Q2)--(Q3)--cycle;
\fill[blue!45,fill opacity=.35] (c)--(P1)--(P2)--cycle;
\draw[gray] (Q1)--(Q2)--(Q3)--cycle;
\draw[blue!50!black] (c)--(P3);
\draw[blue!50!black,thick] (P1)--(P2)--(c)--cycle;
\draw[thick] (P2)--(P3)--(P1);
\draw[blue!50!black,densely dashed,thick] (1.8,0.9)--(5.0,0.9);
\draw[densely dotted] (c)--(2.4,0);
\fill (2.4,0) circle[radius=1.4pt] node[below]{$\mathbf{y}$};
\fill (3.0,0.9) circle[radius=1.4pt] node[above left=-1pt]{$\mathbf{x}$};
\draw[<->] (3.6,1.8)--(3.6,0.02) node[pos=0.62,right=-1pt]{$d_{\mathcal{F}}$};
\draw[->,thick] (5.2,0)--(5.2,-0.8) node[right]{$\bm{\nu}$};
\fill (c) circle[radius=1.4pt] node[above right=-1pt]{$\mathbf{c}_j$};
\node at (1.0,-0.3) {$\mathcal{F}$};
\node at (4.05,1.2) {$\Pi$};
\node at (3.2,2.6) {$\tetra_j$};
\draw[thin] (4.75,0.9)--(7.3,1.1) node[right]{$\lambda_j=\lambda$};
\draw[thin] (5.4,2.0)--(7.3,2.6) node[right]{$1-s\le\lambda_j\le1$};
\end{tikzpicture}
\caption{Two-dimensional representation of the coordinates in the proof of \cref{lem:Fourier_skeleton}.}\label{fig:pyramid}
\end{figure}

Each tetrahedron $\tetra_j$ is the union of four pyramids, which have their apex at the center of mass $\mathbf{c}_j=-\mathbf{z}_j$ and one of the four faces as their base, see \cref{fig:pyramid}. Let $\Pi$ be one of these pyramids, with base the face $\mathcal{F}$, and let $d_{\mathcal{F}}\ge d_{\min}$ be the distance from $\mathbf{c}_j$ to the plane containing $\mathcal{F}$. In the coordinates $\mathbf{x}=\mathbf{c}_j+\lambda(\mathbf{y}-\mathbf{c}_j)$ with $\mathbf{y}\in \mathcal{F}$ and $0\le\lambda\le1$, we have $\lambda_j(\mathbf{x})=\lambda$ and $\dd \mathbf{x}=d_{\mathcal{F}}\lambda^2\dd\lambda\,\dd\Sigma(\mathbf{y})$, hence
\begin{equation}\label{eq:pyramid_coordinates}
\int_\Pi\psi\big(\lambda_j(\mathbf{x})\big)e^{-i\mathbf{k}\cdot \mathbf{x}}\,\dd \mathbf{x}=d_{\mathcal{F}}\,e^{-i\mathbf{k}\cdot \mathbf{c}_j}\int_0^1\psi(\lambda)\lambda^2g_{\mathcal{F},\mathbf{k}}(\lambda)\,\dd\lambda
\end{equation}
with
$$
g_{\mathcal{F},\mathbf{k}}(\lambda):=\int_{\mathcal{F}}e^{-i\lambda \mathbf{k}\cdot(\mathbf{y}-\mathbf{c}_j)}\dd\Sigma(\mathbf{y}).
$$
Let $\bm{\nu}$ be the unit vector normal to $\mathcal{F}$ and $\mathbf{k}_\parallel=\mathbf{k}-(\mathbf{k}\cdot \bm{\nu})\bm{\nu}$. We need three properties of the function $g_{\mathcal{F},\mathbf{k}}$, for $1/2\le\lambda\le1$. The first two are the following.

\begin{claim}
For all $\mathbf{k}\in2\pi\ZZ^3$ and $1/2\le\lambda\le1$,
\begin{equation}\label{eq:properties_g}
|g_{\mathcal{F},\mathbf{k}}'(\lambda)|^2\le C|\mathbf{k}\cdot \bm{\nu}|^2|g_{\mathcal{F},\mathbf{k}}(\lambda)|^2+C,\qquad |g_{\mathcal{F},\mathbf{k}}(\lambda)|^2\le\frac{C}{|\mathbf{k}_\parallel|^2}\quad\text{if }\mathbf{k}_\parallel\neq0,
\end{equation}
with $C=\max\{2,\,2(2|\partial \mathcal{F}|+4|\mathcal{F}|)^2,\,4|\partial \mathcal{F}|^2\}$.
\end{claim}

\begin{proof}
Since $(\mathbf{y}-\mathbf{c}_j)\cdot \bm{\nu}=d_{\mathcal{F}}$ for all $\mathbf{y}\in \mathcal{F}$, we have
$$
g_{\mathcal{F},\mathbf{k}}'(\lambda)=-id_{\mathcal{F}}(\mathbf{k}\cdot \bm{\nu})\,g_{\mathcal{F},\mathbf{k}}(\lambda)-i\int_{\mathcal{F}}\mathbf{k}_\parallel\cdot(\mathbf{y}-\mathbf{c}_j)\,e^{-i\lambda \mathbf{k}\cdot(\mathbf{y}-\mathbf{c}_j)}\dd\Sigma(\mathbf{y}).
$$
To estimate the last integral, we write $u(\mathbf{y}):=e^{-i\lambda \mathbf{k}\cdot(\mathbf{y}-\mathbf{c}_j)}$ for $\mathbf{y}$ in the plane containing $\mathcal{F}$, so that $\grad_{\mathcal{F}} u=-i\lambda\mathbf{k}_\parallel u$, where $\grad_{\mathcal{F}}$ is the gradient in this plane. Let $\bm{\nu}_{\partial \mathcal{F}}$ be the outward unit normal to $\partial \mathcal{F}$ in the plane, $\mathcal{H}^1$ the length measure on $\partial \mathcal{F}$ and $\mathbf{p}:=\mathbf{c}_j+d_{\mathcal{F}}\bm{\nu}$ the orthogonal projection of $\mathbf{c}_j$ onto the plane. Since $\mathbf{k}_\parallel\cdot(\mathbf{y}-\mathbf{c}_j)=\mathbf{k}_\parallel\cdot(\mathbf{y}-\mathbf{p})$ and the divergence of $\mathbf{y}-\mathbf{p}$ in the plane is equal to $2$, the divergence theorem gives
\begin{multline*}
\int_{\mathcal{F}}\mathbf{k}_\parallel\cdot(\mathbf{y}-\mathbf{c}_j)\,u\,\dd\Sigma=\frac{i}{\lambda}\int_{\mathcal{F}}(\mathbf{y}-\mathbf{p})\cdot\grad_{\mathcal{F}} u\,\dd\Sigma\\
=\frac{i}{\lambda}\left(\int_{\partial \mathcal{F}}(\mathbf{y}-\mathbf{p})\cdot\bm{\nu}_{\partial \mathcal{F}}\,u\,\dd\mathcal{H}^1-2\int_{\mathcal{F}} u\,\dd\Sigma\right).
\end{multline*}
Since $|\mathbf{y}-\mathbf{p}|\le|\mathbf{y}-\mathbf{c}_j|\le1$, its modulus is at most $2|\partial \mathcal{F}|+4|\mathcal{F}|$ for $\lambda\ge1/2$. Similarly, if $\mathbf{k}_\parallel\neq0$, then
$$
|\mathbf{k}_\parallel|\,g_{\mathcal{F},\mathbf{k}}(\lambda)=\frac{i}{\lambda}\int_{\mathcal{F}}\frac{\mathbf{k}_\parallel}{|\mathbf{k}_\parallel|}\cdot\grad_{\mathcal{F}} u\,\dd\Sigma=\frac{i}{\lambda}\int_{\partial \mathcal{F}}\frac{\mathbf{k}_\parallel}{|\mathbf{k}_\parallel|}\cdot\bm{\nu}_{\partial \mathcal{F}}\,u\,\dd\mathcal{H}^1,
$$
so that $|\mathbf{k}_\parallel|\,|g_{\mathcal{F},\mathbf{k}}(\lambda)|\le|\partial \mathcal{F}|/\lambda$, where $|\partial \mathcal{F}|$ is the perimeter of $\mathcal{F}$. Since $d_{\mathcal{F}}\le1$ and $\lambda\ge1/2$, this gives~\cref{eq:properties_g}.
\end{proof}

The constant $C$ only depends on the perimeter and the area of $\mathcal{F}$. It is independent of $\mathbf{k}$, of $\lambda\in[1/2,1]$ and of $s$, and it is universal since $\mathcal{F}$ is one of the four faces of the reference tetrahedron $\tetra$, up to a rigid motion. Next, the vertices of the tetrahedra $\tetra_j$ all belong to $(\ZZ/2)^3$, hence $\bm{\nu}$ is proportional to a vector of $\ZZ^3$. This implies that $\mathbf{k}\cdot \bm{\nu}$ takes its values in $\mu\ZZ$ for some $\mu>0$ and that, for every $m\in\ZZ$, the vectors $\mathbf{k}_\parallel$ of the $\mathbf{k}\in2\pi\ZZ^3$ such that $\mathbf{k}\cdot \bm{\nu}=\mu m$ form a translate $\mathbf{a}_m+\mathcal{K}$ of the two-dimensional lattice $\mathcal{K}:=\{\mathbf{k}\in2\pi\ZZ^3:\mathbf{k}\cdot\bm{\nu}=0\}$. Let $A$ be the area of a unit cell of the dual lattice $\mathcal{K}^*:=\{\mathbf{w}\perp\bm{\nu}:\ \mathbf{w}\cdot\mathbf{k}\in2\pi\ZZ\text{ for all }\mathbf{k}\in\mathcal{K}\}$, and let $N$ be the number of points of $\mathcal{K}^*$ in the closed unit disk. The third property of $g_{\mathcal{F},\mathbf{k}}$ is the following.

\begin{claim}
For all $m\in\ZZ$ and $\lambda\in[1/2,1]$,
\begin{equation}\label{eq:Parseval_g}
\sum_{\substack{\mathbf{k}\in2\pi\ZZ^3\\ \mathbf{k}\cdot \bm{\nu}=\mu m}}|g_{\mathcal{F},\mathbf{k}}(\lambda)|^2=\sum_{\mathbf{k}_\parallel\in\mathbf{a}_m+\mathcal{K}}\left|\int_{\mathcal{F}}e^{-i\lambda \mathbf{k}_\parallel\cdot \mathbf{y}}\dd\Sigma(\mathbf{y})\right|^2\le4AN|\mathcal{F}|.
\end{equation}
\end{claim}

\begin{proof}
The equality holds since $|g_{\mathcal{F},\mathbf{k}}(\lambda)|=|\int_{\mathcal{F}}e^{-i\lambda \mathbf{k}_\parallel\cdot \mathbf{y}}\dd\Sigma(\mathbf{y})|$ and $\mathbf{k}\mapsto\mathbf{k}_\parallel$ is one-to-one on each plane $\mathbf{k}\cdot\bm{\nu}=\mu m$. For the inequality, after translating $\mathcal{F}$ into the plane orthogonal to $\bm{\nu}$, we periodize the function $\iden_{\mathcal{F}}(\mathbf{y})e^{-i\lambda\mathbf{a}_m\cdot\mathbf{y}}$ over the lattice $\lambda^{-1}\mathcal{K}^*$. Since $\operatorname{diam}\mathcal{F}\le1$ and $\lambda\le1$, at most $N$ translates of $\mathcal{F}$ overlap, so that the $L^2$ norm squared of the periodized function on a unit cell is at most $N|\mathcal{F}|$. The functions $e^{-i\lambda\mathbf{k}\cdot\mathbf{y}}$ with $\mathbf{k}\in\mathcal{K}$ form an orthogonal basis on this cell, whose area is $\lambda^{-2}A\le4A$, and Parseval's identity gives~\cref{eq:Parseval_g}.
\end{proof}

The numbers $A$, $N$ and $|\mathcal{F}|$ only depend on the face $\mathcal{F}$ of $\tetra$, hence they are universal.

After these preparations, we estimate the contribution of the first term of~\cref{eq:Fourier_two_terms} to the left side of~\cref{eq:Fourier_skeleton}, that is,
$$
\sum_{\substack{\mathbf{k}\in 2\pi\ZZ^3\\ \mathbf{k}\neq0}}\frac{|\wh{\eta}_\star(bs\mathbf{k})|^2\,|\GC_\phi(\mathbf{k})|^2}{|\mathbf{k}|^2}.
$$
The function $\GC_\phi$ is a sum of integrals over the $96$ pyramids of the $24$ tetrahedra. By the Cauchy--Schwarz inequality it suffices to estimate the integral over one pyramid $\Pi$. By~\cref{eq:neutral_skeleton} we can replace $g_{\mathcal{F},\mathbf{k}}(\lambda)$ by $g_{\mathcal{F},\mathbf{k}}(\lambda)-g_{\mathcal{F},\mathbf{k}}(1)$ in~\cref{eq:pyramid_coordinates} for $\psi=\phi$. On the support $[1-s,1]$ of $\phi$, the Cauchy--Schwarz inequality gives $|g_{\mathcal{F},\mathbf{k}}(\lambda)-g_{\mathcal{F},\mathbf{k}}(1)|\le\int_{1-s}^1|g_{\mathcal{F},\mathbf{k}}'|\le s^{1/2}\big(\int_{1-s}^1|g_{\mathcal{F},\mathbf{k}}'|^2\big)^{1/2}$. Since $|\phi|\le\Lambda_s+1\le3$, $\lambda^2\le1$ and $d_{\mathcal{F}}\le1$, integrating over this interval of length $s$ yields
\begin{equation*}
\left|\int_\Pi\phi\big(\lambda_j(\mathbf{x})\big)e^{-i\mathbf{k}\cdot \mathbf{x}}\,\dd \mathbf{x}\right|^2\le 9s^3\int_{1-s}^1|g_{\mathcal{F},\mathbf{k}}'(\lambda)|^2\,\dd\lambda.
\end{equation*}
Since $\eta_\star\in C^\infty_c(\RR^3)$, its Fourier transform decays faster than any power, and since $b$ is fixed and $1+bs|\mathbf{k}|\ge b(1+s|\mathbf{k}|)$, we have $|\wh{\eta}_\star(bs\mathbf{k})|\le C(1+s|\mathbf{k}|)^{-6}$. Using~\cref{eq:properties_g}, \cref{eq:Parseval_g} and $|\mathbf{k}\cdot \bm{\nu}|\le|\mathbf{k}|$, we find
\begin{multline*}
\sum_{\substack{\mathbf{k}\in 2\pi\ZZ^3\\ \mathbf{k}\neq0}}\frac{|g_{\mathcal{F},\mathbf{k}}'(\lambda)|^2}{|\mathbf{k}|^2(1+s|\mathbf{k}|)^{12}}\\
\le C\sum_{m\in\ZZ}\frac{1}{(1+s\mu|m|)^{12}}+C\sum_{\substack{\mathbf{k}\in 2\pi\ZZ^3\\ \mathbf{k}\neq0}}\frac{1}{|\mathbf{k}|^2(1+s|\mathbf{k}|)^{12}}\le \frac{C}{s}.
\end{multline*}
The last inequality follows by comparing the two sums with $\int_0^\infty(1+sx)^{-12}\,\dd x=1/(11s)$, since there are at most $C(1+r^2)$ points $\mathbf{k}\in2\pi\ZZ^3$ with $r\le|\mathbf{k}|<r+1$. After summing over the $96$ pyramids, the contribution of the first term of~\cref{eq:Fourier_two_terms} to the left side of~\cref{eq:Fourier_skeleton} is therefore bounded by $Cs^3\times s\times s^{-1}=Cs^3$. Here $s^3$ comes from the bound on a single pyramid, the factor $s$ from the integration over $[1-s,1]$, and $s^{-1}$ from the sum over $\mathbf{k}$.

We now estimate the contribution of the second term of~\cref{eq:Fourier_two_terms} to the left side of~\cref{eq:Fourier_skeleton}, that is,
$$
\Lambda_s^2\sum_{\substack{\mathbf{k}\in 2\pi\ZZ^3\\ \mathbf{k}\neq0}}\frac{1}{|\mathbf{k}|^2}\left|\wh{\eta}_\star\Big(\frac{bs\mathbf{k}}{2}\Big)-\wh{\eta}_\star(bs\mathbf{k})\right|^2|\GC_{\phi_1}(\mathbf{k})|^2,
$$
where $\Lambda_s\le2$. As for the first term, it suffices to estimate the integral of $\phi_1(\lambda_j)e^{-i\mathbf{k}\cdot\mathbf{x}}$ over one pyramid $\Pi$. Here we do not have the neutrality, but we gain from the difference of the two regularizations. Since $\wh{\eta}_\star$ is smooth and even, we have $\grad\wh{\eta}_\star(0)=0$, hence $\wh{\eta}_\star(\mathbf{q}/2)-\wh{\eta}_\star(\mathbf{q})$ vanishes to second order at $\mathbf{q}=0$. Since $\wh{\eta}_\star$ also decays faster than any polynomial, we have $|\wh{\eta}_\star(\mathbf{q}/2)-\wh{\eta}_\star(\mathbf{q})|\le C|\mathbf{q}|^2(1+|\mathbf{q}|)^{-8}$ and therefore
$$
\left|\wh{\eta}_\star\Big(\frac{bs\mathbf{k}}{2}\Big)-\wh{\eta}_\star(bs\mathbf{k})\right|\le \frac{Cs^2|\mathbf{k}|^2}{(1+s|\mathbf{k}|)^{8}}.
$$
On the other hand, from~\cref{eq:pyramid_coordinates} with $\psi=\phi_1$ and the Cauchy--Schwarz inequality,
$$
\left|\int_\Pi\phi_1\big(\lambda_j(\mathbf{x})\big)e^{-i\mathbf{k}\cdot \mathbf{x}}\,\dd \mathbf{x}\right|^2\le Cs\int_{1-s/2}^1|g_{\mathcal{F},\mathbf{k}}(\lambda)|^2\,\dd\lambda.
$$
Writing $|\mathbf{k}|^2=|\mathbf{k}\cdot \bm{\nu}|^2+|\mathbf{k}_\parallel|^2$ and using again~\cref{eq:properties_g} and~\cref{eq:Parseval_g}, we obtain
$$
\sum_{\substack{\mathbf{k}\in 2\pi\ZZ^3\\ \mathbf{k}\neq0}}\frac{|\mathbf{k}|^2|g_{\mathcal{F},\mathbf{k}}(\lambda)|^2}{(1+s|\mathbf{k}|)^{16}}\le C\sum_{m\in\ZZ}\frac{\mu^2m^2}{(1+s\mu|m|)^{16}}+C\sum_{\mathbf{k}\in 2\pi\ZZ^3}\frac{1}{(1+s|\mathbf{k}|)^{16}}\le \frac{C}{s^3}.
$$
The last inequality follows in the same way, from $\int_0^\infty(1+x^2)(1+sx)^{-16}\,\dd x\le Cs^{-3}$. After summing over the $96$ pyramids, the contribution of the second term of~\cref{eq:Fourier_two_terms} to the left side of~\cref{eq:Fourier_skeleton} is thus bounded by $Cs^4\times s\times s\times s^{-3}=Cs^3$. Here $s^4$ is the square of the factor $s^2$ in the bound on the difference of the two regularizations, the two factors $s$ come from the Cauchy--Schwarz inequality and from the integration over $[1-s/2,1]$, and $s^{-3}$ comes from the sum over $\mathbf{k}$.
\end{proof}

\begin{proof}[Proof of \cref{prop:upper_bound}]
Using~\cref{eq:resolution_identity}, we write
\begin{equation*}
\bm{\rho}(\mathbf{x})=\dashint_{C_\ell}\Bigg(\sum_{\mathbf{z}\in\ZZ^3}\sum_{j=1}^{24} (\iden_{\ell T_j(1-\delta/\ell)\tetra}\ast\eta_\delta)(\mathbf{x}-\ell \mathbf{z}-\bm{\tau})+\Upsilon_{\ell,\delta}(\mathbf{x}-\bm{\tau})\Bigg)\bm{\rho}(\mathbf{x})\,\dd\bm{\tau}.
\end{equation*}
The matrices $(\iden_{\ell T_j(1-\delta/\ell)\tetra}\ast\eta_\delta)(\cdot-\ell \mathbf{z}-\bm{\tau})\bm{\rho}$ and $\Upsilon_{\ell,\delta}(\cdot-\bm{\tau})\bm{\rho}$ satisfy~\cref{eq:representable}, since $\sqrt{\chi \bm{\rho}}=\sqrt\chi\sqrt {\bm{\rho}}$ for a scalar function $\chi\ge0$. As in~\cite{csirik2026non}, for every fixed $\bm{\tau}\in C_\ell$ we pick states having these spin-polarized densities and which minimize the corresponding energies, one for each tile and one for the skeleton. They exist by \cref{thm:LL_attainment}. All these states have disjoint supports and their tensor product can be anti-symmetrized in the standard manner. The so-obtained state $\Gamma_{\bm{\tau}}$ has the spin-polarized density
$$
\bm{\rho}_{\Gamma_{\bm{\tau}}}=\Big(\sum_{\mathbf{z},j}(\iden_{\ell T_j(1-\delta/\ell)\tetra}\ast\eta_\delta)(\cdot-\ell \mathbf{z}-\bm{\tau})+\Upsilon_{\ell,\delta}(\cdot-\bm{\tau})\Big)\bm{\rho}
$$
and its energy is the sum of the energies of the pieces, plus the direct terms between them. Hence
$$
E(\bm{\rho}_{\Gamma_{\bm{\tau}}})\le\sum_{\mathbf{z}\in\ZZ^3}\sum_{j=1}^{24} E\big((\iden_{\ell T_j(1-\delta/\ell)\tetra}\ast\eta_\delta)(\cdot-\ell \mathbf{z}-\bm{\tau})\bm{\rho}\big)+E\big(\Upsilon_{\ell,\delta}(\cdot-\bm{\tau})\bm{\rho}\big).
$$
We do not average the states $\Gamma_{\bm{\tau}}$, which would require choosing them measurably in $\bm{\tau}$. Instead, we use that $F_\LL$ is convex and lower semicontinuous on $L^1(\RR^3)$~\cite{lieb1983density}, hence the supremum of continuous affine functions. Lower semicontinuity also makes the local energies, here and below, measurable in the translation, rotation and dilation parameters of the tilings. Since $\bm{\tau}\mapsto\bm{\rho}_{\Gamma_{\bm{\tau}}}$ is continuous in $L^1$ and $\dashint_{C_\ell}\bm{\rho}_{\Gamma_{\bm{\tau}}}\dd\bm{\tau}=\bm{\rho}$, this gives
$$
F_\LL(\bm{\rho})\le\dashint_{C_\ell}F_\LL(\bm{\rho}_{\Gamma_{\bm{\tau}}})\,\dd\bm{\tau}.
$$
In addition, $\dashint_{C_\ell}D(\rho_{\Gamma_{\bm{\tau}}})\,\dd\bm{\tau}-D(\rho)=\dashint_{C_\ell}D(\rho_{\Gamma_{\bm{\tau}}}-\rho)\,\dd\bm{\tau}$. Combining these, we obtain, see also~\cite[Eq. (37)]{lewin2020local},
\begin{multline*}
E(\bm{\rho})\le\dashint_{C_\ell}\Bigg\{\sum_{\mathbf{z}\in\ZZ^3}\sum_{j=1}^{24} E\big((\iden_{\ell T_j(1-\delta/\ell)\tetra}\ast\eta_\delta)(\cdot-\ell \mathbf{z}-\bm{\tau})\bm{\rho}\big)+E\big(\Upsilon_{\ell,\delta}(\cdot-\bm{\tau})\bm{\rho}\big)\Bigg\}\,\dd\bm{\tau}\\
+\dashint_{C_\ell}D(\rho_{\Gamma_{\bm{\tau}}}-\rho)\,\dd\bm{\tau},
\end{multline*}
The error term only depends on the total density $\rho$. Let $s:=\delta/\ell\le1/2$ and let $f_s$ be the $\ZZ^3$-periodic function defined before \cref{lem:Fourier_skeleton}. It has a vanishing average. Exactly as in~\cite[Eq. (41)]{lewin2020local} we find
\begin{equation*}
\dashint_{C_\ell}D(\rho_{\Gamma_{\bm{\tau}}}-\rho)\,\dd\bm{\tau}
=2\pi\sum_{\substack{\mathbf{k}\in 2\pi\ZZ^3\\ \mathbf{k}\neq0}} \left|\int_{C_1}f_{s}(\mathbf{z})e^{-i\mathbf{k}\cdot \mathbf{z}}\,\dd \mathbf{z}\right|^2\int_{\RR^3}\frac{|\wh{\rho}(\mathbf{p})|^2}{|\mathbf{p}-\mathbf{k}/\ell|^2}\,\dd \mathbf{p}.
\end{equation*}
We then average over the rotations $\mathbf{R}\in \SO(3)$ and the dilations $t\in(1-\alpha,1+\alpha)$ of the tiling, using that $E(\bm{\rho}(\mathbf{R}^{-1}\cdot))=E(\bm{\rho})$ by~\cref{eq:invariance} and that $s$ is independent of $t$. By~\cite[Lemma~14]{lewin2020local} this allows us to replace $|\mathbf{p}-\mathbf{k}/\ell|^{-2}$ by $C\log(\alpha^{-1})\ell^2/|\mathbf{k}|^2$ and we obtain, as in~\cite[Eq. (43)]{lewin2020local}, that the averaged error is bounded by
\begin{equation}\label{eq:very_final_computation_error_term_upper_bound}
C\log(\alpha^{-1})\,\ell^2\Bigg(\sum_{\substack{\mathbf{k}\in 2\pi\ZZ^3\\ \mathbf{k}\neq0}}\frac{1}{|\mathbf{k}|^2}\left|\int_{C_1}f_{s}(\mathbf{z})e^{-i\mathbf{k}\cdot \mathbf{z}}\,\dd \mathbf{z}\right|^2\Bigg)\int_{\RR^3}\rho^2.
\end{equation}
By \cref{lem:Fourier_skeleton}, the error~\cref{eq:very_final_computation_error_term_upper_bound} is bounded by $C\log(\alpha^{-1})\ell^2s^3\int\rho^2$, which is the last term of~\cref{eq:upper_bound_average}.

It remains to bound the energy of the skeleton. By~\cref{cor:upper_bound_E_rho} with $\epsilon=1$ and~\cref{eq:compare_trace_power}, we have for every scalar function $0\le\Upsilon\le2$
$$
E(\Upsilon \bm{\rho})\le C\int_{\RR^3}\rho^{5/3}\Upsilon+C\int_{\RR^3}|\grad\sqrt{\bm{\rho}}|^2\Upsilon+C\int_{\RR^3}\rho|\grad\sqrt\Upsilon|^2.
$$
After averaging over the translations, the rotations and the dilations of the tiling, \cref{eq:skeleton_bounds} shows that the skeleton contributes at most the second and third terms of~\cref{eq:upper_bound_average}.
\end{proof}

\subsection{Spatial decoupling lower bound}\label{sec:lower_bound}
For the lower bound we use the smooth partition of unity~\cref{eq:def_xi_ell_tiling}, which has no holes.

\begin{theorem}[Decoupling lower bound]\label{prop:lower_bound}
There exists a universal constant $C$ such that for any spin-polarized density $\bm{\rho}\ge0$ on $\RR^3$ with $\rho\in L^1(\RR^3)$ and $\sqrt{\bm{\rho}}\in H^1(\RR^3,\CC^{2\times2})$ and any $\delta>0$ with $0<\delta/\ell<1/C$, we have
\begin{multline}\label{eq:lower_bound_average}
E(\bm{\rho})\ge \Big(1-\frac{C\delta}{\ell}\Big)\sum_{\mathbf{z}\in\ZZ^3}\sum_{j=1}^{24}\int_{\SO(3)}\dd\mathbf{R}\,\dashint_{C_\ell}\dd\bm{\tau}\,E\Big((\iden_{\ell\tetra_j}\ast\eta_\delta)(\mathbf{R}\,\cdot\,-\ell \mathbf{z}-\bm{\tau})\bm{\rho}\Big)\\
-\frac{C}{\ell}\int_{\RR^3}\Big(\left(1+\delta^{-1}\right)\rho+\delta^3\rho^2\Big).
\end{multline}
\end{theorem}

\begin{proof}
For a state $\Gamma=\bigoplus_{n\ge0}\Gamma_n$ on Fock space we use the Coulomb and kinetic energies $\Cc(\Gamma)$ and $\TC(\Gamma)$ of~\cref{sec:known_bounds}, and we set $\EC(\Gamma):=\TC(\Gamma)+\Cc(\Gamma)-D(\rho_\Gamma)$. The smeared Graf--Schenker inequality of \cref{thm:GS} and the IMS formula of \cref{thm:IMS} give, as in~\cite[Eq. (54)]{lewin2020local},
\begin{multline*}
\EC(\Gamma)\ge\Big(1-\frac{C\delta}{\ell}\Big)\sum_{\mathbf{z}\in\ZZ^3}\sum_{j=1}^{24}\int_{\SO(3)}\dd\mathbf{R}\,\dashint_{C_\ell}\dd\bm{\tau}\,\EC\left(\Gamma_{|\sqrt{(\iden_{\ell\tetra_j}\ast\eta_\delta)(\mathbf{R}\,\cdot\,-\ell \mathbf{z}-\bm{\tau})}}\right)\\
-\frac{C}{\ell}\left(1+\frac{1}{\delta}\right)\int_{\RR^3}\rho_\Gamma-\frac{C\delta^3}{\ell}\int_{\RR^3}(\rho_\Gamma)^2.
\end{multline*}
Here $\Gamma_{|f}$ is the localized state of~\cref{sec:known_bounds}. This inequality is valid for any state on the fermionic Fock space, independently of the spin. For a real function $f$, the one-particle density matrix of $\Gamma_{|f}$ is $f(\mathbf{x})\bm{\gamma}_\Gamma f(\mathbf{x})$, hence its spin-polarized density is $f^2\bm{\rho}_\Gamma$. Therefore $\EC(\Gamma_{|\sqrt\xi})\ge E(\xi \bm{\rho}_\Gamma)$ and~\cref{eq:lower_bound_average} follows after minimizing over $\Gamma$.
\end{proof}

\subsection{Local spin rotations}\label{sec:deviation}
This section collects the tools that we use to compare a spin-polarized density with a constant matrix or with a spin-spiral wave on a small region. We first collect elementary estimates for functions of Hermitian matrices. The main result is \cref{lem:gauge}, which bounds the energy cost of a position-dependent rotation $\bm{\rho}\mapsto\mathbf{U}\bm{\rho}\,\mathbf{U}^*$ of the spin. Since the internal Hamiltonian does not depend on the spin, such a rotation changes neither the total density nor the Coulomb energy, and only the kinetic energy is affected, by an amount controlled by the gradient of $\mathbf{U}$. Spin-spiral waves are obtained in this way from constant matrices with $\mathbf{U}=\mathbf{U}_{\mathbf{q}}$, and for them the cost is computed exactly in \cref{sec:proof_spirals}.

Let $f$ be a continuous function and $\mathbf{A},\mathbf{B}$ be two Hermitian matrices, with eigenvalues $a_i,b_k$ and corresponding orthonormal eigenvectors $\mathbf{u}_i,\mathbf{v}_k$. Then we have
\begin{equation}\label{eq:HS_functions}
|f(\mathbf{A})-f(\mathbf{B})|^2=\sum_{i,k=1}^2|f(a_i)-f(b_k)|^2|\pscal{\mathbf{u}_i,\mathbf{v}_k}|^2.
\end{equation}
Hence $|f(\mathbf{A})-f(\mathbf{B})|\le L|\mathbf{A}-\mathbf{B}|$ if $f$ is $L$-Lipschitz on an interval containing the eigenvalues of $\mathbf{A}$ and $\mathbf{B}$. Applying this to $f(s)=s^{1/\theta}$, $\mathbf{A}=\bm{\rho}(\mathbf{x})^\theta$ and $\mathbf{B}=\bm{\rho}(\mathbf{y})^\theta$, we find
\begin{equation}\label{eq:Lipschitz_power}
|\bm{\rho}(\mathbf{x})-\bm{\rho}(\mathbf{y})|\le \frac{\max\big(\rho(\mathbf{x}),\rho(\mathbf{y})\big)^{1-\theta}}{\theta}\,\big|\bm{\rho}(\mathbf{x})^\theta-\bm{\rho}(\mathbf{y})^\theta\big|
\end{equation}
for every non-negative matrix field $\bm{\rho}$ and $0<\theta<1$. For a scalar function $f\ge0$ and a spin-polarized density $\bm{\rho}$ with total density $\rho$, differentiation of $\sqrt{f\bm{\rho}}=\sqrt f\sqrt{\bm{\rho}}$ yields
\begin{equation}\label{eq:gradient_product}
|\grad\sqrt{f\bm{\rho}}|^2\le 2\rho|\grad\sqrt f|^2+2f|\grad\sqrt{\bm{\rho}}|^2 .
\end{equation}

The next lemma quantifies the cost of a local rotation of the spin. For a function $\mathbf{U}$ with values in $\SU(2)$ we use the notation
$$
\norm{\grad \mathbf{U}(\mathbf{x})}^2:=\bigg\|\sum_{j=1}^3\partial_{x_j}\mathbf{U}(\mathbf{x})^*\partial_{x_j}\mathbf{U}(\mathbf{x})\bigg\|.
$$
This is the squared operator norm of the map $\mathbf{v}\mapsto\big(\partial_{x_j}\mathbf{U}(\mathbf{x})\mathbf{v}\big)_{j=1,2,3}$ from $\CC^2$ to $(\CC^2)^3$. For $\mathbf{U}_{\mathbf{q}}$ it equals $|\mathbf{q}|^2/4$.

\begin{lemma}[Local rotations of the spin]\label{lem:gauge}
Let $\bm{\rho}\ge0$ be a spin-polarized density on $\RR^3$ with $\rho\in L^1(\RR^3)$ and $\sqrt{\bm{\rho}}\in H^1(\RR^3,\CC^{2\times2})$, and let $\mathbf{U}:\RR^3\to \SU(2)$ be a measurable function which is weakly differentiable in a neighborhood of the support of $\bm{\rho}$, such that $\int_{\RR^3}\rho\norm{\grad \mathbf{U}}^2<\infty$. Then $\mathbf{U}\bm{\rho}\,\mathbf{U}^*$ has the same properties and
\begin{equation*}
E\big(\mathbf{U}\bm{\rho}\,\mathbf{U}^*\big)\le (1+\epsilon)E(\bm{\rho})+\epsilon\, c_\LO\int_{\RR^3}\rho^{4/3}+\left(1+\frac1\epsilon\right)\int_{\RR^3}\rho\norm{\grad \mathbf{U}}^2
\end{equation*}
for every $\epsilon>0$.
\end{lemma}

\begin{proof}
Let $\Gamma$ be a state with $\bm{\rho}_\Gamma=\bm{\rho}$ and one-particle density matrix $\bm{\gamma}=\sum_kn_k|\bm{\Phi}_k\rangle\langle\bm{\Phi}_k|$. Let $\Gamma^{\mathbf{U}}=\{\mathbf{U}^{\otimes n}\Gamma_n(\mathbf{U}^*)^{\otimes n}\}_{n\ge0}$, where $\mathbf{U}$ is the unitary operator $(\mathbf{U}\bm{\Phi})(\mathbf{x})=\mathbf{U}(\mathbf{x})\bm{\Phi}(\mathbf{x})$ on $L^2(\RR^3,\CC^2)$. Then $\bm{\rho}_{\Gamma^{\mathbf{U}}}=\mathbf{U}\bm{\rho}\,\mathbf{U}^*$, $\rho_{\Gamma^{\mathbf{U}}}=\rho_\Gamma$ and $\Cc(\Gamma^{\mathbf{U}})=\Cc(\Gamma)$ since the Coulomb interaction is a multiplication operator which does not depend on the spin. The one-particle density matrix of $\Gamma^{\mathbf{U}}$ is $\mathbf{U}\bm{\gamma} \mathbf{U}^*$. From the pointwise inequality
$$
|\grad(\mathbf{U}\bm{\Phi}_k)|^2=|\mathbf{U}\grad\bm{\Phi}_k+(\grad \mathbf{U})\bm{\Phi}_k|^2\le(1+\epsilon)|\grad\bm{\Phi}_k|^2+\left(1+\frac1\epsilon\right)\sum_{j=1}^3|(\partial_{x_j}\mathbf{U})\bm{\Phi}_k|^2
$$
and $\sum_kn_k\sum_j|(\partial_{x_j}\mathbf{U})\bm{\Phi}_k|^2=\sum_j\Trs\big(\partial_{x_j}\mathbf{U}^*\partial_{x_j}\mathbf{U}\,\bm{\rho}\big)\le\rho\norm{\grad \mathbf{U}}^2$, we obtain
$$
\TC(\Gamma^{\mathbf{U}})\le(1+\epsilon)\TC(\Gamma)+\left(1+\frac1\epsilon\right)\int_{\RR^3}\rho_\Gamma\norm{\grad \mathbf{U}}^2.
$$
Therefore
$$
\EC(\Gamma^{\mathbf{U}})\le (1+\epsilon)\EC(\Gamma)-\epsilon\big(\Cc(\Gamma)-D(\rho_\Gamma)\big)+\left(1+\frac1\epsilon\right)\int_{\RR^3}\rho_\Gamma\norm{\grad \mathbf{U}}^2
$$
and the result follows from the Lieb--Oxford inequality of \cref{thm:LO}, after minimizing over $\Gamma$.
\end{proof}

\subsection{Energy of spin-spiral waves}\label{sec:proof_spirals}
For a unit vector $\mathbf{n}$ write $\mathbf{P}_{\mathbf{n}}=(\iden+\mathbf{n}\cdot\bm{\sigma})/2$. The energy of a fully polarized spin-spiral wave is given in~(iii) below.

\begin{theorem}[Energy of spin-spiral waves]\label{thm:spirals}
Let $\bm{\rho}\ge0$, $\bm{\rho}\neq0$, be a spin-polarized density on $\RR^3$ with total density $\rho\in L^1(\RR^3)$, magnetization $\mathbf{m}=(m_1,m_2,m_3)$ and $\sqrt{\bm{\rho}}\in H^1(\RR^3,\CC^{2\times2})$, and let $\mathbf{U}_{\mathbf{q}}(\mathbf{x})=e^{-\frac{i}2(\mathbf{q}\cdot\mathbf{x})\bm{\sigma}_3}$.
\begin{itemize}
\item[(i)] \emph{(Concavity)}. The function
$$
\mathbf{q}\in\RR^3\mapsto E\big(\mathbf{U}_{\mathbf{q}}\bm{\rho}\,\mathbf{U}_{\mathbf{q}}^*\big)-\frac{|\mathbf{q}|^2}{4}\int_{\RR^3}\rho
$$
is concave.
\item[(ii)] \emph{(Lower and upper bounds)}. For every $\mathbf{q}\in\RR^3$ we have
\begin{equation}\label{eq:spiral_lower}
E\begin{pmatrix}\rho^{\spu\spu}&0\\0&\rho^{\spd\spd}\end{pmatrix}\le E\big(\mathbf{U}_{\mathbf{q}}\bm{\rho}\,\mathbf{U}_{\mathbf{q}}^*\big).
\end{equation}
If in addition $\mathbf{m}(\mathbf{x})$ belongs to a fixed plane containing $\mathbf{e}_3$ for almost every $\mathbf{x}$, for instance if $\bm{\rho}=\bm{\rho}_0f$ with a constant matrix $\bm{\rho}_0$ and a scalar function $f$, then
\begin{equation}\label{eq:spiral_upper}
E\big(\mathbf{U}_{\mathbf{q}}\bm{\rho}\,\mathbf{U}_{\mathbf{q}}^*\big)\le E(\bm{\rho})+\frac{|\mathbf{q}|^2}{4}\left(\int_{\RR^3}\rho-\frac{\big(\int_{\RR^3}m_3\big)^2}{\int_{\RR^3}\rho}\right).
\end{equation}
\item[(iii)] \emph{(Full polarization)}. If $\bm{\rho}=\rho\,\mathbf{P}_{\mathbf{n}_0}$ with a constant unit vector $\mathbf{n}_0\in\RR^3$, where $\mathbf{P}_{\mathbf{n}_0}=(\iden+\mathbf{n}_0\cdot\bm{\sigma})/2$, then
\begin{equation}\label{eq:spiral_exact}
E\big(\mathbf{U}_{\mathbf{q}}\bm{\rho}\,\mathbf{U}_{\mathbf{q}}^*\big)= E(\bm{\rho})+\frac{|\mathbf{q}|^2}{4}\Big(1-(\mathbf{n}_0\cdot\mathbf{e}_3)^2\Big)\int_{\RR^3}\rho
\end{equation}
for every $\mathbf{q}\in\RR^3$.
\end{itemize}
\end{theorem}

In (iii) the matrix $\mathbf{U}_{\mathbf{q}}\bm{\rho}\,\mathbf{U}_{\mathbf{q}}^*$ is equal to $\rho\,\mathbf{P}_{\mathbf{n}}$, where $\mathbf{n}(\mathbf{x})$ is the rotation of $\mathbf{n}_0$ by the angle $\mathbf{q}\cdot\mathbf{x}$ about $\mathbf{e}_3$. Since $|\grad\mathbf{n}|^2=|\mathbf{q}|^2(1-(\mathbf{n}_0\cdot\mathbf{e}_3)^2)$, the equality~\cref{eq:spiral_exact} reads
$$
E\big(\rho\,\mathbf{P}_{\mathbf{n}}\big)=E\big(\rho\,\mathbf{P}_{\mathbf{n}_0}\big)+\frac14\int_{\RR^3}\rho\,|\grad\mathbf{n}|^2.
$$
For a fully polarized spin-spiral wave, the rotation of the magnetization adds exactly $\frac14\int\rho|\grad\mathbf{n}|^2$ to the error of the LSDA. By~\cref{eq:formula_gradient_sqrt_R} this is one half of the contribution of the rotation to $\int|\grad\sqrt{\bm{\rho}}|^2$, in agreement with the optimal constant of \cref{lem:HO}. The upper bound~\cref{eq:spiral_upper} is an equality in this case. On the contrary it is far from optimal at low polarization, since $\mathbf{U}_{\mathbf{q}}\bm{\rho}\,\mathbf{U}_{\mathbf{q}}^*=\bm{\rho}$ for an unpolarized $\bm{\rho}$.

\begin{proof}
For a state $\Gamma$ with a finite kinetic energy and the one-particle density matrix $\bm{\gamma}=\sum_kn_k|\bm{\Phi}_k\rangle\langle\bm{\Phi}_k|$, we call
$$
\mathbf{p}_\Gamma:=\sum_kn_k\pscal{\bm{\Phi}_k,-i\grad\bm{\Phi}_k}\in\RR^3,\qquad \mathbf{j}_\Gamma:=\sum_kn_k\pscal{\bm{\Phi}_k,\bm{\sigma}_3(-i\grad)\bm{\Phi}_k}\in\RR^3
$$
its total momentum and the total current of the third component of its spin. Both are linear in $\Gamma$. Let $\mathbf{k},\mathbf{q}\in\RR^3$ and let $\mathbf{V}$ be the unitary operator on $L^2(\RR^3,\CC^2)$ defined by $(\mathbf{V}\bm{\Phi})(\mathbf{x})=e^{i\mathbf{k}\cdot\mathbf{x}}\mathbf{U}_{\mathbf{q}}(\mathbf{x})\bm{\Phi}(\mathbf{x})$. As in the proof of \cref{lem:gauge}, the state $\Gamma^{\mathbf{V}}=\{\mathbf{V}^{\otimes n}\Gamma_n(\mathbf{V}^*)^{\otimes n}\}_{n\ge0}$ satisfies $\bm{\rho}_{\Gamma^{\mathbf{V}}}=\mathbf{U}_{\mathbf{q}}\bm{\rho}_\Gamma\mathbf{U}_{\mathbf{q}}^*$ and $\Cc(\Gamma^{\mathbf{V}})=\Cc(\Gamma)$. Since $\grad(\mathbf{V}\bm{\Phi})=\mathbf{V}\big(\grad+i\mathbf{k}-\frac i2\mathbf{q}\,\bm{\sigma}_3\big)\bm{\Phi}$, expanding the square gives
\begin{equation}\label{eq:kinetic_spiral}
\TC(\Gamma^{\mathbf{V}})=\TC(\Gamma)+2\mathbf{k}\cdot\mathbf{p}_\Gamma-\mathbf{q}\cdot\mathbf{j}_\Gamma+\Big(|\mathbf{k}|^2+\frac{|\mathbf{q}|^2}4\Big)\int_{\RR^3}\rho_\Gamma-\mathbf{k}\cdot\mathbf{q}\int_{\RR^3}m_3,
\end{equation}
where $m_3=\Trs(\bm{\sigma}_3\bm{\rho}_\Gamma)=\rho_\Gamma^{\spu\spu}-\rho_\Gamma^{\spd\spd}$. We also have $\mathbf{p}_{\Gamma^{\mathbf{V}}}=\mathbf{p}_\Gamma+\mathbf{k}\int\rho_\Gamma$ when $\mathbf{q}=0$.

\medskip

\noindent\textbf{Part (i).} We take $\mathbf{k}=0$. The map $\Gamma\mapsto\Gamma^{\mathbf{V}}$ is a bijection from the states with spin-polarized density $\bm{\rho}$ onto the states with spin-polarized density $\mathbf{U}_{\mathbf{q}}\bm{\rho}\,\mathbf{U}_{\mathbf{q}}^*$, which preserves the total density and the finiteness of the kinetic energy. From~\cref{eq:kinetic_spiral} we thus obtain
\begin{equation}\label{eq:spiral_formula}
E\big(\mathbf{U}_{\mathbf{q}}\bm{\rho}\,\mathbf{U}_{\mathbf{q}}^*\big)-\frac{|\mathbf{q}|^2}{4}\int_{\RR^3}\rho=\inf_{\bm{\rho}_\Gamma=\bm{\rho}}\Big\{\EC(\Gamma)-\mathbf{q}\cdot\mathbf{j}_\Gamma\Big\},
\end{equation}
where the infimum is taken over the states of finite kinetic energy. The right side is an infimum of affine functions of $\mathbf{q}$, hence it is concave.

\medskip

\noindent\textbf{Part (ii).} We start with the lower bound~\cref{eq:spiral_lower}. Let $\Gamma$ be a state with $\bm{\rho}_\Gamma=\mathbf{U}_{\mathbf{q}}\bm{\rho}\,\mathbf{U}_{\mathbf{q}}^*$ and let $\mathbf{W}=\diag(i,-i)$. As in \cref{rmk:collinear}, the state $\frac12(\Gamma+\GC_{\mathbf{W}}\Gamma \GC_{\mathbf{W}}^*)$ has the same energy as $\Gamma$ and its spin-polarized density is the diagonal part of $\mathbf{U}_{\mathbf{q}}\bm{\rho}\,\mathbf{U}_{\mathbf{q}}^*$. Since $\mathbf{U}_{\mathbf{q}}$ is diagonal, this is $\diag(\rho^{\spu\spu},\rho^{\spd\spd})$, which gives~\cref{eq:spiral_lower} after minimizing over $\Gamma$.

For the upper bound~\cref{eq:spiral_upper}, we first assume that $m_2\equiv0$, that is, the entries of $\bm{\rho}$ are real. Let $\Gamma$ be a state with $\bm{\rho}_\Gamma=\bm{\rho}$ and let $\ol\Gamma$ be the state obtained by taking the complex conjugate of the kernels of the $\Gamma_n$. Since the Hamiltonian is real, we have $\EC(\ol\Gamma)=\EC(\Gamma)$. The spin-polarized density of $\ol\Gamma$ is the complex conjugate of $\bm{\rho}$, hence it is equal to $\bm{\rho}$, whereas $\mathbf{p}_{\ol\Gamma}=-\mathbf{p}_\Gamma$ and $\mathbf{j}_{\ol\Gamma}=-\mathbf{j}_\Gamma$. The state $\Gamma_{\rm s}:=\frac12(\Gamma+\ol\Gamma)$ therefore satisfies $\bm{\rho}_{\Gamma_{\rm s}}=\bm{\rho}$, $\EC(\Gamma_{\rm s})=\EC(\Gamma)$ and $\mathbf{p}_{\Gamma_{\rm s}}=\mathbf{j}_{\Gamma_{\rm s}}=0$. Using $(\Gamma_{\rm s})^{\mathbf{V}}$ as a trial state for $\mathbf{U}_{\mathbf{q}}\bm{\rho}\,\mathbf{U}_{\mathbf{q}}^*$, we deduce from~\cref{eq:kinetic_spiral} that
$$
E\big(\mathbf{U}_{\mathbf{q}}\bm{\rho}\,\mathbf{U}_{\mathbf{q}}^*\big)\le \EC(\Gamma)+\Big(|\mathbf{k}|^2+\frac{|\mathbf{q}|^2}4\Big)\int_{\RR^3}\rho-\mathbf{k}\cdot\mathbf{q}\int_{\RR^3}m_3
$$
for all $\mathbf{k}\in\RR^3$. We minimize over $\Gamma$ and choose $\mathbf{k}=\mathbf{q}\int m_3/(2\int\rho)$, which gives~\cref{eq:spiral_upper}. If $\mathbf{m}$ belongs to a fixed plane containing $\mathbf{e}_3$, there is a constant rotation $\mathbf{W}_0=e^{-\frac i2\varphi_0\bm{\sigma}_3}$ about $\mathbf{e}_3$ such that $\mathbf{W}_0\bm{\rho}\,\mathbf{W}_0^*$ has real entries. This rotation commutes with $\mathbf{U}_{\mathbf{q}}$, it does not change $m_3$ and it does not change the energy by~\cref{eq:invariance}. Hence~\cref{eq:spiral_upper} holds in this case as well.

\medskip

\noindent\textbf{Part (iii).} Let $\mathbf{u}_0\in\CC^2$ be a unit vector such that $\mathbf{P}_{\mathbf{n}_0}\mathbf{u}_0=\mathbf{u}_0$ and let $c:=\mathbf{n}_0\cdot\mathbf{e}_3=\pscal{\mathbf{u}_0,\bm{\sigma}_3\mathbf{u}_0}$. Let $\Gamma$ be a state with $\bm{\rho}_\Gamma=\rho\,\mathbf{P}_{\mathbf{n}_0}$ and one-particle density matrix $\bm{\gamma}$. Then $\Tr\big((\iden-\mathbf{P}_{\mathbf{n}_0})\bm{\gamma}\big)=\int\Trs\big((\iden-\mathbf{P}_{\mathbf{n}_0})\bm{\rho}_\Gamma\big)=0$, where $\mathbf{P}_{\mathbf{n}_0}$ acts on the spin variable. Since $\bm{\gamma}\ge0$, this implies $(\iden-\mathbf{P}_{\mathbf{n}_0})\bm{\gamma}=0$, hence the eigenfunctions of $\bm{\gamma}$ are of the form $\bm{\Phi}_k=\phi_k\mathbf{u}_0$ with $\phi_k\in L^2(\RR^3,\CC)$. We deduce that
$$
\mathbf{j}_\Gamma=c\,\mathbf{p}_\Gamma
$$
for all such states. By~\cref{eq:spiral_formula}, we have
$$
E\big(\mathbf{U}_{\mathbf{q}}\bm{\rho}\,\mathbf{U}_{\mathbf{q}}^*\big)-\frac{|\mathbf{q}|^2}{4}\int_{\RR^3}\rho=\inf_{\bm{\rho}_\Gamma=\bm{\rho}}\Big\{\EC(\Gamma)-c\,\mathbf{q}\cdot\mathbf{p}_\Gamma\Big\}.
$$
The term involving $\mathbf{p}_\Gamma$ is removed by a Galilean transformation. We apply~\cref{eq:kinetic_spiral} with $\mathbf{q}$ replaced by $0$ and $\mathbf{k}=c\,\mathbf{q}/2$, in which case $\Gamma\mapsto\Gamma^{\mathbf{V}}$ is a bijection of the set of the states with spin-polarized density $\bm{\rho}$. We find
$$
\EC(\Gamma^{\mathbf{V}})-c\,\mathbf{q}\cdot\mathbf{p}_{\Gamma^{\mathbf{V}}}=\EC(\Gamma)+(2\mathbf{k}-c\,\mathbf{q})\cdot\mathbf{p}_\Gamma+\big(|\mathbf{k}|^2-c\,\mathbf{q}\cdot\mathbf{k}\big)\int_{\RR^3}\rho=\EC(\Gamma)-\frac{c^2|\mathbf{q}|^2}{4}\int_{\RR^3}\rho.
$$
Taking the infimum over $\Gamma$ gives $\inf_{\bm{\rho}_\Gamma=\bm{\rho}}\{\EC(\Gamma)-c\,\mathbf{q}\cdot\mathbf{p}_\Gamma\}=E(\bm{\rho})-\frac14c^2|\mathbf{q}|^2\int\rho$, which is~\cref{eq:spiral_exact}.
\end{proof}

\subsection{Local spin-polarized density replacement}\label{sec:deviation_spiral}

For a continuous function $\mathbf{F}$ on a set $K$ we write
$$
\operatorname{osc}_K(\mathbf{F}):=\max_{\mathbf{x},\mathbf{y}\in K}|\mathbf{F}(\mathbf{x})-\mathbf{F}(\mathbf{y})|.
$$
We first collect elementary estimates relating the oscillation of a matrix power on a ball to its gradient.

\begin{lemma}[Oscillation estimates]\label{lem:oscillation}
Let $p>3$, $0<\theta<1$ and $\ell>0$, let $K$ be a closed ball of radius at most $3\ell$ and let $\bm{\rho}\ge0$ be a spin-polarized density on $K$ with $\bm{\rho}^\theta\in W^{1,p}(K)$. We use the continuous representative of $\bm{\rho}^\theta$ given by Morrey's inequality of \cref{thm:Morrey}. Then the following holds with $C=C(p,\theta)$.
\begin{itemize}
\item[(i)] \emph{(Morrey bound)}. We have
\begin{equation}\label{eq:Sobolev_domain}
\operatorname{osc}_K(\bm{\rho}^\theta)^p\le C\ell^{p-3}\int_K|\grad\bm{\rho}^\theta|^p .
\end{equation}
\item[(ii)] \emph{(Large oscillation)}. For every $\kappa>0$, if $\operatorname{osc}_K(\bm{\rho}^\theta)>\kappa\big(\max_K\rho\big)^\theta$, then
\begin{equation}\label{eq:large_oscillation}
\big(\max_K\rho\big)^{p\theta}\le C\kappa^{-p}\ell^{p-3}\int_K|\grad\bm{\rho}^\theta|^p .
\end{equation}
\item[(iii)] \emph{(Small oscillation)}. If $\operatorname{osc}_K(\bm{\rho}^\theta)\le\frac\theta4\big(\max_K\rho\big)^\theta$, then
\begin{equation}\label{eq:comparable_density}
\rho(\mathbf{x})\ge\frac12\max_K\rho\qquad\text{for all $\mathbf{x}\in K$.}
\end{equation}
\end{itemize}
\end{lemma}

\begin{proof}
Since $(\diam K)^3/|K|=6/\pi$ for every ball, the constant in Morrey's inequality of \cref{thm:Morrey} only depends on $p$. Taking the maximum over $\mathbf{x},\mathbf{y}\in K$ in Morrey's inequality and using $\diam K\le6\ell$ gives~\cref{eq:Sobolev_domain}. Under the assumption of (ii), $(\max_K\rho)^{p\theta}<\kappa^{-p}\operatorname{osc}_K(\bm{\rho}^\theta)^p$, and~\cref{eq:large_oscillation} follows from~\cref{eq:Sobolev_domain}. Next,~\cref{eq:Lipschitz_power} and $|\Trs\mathbf{A}|\le\sqrt2|\mathbf{A}|$ give
$$
\big|\rho(\mathbf{x})-\rho(\mathbf{y})\big|\le\frac{\sqrt2}{\theta}\Big(\max_K\rho\Big)^{1-\theta}\operatorname{osc}_K(\bm{\rho}^\theta)
$$
for all $\mathbf{x},\mathbf{y}\in K$. Under the assumption of (iii) the right side is at most $\frac12\max_K\rho$, and~\cref{eq:comparable_density} follows by taking for $\mathbf{y}$ a point where the continuous function $\rho$ attains its maximum on $K$.
\end{proof}

In the next proposition the comparison densities are chosen from the density on a ball containing the tile, and are therefore independent of the orientation of the tile.

\begin{proposition}[Local spin-polarized density replacement]\label{prop:replace_spiral}
Let $p>3$, $2/p\le\theta<1$, $0<\delta\le\ell/C$, $\mathbf{R}\in\SO(3)$ and $\mathbf{z}\in\RR^3$. Let $K$ be a closed ball of radius at most $6\ell$ containing the support of $(\iden_{\ell\tetra}\ast\eta_\delta)(\mathbf{R}\,\cdot\,-\mathbf{z})$ in its interior, and let $\bm{\rho}\ge0$ be a spin-polarized density on $\RR^3$ with total density $\rho\in L^1(\RR^3)$, $\sqrt{\bm{\rho}}\in H^1(\RR^3,\CC^{2\times2})$ and $\grad\bm{\rho}^\theta\in L^p(K)$. There exists $\kappa=\kappa(p,\theta)\in(0,\theta/64]$ such that, if $\operatorname{osc}_K(\bm{\rho}^\theta)\le\kappa(\max_K\rho)^\theta$, two constant matrices $\ul{\bm{\rho}},\ol{\bm{\rho}}\ge0$, which only depend on $p$, $\theta$, $K$ and the restriction of $\bm{\rho}$ to $K$, satisfy, for $\bm{\rho}_\star\in\{\ul{\bm{\rho}},\ol{\bm{\rho}}\}$ and every $\mathbf{x}\in K$,
\begin{equation}\label{eq:spiral_comparison_rotation}
\begin{aligned}
|\bm{\rho}_\star^{1/p}-\bm{\rho}(\mathbf{x})^{1/p}|&\le C\operatorname{osc}_K(\bm{\rho}^{1/p}),\qquad
|\bm{\rho}_\star^\theta-\bm{\rho}(\mathbf{x})^\theta|\le C\operatorname{osc}_K(\bm{\rho}^\theta),\\
|\bm{\rho}_\star-\bm{\rho}(\mathbf{x})|&\le C\Big(\max_K\rho\Big)^{1-\theta}\operatorname{osc}_K(\bm{\rho}^\theta).
\end{aligned}
\end{equation}
Their traces satisfy $\ul{\rho}\le\rho(\mathbf{x})\le\ol{\rho}\le C\max_K\rho$.

For every $\mathbf{q}\in\RR^3$ and $0<\epsilon\le1/2$ we have
\begin{align}
E\big(\mathbf{U}_{\mathbf{q}}\bm{\rho}\,\mathbf{U}_{\mathbf{q}}^*(\iden_{\ell\tetra}\ast\eta_\delta)(\mathbf{R}\,\cdot\,-\mathbf{z})\big)&\le E\big(\mathbf{U}_{\mathbf{q}}\ul{\bm{\rho}}\,\mathbf{U}_{\mathbf{q}}^*(\iden_{\ell\tetra}\ast\eta_\delta)(\mathbf{R}\,\cdot\,-\mathbf{z})\big)+\mathcal{R},\label{eq:replace_spiral_upper}\\
E\big(\mathbf{U}_{\mathbf{q}}\bm{\rho}\,\mathbf{U}_{\mathbf{q}}^*(\iden_{\ell\tetra}\ast\eta_\delta)(\mathbf{R}\,\cdot\,-\mathbf{z})\big)&\ge E\big(\mathbf{U}_{\mathbf{q}}\ol{\bm{\rho}}\,\mathbf{U}_{\mathbf{q}}^*(\iden_{\ell\tetra}\ast\eta_\delta)(\mathbf{R}\,\cdot\,-\mathbf{z})\big)-\mathcal{R},\label{eq:replace_spiral_lower}
\end{align}
where
\begin{equation}\label{eq:def_s}
s:=\max\{p,3\theta p-3\}-1
\end{equation}
and
\begin{multline}\label{eq:spiral_rotation_error}
\mathcal{R}=C\int_{\RR^3}\Big(\epsilon(\rho+\rho^2)+\frac{|\grad\sqrt{\bm{\rho}}|^2}{\epsilon}\Big)(\iden_{\ell\tetra}\ast\eta_\delta)(\mathbf{R}\,\cdot\,-\mathbf{z})+C\epsilon|\mathbf{q}|^2\ell^3\max_K\rho\\
+C\int_{\RR^3}\rho\,\Big|\grad\sqrt{(\iden_{\ell\tetra}\ast\eta_\delta)(\mathbf{R}\,\cdot\,-\mathbf{z})}\Big|^2+\frac{C\ell^p}{\epsilon^s}\int_K|\grad\bm{\rho}^\theta|^p\\
+C|\mathbf{q}|^2\ell^3\bigg(\Big(\max_K\rho\Big)^{1-1/p}\operatorname{osc}_K(\bm{\rho}^{1/p})+\frac{(\max_K\rho)^{1-2/p}}{\epsilon}\operatorname{osc}_K(\bm{\rho}^{1/p})^2\bigg)
\end{multline}
with $C=C(p,\theta)$.
\end{proposition}

\begin{proof}
We write
$$
\rho_{\max}:=\max_K\rho,\qquad\omega:=\operatorname{osc}_K(\bm{\rho}^\theta),\qquad d:=\frac{\operatorname{osc}_K(\bm{\rho}^{1/p})}{\rho_{\max}^{1/p}},
$$
with $d=0$ if $\rho_{\max}=0$. The terms of~\cref{eq:spiral_rotation_error} containing $\mathbf{q}$ are then $C|\mathbf{q}|^2\ell^3\rho_{\max}(\epsilon+d+d^2/\epsilon)$. Since $\theta<1$ and $\rho$ is integrable, $\bm{\rho}^\theta\in L^1(K)$. Poincar\'e's inequality and $\grad\bm{\rho}^\theta\in L^p(K)$ then give $\bm{\rho}^\theta\in W^{1,p}(K)$. We take its continuous representative from Morrey's inequality of \cref{thm:Morrey}, so all positive powers of $\bm{\rho}$ are continuous on $K$. Suppose $\rho_{\max}>0$, since both matrices may be taken to vanish when $\rho_{\max}=0$. Let $\lambda_\pm$ denote the ordered eigenvalues of $\bm{\rho}$. By~\cref{eq:Lipschitz_power} and the same inequality applied to the power $p\theta$ of $\bm{\rho}^{1/p}$,
\[
 \operatorname{osc}_K\bm{\rho}\le \theta^{-1}\rho_{\max}^{1-\theta}\omega=:\eta,\qquad
 \omega\le p\theta\,\rho_{\max}^\theta d .
\]
Choose $\kappa\le\theta/64$, so that $\eta\le \rho_{\max}/64$. Let $\mathbf{x}_0$ be the center of $K$.

\noindent\emph{Case I.} The smaller eigenvalue of $\bm{\rho}$ is comparable to $\rho_{\max}$ on $K$, that is, $\min_K\lambda_-\ge \rho_{\max}/8$. We take $\mathbf{V}=\iden$, so that $\bm{\rho}_{\mathbf{V}}=\bm{\rho}$, and
\[
 \ul{\bm{\rho}}=\bm{\rho}(\mathbf{x}_0)-\eta\iden,\qquad
 \ol{\bm{\rho}}=\bm{\rho}(\mathbf{x}_0)+\eta\iden .
\]
They satisfy $\ul{\bm{\rho}}\le\bm{\rho}_{\mathbf{V}}\le\ol{\bm{\rho}}$ and their spectra lie in $[\rho_{\max}/16,2\rho_{\max}]$. Applying the matrix Lipschitz bound~\cref{eq:HS_functions} to the powers $\theta$ and $1/p$ on this interval gives~\cref{eq:spiral_comparison_rotation}.

\noindent\emph{Case II.} The smaller eigenvalue of $\bm{\rho}$ is small somewhere on $K$, that is, $\min_K\lambda_-<\rho_{\max}/8$. Then $|\rho(\mathbf{x})-\rho_{\max}|\le\sqrt2\,\eta$ and $\lambda_-\le \rho_{\max}/8+\eta$. Thus $\lambda_+-\lambda_-\ge \rho_{\max}/2$ throughout $K$, and $g_u=\lambda_+^u-\lambda_-^u\ge c_u \rho_{\max}^u$ for $u=1,\theta,1/p$. Writing $\mathbf{n}$ for the magnetization direction and $\mathbf{n}_0=\mathbf{n}(\mathbf{x}_0)$, the traceless part of $\bm{\rho}^u$ is $g_u\mathbf{n}\cdot\bm{\sigma}/2$. Taking the traceless part is the orthogonal projection onto the complement of $\iden$ for the Hilbert--Schmidt scalar product, so it does not increase distances. Since $|\mathbf{v}\cdot\bm{\sigma}|=\sqrt2|\mathbf{v}|$ for $\mathbf{v}\in\RR^3$, we obtain for every $\mathbf{x}\in K$
\[
|g_u(\mathbf{x})\mathbf{n}(\mathbf{x})-g_u(\mathbf{x}_0)\mathbf{n}_0|\le\sqrt2\operatorname{osc}_K(\bm{\rho}^u).
\]
For unit vectors $\mathbf{n},\mathbf{n}_0$ and $g,g_0>0$ we have $|\mathbf{n}-\mathbf{n}_0|\le2|g\mathbf{n}-g_0\mathbf{n}_0|/\max(g,g_0)$. Together with $g_u\ge c_u\rho_{\max}^u$ this gives
\[
|\mathbf{n}(\mathbf{x})-\mathbf{n}_0|\le C_u \rho_{\max}^{-u}\operatorname{osc}_K(\bm{\rho}^u).
\]
For $u=1$ the right side can be made less than $1/4$ by decreasing $\kappa$. The $\SU(2)$-valued rotation
\[
\mathbf{V}=\frac{\iden+(\mathbf{n}\cdot\bm{\sigma})(\mathbf{n}_0\cdot\bm{\sigma})}
 {\sqrt{2(1+\mathbf{n}\cdot\mathbf{n}_0)}}
\]
maps $\mathbf{P}_{\mathbf{n}_0}$ to $\mathbf{P}_{\mathbf{n}}$ by conjugation. When $\mathbf{n}\ne\mathbf{n}_0$, it rotates through the angle between the two directions about the axis $\mathbf{n}_0\times\mathbf{n}$. Recall that $\mathbf{P}_{\mathbf{n}}=(\iden+\mathbf{n}\cdot\bm{\sigma})/2$ is the projection onto the spin state pointing in the direction $\mathbf{n}$, and that $\mathbf{P}_{\mathbf{n}}+\mathbf{P}_{-\mathbf{n}}=\iden$. The rotation $\mathbf{V}$ satisfies $\|\mathbf{V}-\iden\|\le C\min\{\rho_{\max}^{-\theta}\omega,d\}$ and $\|\grad\mathbf{V}\|\le C|\grad\mathbf{n}|$. Since the gap of the square roots is at least $c\sqrt{\rho_{\max}}$,~\cref{eq:formula_gradient_sqrt_R} gives $\rho_{\max}\|\grad\mathbf{V}\|^2\le C|\grad\sqrt{\bm{\rho}}|^2$.
Set
\begin{equation*}
\bm{\rho}_{\mathbf{V}}:=\mathbf{V}^*\bm{\rho}\,\mathbf{V}=\sum_{\pm}\lambda_\pm\mathbf{P}_{\pm\mathbf{n}_0}.
\end{equation*}
In this fixed basis take
\[
\ul{\bm{\rho}}=\sum_{\pm}(\min_K\lambda_\pm)\mathbf{P}_{\pm\mathbf{n}_0},\qquad
\ol{\bm{\rho}}=\sum_{\pm}(\max_K\lambda_\pm)\mathbf{P}_{\pm\mathbf{n}_0}.
\]
Then $\ul{\bm{\rho}}\le\bm{\rho}_{\mathbf{V}}\le\ol{\bm{\rho}}$. For $u=1,\theta,1/p$, each eigenvalue of $\bm{\rho}^u$ has oscillation at most $\operatorname{osc}_K(\bm{\rho}^u)$. Hence $|\bm{\rho}_\star^u-\bm{\rho}_{\mathbf{V}}^u|\le\sqrt2\,\operatorname{osc}_K(\bm{\rho}^u)$, and the bound on $\|\mathbf{V}-\iden\|$ yields~\cref{eq:spiral_comparison_rotation}.

It remains to prove the trace bounds. In both cases, conjugation by $\mathbf{V}$ does not change the trace, so $\Trs\bm{\rho}_{\mathbf{V}}(\mathbf{x})=\rho(\mathbf{x})$, and the trace is monotone for the matrix order. Hence $\ul{\bm{\rho}}\le\bm{\rho}_{\mathbf{V}}\le\ol{\bm{\rho}}$ gives $\ul{\rho}\le\rho(\mathbf{x})\le\ol{\rho}$. Since $|\Trs\mathbf{A}|\le\sqrt2|\mathbf{A}|$ for every $2\times2$ matrix $\mathbf{A}$, the last bound of~\cref{eq:spiral_comparison_rotation} and $\omega\le\kappa\rho_{\max}^\theta$ give
\[
\ol{\rho}\le\rho(\mathbf{x})+\sqrt2\,|\ol{\bm{\rho}}-\bm{\rho}(\mathbf{x})|\le\rho_{\max}+C\rho_{\max}^{1-\theta}\omega\le C\rho_{\max}.
\]

We apply \cref{lem:subadditivity_worse}, in the form~\cref{eq:subadditivity_lambda}, with
$$
\bm{\rho}_1=\mathbf{U}_{\mathbf{q}}\mathbf{A}\,\mathbf{U}_{\mathbf{q}}^*(\iden_{\ell\tetra}\ast\eta_\delta)(\mathbf{R}\,\cdot\,-\mathbf{z}),\qquad
\bm{\rho}_2=\mathbf{U}_{\mathbf{q}}\mathbf{H}\,\mathbf{U}_{\mathbf{q}}^*(\iden_{\ell\tetra}\ast\eta_\delta)(\mathbf{R}\,\cdot\,-\mathbf{z}),
$$
where $\mathbf{A}=\ul{\bm{\rho}}$ and $\mathbf{H}=\bm{\rho}_{\mathbf{V}}-\ul{\bm{\rho}}$ for the upper bound~\cref{eq:replace_spiral_upper}, and $\mathbf{A}=\bm{\rho}_{\mathbf{V}}$ and $\mathbf{H}=\ol{\bm{\rho}}-\bm{\rho}_{\mathbf{V}}$ for the lower bound~\cref{eq:replace_spiral_lower}. In both cases $\mathbf{H}\ge0$ and the total density of $\bm{\rho}_2$ satisfies $\rho_2\le\Trs\mathbf{H}\le\lambda:=C\rho_{\max}^{1-\theta}\omega$. For the upper bound, since $\ul{\bm{\rho}}$ is constant and $0\le(\iden_{\ell\tetra}\ast\eta_\delta)(\mathbf{R}\,\cdot\,-\mathbf{z})\le1$, the estimate~\cref{eq:subadditivity_lambda} of \cref{lem:subadditivity_worse} reads
\begin{align*}
&E\big(\mathbf{U}_{\mathbf{q}}\bm{\rho}_{\mathbf{V}}\mathbf{U}_{\mathbf{q}}^*(\iden_{\ell\tetra}\ast\eta_\delta)(\mathbf{R}\,\cdot\,-\mathbf{z})\big)\le E\big(\mathbf{U}_{\mathbf{q}}\ul{\bm{\rho}}\,\mathbf{U}_{\mathbf{q}}^*(\iden_{\ell\tetra}\ast\eta_\delta)(\mathbf{R}\,\cdot\,-\mathbf{z})\big)\\
&\qquad+C\epsilon\big(\ul{\rho}+\ul{\rho}^2\big)\int_{\RR^3}(\iden_{\ell\tetra}\ast\eta_\delta)(\mathbf{R}\,\cdot\,-\mathbf{z})\\
&\qquad+C\epsilon\int_{\RR^3}\Big|\grad\sqrt{\mathbf{U}_{\mathbf{q}}\ul{\bm{\rho}}\,\mathbf{U}_{\mathbf{q}}^*(\iden_{\ell\tetra}\ast\eta_\delta)(\mathbf{R}\,\cdot\,-\mathbf{z})}\Big|^2\\
&\qquad+C\int_{\RR^3}\Big|\grad\sqrt{\mathbf{U}_{\mathbf{q}}\big(\bm{\rho}_{\mathbf{V}}-(1-\epsilon)\ul{\bm{\rho}}\big)\mathbf{U}_{\mathbf{q}}^*(\iden_{\ell\tetra}\ast\eta_\delta)(\mathbf{R}\,\cdot\,-\mathbf{z})}\Big|^2\\
&\qquad+C\bigg(\epsilon+\Big(\frac{\lambda}{\epsilon}\Big)^{1/2}+\frac{\lambda}{\epsilon}\bigg)\int_{\RR^3}(\rho-\ul{\rho})(\iden_{\ell\tetra}\ast\eta_\delta)(\mathbf{R}\,\cdot\,-\mathbf{z}).
\end{align*}
For the lower bound, the same inequality holds with $\ul{\bm{\rho}}$ and $\bm{\rho}_{\mathbf{V}}$ replaced by $\bm{\rho}_{\mathbf{V}}$ and $\ol{\bm{\rho}}$, respectively. In both cases
\[
|\grad\sqrt{\mathbf{A}}|^2\le|\grad\sqrt{\bm{\rho}}|^2,\qquad
|\grad\sqrt{\mathbf{H}+\epsilon\mathbf{A}}|^2\le\epsilon^{-1}|\grad\sqrt{\bm{\rho}}|^2 .
\]
Indeed $|\grad\sqrt{\bm{\rho}_{\mathbf{V}}}|\le|\grad\sqrt{\bm{\rho}}|$ by~\cref{eq:formula_gradient_sqrt_R}. For the upper bound $\mathbf{H}+\epsilon\mathbf{A}=\bm{\rho}_{\mathbf{V}}-(1-\epsilon)\ul{\bm{\rho}}\ge\epsilon\bm{\rho}_{\mathbf{V}}$, with derivative $\grad\bm{\rho}_{\mathbf{V}}$. For the lower bound it equals $\ol{\bm{\rho}}-(1-\epsilon)\bm{\rho}_{\mathbf{V}}\ge\epsilon\bm{\rho}_{\mathbf{V}}$ with derivative $-(1-\epsilon)\grad\bm{\rho}_{\mathbf{V}}$. The bound follows from \cref{lem:convexity}~(ii) and~(iii) and~\cref{eq:gradient_Q}.

Differentiation of $\mathbf{U}_{\mathbf{q}}\sqrt{\mathbf{A}}\,\mathbf{U}_{\mathbf{q}}^*$ and the inequality $|[\bm{\sigma}_3,\sqrt{\mathbf{A}}]|\le2|\sqrt{\mathbf{A}}|$ give
\begin{equation}\label{eq:gradient_spiral}
|\grad\sqrt{\mathbf{U}_{\mathbf{q}}\mathbf{A}\mathbf{U}_{\mathbf{q}}^*}|^2\le
2|\grad\sqrt{\mathbf{A}}|^2+2|\mathbf{q}|^2\Trs\mathbf{A}.
\end{equation}
Together with~\cref{eq:gradient_product} this bounds the two gradient terms in~\cref{eq:subadditivity_lambda} of \cref{lem:subadditivity_worse} by
\begin{multline*}
 C\epsilon^{-1}\int|\grad\sqrt{\bm{\rho}}|^2(\iden_{\ell\tetra}\ast\eta_\delta)(\mathbf{R}\,\cdot\,-\mathbf{z})
 + C\int\rho\,\Big|\grad\sqrt{(\iden_{\ell\tetra}\ast\eta_\delta)(\mathbf{R}\,\cdot\,-\mathbf{z})}\Big|^2\\
 +C|\mathbf{q}|^2\ell^3(\epsilon \rho_{\max}+\lambda).
\end{multline*}
Here all traces are bounded by $C\rho_{\max}\le C\rho$ on $K$ and $\Trs(\mathbf{H}+\epsilon\mathbf{A})\le\lambda+C\epsilon \rho_{\max}$. The last term is bounded by the wavevector term of~\cref{eq:spiral_rotation_error} since $\lambda\le C\rho_{\max}d$.

In the remaining terms of~\cref{eq:subadditivity_lambda} in \cref{lem:subadditivity_worse}, it suffices to bound $\ell^3\epsilon^{1-m}\lambda^m$ for $m=3/2,2$. Take $\alpha=mp/(s+1)\le m$ and $\beta=(m-\theta\alpha)/(1-\alpha/p)$. Then $1\le\beta\le2$ since $p\theta\ge2$ and $s+1\ge3p\theta-3$. Since $\omega\le \rho_{\max}^\theta$, we have $\lambda^m\le C\omega^\alpha \rho_{\max}^{m-\theta\alpha}$. Young's inequality and \cref{lem:oscillation}~(i), applied with $2\ell$ in place of $\ell$, give
\begin{multline*}
 \ell^3\epsilon^{1-m}\lambda^m
 \le C\epsilon^{-s}\ell^3\omega^p+C\epsilon\ell^3\rho_{\max}^\beta\\
 \le C\epsilon^{-s}\ell^p\int_K|\grad\bm{\rho}^\theta|^p+C\epsilon\int(\rho+\rho^2)(\iden_{\ell\tetra}\ast\eta_\delta)(\mathbf{R}\,\cdot\,-\mathbf{z}) ,
\end{multline*}
where $\int(\iden_{\ell\tetra}\ast\eta_\delta)(\mathbf{R}\,\cdot\,-\mathbf{z})=|\ell\tetra|$ and $\rho\ge \rho_{\max}/2$ on $K$. This proves the two energy comparisons with $\bm{\rho}_{\mathbf{V}}$ in place of $\bm{\rho}$.

Finally $\mathbf{W}=\mathbf{U}_{\mathbf{q}}\mathbf{V}\mathbf{U}_{\mathbf{q}}^*$ maps the spin-spiral wave of $\bm{\rho}_{\mathbf{V}}(\iden_{\ell\tetra}\ast\eta_\delta)(\mathbf{R}\,\cdot\,-\mathbf{z})$ to that of $\bm{\rho}(\iden_{\ell\tetra}\ast\eta_\delta)(\mathbf{R}\,\cdot\,-\mathbf{z})$. Since $\|[\bm{\sigma}_3,\mathbf{V}]\|\le2\|\mathbf{V}-\iden\|$, we have
\[
 \|\grad\mathbf{W}\|^2\le C\|\grad\mathbf{V}\|^2+C|\mathbf{q}|^2 d^2 .
\]
Applying \cref{lem:gauge} to $\mathbf{W}$ and $\mathbf{W}^*$ gives the rotation cost $C\epsilon^{-1}\int |\grad\sqrt{\bm{\rho}}|^2(\iden_{\ell\tetra}\ast\eta_\delta)(\mathbf{R}\,\cdot\,-\mathbf{z})+C\epsilon^{-1}|\mathbf{q}|^2\ell^3\rho_{\max} d^2$. The factor $1+\epsilon$ multiplying the energy in \cref{lem:gauge} contributes at most
\begin{multline*}
 C\epsilon\int(\rho^{4/3}+\rho^{5/3}+\rho|\mathbf{q}|^2+|\grad\sqrt{\bm{\rho}}|^2)(\iden_{\ell\tetra}\ast\eta_\delta)(\mathbf{R}\,\cdot\,-\mathbf{z})\\
 +C\epsilon\int\rho\,\Big|\grad\sqrt{(\iden_{\ell\tetra}\ast\eta_\delta)(\mathbf{R}\,\cdot\,-\mathbf{z})}\Big|^2
\end{multline*}
including the Lieb--Oxford term, by~\cref{cor:upper_bound_E_rho} and~\cref{eq:gradient_product,eq:gradient_spiral}. These terms are bounded by $\mathcal{R}$, giving~\cref{eq:replace_spiral_upper,eq:replace_spiral_lower}.
\end{proof}

\cref{prop:replace_spiral} applies to spin-spiral waves with a constant wavevector $\mathbf{q}$. To use it for a general phase $\varphi$, the next lemma replaces $\varphi$ by the linear phase $\mathbf{q}\cdot\mathbf{x}$ on the support of a cutoff $\xi$, at a cost controlled by $\int\rho|\grad\varphi-\mathbf{q}|^2\xi$.

\begin{lemma}[Local wavevector replacement]\label{lem:constant_wavevector}
Let $\varphi\in W^{2,p}_{\rm loc}(\RR^3)$ be real with $p>3$, let $\bm{\rho}\ge0$ be a spin-polarized density on $\RR^3$ with total density $\rho\in L^1(\RR^3)$ and $\sqrt{\bm{\rho}}\in H^1(\RR^3,\CC^{2\times2})$, and let $0\le\xi\le1$ have compact support and $\grad\sqrt\xi$ bounded. For every $\mathbf{q}\in\RR^3$ and $0<\epsilon\le1$,
\begin{equation*}
E\big(\mathbf{U}_\varphi\bm{\rho}\,\mathbf{U}_\varphi^*\xi\big)
\le(1+\epsilon)E\big(\mathbf{U}_{\mathbf{q}}\bm{\rho}\,\mathbf{U}_{\mathbf{q}}^*\xi\big)
+\epsilon\,c_\LO\int_{\RR^3}\rho^{4/3}\xi
+\frac1{2\epsilon}\int_{\RR^3}\rho\,|\grad\varphi-\mathbf{q}|^2\xi .
\end{equation*}
The same inequality holds with the two energies interchanged.
\end{lemma}

\begin{proof}
The $\SU(2)$-valued rotation $\mathbf{U}_{\varphi-\mathbf{q}\cdot\mathbf{x}}$ maps the second density to the first and its squared gradient norm is $|\grad\varphi-\mathbf{q}|^2/4$. Apply \cref{lem:gauge} to this rotation and its adjoint, using $(\rho\xi)^{4/3}\le\rho^{4/3}\xi$ and $(1+\epsilon^{-1})/4\le1/(2\epsilon)$.
\end{proof}

\subsection{Estimates on a ball}\label{sec:ball_estimates}
The following choice of constant wavevector controls the error in \cref{lem:constant_wavevector} on every ball.

\begin{lemma}[Freezing the wavevector on a ball]\label{lem:weighted_wavevector}
Let $p>3$, $\ell>0$ and let $K$ be a closed ball of center $\mathbf{c}$ and radius at most $3\ell$. Let $\varphi\in W^{2,p}(K)$ be real and let $\bm{\rho}\ge0$ have trace $\rho\in L^1(K)$, with $\bm{\rho}^{1/p}\in W^{1,p}_{\rm loc}(\Int K\cap\{\grad\varphi\ne0\})$. Choose $\mathbf{q}_{\mathbf{c}}$ to be the value of $\grad\varphi$ at a point where $|\grad\varphi|$ attains its minimum on $K$. For every measurable $0\le\xi\le1$ on $K$ and every $0<\epsilon\le1$,
\begin{multline}\label{eq:weighted_wavevector}
\frac1\epsilon\int_K\rho\,|\grad\varphi-\mathbf{q}_{\mathbf{c}}|^2\xi
\le C\epsilon\int_K\rho\,\xi\\
+\frac{C\ell^p}{\epsilon}\int_{K\cap\{\grad\varphi\ne0\}}|\grad\varphi|^2\,|\grad\bm{\rho}^{1/p}|^p
+\frac{C\ell^p}{\epsilon^{p-1}}\int_K\rho\,|\mathbf{D}^2\varphi|^p ,
\end{multline}
with $C=C(p)$.
\end{lemma}

\begin{proof}
Morrey's inequality gives a continuous representative of $\grad\varphi$ on $K$. By the choice of $\mathbf{q}_{\mathbf{c}}$ we have $|\grad\varphi-\mathbf{q}_{\mathbf{c}}|\le2|\grad\varphi|$. Set
\[
g=\rho^{1/p}\big(|\grad\varphi-\mathbf{q}_{\mathbf{c}}|^{2/p}-\epsilon^{2/p}\big)_+ .
\]
This function vanishes where $|\grad\varphi-\mathbf{q}_{\mathbf{c}}|<\epsilon$. Where $|\grad\varphi-\mathbf{q}_{\mathbf{c}}|>\epsilon/2$, the root $\bm{\rho}^{1/p}$ is locally Sobolev and $|\grad\rho^{1/p}|\le|\grad\bm{\rho}^{1/p}|$, since $\rho^{1/p}$ is its Schatten $p$-norm. The function $t\mapsto(t^{2/p}-\epsilon^{2/p})_+$ has Lipschitz constant $(2/p)\epsilon^{2/p-1}$. The product rule on these two open sets gives
\[
\int_K|\grad g|^p\le C\int_{K\cap\{\grad\varphi\ne0\}}|\grad\varphi|^2\,|\grad\bm{\rho}^{1/p}|^p
+C\epsilon^{2-p}\int_K\rho\,|\mathbf{D}^2\varphi|^p .
\]
We may assume that the right side is finite. Since $\rho\in L^1(K)$ and $\grad\varphi$ is bounded on $K$, we have $g\in W^{1,p}(K)$. Its continuous representative vanishes at the chosen minimum point, so \cref{thm:Morrey} gives $\int_K g^p\le C\ell^p\int_K|\grad g|^p$. Finally,
$\rho|\grad\varphi-\mathbf{q}_{\mathbf{c}}|^2\xi\le 2^{p-1}(\epsilon^2\rho\xi+g^p)$, which proves~\cref{eq:weighted_wavevector}.
\end{proof}

The following comparison also controls the cost of replacing the wavevector in the gas energy.

\begin{lemma}[Comparison of SSW gas energies]\label{lem:compare_SSW}
Let $p\ge3$, let $\bm{\rho}_1,\bm{\rho}_2\ge0$ be Hermitian $2\times2$ matrices with traces $\rho_1$ and $\rho_2$, and let $r>0$ be such that $\rho_1\le r$ and $\rho_2\le r$. For all $\mathbf{q},\mathbf{q}'\in\RR^3$ and all $0<\epsilon\le1$ we have
\begin{multline}\label{eq:compare_SSW}
\big|e_\SSW(\bm{\rho}_1,\mathbf{q})-e_\SSW(\bm{\rho}_2,\mathbf{q}')\big|\le C\big(r^{4/3}+r^{5/3}+r|\mathbf{q}|^2\big)\frac{|\bm{\rho}_1^{1/p}-\bm{\rho}_2^{1/p}|}{r^{1/p}}\\
+C\big(r+r^{4/3}\big)|\mathbf{q}-\mathbf{q}'|+\frac{Cr}{\epsilon}|\mathbf{q}-\mathbf{q}'|^2+C\epsilon r|\mathbf{q}|^2
\end{multline}
with $C=C(p)$.
\end{lemma}

\begin{proof}
Let $r'=\max\{\rho_1,\rho_2\}\le r$. If $r'=0$, both matrices vanish and the left side is zero. Otherwise \cref{prop:Lipschitz_SSW}~(vi) gives
$$
\big|e_\SSW(\bm{\rho}_1,\mathbf{q})-e_\SSW(\bm{\rho}_2,\mathbf{q})\big|\le C\big(r'^{4/3}+r'^{5/3}+r'|\mathbf{q}|^2\big)\frac{|\bm{\rho}_1^{1/p}-\bm{\rho}_2^{1/p}|}{r'^{1/p}},
$$
which is bounded by the first term on the right of~\cref{eq:compare_SSW}, since $t\mapsto t^{m-1/p}$ is non-decreasing for $m\ge1$. Next, \cref{prop:Lipschitz_SSW}~(iii) at the matrix $\bm{\rho}_2$, together with $\max(|\mathbf{q}|,|\mathbf{q}'|)\le|\mathbf{q}|+|\mathbf{q}-\mathbf{q}'|$, gives
$$
\big|e_\SSW(\bm{\rho}_2,\mathbf{q})-e_\SSW(\bm{\rho}_2,\mathbf{q}')\big|\le C\big(r^{7/6}+r^{4/3}+r|\mathbf{q}|+r|\mathbf{q}-\mathbf{q}'|\big)|\mathbf{q}-\mathbf{q}'|.
$$
Since $r^{7/6}\le r+r^{4/3}$, $r|\mathbf{q}||\mathbf{q}-\mathbf{q}'|\le\epsilon r|\mathbf{q}|^2+r|\mathbf{q}-\mathbf{q}'|^2/(4\epsilon)$ and $\epsilon\le1$, this is bounded by the last three terms of~\cref{eq:compare_SSW}.
\end{proof}

\subsection{Proof of \texorpdfstring{\cref{thm:LSSWA}}{the fixed-axis TSGA estimate}}\label{sec:proof_LSSWA}

We are now ready to prove our main result, which we do in several steps. We prove~\cref{eq:LSSWA_estim} in its two forms together: in general, and with the bulk term replaced by~\cref{eq:LSSWA_sharp_bulk}.

\medskip

\noindent\textbf{Step 1. (Reductions and a priori bound)} For~\cref{eq:LSSWA_sharp_bulk} we may first reduce to $\theta=2/p$. Indeed, for any $k=p\theta\ge2$ the function $F$ on $[0,\infty)$ defined by $F(t+t^k)=t^2$ is $2$-Lipschitz, since $2t/(1+kt^{k-1})\le2$. Applying the Hilbert--Schmidt Lipschitz estimate used in~\cref{eq:Lipschitz_power} to $F$ and taking difference quotients gives
\[
|\grad\wt{\bm{\rho}}^{2/p}|\le2\big(|\grad\wt{\bm{\rho}}^{1/p}|+|\grad\wt{\bm{\rho}}^\theta|\big).
\]
We therefore use $\theta=2/p$ throughout the proof of~\cref{eq:LSSWA_sharp_bulk} and keep the prescribed $\theta$ for~\cref{eq:LSSWA_estim}. For the density comparison choose
\begin{equation}\label{eq:def_a_b}
(\beta,a)=
\begin{cases}
\big(\theta,\min\{1,1/(3\theta)\}\big) & \text{in~\cref{eq:LSSWA_estim}},\\
(1/p,1) & \text{in~\cref{eq:LSSWA_sharp_bulk}}.
\end{cases}
\end{equation}
Then $b=p+p/a-1$ gives the two exponents in the theorem. The replacement exponent $s$ in~\cref{eq:def_s} of \cref{prop:replace_spiral} satisfies $p+s\le b$ in each case. For~\cref{eq:LSSWA_sharp_bulk}, the reduction above and the choice $\theta=2/p$ give constants depending only on $p$.

The density $\bm{\rho}=\mathbf{U}_\varphi\wt{\bm{\rho}}\,\mathbf{U}_\varphi^*$ satisfies~\cref{eq:representable} by \cref{lem:gauge} and~\cref{eq:hyp_phi}. Morrey's inequality of \cref{thm:Morrey} applied on balls to $\wt{\bm{\rho}}^\theta$ and $\grad\varphi$ gives continuous representatives of $\wt{\bm{\rho}}$ and $\grad\varphi$. By \cref{prop:Lipschitz_SSW} the function $\mathbf{x}\mapsto e_\SSW(\wt{\bm{\rho}}(\mathbf{x}),\grad\varphi(\mathbf{x}))$ is therefore continuous, and it is integrable since it is bounded by $C(\rho^{4/3}+\rho^{5/3}+\rho|\grad\varphi|^2)$, by~\cref{eq:bounds_SSW} of \cref{prop:Lipschitz_SSW}~(i). From~\cref{eq:gradient_product} and the same computation as for~\cref{eq:gradient_spiral} we have, for every scalar function $0\le f\le1$,
\begin{equation}\label{eq:a_priori_LSSWA}
\big|\grad\sqrt{\bm{\rho}f}\big|^2\le2\rho|\grad\sqrt f|^2+4f|\grad\sqrt{\wt{\bm{\rho}}}|^2+4\rho|\grad\varphi|^2 f.
\end{equation}
Together with~\cref{cor:upper_bound_E_rho,cor:lower_bound_E_rho} and~\cref{eq:bounds_SSW} of \cref{prop:Lipschitz_SSW}~(i), this gives the \emph{a priori} bound
\begin{multline}\label{eq:a_priori_LSSWA_2}
\left|E(\bm{\rho}f)-\int_{\RR^3}e_\SSW\big(\wt{\bm{\rho}}(\mathbf{x}),\grad\varphi(\mathbf{x})\big)f(\mathbf{x})\,\dd\mathbf{x}\right|\le C\int_{\RR^3}\big(\rho^{4/3}+\rho^{5/3}\big)f\\
+C\int_{\RR^3}\rho|\grad\varphi|^2 f+C\int_{\RR^3}|\grad\sqrt{\wt{\bm{\rho}}}|^2f+C\int_{\RR^3}\rho|\grad\sqrt f|^2.
\end{multline}
We use it on balls of large oscillation (defined below), where no comparison with a constant matrix is available, and for large $\epsilon$. The second term on the right comes from the kinetic energy of the spin-spiral wave. Even at low density its size relative to $\int\rho f$ depends on the wavevector, so it requires a separate estimate.

\medskip

\noindent\textbf{Step 2. (Choice of scales)} Let $0<\epsilon_0\le1/3$ be such that, for $0<\epsilon\le\epsilon_0$, the choice $\delta=1$ and $\ell=\epsilon^{-1}$ fulfills the conditions on $\delta/\ell$ in the spatial decoupling estimates of \cref{prop:upper_bound,prop:lower_bound}, in \cref{prop:replace_spiral} and in \cref{prop:rate_SSW}, also when $\ell$ is replaced by $t\ell$ or $t(\ell-\delta)$ with $t\in[1/2,3/2]$. Here $\epsilon_0$ only depends on $p$ and $\theta$. Until the last paragraph of the proof we assume $0<\epsilon\le\epsilon_0$, and we prove the bounds with $C\epsilon$ in place of $\epsilon$ in front of the first integral. The other values of $\epsilon$ are treated at the end. The localization errors in~\cref{prop:upper_bound,prop:lower_bound} have coefficient at most $C(1+\delta^{-1}+\delta^3)/\ell$ on $\rho+\rho^2$. We take
$$
\delta=1,\qquad \ell=\epsilon^{-1}.
$$

For $\mathbf{c}\in\RR^3$ let $K_{\mathbf{c}}$ be the closed ball of center $\mathbf{c}$ and radius $2\ell$. We choose $\mathbf{q}_{\mathbf{c}}$ as in \cref{lem:weighted_wavevector}, before averaging over the orientations or sizes of the tiles. Since $\ell\ge3$ and $\tetra$ has diameter one, all tile supports lie inside $K_{\mathbf{c}}$ and $2\ell\le6t(\ell-\delta)$ for every $1/2\le t\le3/2$, as required by \cref{prop:replace_spiral}. Each cutoff has integral at least $\ell^3/C$.

\medskip

\noindent\textbf{Step 3. (Freezing the wavevector)} On each ball $K_{\mathbf{c}}$ we replace $\varphi$ by $\mathbf{q}_{\mathbf{c}}\cdot\mathbf{x}$ in the localized energy and $\grad\varphi$ by $\mathbf{q}_{\mathbf{c}}$ in the gas energy. For a cutoff $\xi$ with support in $K_{\mathbf{c}}$, \cref{lem:constant_wavevector}, applied to $\wt{\bm{\rho}}$ in both directions, gives
\begin{multline*}
\Big|E\big(\mathbf{U}_\varphi\wt{\bm{\rho}}\,\mathbf{U}_\varphi^*\xi\big)-E\big(\mathbf{U}_{\mathbf{q}_{\mathbf{c}}}\wt{\bm{\rho}}\,\mathbf{U}_{\mathbf{q}_{\mathbf{c}}}^*\xi\big)\Big|
\le C\epsilon\int_{K_{\mathbf{c}}}\big(\rho^{4/3}+\rho^{5/3}+\rho|\grad\varphi|^2+|\grad\sqrt{\wt{\bm{\rho}}}|^2\big)\xi\\
+C\epsilon\int_{K_{\mathbf{c}}}\rho|\grad\sqrt\xi|^2+\frac1{2\epsilon}\int_{K_{\mathbf{c}}}\rho\,|\grad\varphi-\mathbf{q}_{\mathbf{c}}|^2\xi.
\end{multline*}
Here the terms coming from the factor $1+\epsilon$ are bounded by \cref{cor:lower_bound_E_rho,cor:upper_bound_E_rho} and~\cref{eq:a_priori_LSSWA}, since $|\mathbf{q}_{\mathbf{c}}|\le|\grad\varphi|$ throughout the ball. For the gas energy, we apply \cref{lem:compare_SSW} with both matrices equal to $\wt{\bm{\rho}}$ and $r=\rho$ wherever $\rho>0$. Young's inequality then gives
\[
\big|e_\SSW(\wt{\bm{\rho}},\grad\varphi)-e_\SSW(\wt{\bm{\rho}},\mathbf{q}_{\mathbf{c}})\big|
\le C\epsilon\big(\rho+\rho^2+\rho|\grad\varphi|^2\big)
+\frac C\epsilon\rho|\grad\varphi-\mathbf{q}_{\mathbf{c}}|^2 .
\]
For both replacements, \cref{lem:weighted_wavevector}, applied to $\wt{\bm{\rho}}$ on $K_{\mathbf{c}}$, bounds the integrated wavevector difference. With $\ell=\epsilon^{-1}$ its two gradient coefficients are at most $C\epsilon^{1-2p}$. These estimates apply under~\cref{eq:hyp_phi}, including on balls that meet $\{\grad\varphi=0\}$.

\medskip

\noindent\textbf{Step 4. (Small and large oscillation)} Let
$$
\omega_{\mathbf{c}}=\operatorname{osc}_{K_{\mathbf{c}}}\big(\wt{\bm{\rho}}^\theta\big).
$$
We omit the balls on which $\max_{K_{\mathbf{c}}}\rho=0$. Let $\kappa=\kappa(p,\theta)$ be the constant of \cref{prop:replace_spiral}. We say that $K_{\mathbf{c}}$ has \emph{small oscillation} if $\omega_{\mathbf{c}}\le\kappa(\max_{K_{\mathbf{c}}}\rho)^\theta$, and \emph{large oscillation} otherwise. On a ball of small oscillation we take the comparison densities $\ul{\bm{\rho}}_{\mathbf{c}},\ol{\bm{\rho}}_{\mathbf{c}}$ of \cref{prop:replace_spiral}, applied to $\wt{\bm{\rho}}$ on $K_{\mathbf{c}}$. These too are independent of the orientation and size of the tile.

Let $\omega'_{\mathbf{c}}$ denote the oscillation of $\wt{\bm{\rho}}^{1/p}$ on the ball. We use its relative size
\[
d_{\mathbf{c}}:=\frac{\omega'_{\mathbf{c}}}{(\max_{K_{\mathbf{c}}}\rho)^{1/p}},\qquad
L_{\mathbf{c}}:=\epsilon+d_{\mathbf{c}}+\epsilon^{-1}d_{\mathbf{c}}^2.
\]
On a ball of small oscillation, \cref{prop:replace_spiral}, applied to $\wt{\bm{\rho}}$ on $K_{\mathbf{c}}$ with $\mathbf{q}=\mathbf{q}_{\mathbf{c}}$, gives for every tile cutoff $\xi$ with support in $K_{\mathbf{c}}$
\begin{align*}
E\big(\mathbf{U}_{\mathbf{q}_{\mathbf{c}}}\wt{\bm{\rho}}\,\mathbf{U}_{\mathbf{q}_{\mathbf{c}}}^*\xi\big)&\le E\big(\mathbf{U}_{\mathbf{q}_{\mathbf{c}}}\ul{\bm{\rho}}_{\mathbf{c}}\mathbf{U}_{\mathbf{q}_{\mathbf{c}}}^*\xi\big)+\mathcal{R}_{\mathbf{c}},\\
E\big(\mathbf{U}_{\mathbf{q}_{\mathbf{c}}}\wt{\bm{\rho}}\,\mathbf{U}_{\mathbf{q}_{\mathbf{c}}}^*\xi\big)&\ge E\big(\mathbf{U}_{\mathbf{q}_{\mathbf{c}}}\ol{\bm{\rho}}_{\mathbf{c}}\mathbf{U}_{\mathbf{q}_{\mathbf{c}}}^*\xi\big)-\mathcal{R}_{\mathbf{c}},
\end{align*}
where, since $\operatorname{osc}_{K_{\mathbf{c}}}(\wt{\bm{\rho}}^{1/p})=(\max_{K_{\mathbf{c}}}\rho)^{1/p}d_{\mathbf{c}}$,
\begin{multline*}
\mathcal{R}_{\mathbf{c}}=C\int_{\RR^3}\Big(\epsilon(\rho+\rho^2)+\frac{|\grad\sqrt{\wt{\bm{\rho}}}|^2}{\epsilon}\Big)\xi+C\int_{\RR^3}\rho\,|\grad\sqrt\xi|^2\\
+\frac{C\ell^p}{\epsilon^s}\int_{K_{\mathbf{c}}}|\grad\wt{\bm{\rho}}^\theta|^p+C|\mathbf{q}_{\mathbf{c}}|^2\ell^3\Big(\max_{K_{\mathbf{c}}}\rho\Big)L_{\mathbf{c}}.
\end{multline*}
The matrix power estimate~\cref{eq:Lipschitz_power} gives $\omega_{\mathbf{c}}\le p\theta\,(\max_{K_{\mathbf{c}}}\rho)^\theta d_{\mathbf{c}}$. Moreover,
\begin{equation}\label{eq:bound_q}
|\mathbf{q}_{\mathbf{c}}|^2\ell^3(\max_{K_{\mathbf{c}}}\rho)d_{\mathbf{c}}^p
\le C\ell^p\int_{K_{\mathbf{c}}\cap\{\grad\varphi\ne0\}}|\grad\varphi|^2\,|\grad\wt{\bm{\rho}}^{1/p}|^p .
\end{equation}
If $\mathbf{q}_{\mathbf{c}}=0$ the left side vanishes. Otherwise $K_{\mathbf{c}}\subset\{\grad\varphi\ne0\}$. Since $(\max_{K_{\mathbf{c}}}\rho)d_{\mathbf{c}}^p=(\omega'_{\mathbf{c}})^p$, \cref{lem:oscillation}~(i) applied to $\wt{\bm{\rho}}^{1/p}$ gives~\cref{eq:bound_q} upon multiplication by $|\mathbf{q}_{\mathbf{c}}|^2=\min_{K_{\mathbf{c}}}|\grad\varphi|^2$.

On balls of small oscillation, \cref{lem:oscillation}~(iii) gives $\rho\ge(\max_{K_{\mathbf{c}}}\rho)/2$ on $K_{\mathbf{c}}$. Since moreover $|\mathbf{q}_{\mathbf{c}}|\le|\grad\varphi|$ on $K_{\mathbf{c}}$ and each tile cutoff has integral at least $\ell^3/C$, we obtain
$$
|\mathbf{q}_{\mathbf{c}}|^2\ell^3\max_{K_{\mathbf{c}}}\rho\le C\int_{\RR^3}\rho|\grad\varphi|^2\xi
$$
for each tile cutoff $\xi$. Since $L_{\mathbf{c}}\le C(\epsilon+\epsilon^{1-p}d_{\mathbf{c}}^p)$, this and~\cref{eq:bound_q} control the last term of $\mathcal{R}_{\mathbf{c}}$. On balls of large oscillation $d_{\mathbf{c}}>\kappa/(p\theta)$, so~\cref{eq:bound_q} controls the entire kinetic term $|\mathbf{q}_{\mathbf{c}}|^2\int\rho\xi$.

\medskip

\noindent\textbf{Step 5. (Density comparison)} In this step we bound the errors caused by the variation of the density on $K_{\mathbf{c}}$. On a ball of small oscillation they come from replacing $\wt{\bm{\rho}}(\mathbf{x})$ by the comparison densities in the gas energy, and on a ball of large oscillation from the density powers in~\cref{eq:a_priori_LSSWA_2}. In both cases we reduce them to the expression $\rho^{m-\beta a}\operatorname{osc}_{K_{\mathbf{c}}}(\wt{\bm{\rho}}^\beta)^a$ with $m=4/3,5/3$, which we then bound by Young's inequality and Morrey's inequality. We write
\[
d_{\mathbf{c},\beta}:=\frac{\operatorname{osc}_{K_{\mathbf{c}}}(\wt{\bm{\rho}}^\beta)}{(\max_{K_{\mathbf{c}}}\rho)^\beta},
\]
so that $d_{\mathbf{c}}=d_{\mathbf{c},1/p}$.

\noindent\emph{Small oscillation.} Let $\bm{\rho}_\star\in\{\ul{\bm{\rho}}_{\mathbf{c}},\ol{\bm{\rho}}_{\mathbf{c}}\}$. By~\cref{eq:spiral_comparison_rotation} of \cref{prop:replace_spiral} and \cref{prop:Lipschitz_SSW}~(vi), we have for every $\mathbf{x}\in K_{\mathbf{c}}$
\[
\big|e_\SSW\big(\bm{\rho}_\star,\mathbf{q}_{\mathbf{c}}\big)-e_\SSW\big(\wt{\bm{\rho}}(\mathbf{x}),\mathbf{q}_{\mathbf{c}}\big)\big|\le C\big((\max_{K_{\mathbf{c}}}\rho)^{4/3}+(\max_{K_{\mathbf{c}}}\rho)^{5/3}+(\max_{K_{\mathbf{c}}}\rho)|\mathbf{q}_{\mathbf{c}}|^2\big)d_{\mathbf{c}} .
\]
The same bound holds with $d_{\mathbf{c},\beta}^a$ in place of $d_{\mathbf{c}}$, for both choices in~\cref{eq:def_a_b}:
\begin{itemize}
\item in~\cref{eq:LSSWA_estim}, we apply~\cref{eq:theta_to_cube_root} to the $\theta$-power comparison in~\cref{eq:spiral_comparison_rotation} of \cref{prop:replace_spiral}, then use \cref{prop:Lipschitz_SSW}~(vi) with $p=3$;
\item in~\cref{eq:LSSWA_sharp_bulk}, $d_{\mathbf{c},\beta}^a=d_{\mathbf{c}}$.
\end{itemize}
Since the left side is bounded by both expressions, we use the factor $d_{\mathbf{c}}$ for the wavevector term and $d_{\mathbf{c},\beta}^a$ for the density powers. The wavevector contribution is at most $C(\max_{K_{\mathbf{c}}}\rho)|\mathbf{q}_{\mathbf{c}}|^2d_{\mathbf{c}}$ and is controlled after integration by~\cref{eq:bound_q} as above. Since $\rho(\mathbf{x})\ge(\max_{K_{\mathbf{c}}}\rho)/2$ on this ball, each density power $(\max_{K_{\mathbf{c}}}\rho)^m d_{\mathbf{c},\beta}^a$ is bounded by
\[
C\rho(\mathbf{x})^{m-\beta a}\operatorname{osc}_{K_{\mathbf{c}}}(\wt{\bm{\rho}}^\beta)^a .
\]

\noindent\emph{Large oscillation.} On a ball of large oscillation the same expression bounds the full density power $\rho(\mathbf{x})^m$ in~\cref{eq:a_priori_LSSWA_2}. Indeed, $\operatorname{osc}_{K_{\mathbf{c}}}(\wt{\bm{\rho}}^\beta)\ge c(\max_{K_{\mathbf{c}}}\rho)^\beta$ by the definition of large oscillation when $\beta=\theta$, and by $d_{\mathbf{c}}>\kappa/(p\theta)$ when $\beta=1/p$.

For either class of balls, Young's inequality gives
\[
\rho^{m-\beta a}\operatorname{osc}_{K_{\mathbf{c}}}(\wt{\bm{\rho}}^\beta)^a
\le \epsilon\rho^{\gamma_m}
+C\epsilon^{1-p/a}\operatorname{osc}_{K_{\mathbf{c}}}(\wt{\bm{\rho}}^\beta)^p,
\qquad \gamma_m=\frac{p(m-\beta a)}{p-a}.
\]
Both $\gamma_{4/3}$ and $\gamma_{5/3}$ lie in $[1,2]$ precisely when
$2-p/(3a)\le p\beta\le1+p/(3a)$. For $\beta=\theta$ these inequalities follow from $p\theta\ge2$ and~\cref{eq:def_a_b}. For $(\beta,a)=(1/p,1)$ they follow from $p>3$. Multiplying by the tile cutoff $\xi$ and integrating, \cref{lem:oscillation}~(i) and $\int\xi\le C\ell^3$ therefore give
\begin{equation}\label{eq:density_interpolation}
\int_{K_{\mathbf{c}}}\rho^{m-\beta a}\operatorname{osc}_{K_{\mathbf{c}}}(\wt{\bm{\rho}}^\beta)^a\xi
\le C\epsilon\int_{K_{\mathbf{c}}}(\rho+\rho^2)\xi
+\frac{C}{\epsilon^b}\int_{K_{\mathbf{c}}}|\grad\wt{\bm{\rho}}^\beta|^p,
\end{equation}
since $\epsilon^{1-p/a}\ell^p=\epsilon^{-b}$.

\medskip

\noindent\textbf{Step 6. (Upper bound)} We use the upper bound of \cref{prop:upper_bound} in the limit $\alpha\uparrow1/2$. The translated tile energies are integrable uniformly in $t\in[1/2,3/2]$ by~\cref{eq:a_priori_LSSWA_2} and the cutoff gradient bound. By~\cref{eq:a_priori_LSSWA} with $f=1$ and $\rho^{5/3}\le\rho+\rho^2$, its last three terms are bounded by $C\epsilon\int(\rho+\rho^2+\rho|\grad\varphi|^2+|\grad\sqrt{\wt{\bm{\rho}}}|^2)$. Let $t\in[1/2,3/2]$. The function
$$
\chi:=(\iden_{t\ell T_j(1-\delta/\ell)\tetra}\ast\eta_{t\delta})(\mathbf{R}\,\cdot\,-t\ell\mathbf{z}-\bm{\tau})
$$
is a smeared tetrahedron of size $t(\ell-\delta)$, of integral $(1-\delta/\ell)^3|t\ell\tetra|$. We call $\mathbf{c}$ its center of mass.

\begin{claim}
For every fixed $\mathbf{c}$, $t$ and $j$,
\begin{multline}\label{eq:estim_local_upper_SSW}
\int_{\SO(3)}\dd\mathbf{R}\,E(\bm{\rho}\,\chi)\le\int_{\SO(3)}\dd\mathbf{R}\,\bigg\{\int_{\RR^3}e_\SSW\big(\wt{\bm{\rho}},\grad\varphi\big)\chi+C\epsilon\int_{\RR^3}\big(\rho+\rho^2+\rho|\grad\varphi|^2\big)\chi\\
+\frac C\epsilon\int_{\RR^3}|\grad\sqrt{\wt{\bm{\rho}}}|^2\chi+C\int_{\RR^3}\rho|\grad\sqrt\chi|^2\bigg\}\\
+\frac{C}{\epsilon^b}\int_{K_{\mathbf{c}}}\Big(|\grad\wt{\bm{\rho}}^\theta|^p+|\grad\wt{\bm{\rho}}^\beta|^p\Big)\\
+\frac{C}{\epsilon^{2p-1}}\left(\int_{K_{\mathbf{c}}}\rho\,|\mathbf{D}^2\varphi|^p
+\int_{K_{\mathbf{c}}\cap\{\grad\varphi\ne0\}}|\grad\varphi|^2\,|\grad\wt{\bm{\rho}}^{1/p}|^p\right),
\end{multline}
where the integral is over the orientations of the tile at fixed center of mass.
\end{claim}

\begin{proof}
\emph{Case I.} The ball $K_{\mathbf{c}}$ has large oscillation, that is, $\omega_{\mathbf{c}}>\kappa(\max_{K_{\mathbf{c}}}\rho)^\theta$. Then \cref{prop:replace_spiral} does not apply, and we use~\cref{eq:a_priori_LSSWA_2} at the frozen wavevector instead. We fix $\mathbf{R}$. Step~3 replaces $\varphi$ by $\mathbf{q}_{\mathbf{c}}\cdot\mathbf{x}$ in $E(\bm{\rho}\,\chi)$. Then~\cref{eq:a_priori_LSSWA_2}, with $f=\chi$ and the constant wavevector $\mathbf{q}_{\mathbf{c}}$ in place of $\grad\varphi$, compares this energy with the gas energy at $\mathbf{q}_{\mathbf{c}}$, and Step~3 replaces $\mathbf{q}_{\mathbf{c}}$ by $\grad\varphi$ in the gas energy. Since $\epsilon\le1$, this gives
\begin{multline*}
E(\bm{\rho}\,\chi)\le\int_{\RR^3}e_\SSW\big(\wt{\bm{\rho}},\grad\varphi\big)\chi+C\epsilon\int_{\RR^3}\big(\rho+\rho^2+\rho|\grad\varphi|^2\big)\chi\\
+\frac C\epsilon\int_{\RR^3}|\grad\sqrt{\wt{\bm{\rho}}}|^2\chi+C\int_{\RR^3}\rho|\grad\sqrt\chi|^2\\
+C\int_{\RR^3}\big(\rho^{4/3}+\rho^{5/3}\big)\chi+C|\mathbf{q}_{\mathbf{c}}|^2\int_{\RR^3}\rho\,\chi+\frac C\epsilon\int_{\RR^3}\rho\,|\grad\varphi-\mathbf{q}_{\mathbf{c}}|^2\chi.
\end{multline*}
The terms on the first two lines occur in~\cref{eq:estim_local_upper_SSW}. By Step~5, $\rho^m\le C\rho^{m-\beta a}\operatorname{osc}_{K_{\mathbf{c}}}(\wt{\bm{\rho}}^\beta)^a$ on this ball for $m=4/3,5/3$, so~\cref{eq:density_interpolation} bounds the first term on the last line by $C\epsilon\int(\rho+\rho^2)\chi+C\epsilon^{-b}\int_{K_{\mathbf{c}}}|\grad\wt{\bm{\rho}}^\beta|^p$. For the second term on the last line, $\int\rho\,\chi\le C\ell^3\max_{K_{\mathbf{c}}}\rho$ and $d_{\mathbf{c}}>\kappa/(p\theta)$, so~\cref{eq:bound_q} gives
$$
C|\mathbf{q}_{\mathbf{c}}|^2\int_{\RR^3}\rho\,\chi\le C|\mathbf{q}_{\mathbf{c}}|^2\ell^3\Big(\max_{K_{\mathbf{c}}}\rho\Big)d_{\mathbf{c}}^p\le C\ell^p\int_{K_{\mathbf{c}}\cap\{\grad\varphi\ne0\}}|\grad\varphi|^2\,|\grad\wt{\bm{\rho}}^{1/p}|^p,
$$
and $\ell^p=\epsilon^{-p}\le\epsilon^{1-2p}$. Finally, \cref{lem:weighted_wavevector} with $\xi=\chi$ bounds the last term by $C\epsilon\int\rho\,\chi$ plus the last two terms of~\cref{eq:estim_local_upper_SSW}, since $\ell^p/\epsilon\le\epsilon^{1-2p}$ and $\ell^p/\epsilon^{p-1}=\epsilon^{1-2p}$. Integrating over $\mathbf{R}$ gives~\cref{eq:estim_local_upper_SSW}.

\noindent\emph{Case II.} The ball $K_{\mathbf{c}}$ has small oscillation, that is, $\omega_{\mathbf{c}}\le\kappa(\max_{K_{\mathbf{c}}}\rho)^\theta$, so that the comparison densities $\ul{\bm{\rho}}_{\mathbf{c}},\ol{\bm{\rho}}_{\mathbf{c}}$ of \cref{prop:replace_spiral} are available. We fix $\mathbf{R}$. Step~3 replaces $\varphi$ by $\mathbf{q}_{\mathbf{c}}\cdot\mathbf{x}$ in $E(\bm{\rho}\,\chi)$, and the upper bound of Step~4 replaces $\wt{\bm{\rho}}$ by $\ul{\bm{\rho}}_{\mathbf{c}}$. Since $\epsilon\le1$, this gives
\begin{multline*}
E(\bm{\rho}\,\chi)\le E\big(\mathbf{U}_{\mathbf{q}_{\mathbf{c}}}\ul{\bm{\rho}}_{\mathbf{c}}\mathbf{U}_{\mathbf{q}_{\mathbf{c}}}^*\chi\big)+C\epsilon\int_{\RR^3}\big(\rho+\rho^2+\rho|\grad\varphi|^2\big)\chi\\
+\frac C\epsilon\int_{\RR^3}|\grad\sqrt{\wt{\bm{\rho}}}|^2\chi+C\int_{\RR^3}\rho|\grad\sqrt\chi|^2\\
+\frac{C\ell^p}{\epsilon^s}\int_{K_{\mathbf{c}}}|\grad\wt{\bm{\rho}}^\theta|^p+C|\mathbf{q}_{\mathbf{c}}|^2\ell^3\Big(\max_{K_{\mathbf{c}}}\rho\Big)L_{\mathbf{c}}+\frac C\epsilon\int_{\RR^3}\rho\,|\grad\varphi-\mathbf{q}_{\mathbf{c}}|^2\chi.
\end{multline*}
Apart from the first one, the terms on the first two lines occur in~\cref{eq:estim_local_upper_SSW}. On the last line, $\ell^p\epsilon^{-s}\le\epsilon^{-b}$ since $p+s\le b$, the second term was bounded in Step~4, and \cref{lem:weighted_wavevector} bounds the last term as in Case~I.

Next we integrate over $\mathbf{R}$. By~\cref{eq:covariance_spiral}, the energy of the comparison density is
$$
\Big(1-\frac\delta\ell\Big)^3|t\ell\tetra|\,e_{t(\ell-\delta)\tetra,\eta_{t\delta}}\big(\ul{\bm{\rho}}_{\mathbf{c}},\mathbf{R}_j^{-1}\mathbf{R}\mathbf{q}_{\mathbf{c}}\big).
$$
The density $\ul{\bm{\rho}}_{\mathbf{c}}$ and the vector $\mathbf{q}_{\mathbf{c}}$ do not depend on $\mathbf{R}$. Integrating over $\mathbf{R}$ we therefore obtain the averaged energy $\ol{e}_{t(\ell-\delta)\tetra,\eta_{t\delta}}(\ul{\bm{\rho}}_{\mathbf{c}},\mathbf{q}_{\mathbf{c}})$, to which we apply the upper bound~\cref{eq:rate_SSW_upper} of \cref{prop:rate_SSW}. Since $\ul{\rho}_{\mathbf{c}}\le\rho$ on $K_{\mathbf{c}}$, we find
\begin{multline*}
\int_{\SO(3)}\dd\mathbf{R}\,E\big(\mathbf{U}_{\mathbf{q}_{\mathbf{c}}}\ul{\bm{\rho}}_{\mathbf{c}}\mathbf{U}_{\mathbf{q}_{\mathbf{c}}}^*\chi\big)\\
\le\int_{\SO(3)}\dd\mathbf{R}\int_{\RR^3}\Big(e_\SSW\big(\ul{\bm{\rho}}_{\mathbf{c}},\mathbf{q}_{\mathbf{c}}\big)+C\epsilon\big(\rho+\rho^2\big)\Big)\chi+C\epsilon|\mathbf{q}_{\mathbf{c}}|^2\ell^3\max_{K_{\mathbf{c}}}\rho,
\end{multline*}
where the last term is at most $C\epsilon\int\rho|\grad\varphi|^2\chi$ by Step~4.

It remains to return to the gas energy of $\wt{\bm{\rho}}$ at $\grad\varphi$. By Step~5, for every $\mathbf{x}\in K_{\mathbf{c}}$,
\begin{multline*}
e_\SSW\big(\ul{\bm{\rho}}_{\mathbf{c}},\mathbf{q}_{\mathbf{c}}\big)\le e_\SSW\big(\wt{\bm{\rho}}(\mathbf{x}),\mathbf{q}_{\mathbf{c}}\big)+C\Big(\big(\max_{K_{\mathbf{c}}}\rho\big)^{4/3}+\big(\max_{K_{\mathbf{c}}}\rho\big)^{5/3}\Big)d_{\mathbf{c},\beta}^a\\
+C\Big(\max_{K_{\mathbf{c}}}\rho\Big)|\mathbf{q}_{\mathbf{c}}|^2d_{\mathbf{c}}.
\end{multline*}
For the density powers, Step~5 and~\cref{eq:density_interpolation} give, for $m=4/3,5/3$,
\begin{align*}
\Big(\max_{K_{\mathbf{c}}}\rho\Big)^m d_{\mathbf{c},\beta}^a\int_{\RR^3}\chi
&\le C\int_{K_{\mathbf{c}}}\rho^{m-\beta a}\operatorname{osc}_{K_{\mathbf{c}}}(\wt{\bm{\rho}}^\beta)^a\chi\\
&\le C\epsilon\int_{K_{\mathbf{c}}}(\rho+\rho^2)\chi+\frac{C}{\epsilon^b}\int_{K_{\mathbf{c}}}|\grad\wt{\bm{\rho}}^\beta|^p.
\end{align*}
Since $\int\chi\le C\ell^3$ and $d_{\mathbf{c}}\le L_{\mathbf{c}}$, the last term gives at most $C|\mathbf{q}_{\mathbf{c}}|^2\ell^3(\max_{K_{\mathbf{c}}}\rho)L_{\mathbf{c}}$, which was bounded in Step~4. Finally, Step~3 replaces $\mathbf{q}_{\mathbf{c}}$ by $\grad\varphi$ in the gas energy, with an error bounded as in Case~I. Collecting the terms gives~\cref{eq:estim_local_upper_SSW}.
\end{proof}

It remains to sum over the tiles. We fix $t\in[1/2,3/2]$ and $j$. For fixed $\mathbf{R}$ the center of mass of $\chi$ is $\mathbf{c}=\mathbf{R}^{-1}(t\ell(\mathbf{z}-\mathbf{z}_j)+\bm{\tau})$, and
$$
\chi(\mathbf{x})=(\iden_{t\ell T_j(1-\delta/\ell)\tetra}\ast\eta_{t\delta})\big(\mathbf{R}(\mathbf{x}-\mathbf{c})-t\ell\mathbf{z}_j\big)
$$
depends on $\mathbf{z}$ and $\bm{\tau}$ only through $\mathbf{c}$. The map $(\mathbf{z},\bm{\tau})\mapsto\mathbf{c}$ is a bijection from $\ZZ^3\times C_{t\ell}$ onto $\RR^3$ which preserves the Lebesgue measure. The sum over $\mathbf{z}$ and the average over $\bm{\tau}$ in \cref{prop:upper_bound} therefore become an integral over the center of mass,
$$
\int_{\SO(3)}\dd\mathbf{R}\,\dashint_{C_{t\ell}}\dd\bm{\tau}\sum_{\mathbf{z}\in\ZZ^3}E(\bm{\rho}\,\chi)
=\frac{1}{(t\ell)^3}\int_{\RR^3}\dd\mathbf{c}\int_{\SO(3)}\dd\mathbf{R}\,E(\bm{\rho}\,\chi).
$$
Here we have exchanged the integrals over $\mathbf{R}$ and $\mathbf{c}$. This is allowed by Tonelli's theorem applied to $E(\bm{\rho}\,\chi)+c_\LO\int\rho^{4/3}\chi\ge0$, since the integral of $\int\rho^{4/3}\chi$ over $\mathbf{c}$ is $(1-\delta/\ell)^3|t\ell\tetra|\int\rho^{4/3}<\infty$.

We apply the Claim at every $\mathbf{c}$ and integrate its right side over $\mathbf{c}$. For every $\mathbf{x}\in\RR^3$ and $\mathbf{R}\in\SO(3)$, the change of variables $\mathbf{y}=\mathbf{R}(\mathbf{x}-\mathbf{c})-t\ell\mathbf{z}_j$ and the definition of $K_{\mathbf{c}}$ give
\begin{gather*}
\int_{\RR^3}\chi(\mathbf{x})\,\dd\mathbf{c}=\Big(1-\frac\delta\ell\Big)^3|t\ell\tetra|,\qquad
\int_{\RR^3}|\grad\sqrt\chi(\mathbf{x})|^2\,\dd\mathbf{c}\le\frac{C\ell^2}{\delta},\\
\int_{\RR^3}\iden_{K_{\mathbf{c}}}(\mathbf{x})\,\dd\mathbf{c}=\frac{32\pi}{3}\ell^3.
\end{gather*}
For the second bound, \cref{thm:Glaeser} gives $|\grad\sqrt\chi|^2\le C\delta^{-2}$, and $\grad\chi$ is supported in a layer of volume $C\ell^2\delta$ around the boundary of the tile. We use $|t\ell\tetra|=(t\ell)^3/24$ and $t\ge1/2$, sum over $j$ and divide by $(t\ell)^3$. First, every integral against $\chi$ in~\cref{eq:estim_local_upper_SSW} becomes the same integral over $\RR^3$, multiplied by $(1-\delta/\ell)^3\le1$. Second, the cutoff term $C\int\rho|\grad\sqrt\chi|^2$ becomes at most $C(\ell\delta)^{-1}\int\rho=C\epsilon\int\rho$. Third, by Tonelli's theorem and the third identity above, for every function $f\ge0$,
$$
\frac{1}{\ell^3}\int_{\RR^3}\dd\mathbf{c}\int_{K_{\mathbf{c}}}f=\frac{1}{\ell^3}\int_{\RR^3}\dd\mathbf{x}\,f(\mathbf{x})\int_{\RR^3}\dd\mathbf{c}\,\iden_{K_{\mathbf{c}}}(\mathbf{x})=\frac{32\pi}{3}\int_{\RR^3}f,
$$
and the same holds with $K_{\mathbf{c}}\cap\{\grad\varphi\ne0\}$ and $\{\grad\varphi\ne0\}$ in place of $K_{\mathbf{c}}$ and $\RR^3$. Hence each integral over $K_{\mathbf{c}}$ becomes the corresponding integral over $\RR^3$, multiplied by a constant and with the same power of $\epsilon$. For the gas energy we use $1-(1-\delta/\ell)^3\le3\delta/\ell=3\epsilon$ and $-e_\SSW(\wt{\bm{\rho}},\grad\varphi)\le c_\LO\rho^{4/3}\le c_\LO(\rho+\rho^2)$, so that
$$
\Big(1-\frac\delta\ell\Big)^3\int_{\RR^3}e_\SSW\big(\wt{\bm{\rho}},\grad\varphi\big)\le\int_{\RR^3}e_\SSW\big(\wt{\bm{\rho}},\grad\varphi\big)+3c_\LO\,\epsilon\int_{\RR^3}\big(\rho+\rho^2\big).
$$
These bounds are uniform in $t$, and the weight $\dd t/t^4$ is normalized, so the average over $t$ does not change them. Adding the error terms of \cref{prop:upper_bound}, which were bounded at the beginning of Step~6, we obtain
\begin{multline*}
E(\bm{\rho})\le\int_{\RR^3}e_\SSW\big(\wt{\bm{\rho}},\grad\varphi\big)+C\epsilon\int_{\RR^3}\big(\rho+\rho^2+\rho|\grad\varphi|^2\big)+\frac{C}{\epsilon}\int_{\RR^3}|\grad\sqrt{\wt{\bm{\rho}}}|^2\\
+\frac{C}{\epsilon^b}\int_{\RR^3}\Big(|\grad\wt{\bm{\rho}}^\theta|^p+|\grad\wt{\bm{\rho}}^\beta|^p\Big)
+\frac{C}{\epsilon^{2p-1}}\left(\int_{\RR^3}\rho\,|\mathbf{D}^2\varphi|^p+\int_{\{\grad\varphi\ne0\}}|\grad\varphi|^2\,|\grad\wt{\bm{\rho}}^{1/p}|^p\right).
\end{multline*}
For~\cref{eq:LSSWA_estim} we have $\beta=\theta$, so the two density gradients coincide. For~\cref{eq:LSSWA_sharp_bulk} they are the gradients of the $2/p$ and $1/p$ powers, and the reduction at the start of the proof gives the stated bound.

\medskip

\noindent\textbf{Step 7. (Lower bound)} For $t\in[1/2,3/2]$ we apply \cref{prop:lower_bound} with $t\ell$ and $t\delta$ in place of $\ell$ and $\delta$, and we average over $t$ with the weighted mean $\dashint_{1/2}^{3/2}\frac{\dd t}{t^4}$. The tiles are now
$$
\xi_t:=(\iden_{t\ell\tetra_j}\ast\eta_{t\delta})(\mathbf{R}\,\cdot\,-t\ell\mathbf{z}-\bm{\tau}).
$$
They have integral $|t\ell\tetra|=(t\ell)^3/24$. As in Step~6 we call $\mathbf{c}=\mathbf{R}^{-1}(t\ell(\mathbf{z}-\mathbf{z}_j)+\bm{\tau})$ the center of mass of $\xi_t$. For fixed $t$, $\mathbf{R}$ and $j$ the sum over $\mathbf{z}$ and the average over $\bm{\tau}$ become $(t\ell)^{-3}\int_{\RR^3}\dd\mathbf{c}$, and by Tonelli's theorem, applied as in Step~6, we may exchange the integrals over $t$, $\mathbf{R}$ and $\mathbf{c}$. Since $\delta=1$ and $\ell=\epsilon^{-1}$, the last term in \cref{prop:lower_bound} is at most $C\epsilon\int(\rho+\rho^2)$ for every $t\in[1/2,3/2]$. We obtain
$$
E(\bm{\rho})\ge(1-C\epsilon)\sum_{j=1}^{24}\int_{\RR^3}\dd\mathbf{c}\,\dashint_{1/2}^{3/2}\frac{\dd t}{t^4}\int_{\SO(3)}\dd\mathbf{R}\,\frac{E(\bm{\rho}\,\xi_t)}{(t\ell)^3}-C\epsilon\int_{\RR^3}(\rho+\rho^2).
$$
By \cref{cor:upper_bound_E_rho}, with parameter $1$ in place of $\epsilon$, and~\cref{eq:a_priori_LSSWA} with $f=\xi_t$, we have
$$
E(\bm{\rho}\,\xi_t)\le C\int_{\RR^3}\big(\rho^{5/3}+\rho|\grad\varphi|^2+|\grad\sqrt{\wt{\bm{\rho}}}|^2\big)\xi_t+C\int_{\RR^3}\rho|\grad\sqrt{\xi_t}|^2 .
$$
Summing over the tiles as below, the sum on the right of the lower bound for $E(\bm{\rho})$ is therefore at most $C\int(\rho+\rho^2+\rho|\grad\varphi|^2+|\grad\sqrt{\wt{\bm{\rho}}}|^2)$. Hence replacing the factor $1-C\epsilon$ by $1$ costs at most this bound times $C\epsilon$, and
\begin{multline*}
E(\bm{\rho})\ge\sum_{j=1}^{24}\int_{\RR^3}\dd\mathbf{c}\,\dashint_{1/2}^{3/2}\frac{\dd t}{t^4}\int_{\SO(3)}\dd\mathbf{R}\,\frac{E(\bm{\rho}\,\xi_t)}{(t\ell)^3}\\
-C\epsilon\int_{\RR^3}\big(\rho+\rho^2+\rho|\grad\varphi|^2+|\grad\sqrt{\wt{\bm{\rho}}}|^2\big).
\end{multline*}
The lower bound of \cref{prop:rate_SSW} only holds for the average over the sizes of the tiles. Unlike in~\cref{eq:estim_local_upper_SSW}, we therefore divide by $(t\ell)^3$ and average over $t$ before comparing with the gas energy.

\begin{claim}
For every fixed $\mathbf{c}$ and $j$,
\begin{multline}\label{eq:estim_local_lower_SSW}
\dashint_{1/2}^{3/2}\frac{\dd t}{t^4}\int_{\SO(3)}\dd\mathbf{R}\,\frac{E(\bm{\rho}\,\xi_t)}{(t\ell)^3}\\
\ge\dashint_{1/2}^{3/2}\frac{\dd t}{t^4}\int_{\SO(3)}\dd\mathbf{R}\,\frac{1}{(t\ell)^3}\bigg\{\int_{\RR^3}e_\SSW\big(\wt{\bm{\rho}},\grad\varphi\big)\xi_t-C\epsilon\int_{\RR^3}\big(\rho+\rho^2+\rho|\grad\varphi|^2\big)\xi_t\\
-\frac C\epsilon\int_{\RR^3}|\grad\sqrt{\wt{\bm{\rho}}}|^2\xi_t-C\int_{\RR^3}\rho|\grad\sqrt{\xi_t}|^2\bigg\}\\
-\frac{C}{\ell^3\epsilon^b}\int_{K_{\mathbf{c}}}\Big(|\grad\wt{\bm{\rho}}^\theta|^p+|\grad\wt{\bm{\rho}}^\beta|^p\Big)\\
-\frac{C}{\ell^3\epsilon^{2p-1}}\left(\int_{K_{\mathbf{c}}}\rho\,|\mathbf{D}^2\varphi|^p
+\int_{K_{\mathbf{c}}\cap\{\grad\varphi\ne0\}}|\grad\varphi|^2\,|\grad\wt{\bm{\rho}}^{1/p}|^p\right),
\end{multline}
where $\xi_t$ is the tile of size $t\ell$, orientation $\mathbf{R}$ and center of mass $\mathbf{c}$.
\end{claim}

\begin{proof}
\emph{Case I.} The ball $K_{\mathbf{c}}$ has large oscillation, that is, $\omega_{\mathbf{c}}>\kappa(\max_{K_{\mathbf{c}}}\rho)^\theta$. We fix $t$ and $\mathbf{R}$ and argue as in Case~I of Step~6, with lower instead of upper bounds. Step~3 replaces $\varphi$ by $\mathbf{q}_{\mathbf{c}}\cdot\mathbf{x}$ in $E(\bm{\rho}\,\xi_t)$. Then the lower bound in~\cref{eq:a_priori_LSSWA_2}, with $f=\xi_t$ and the constant wavevector $\mathbf{q}_{\mathbf{c}}$ in place of $\grad\varphi$, compares this energy with the gas energy at $\mathbf{q}_{\mathbf{c}}$, and Step~3 replaces $\mathbf{q}_{\mathbf{c}}$ by $\grad\varphi$ in the gas energy. Since $\epsilon\le1$, this gives
\begin{multline*}
E(\bm{\rho}\,\xi_t)\ge\int_{\RR^3}e_\SSW\big(\wt{\bm{\rho}},\grad\varphi\big)\xi_t-C\epsilon\int_{\RR^3}\big(\rho+\rho^2+\rho|\grad\varphi|^2\big)\xi_t\\
-\frac C\epsilon\int_{\RR^3}|\grad\sqrt{\wt{\bm{\rho}}}|^2\xi_t-C\int_{\RR^3}\rho|\grad\sqrt{\xi_t}|^2\\
-C\int_{\RR^3}\big(\rho^{4/3}+\rho^{5/3}\big)\xi_t-C|\mathbf{q}_{\mathbf{c}}|^2\int_{\RR^3}\rho\,\xi_t-\frac C\epsilon\int_{\RR^3}\rho\,|\grad\varphi-\mathbf{q}_{\mathbf{c}}|^2\xi_t.
\end{multline*}
The terms on the first two lines occur in the braces of~\cref{eq:estim_local_lower_SSW}. The terms on the last line are bounded as in Case~I of Step~6, by~\cref{eq:density_interpolation}, by~\cref{eq:bound_q} since $d_{\mathbf{c}}>\kappa/(p\theta)$, and by \cref{lem:weighted_wavevector} with $\xi=\xi_t$. Thus the last line is at least
\begin{multline*}
-C\epsilon\int_{\RR^3}(\rho+\rho^2)\xi_t-\frac{C}{\epsilon^b}\int_{K_{\mathbf{c}}}|\grad\wt{\bm{\rho}}^\beta|^p\\
-\frac{C}{\epsilon^{2p-1}}\bigg(\int_{K_{\mathbf{c}}}\rho\,|\mathbf{D}^2\varphi|^p+\int_{K_{\mathbf{c}}\cap\{\grad\varphi\ne0\}}|\grad\varphi|^2\,|\grad\wt{\bm{\rho}}^{1/p}|^p\bigg).
\end{multline*}
No average over $t$ is needed here. We divide by $(t\ell)^3$, use $(t\ell)^{-3}\le8\ell^{-3}$ for the integrals over $K_{\mathbf{c}}$ and average over $\mathbf{R}$ and $t$.

\noindent\emph{Case II.} The ball $K_{\mathbf{c}}$ has small oscillation, that is, $\omega_{\mathbf{c}}\le\kappa(\max_{K_{\mathbf{c}}}\rho)^\theta$. We fix $t$ and $\mathbf{R}$. Step~3 replaces $\varphi$ by $\mathbf{q}_{\mathbf{c}}\cdot\mathbf{x}$ in $E(\bm{\rho}\,\xi_t)$, and the lower bound of Step~4 replaces $\wt{\bm{\rho}}$ by $\ol{\bm{\rho}}_{\mathbf{c}}$. Since $\epsilon\le1$, this gives
\begin{multline*}
E(\bm{\rho}\,\xi_t)\ge E\big(\mathbf{U}_{\mathbf{q}_{\mathbf{c}}}\ol{\bm{\rho}}_{\mathbf{c}}\mathbf{U}_{\mathbf{q}_{\mathbf{c}}}^*\xi_t\big)-C\epsilon\int_{\RR^3}\big(\rho+\rho^2+\rho|\grad\varphi|^2\big)\xi_t\\
-\frac C\epsilon\int_{\RR^3}|\grad\sqrt{\wt{\bm{\rho}}}|^2\xi_t-C\int_{\RR^3}\rho|\grad\sqrt{\xi_t}|^2\\
-\frac{C\ell^p}{\epsilon^s}\int_{K_{\mathbf{c}}}|\grad\wt{\bm{\rho}}^\theta|^p-C|\mathbf{q}_{\mathbf{c}}|^2\ell^3\Big(\max_{K_{\mathbf{c}}}\rho\Big)L_{\mathbf{c}}-\frac C\epsilon\int_{\RR^3}\rho\,|\grad\varphi-\mathbf{q}_{\mathbf{c}}|^2\xi_t.
\end{multline*}
Apart from the first one, the terms on the first two lines occur in the braces of~\cref{eq:estim_local_lower_SSW}. On the last line, $\ell^p\epsilon^{-s}\le\epsilon^{-b}$ since $p+s\le b$, the second term was bounded in Step~4, and \cref{lem:weighted_wavevector} bounds the last term as in Case~I.

Next we integrate over $\mathbf{R}$ and average over $t$. By~\cref{eq:covariance_spiral}, the energy of the comparison density is
$$
|t\ell\tetra|\,e_{t\ell\tetra,\eta_{t\delta}}\big(\ol{\bm{\rho}}_{\mathbf{c}},\mathbf{R}_j^{-1}\mathbf{R}\mathbf{q}_{\mathbf{c}}\big).
$$
The density $\ol{\bm{\rho}}_{\mathbf{c}}$ and the vector $\mathbf{q}_{\mathbf{c}}$ depend neither on $\mathbf{R}$ nor on $t$. Integrating over $\mathbf{R}$ therefore gives $|t\ell\tetra|\,\ol{e}_{t\ell\tetra,\eta_{t\delta}}(\ol{\bm{\rho}}_{\mathbf{c}},\mathbf{q}_{\mathbf{c}})$, and after division by $(t\ell)^3$ the prefactor $|t\ell\tetra|/(t\ell)^3=1/24$ does not depend on $t$. We now average over $t$ and use the lower bound~\cref{eq:rate_SSW_lower} of \cref{prop:rate_SSW}. Since $\ol{\rho}_{\mathbf{c}}\le C\max_{K_{\mathbf{c}}}\rho$, we find
$$
\dashint_{1/2}^{3/2}\ol{e}_{t\ell\tetra,\eta_{t\delta}}\big(\ol{\bm{\rho}}_{\mathbf{c}},\mathbf{q}_{\mathbf{c}}\big)\,\frac{\dd t}{t^4}
\ge e_\SSW\big(\ol{\bm{\rho}}_{\mathbf{c}},\mathbf{q}_{\mathbf{c}}\big)-C\epsilon\big(\max_{K_{\mathbf{c}}}\rho+(\max_{K_{\mathbf{c}}}\rho)^2\big)-C\epsilon|\mathbf{q}_{\mathbf{c}}|^2\max_{K_{\mathbf{c}}}\rho.
$$
Each term on the right is a constant, hence equal to its average against $\xi_t/|t\ell\tetra|$ for every $t$ and $\mathbf{R}$. In the last two terms we use $\rho\ge(\max_{K_{\mathbf{c}}}\rho)/2$ on $K_{\mathbf{c}}$, by \cref{lem:oscillation}~(iii), and $|\mathbf{q}_{\mathbf{c}}|\le|\grad\varphi|$. This bounds them by the averages of $C\epsilon(\rho+\rho^2+\rho|\grad\varphi|^2)$.

It remains to return to the gas energy of $\wt{\bm{\rho}}$ at $\grad\varphi$. By Step~5, for every $\mathbf{x}\in K_{\mathbf{c}}$,
\begin{multline*}
e_\SSW\big(\ol{\bm{\rho}}_{\mathbf{c}},\mathbf{q}_{\mathbf{c}}\big)\ge e_\SSW\big(\wt{\bm{\rho}}(\mathbf{x}),\mathbf{q}_{\mathbf{c}}\big)-C\Big(\big(\max_{K_{\mathbf{c}}}\rho\big)^{4/3}+\big(\max_{K_{\mathbf{c}}}\rho\big)^{5/3}\Big)d_{\mathbf{c},\beta}^a\\
-C\Big(\max_{K_{\mathbf{c}}}\rho\Big)|\mathbf{q}_{\mathbf{c}}|^2d_{\mathbf{c}}.
\end{multline*}
The density powers are bounded as in Case~II of Step~6, with $\xi_t$ in place of $\chi$. Since $\int\xi_t\le C\ell^3$ and $d_{\mathbf{c}}\le L_{\mathbf{c}}$, the last term gives at most $C|\mathbf{q}_{\mathbf{c}}|^2\ell^3(\max_{K_{\mathbf{c}}}\rho)L_{\mathbf{c}}$, which was bounded in Step~4. Finally, Step~3 replaces $\mathbf{q}_{\mathbf{c}}$ by $\grad\varphi$ in the gas energy, with an error bounded as in Case~I. We divide by $(t\ell)^3$ and use $(t\ell)^{-3}\le8\ell^{-3}$ for the integrals over $K_{\mathbf{c}}$. Collecting the terms gives~\cref{eq:estim_local_lower_SSW}.
\end{proof}

It remains to sum over the tiles. We apply the Claim at every $\mathbf{c}$, integrate over $\mathbf{c}$ and sum over $j$. Since the tiles $\xi_t$ are not shrunk, the change of variables $\mathbf{y}=\mathbf{R}(\mathbf{x}-\mathbf{c})-t\ell\mathbf{z}_j$ of Step~6 now gives
$$
\int_{\RR^3}\xi_t(\mathbf{x})\,\dd\mathbf{c}=|t\ell\tetra|=\frac{(t\ell)^3}{24},\qquad
\int_{\RR^3}|\grad\sqrt{\xi_t}(\mathbf{x})|^2\,\dd\mathbf{c}\le\frac{C\ell^2}{\delta},
$$
while $\int\iden_{K_{\mathbf{c}}}(\mathbf{x})\,\dd\mathbf{c}=\frac{32\pi}{3}\ell^3$ as before. Using the lower bound for $E(\bm{\rho})$ above, and then~\cref{eq:estim_local_lower_SSW} integrated over $\mathbf{c}$ and summed over $j$, we obtain
\begin{align*}
E(\bm{\rho})&\ge\sum_{j=1}^{24}\int_{\RR^3}\dd\mathbf{c}\,\dashint_{1/2}^{3/2}\frac{\dd t}{t^4}\int_{\SO(3)}\dd\mathbf{R}\,\frac{E(\bm{\rho}\,\xi_t)}{(t\ell)^3}\\
&\qquad-C\epsilon\int_{\RR^3}\big(\rho+\rho^2+\rho|\grad\varphi|^2+|\grad\sqrt{\wt{\bm{\rho}}}|^2\big)\\
&\ge\dashint_{1/2}^{3/2}\frac{\dd t}{t^4}\int_{\SO(3)}\dd\mathbf{R}\sum_{j=1}^{24}\frac{1}{(t\ell)^3}\int_{\RR^3}\dd\mathbf{c}\,\bigg\{\int_{\RR^3}e_\SSW\big(\wt{\bm{\rho}},\grad\varphi\big)\xi_t\\
&\qquad\qquad-C\epsilon\int_{\RR^3}\big(\rho+\rho^2+\rho|\grad\varphi|^2\big)\xi_t-\frac C\epsilon\int_{\RR^3}|\grad\sqrt{\wt{\bm{\rho}}}|^2\xi_t\\
&\qquad\qquad-C\int_{\RR^3}\rho|\grad\sqrt{\xi_t}|^2\bigg\}\\
&\qquad-\frac{C}{\epsilon^b}\sum_{j=1}^{24}\frac{1}{\ell^3}\int_{\RR^3}\dd\mathbf{c}\int_{K_{\mathbf{c}}}\Big(|\grad\wt{\bm{\rho}}^\theta|^p+|\grad\wt{\bm{\rho}}^\beta|^p\Big)\\
&\qquad-\frac{C}{\epsilon^{2p-1}}\sum_{j=1}^{24}\frac{1}{\ell^3}\int_{\RR^3}\dd\mathbf{c}\bigg(\int_{K_{\mathbf{c}}}\rho\,|\mathbf{D}^2\varphi|^p+\int_{K_{\mathbf{c}}\cap\{\grad\varphi\ne0\}}|\grad\varphi|^2\,|\grad\wt{\bm{\rho}}^{1/p}|^p\bigg)\\
&\qquad-C\epsilon\int_{\RR^3}\big(\rho+\rho^2+\rho|\grad\varphi|^2+|\grad\sqrt{\wt{\bm{\rho}}}|^2\big).
\end{align*}
In the last expression, first, after the sum over $j$, every integral against $\xi_t/(t\ell)^3$ becomes the same integral over $\RR^3$, without any prefactor, since the $24$ tiles sum to one after the translation average. In particular, the gas energy term becomes exactly $\int e_\SSW(\wt{\bm{\rho}},\grad\varphi)$, and no correction as in Step~6 is needed. Second, using $t\ge1/2$, the cutoff term $C(t\ell)^{-3}\int\rho|\grad\sqrt{\xi_t}|^2$ becomes at most $C(\ell\delta)^{-1}\int\rho=C\epsilon\int\rho$. Third, by the identity for the integrals over $K_{\mathbf{c}}$ in Step~6, each such integral, which carries the factor $\ell^{-3}$ in~\cref{eq:estim_local_lower_SSW}, becomes the corresponding integral over $\RR^3$ multiplied by a constant, with the same power of $\epsilon$. These bounds are uniform in $t$, and the weight $\dd t/t^4$ is normalized, so the average over $t$ does not change them. Since $\epsilon\le1$, the last term is at most $C\epsilon\int(\rho+\rho^2+\rho|\grad\varphi|^2)+C\epsilon^{-1}\int|\grad\sqrt{\wt{\bm{\rho}}}|^2$, and we obtain
\begin{multline*}
E(\bm{\rho})\ge\int_{\RR^3}e_\SSW\big(\wt{\bm{\rho}},\grad\varphi\big)-C\epsilon\int_{\RR^3}\big(\rho+\rho^2+\rho|\grad\varphi|^2\big)-\frac{C}{\epsilon}\int_{\RR^3}|\grad\sqrt{\wt{\bm{\rho}}}|^2\\
-\frac{C}{\epsilon^b}\int_{\RR^3}\Big(|\grad\wt{\bm{\rho}}^\theta|^p+|\grad\wt{\bm{\rho}}^\beta|^p\Big)
-\frac{C}{\epsilon^{2p-1}}\left(\int_{\RR^3}\rho\,|\mathbf{D}^2\varphi|^p+\int_{\{\grad\varphi\ne0\}}|\grad\varphi|^2\,|\grad\wt{\bm{\rho}}^{1/p}|^p\right).
\end{multline*}
As at the end of Step~6, this gives the lower bounds in~\cref{eq:LSSWA_estim,eq:LSSWA_sharp_bulk} for $0<\epsilon\le\epsilon_0$.

\medskip

\noindent\textbf{Step 8. (Remaining values of $\epsilon$)} We have proved~\cref{eq:LSSWA_estim}, and~\cref{eq:LSSWA_sharp_bulk} under its additional assumption, for $0<\epsilon\le\epsilon_0$ with $C\epsilon$ in place of $\epsilon$ in front of the first integral. Changing $\epsilon$ into $\epsilon/C$ gives both bounds for $0<\epsilon\le\epsilon_1:=C\epsilon_0$, with larger constants in front of the other terms. On the other hand,~\cref{eq:a_priori_LSSWA_2} with $f=1$ and $\rho^{4/3}+\rho^{5/3}\le2(\rho+\rho^2)$ bounds the error by $C_1\int(\rho+\rho^2+\rho|\grad\varphi|^2)+C_1\int|\grad\sqrt{\wt{\bm{\rho}}}|^2$ with a universal constant $C_1$. Since $C(1+\epsilon)/\epsilon\ge C$, this gives both bounds for $\epsilon\ge C_1$ once $C\ge C_1$. Finally, for $\epsilon_1\le\epsilon\le C_1$ we use the bound at $\epsilon_1$. Its first term is at most $\epsilon\int(\rho+\rho^2+\rho|\grad\varphi|^2)$, and its other coefficients are larger than those at $\epsilon$ by at most a factor depending only on $\epsilon_1$ and $C_1$. After increasing $C$, this proves both bounds for every $\epsilon>0$.
\qed

\subsection{Proof of \texorpdfstring{\cref{cor:gradient}}{the azimuthal gradient form}}\label{sec:proof_gradient}
The matrix $\wt{\bm{\rho}}$ is real, it has the total density $\rho$ and the magnetization $(a,0,m_3)$. By~\cref{eq:canonical_phase}, the magnetization of $\mathbf{U}_\varphi\wt{\bm{\rho}}\,\mathbf{U}_\varphi^*$ is therefore $\mathbf{m}$, that is, $\bm{\rho}=\mathbf{U}_\varphi\wt{\bm{\rho}}\,\mathbf{U}_\varphi^*$. Apply \cref{thm:LSSWA} to $\wt{\bm{\rho}}$ and $\varphi$. For every $\mathbf{x}$, the matrix $\bm{\rho}(\mathbf{x})$ is obtained from $\wt{\bm{\rho}}(\mathbf{x})$ by a rotation $\mathbf{W}$ of the spin about $\mathbf{e}_3$. Such a rotation commutes with $\mathbf{U}_{\mathbf{q}}$, hence $E(\mathbf{U}_{\mathbf{q}}\mathbf{W}\bm{\rho}_0\mathbf{W}^*\mathbf{U}_{\mathbf{q}}^*f)=E(\mathbf{U}_{\mathbf{q}}\bm{\rho}_0\mathbf{U}_{\mathbf{q}}^*f)$ by~\cref{eq:invariance} and $e_\SSW(\mathbf{W}\bm{\rho}_0\mathbf{W}^*,\mathbf{q})=e_\SSW(\bm{\rho}_0,\mathbf{q})$. Therefore $e_\SSW(\wt{\bm{\rho}}(\mathbf{x}),\grad\varphi(\mathbf{x}))=e_\SSW(\bm{\rho}(\mathbf{x}),\grad\varphi(\mathbf{x}))$, which gives~\cref{eq:gradient_approx}.

Let $\mu:=m_1+im_2$. Differentiating~\cref{eq:canonical_phase} we find $\ol\mu\,\grad\mu=a\grad a+ia^2\grad\varphi$, whose imaginary part is the first formula in~\cref{eq:azimuthal_gradient} since $a^2=|\mu|^2$. For the second formula we differentiate $\sqrt{\bm{\rho}}=\mathbf{U}_\varphi\sqrt{\wt{\bm{\rho}}}\,\mathbf{U}_\varphi^*$, which gives
$$
\partial_{x_j}\sqrt{\bm{\rho}}=\mathbf{U}_\varphi\Big(\partial_{x_j}\sqrt{\wt{\bm{\rho}}}-\frac i2\partial_{x_j}\varphi\,\big[\bm{\sigma}_3,\sqrt{\wt{\bm{\rho}}}\big]\Big)\mathbf{U}_\varphi^*.
$$
The matrix $\sqrt{\wt{\bm{\rho}}}$ is real and symmetric. The first matrix in the parenthesis is therefore real and symmetric, whereas the second is Hermitian with purely imaginary entries. Two such matrices are orthogonal for the Hilbert--Schmidt scalar product, hence $|\grad\sqrt{\bm{\rho}}|^2=|\grad\sqrt{\wt{\bm{\rho}}}|^2+\frac14|\grad\varphi|^2\,|[\bm{\sigma}_3,\sqrt{\wt{\bm{\rho}}}]|^2$. The off-diagonal entry of $\sqrt{\wt{\bm{\rho}}}$ is equal to $\pm\frac12(\sqrt{\rho^+}-\sqrt{\rho^-})\sin\vartheta$, so that $|[\bm{\sigma}_3,\sqrt{\wt{\bm{\rho}}}]|^2=2(\sqrt{\rho^+}-\sqrt{\rho^-})^2\sin^2\vartheta$, which is the second formula.
\qed

\appendix
\crefalias{section}{appendix}
\crefalias{subsection}{appendix}

\section{Proof of \texorpdfstring{\cref{lem:HO}}{the matrix Hoffmann--Ostenhof inequality}}\label{app:HO}
In the scalar case the Hoffmann-Ostenhof inequality follows from the Cauchy--Schwarz inequality. For the matrix version in \cref{lem:HO} we use instead the convexity of $\bm{\rho}\mapsto|\grad\sqrt{\bm{\rho}}|^2$, the analogue of the convexity of $\rho\mapsto|\grad\sqrt\rho|^2$ which was used several times in~\cite{lewin2020local}.

For a Hermitian matrix $\mathbf{A}>0$ and a Hermitian matrix $\mathbf{Y}$, we denote by $\mathbf{X}$ the unique Hermitian solution of the equation $\sqrt{\mathbf{A}}\mathbf{X}+\mathbf{X}\sqrt{\mathbf{A}}=\mathbf{Y}$ and we define
\begin{equation}\label{eq:def_Q}
\QC(\mathbf{A},\mathbf{Y}):=\Trs \mathbf{X}^2=\sum_{i,k=1}^2\frac{|\pscal{\mathbf{u}_i,\mathbf{Y}\mathbf{u}_k}|^2}{\big(\sqrt{a_i}+\sqrt{a_k}\big)^2},
\end{equation}
where $a_1,a_2$ are the eigenvalues of $\mathbf{A}$ and $\mathbf{u}_1,\mathbf{u}_2$ are corresponding orthonormal eigenvectors. For $\mathbf{A}\ge0$ we set $\QC(\mathbf{A},\mathbf{Y}):=\lim_{\eta\to0^+}\QC(\mathbf{A}+\eta,\mathbf{Y})\in[0,\infty]$. This is again given by the right side of~\cref{eq:def_Q}, with the convention that the terms for which $a_i=a_k=0$ vanish when $\pscal{\mathbf{u}_i,\mathbf{Y}\mathbf{u}_k}=0$ and are infinite otherwise. If $\sqrt{\bm{\rho}}\in H^1_{\mathrm{loc}}(\RR^d,\CC^{2\times2})$, differentiating the relation $(\sqrt{\bm{\rho}})^2=\bm{\rho}$ we find $\sqrt{\bm{\rho}}\,\partial_{x_j}\sqrt{\bm{\rho}}+\partial_{x_j}\sqrt{\bm{\rho}}\,\sqrt{\bm{\rho}}=\partial_{x_j}\bm{\rho}$, hence
\begin{equation}\label{eq:gradient_Q}
|\grad\sqrt{\bm{\rho}}(\mathbf{x})|^2=\sum_{j=1}^d\QC\big(\bm{\rho}(\mathbf{x}),\partial_{x_j}\bm{\rho}(\mathbf{x})\big)
\end{equation}
for almost every $\mathbf{x}\in\RR^d$. At a point where $\bm{\rho}(\mathbf{x})$ is not invertible, we use here that $\pscal{\mathbf{v},\sqrt{\bm{\rho}}(\cdot)\mathbf{v}}\ge0$ attains its minimum at $\mathbf{x}$ for every $\mathbf{v}\in\ker \bm{\rho}(\mathbf{x})$, so that $\partial_{x_j}\sqrt{\bm{\rho}}(\mathbf{x})$ vanishes on $\ker \bm{\rho}(\mathbf{x})$ in the sense of quadratic forms. Up to a factor $4$, $\QC$ is called the Wigner--Yanase metric in quantum information theory.

\begin{lemma}[Convexity and monotonicity]\label{lem:convexity}
The function $\QC$ has the following properties.
\begin{itemize}
\item[(i)] It is jointly convex and homogeneous of degree one, that is, $\QC(t\mathbf{A},t\mathbf{Y})=t\,\QC(\mathbf{A},\mathbf{Y})$ for $t>0$. In particular $\bm{\rho}\mapsto \int_{\RR^d}|\grad\sqrt{\bm{\rho}}|^2$ is convex and
\begin{equation*}
\big|\grad\sqrt{\bm{\rho}_1+\bm{\rho}_2}\big|^2\le \big|\grad\sqrt{\bm{\rho}_1}\big|^2+\big|\grad\sqrt{\bm{\rho}_2}\big|^2\qquad\text{a.e.}
\end{equation*}
\item[(ii)] If $\mathbf{A}'\ge c\mathbf{A}$ for some $c>0$, then $\QC(\mathbf{A}',\mathbf{Y})\le c^{-1}\QC(\mathbf{A},\mathbf{Y})$.
\item[(iii)] Let $\mathbf{M}\ge0$ be a function on $\RR^d$ with values in the Hermitian $2\times2$ matrices and entries in $W^{1,1}_{\mathrm{loc}}(\RR^d)$, such that $\sum_{j=1}^d\QC(\mathbf{M},\partial_{x_j}\mathbf{M})\in L^1_{\mathrm{loc}}(\RR^d)$. Then $\sqrt{\mathbf{M}}\in H^1_{\mathrm{loc}}(\RR^d,\CC^{2\times2})$ and~\cref{eq:gradient_Q} holds for $\bm{\rho}=\mathbf{M}$.
\end{itemize}
\end{lemma}

\begin{proof}
It suffices to consider positive definite matrices. On the space of $2\times2$ matrices endowed with the Hilbert--Schmidt scalar product, we introduce the two commuting positive operators $\LC_{\mathbf{A}}\mathbf{Y}=\mathbf{A}\mathbf{Y}$ and $\mathcal{R}_{\mathbf{A}}\mathbf{Y}=\mathbf{Y}\mathbf{A}$. Then $\sqrt{\mathbf{A}}\mathbf{X}+\mathbf{X}\sqrt{\mathbf{A}}=(\LC_{\mathbf{A}}^{1/2}+\mathcal{R}_{\mathbf{A}}^{1/2})\mathbf{X}$ and therefore
$$
\QC(\mathbf{A},\mathbf{Y})=\pscal{\mathbf{Y},\big(\LC_{\mathbf{A}}^{1/2}+\mathcal{R}_{\mathbf{A}}^{1/2}\big)^{-2}\mathbf{Y}}=\pscal{\mathbf{Y},\big(\LC_{\mathbf{A}}+\mathcal{R}_{\mathbf{A}}+2\,\LC_{\mathbf{A}}\#\mathcal{R}_{\mathbf{A}}\big)^{-1}\mathbf{Y}}
$$
where $S\#T$ denotes the geometric mean of two positive operators, which equals $S^{1/2}T^{1/2}$ when they commute. The maps $\mathbf{A}\mapsto\LC_{\mathbf{A}}$ and $\mathbf{A}\mapsto\mathcal{R}_{\mathbf{A}}$ are linear and order-preserving, whereas $(S,T)\mapsto S\#T$ is jointly concave and monotone~\cite{ando1979concavity,kubo1980means}. Hence $\mathbf{A}\mapsto \LC_{\mathbf{A}}+\mathcal{R}_{\mathbf{A}}+2\,\LC_{\mathbf{A}}\#\mathcal{R}_{\mathbf{A}}$ is concave, monotone and homogeneous of degree one. Since $(S,\mathbf{Y})\mapsto\pscal{\mathbf{Y},S^{-1}\mathbf{Y}}$ is jointly convex and non-increasing in $S$, we obtain (ii) and the properties of $\QC$ in (i).

For (iii), let $\eta>0$. The matrix square root is smooth with bounded derivative on the Hermitian matrices $\ge\eta$. By the chain rule, $\sqrt{\mathbf{M}+\eta}$ has entries in $W^{1,1}_{\mathrm{loc}}(\RR^d)$, and differentiating $(\sqrt{\mathbf{M}+\eta})^2=\mathbf{M}+\eta$ gives
$$
\big|\partial_{x_j}\sqrt{\mathbf{M}+\eta}\big|^2=\QC\big(\mathbf{M}+\eta,\partial_{x_j}\mathbf{M}\big)\le\QC\big(\mathbf{M},\partial_{x_j}\mathbf{M}\big)
$$
by (ii). In addition $0\le\sqrt{\mathbf{M}+\eta}-\sqrt{\mathbf{M}}\le\sqrt\eta$, since the square root is operator monotone and $(\sqrt{\mathbf{M}}+\sqrt\eta)^2\ge\mathbf{M}+\eta$. Hence $\sqrt{\mathbf{M}+\eta}\to\sqrt{\mathbf{M}}$ uniformly as $\eta\to0^+$, with gradients bounded in $L^2_{\mathrm{loc}}$. Thus $\sqrt{\mathbf{M}}\in H^1_{\mathrm{loc}}$, and~\cref{eq:gradient_Q} holds as shown above.

The statements on $|\grad\sqrt{\bm{\rho}}|^2$ in (i) follow from (iii) and~\cref{eq:gradient_Q}, since the joint convexity and the homogeneity imply $\QC(\mathbf{A}_1+\mathbf{A}_2,\mathbf{Y}_1+\mathbf{Y}_2)\le\QC(\mathbf{A}_1,\mathbf{Y}_1)+\QC(\mathbf{A}_2,\mathbf{Y}_2)$.
\end{proof}

We can now prove \cref{lem:HO}. We write
$$
\bm{\gamma}=\sum_kn_k|\bm{\Phi}_k\rangle\langle\bm{\Phi}_k|
$$
with $\bm{\Phi}_k\in H^1(\RR^d,\CC^2)$, so that
$$
\bm{\rho}_{\bm{\gamma}}=\sum_kn_k\bm{\Phi}_k\bm{\Phi}_k^*.
$$
We first prove the pointwise inequality
$$
\QC\big(\bm{\Phi}\bm{\Phi}^*,\partial_{x_j}(\bm{\Phi}\bm{\Phi}^*)\big)\le 2|\partial_{x_j}\bm{\Phi}|^2
$$
for $\bm{\Phi}\in H^1(\RR^d,\CC^2)$ and almost every $\mathbf{x}$. At a point where $\bm{\Phi}(\mathbf{x})\neq0$, we take $\mathbf{u}_1=\bm{\Phi}/|\bm{\Phi}|$ and $\mathbf{u}_2$ a normalized vector orthogonal to $\bm{\Phi}$. The matrix $\mathbf{Y}=\partial_{x_j}\bm{\Phi}\,\bm{\Phi}^*+\bm{\Phi}\,\partial_{x_j}\bm{\Phi}^*$ satisfies $\pscal{\mathbf{u}_1,\mathbf{Y}\mathbf{u}_1}=2|\bm{\Phi}|\Real\pscal{\mathbf{u}_1,\partial_{x_j}\bm{\Phi}}$, $\pscal{\mathbf{u}_2,\mathbf{Y}\mathbf{u}_1}=|\bm{\Phi}|\pscal{\mathbf{u}_2,\partial_{x_j}\bm{\Phi}}$ and $\pscal{\mathbf{u}_2,\mathbf{Y}\mathbf{u}_2}=0$. From~\cref{eq:def_Q} we find
$$
\QC\big(\bm{\Phi}\bm{\Phi}^*,\partial_{x_j}(\bm{\Phi}\bm{\Phi}^*)\big)=\big(\Real\pscal{\mathbf{u}_1,\partial_{x_j}\bm{\Phi}}\big)^2+2\big|\pscal{\mathbf{u}_2,\partial_{x_j}\bm{\Phi}}\big|^2\le 2|\partial_{x_j}\bm{\Phi}|^2.
$$
Let now $\bm{\rho}_N:=\sum_{k\le N}n_k\bm{\Phi}_k\bm{\Phi}_k^*$, whose entries are in $W^{1,1}(\RR^d)$. By \cref{lem:convexity}~(i), $\QC$ is subadditive, hence
$$
\sum_{j=1}^d\QC\big(\bm{\rho}_N,\partial_{x_j}\bm{\rho}_N\big)\le2\sum_{k\le N}n_k|\grad\bm{\Phi}_k|^2.
$$
By \cref{lem:convexity}~(iii), $\sqrt{\bm{\rho}_N}\in H^1_{\mathrm{loc}}$ and $\int|\grad\sqrt{\bm{\rho}_N}|^2\le2\Tr(-\lapl)\bm{\gamma}$. If $\bm{\gamma}$ has infinite rank, we use that $0\le\sqrt{\bm{\rho}_N}\le\sqrt{\bm{\rho}_{\bm{\gamma}}}$ by operator monotonicity. Since $\sqrt{\bm{\rho}_{\bm{\gamma}}}+\sqrt{\bm{\rho}_N}\ge\sqrt{\bm{\rho}_{\bm{\gamma}}}-\sqrt{\bm{\rho}_N}\ge0$, this gives
$$
\Trs\big(\sqrt{\bm{\rho}_{\bm{\gamma}}}-\sqrt{\bm{\rho}_N}\big)^2\le\Trs\Big(\big(\sqrt{\bm{\rho}_{\bm{\gamma}}}-\sqrt{\bm{\rho}_N}\big)\big(\sqrt{\bm{\rho}_{\bm{\gamma}}}+\sqrt{\bm{\rho}_N}\big)\Big)=\Trs\big(\bm{\rho}_{\bm{\gamma}}-\bm{\rho}_N\big),
$$
which tends to $0$ in $L^1_{\mathrm{loc}}$ by dominated convergence. Thus $\sqrt{\bm{\rho}_N}\to\sqrt{\bm{\rho}_{\bm{\gamma}}}$ in $L^2_{\mathrm{loc}}$, and~\cref{eq:Hoffmann-Ostenhof} follows from the weak lower semicontinuity of $\int|\grad\cdot|^2$. Since $\int\Trs\bm{\rho}_{\bm{\gamma}}=\Tr\bm{\gamma}<\infty$, the square root is in $L^2$ and hence in $H^1$. To see that $1/2$ is optimal, fix $0\le f\in C_c^\infty(\RR^d)$ with $\int f^2=1$ and take $\bm{\Phi}_k(\mathbf{x})=f(\mathbf{x})(\cos(kx_1),\sin(kx_1))$. With $A=\int|\grad f|^2$, the orbital kinetic energy is $A+k^2$ and the squared $L^2$ norm of the gradient of $\sqrt{\bm{\Phi}_k\bm{\Phi}_k^*}$ is $A+2k^2$. Their ratio tends to $1/2$ as $k\to\infty$.\qed

\section{Proof of \texorpdfstring{\cref{thm:upper_bound_T}}{the kinetic energy upper bound}}\label{app:upper_bound_T}
The free Fermi sea
$$
P_r=\iden\left(-\lapl\le \frac{d+2}d c_\TF r^{2/d}\right)\otimes\iden_{\CC^2}
$$
on $L^2(\RR^d,\CC^2)$ has the constant spin-polarized density $\bm{\rho}_{P_r}=r\iden_{\CC^2}$ and the constant kinetic energy density $2c_\TF r^{1+2/d}$. If we take
$\bm{\gamma}=\mathbf{F}(\mathbf{x})\,P_r\,\mathbf{F}(\mathbf{x})$
where $\mathbf{F}$ is a function with values in the Hermitian $2\times2$ matrices, we obtain
\begin{equation}\label{eq:computation_simple_trial_state}
\begin{gathered}
\bm{\rho}_{\bm{\gamma}}=r\mathbf{F}(\mathbf{x})^2,\\
\Tr(-\lapl)\bm{\gamma}=r\int_{\RR^d}|\grad \mathbf{F}(\mathbf{x})|^2\,\dd \mathbf{x}+c_\TF r^{1+2/d}\,\int_{\RR^d}\Trs \mathbf{F}(\mathbf{x})^2\,\dd \mathbf{x}.
\end{gathered}
\end{equation}
Indeed, the kernel of $\bm{\gamma}$ is $\mathbf{F}(\mathbf{x})P_r(\mathbf{x}-\mathbf{y})\mathbf{F}(\mathbf{y})$ where $P_r(\mathbf{x}-\mathbf{y})$ is scalar and even, and $\Tr(-\lapl)\bm{\gamma}$ is the integral of $\Trs\grad_{\mathbf{x}}\cdot\grad_{\mathbf{y}}\bm{\gamma}(\mathbf{x},\mathbf{y})$ on the diagonal $\mathbf{y}=\mathbf{x}$. The cross terms vanish since $\grad P_r(0)=0$, even though $\mathbf{F}(\mathbf{x})$ and $\grad \mathbf{F}(\mathbf{x})$ need not commute. In addition $0\le\bm{\gamma}\le \mathbf{F}(\mathbf{x})^2$ in the sense of operators.

Let now $\eta\ge0$ be a non-negative function with compact support in $(0,\infty)$, such that $\sqrt\eta$ is Lipschitz and
\begin{equation}\label{eq:cond_eta}
\int_0^\infty\eta(t)\,\dd t=1,\qquad \int_0^\infty\eta(t)\,\frac{\dd t}{t}\le1.
\end{equation}
For every $t>0$ we define the matrix $\eta(t/\bm{\rho}(\mathbf{x}))$ by the functional calculus, that is,
$$
\eta\left(\frac{t}{\bm{\rho}(\mathbf{x})}\right)=\eta\left(\frac{t}{\rho^+(\mathbf{x})}\right)|\mathbf{u}_+(\mathbf{x})\rangle\langle \mathbf{u}_+(\mathbf{x})|+\eta\left(\frac{t}{\rho^-(\mathbf{x})}\right)|\mathbf{u}_-(\mathbf{x})\rangle\langle \mathbf{u}_-(\mathbf{x})|
$$
where $\mathbf{u}_\pm(\mathbf{x})$ are normalized eigenvectors of $\bm{\rho}(\mathbf{x})$ and with the convention that $\eta(t/0)=0$. The smooth layer cake principle now reads $\bm{\rho}(\mathbf{x})=\int_0^\infty \eta(t/\bm{\rho}(\mathbf{x}))\,\dd t$ and we introduce the trial state
$$
\bm{\gamma}=\int_0^\infty  \sqrt{\eta\left(\frac{t}{\bm{\rho}(\mathbf{x})}\right)}\;P_t\;\sqrt{\eta\left(\frac{t}{\bm{\rho}(\mathbf{x})}\right)}\;\frac{\dd t}{t}.
$$
In the sense of operators, we have
$$
0\le \bm{\gamma}\le \int_0^\infty  \eta\left(\frac{t}{\bm{\rho}(\mathbf{x})}\right)\;\frac{\dd t}{t}\le\int_0^\infty\eta(t)\,\frac{\dd t}{t}\le1
$$
and $\bm{\gamma}$ has the required spin-polarized density
$$
\bm{\rho}_{\bm{\gamma}}(\mathbf{x})=\int_0^\infty  \eta\left(\frac{t}{\bm{\rho}(\mathbf{x})}\right)\,\dd t=\bm{\rho}(\mathbf{x}).
$$
From~\cref{eq:computation_simple_trial_state}, its kinetic energy is
\begin{multline}\label{eq:kinetic_trial}
\Tr(-\lapl)\bm{\gamma}=\int_{\RR^d}\dd \mathbf{x}\int_0^\infty \dd t  \left|\grad_{\mathbf{x}}\sqrt{\eta\left(\frac{t}{\bm{\rho}(\mathbf{x})}\right)}\right|^2\\
+c_\TF\int_{\RR^d}\Trs \bm{\rho}(\mathbf{x})^{1+2/d}\dd \mathbf{x}\int_0^\infty \eta(t)t^{2/d}\,\dd t.
\end{multline}
The difference with the scalar case is that the first term is not exactly proportional to $|\grad\sqrt{\bm{\rho}}|^2$, because the eigenvectors $\mathbf{u}_\pm(\mathbf{x})$ depend on $\mathbf{x}$. We claim that
\begin{equation}\label{eq:claim_gradient}
\int_0^\infty \left|\grad_{\mathbf{x}}\sqrt{\eta\left(\frac{t}{\bm{\rho}(\mathbf{x})}\right)}\right|^2\dd t\le \big(1+\sqrt{K_\eta}\big)^2\,|\grad\sqrt{\bm{\rho}}(\mathbf{x})|^2
\end{equation}
for almost every $\mathbf{x}$, where
$$
K_\eta:=\int_0^\infty \frac{t^2\eta'(t)^2}{\eta(t)}\,\dd t.
$$
Let $\phi=\sqrt\eta$ and $g_t(\lambda)=\phi(t/\lambda)$, so that $\sqrt{\eta(t/\bm{\rho}(\mathbf{x}))}=g_t(\bm{\rho}(\mathbf{x}))$. By the Daleckii--Krein formula~\cite[Chapter V]{bhatia1997matrix} (we may assume by density that $\phi$ is $C^1$), we have in the basis $\mathbf{u}_1=\mathbf{u}_+(\mathbf{x})$, $\mathbf{u}_2=\mathbf{u}_-(\mathbf{x})$ with $\lambda_1=\rho^+(\mathbf{x})$ and $\lambda_2=\rho^-(\mathbf{x})$,
$$
\left|\partial_{x_j}g_t\big(\bm{\rho}(\mathbf{x})\big)\right|^2=\sum_{i,k=1}^2\left|\frac{g_t(\lambda_i)-g_t(\lambda_k)}{\lambda_i-\lambda_k}\right|^2\big|\pscal{\mathbf{u}_i,\partial_{x_j}\bm{\rho}(\mathbf{x})\mathbf{u}_k}\big|^2
$$
where the quotient is interpreted as $g_t'(\lambda_i)$ when $\lambda_i=\lambda_k$. Comparing with~\cref{eq:def_Q} and~\cref{eq:gradient_Q}, we see that~\cref{eq:claim_gradient} follows from the inequality
\begin{equation}\label{eq:dilation_estimate}
\int_0^\infty\left|\phi\left(\frac{t}{\lambda_1}\right)-\phi\left(\frac{t}{\lambda_2}\right)\right|^2\dd t\le \big(1+\sqrt{K_\eta}\big)^2\big(\sqrt{\lambda_1}-\sqrt{\lambda_2}\big)^2
\end{equation}
for all $\lambda_1\ge\lambda_2\ge0$. To prove~\cref{eq:dilation_estimate}, we use the unitary group of dilations $(V_s\phi)(t)=e^{-s/2}\phi(e^{-s}t)$ on $L^2(0,\infty)$, for which $\phi(t/\lambda)=\sqrt\lambda\,(V_{\log\lambda}\phi)(t)$. Its generator satisfies
\begin{multline*}
\norm{\frac{\dd}{\dd s}V_s\phi}_{L^2}^2=\int_0^\infty\left(\frac{\phi(t)}2+t\phi'(t)\right)^2\dd t\\
=\int_0^\infty t^2\phi'(t)^2\,\dd t-\frac14\int_0^\infty\phi(t)^2\,\dd t=\frac{K_\eta-1}{4}.
\end{multline*}
Since $\norm{\phi}_{L^2}=1$, we obtain for $\lambda_1\ge\lambda_2>0$
\begin{align*}
\norm{\phi(\cdot/\lambda_1)-\phi(\cdot/\lambda_2)}_{L^2}&\le \sqrt{\lambda_1}-\sqrt{\lambda_2}+\sqrt{\lambda_2}\norm{V_{\log\lambda_1}\phi-V_{\log\lambda_2}\phi}_{L^2}\\
&\le \sqrt{\lambda_1}-\sqrt{\lambda_2}+\sqrt{K_\eta}\sqrt{\lambda_2}\log\sqrt{\frac{\lambda_1}{\lambda_2}}
\end{align*}
and~\cref{eq:dilation_estimate} follows from $\log x\le x-1$. The case $\lambda_2=0$ is immediate.

Inserting~\cref{eq:claim_gradient} in~\cref{eq:kinetic_trial} we have proved that
\begin{multline*}
T(\bm{\rho})\le \big(1+\sqrt{K_\eta}\big)^2\int_{\RR^d}|\grad\sqrt{\bm{\rho}}(\mathbf{x})|^2\,\dd \mathbf{x}\\
+c_\TF\int_{\RR^d}\Trs \bm{\rho}(\mathbf{x})^{1+2/d}\dd \mathbf{x}\int_0^\infty \eta(t)t^{2/d}\,\dd t
\end{multline*}
for all $\eta$ satisfying~\cref{eq:cond_eta}. In the proof of~\cite[Theorem~8]{lewin2020local}, explicit functions $\eta$ are constructed, for which $\int_0^\infty\eta(t)t^{2/d}\dd t\le 1+\kappa_1\epsilon$ and $K_\eta\le\kappa_2(1+\sqrt\epsilon)^2/\epsilon$. Using $(1+\sqrt{K_\eta})^2\le 2+2K_\eta$ we obtain~\cref{eq:upper_bound_T_kappa} with $\kappa_2$ replaced by $2(1+\kappa_2)$.\qed

\section{Subadditivity estimate with local errors}\label{app:subadditivity}
In this appendix we prove the following subadditivity estimate, which is used in~\cref{sec:proof}. In~\cite[Lemma~19]{lewin2020local}, an upper bound on $E(\rho_1+\rho_2)$ was derived in the spinless case by mixing two optimal states for $\rho_1$ and $\rho_1+\rho_2/\epsilon$, with the weights $1-\epsilon$ and $\epsilon$. This produces the error term $(1-\epsilon)D(\rho_2)/\epsilon$, which is non-local and grows much faster than the volume. It was responsible for the large power of $\epsilon$ in the main estimate of~\cite{lewin2020local}, where it was conjectured that it can be removed. Here we prove a similar estimate in which this term is replaced by \emph{local} terms.

\begin{lemma}[Subadditivity estimate with local errors]\label{lem:subadditivity_worse}
Let $\bm{\rho}_1\ge0$ be a spin-polarized density on $\RR^3$ with total density $\rho_1\in L^1(\RR^3)$ and $\sqrt{\bm{\rho}_1}\in H^1(\RR^3,\CC^{2\times2})$. Let $\bm{\rho}_2\ge0$ be a measurable function with values in the Hermitian $2\times2$ matrices, such that $\bm{\rho}_1+\bm{\rho}_2$ has the same properties. Assume that $\rho_1$ and the total density $\rho_2$ of $\bm{\rho}_2$ are in $L^2(\RR^3)$. Then
\begin{multline}\label{eq:subadditivity_worse}
 E(\bm{\rho}_1+\bm{\rho}_2)\le E(\bm{\rho}_1)+C\epsilon\int_{\RR^3}\left(\rho_1+\rho_1^{2}\right)+C\epsilon\int_{\RR^3}\left|\grad\sqrt{\bm{\rho}_1}\right|^2+C\epsilon^{-2/3}\int_{\RR^3}\rho_2^{5/3}\\
 +C\int_{\RR^3}\left|\grad\sqrt{\bm{\rho}_2+\epsilon \bm{\rho}_1}\right|^2+\frac{C}{r^2}\int_{\RR^3}\rho_2+\frac{Cr^2}{\epsilon}\int_{\RR^3}\rho_2^2
\end{multline}
for all $0<\epsilon\le1$ and all $1\le r\le\epsilon^{-1/2}$. If in addition $\rho_2\le\lambda$ almost everywhere for some constant $\lambda\ge0$, then
\begin{multline}\label{eq:subadditivity_lambda}
 E(\bm{\rho}_1+\bm{\rho}_2)\le E(\bm{\rho}_1)+C\epsilon\int_{\RR^3}\left(\rho_1+\rho_1^{2}\right)+C\epsilon\int_{\RR^3}\left|\grad\sqrt{\bm{\rho}_1}\right|^2\\
 +C\int_{\RR^3}\left|\grad\sqrt{\bm{\rho}_2+\epsilon \bm{\rho}_1}\right|^2+C\left(\epsilon+\Big(\frac{\lambda}{\epsilon}\Big)^{1/2}+\frac{\lambda}{\epsilon}\right)\int_{\RR^3}\rho_2
\end{multline}
for all $0<\epsilon\le1$.
\end{lemma}

The idea of the proof is the following. Instead of mixing two states whose densities differ everywhere, we put the additional electrons in small islands of size $r$, which occupy a fraction $\epsilon$ of the space and are placed at random. The optimal state for $\bm{\rho}_1$ is kept outside of the islands. The fluctuations of the density then have a finite range, and so has their electrostatic energy. That the state for $\bm{\rho}_1$ can be removed from the islands at a small cost follows from the Graf--Schenker inequality. The last two terms of~\cref{eq:subadditivity_worse} are the kinetic energy which is necessary to confine the additional electrons in the islands, and the electrostatic energy of the islands. The best size $r$ therefore depends on the number of additional electrons, and optimizing over $r$ gives~\cref{eq:subadditivity_lambda}. The choice $r=\epsilon^{-1/2}$, for which the last two terms of~\cref{eq:subadditivity_worse} are $C\epsilon\int\rho_2+C\epsilon^{-2}\int\rho_2^2$, is only the best one when $\lambda\le\epsilon^3$.

Note that we do not require that $\sqrt{\bm{\rho}_2}\in H^1(\RR^3)$, which is not the case in our application in~\cref{sec:proof}. The gradient terms involving $\bm{\rho}_2$ are nevertheless finite. Indeed, for $a>0$ the matrix $a\bm{\rho}_1+\bm{\rho}_2$ has entries in $W^{1,1}_{\mathrm{loc}}(\RR^3)$, its derivatives are $\partial_{x_j}(\bm{\rho}_1+\bm{\rho}_2)-(1-a)\partial_{x_j}\bm{\rho}_1$, and it is bounded from below by $\min(a,1)(\bm{\rho}_1+\bm{\rho}_2)$ and by $a\bm{\rho}_1$. Hence \cref{lem:convexity}~(ii) and~(iii) give $\sqrt{a\bm{\rho}_1+\bm{\rho}_2}\in H^1(\RR^3)$ for every $a>0$ and, for $a=\epsilon$,
$$
\big|\grad\sqrt{\bm{\rho}_2+\epsilon\bm{\rho}_1}\big|^2\le\frac2\epsilon\Big(\big|\grad\sqrt{\bm{\rho}_1+\bm{\rho}_2}\big|^2+\big|\grad\sqrt{\bm{\rho}_1}\big|^2\Big).
$$

\begin{proof}[Proof of \cref{lem:subadditivity_worse}]
Let $0<v\le1/2$, $r>0$ and $0<\delta\le r/C$ be three parameters, to be chosen at the end of the proof.

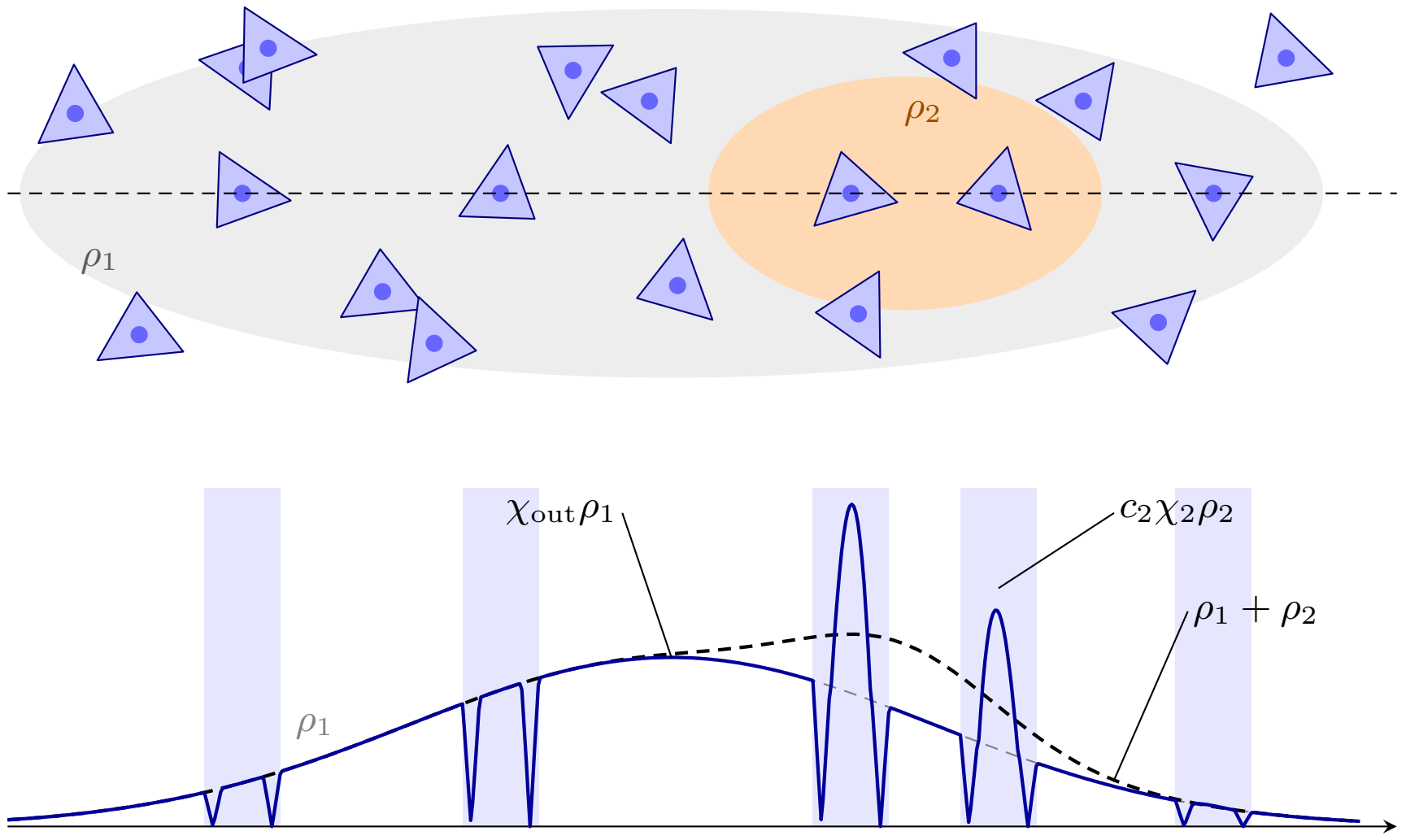
\begin{figure}[t]
\centering
\begin{tikzpicture}[>=stealth,line join=round]
\def\isl{(-0.29,-0.19)--(0.35,-0.19)--(-0.06,0.38)--cycle}
\fill[gray!14] (5.4,1.6) ellipse[x radius=5.3,y radius=1.5];
\fill[orange!30] (7.3,1.6) ellipse[x radius=1.6,y radius=0.95];
\foreach \x/\y/\a in {1.91/1.6/20,4.01/1.6/110,6.86/1.6/250,8.06/1.6/340,9.81/1.6/170,
  1.07/0.45/240,3.05/0.8/240,6.92/0.62/325,1.95/2.62/199,2.12/2.78/21,3.47/0.38/137,5.22/2.35/198,
  7.68/2.7/202,9.36/0.55/69,5.45/0.85/344,4.6/2.6/293,8.75/2.35/80,10.4/2.7/10,0.55/2.25/120}{
  \begin{scope}[shift={(\x,\y)},rotate=\a]
  \fill[blue!22,draw=blue!50!black] \isl; \fill[blue!60] (0,0.0) circle[radius=0.07];
  \end{scope}}
\draw[densely dashed] (0,1.6)--(11.3,1.6);
\node[gray!70!black] at (0.75,1.05) {\small $\rho_1$};
\node[orange!60!black] at (7.45,2.25) {\small $\rho_2$};
\begin{scope}[shift={(0,-3.55)}]
\fill[blue!10] (1.60,0) rectangle (2.22,2.75);
\fill[blue!10] (3.70,0) rectangle (4.32,2.75);
\fill[blue!10] (6.55,0) rectangle (7.17,2.75);
\fill[blue!10] (7.75,0) rectangle (8.37,2.75);
\fill[blue!10] (9.50,0) rectangle (10.12,2.75);
\draw[->] (0,0)--(11.3,0);
\draw[gray,densely dashed] plot coordinates {(0.000,0.054) (0.100,0.061) (0.200,0.068) (0.300,0.076) (0.400,0.085) (0.500,0.095) (0.600,0.106) (0.700,0.118) (0.800,0.131) (0.900,0.145) (1.000,0.160) (1.100,0.176) (1.200,0.194) (1.300,0.212) (1.400,0.232) (1.500,0.254) (1.600,0.276) (1.700,0.300) (1.800,0.326) (1.900,0.353) (2.000,0.381) (2.100,0.410) (2.200,0.441) (2.300,0.473) (2.400,0.506) (2.500,0.540) (2.600,0.575) (2.700,0.612) (2.800,0.649) (2.900,0.687) (3.000,0.725) (3.100,0.764) (3.200,0.803) (3.300,0.842) (3.400,0.882) (3.500,0.921) (3.600,0.959) (3.700,0.997) (3.800,1.035) (3.900,1.071) (4.000,1.106) (4.100,1.140) (4.200,1.172) (4.300,1.202) (4.400,1.230) (4.500,1.257) (4.600,1.281) (4.700,1.302) (4.800,1.321) (4.900,1.337) (5.000,1.351) (5.100,1.361) (5.200,1.369) (5.300,1.373) (5.400,1.375) (5.500,1.373) (5.600,1.369) (5.700,1.361) (5.800,1.351) (5.900,1.337) (6.000,1.321) (6.100,1.302) (6.200,1.281) (6.300,1.257) (6.400,1.230) (6.500,1.202) (6.600,1.172) (6.700,1.140) (6.800,1.106) (6.900,1.071) (7.000,1.035) (7.100,0.997) (7.200,0.959) (7.300,0.921) (7.400,0.882) (7.500,0.842) (7.600,0.803) (7.700,0.764) (7.800,0.725) (7.900,0.687) (8.000,0.649) (8.100,0.612) (8.200,0.575) (8.300,0.540) (8.400,0.506) (8.500,0.473) (8.600,0.441) (8.700,0.410) (8.800,0.381) (8.900,0.353) (9.000,0.326) (9.100,0.300) (9.200,0.276) (9.300,0.254) (9.400,0.232) (9.500,0.212) (9.600,0.194) (9.700,0.176) (9.800,0.160) (9.900,0.145) (10.000,0.131) (10.100,0.118) (10.200,0.106) (10.300,0.095) (10.400,0.085) (10.500,0.076) (10.600,0.068) (10.700,0.061) (10.800,0.054) (10.900,0.048) (11.000,0.042)};
\draw[densely dashed,thick] plot coordinates {(0.000,0.054) (0.100,0.061) (0.200,0.068) (0.300,0.076) (0.400,0.085) (0.500,0.095) (0.600,0.106) (0.700,0.118) (0.800,0.131) (0.900,0.145) (1.000,0.160) (1.100,0.176) (1.200,0.194) (1.300,0.212) (1.400,0.232) (1.500,0.254) (1.600,0.276) (1.700,0.300) (1.800,0.326) (1.900,0.353) (2.000,0.381) (2.100,0.410) (2.200,0.441) (2.300,0.473) (2.400,0.506) (2.500,0.540) (2.600,0.575) (2.700,0.612) (2.800,0.649) (2.900,0.687) (3.000,0.725) (3.100,0.764) (3.200,0.803) (3.300,0.842) (3.400,0.882) (3.500,0.921) (3.600,0.959) (3.700,0.997) (3.800,1.035) (3.900,1.071) (4.000,1.106) (4.100,1.140) (4.200,1.172) (4.300,1.202) (4.400,1.231) (4.500,1.258) (4.600,1.282) (4.700,1.304) (4.800,1.324) (4.900,1.342) (5.000,1.358) (5.100,1.372) (5.200,1.384) (5.300,1.394) (5.400,1.403) (5.500,1.412) (5.600,1.421) (5.700,1.429) (5.800,1.438) (5.900,1.449) (6.000,1.460) (6.100,1.473) (6.200,1.488) (6.300,1.503) (6.400,1.518) (6.500,1.533) (6.600,1.547) (6.700,1.557) (6.800,1.563) (6.900,1.564) (7.000,1.557) (7.100,1.542) (7.200,1.517) (7.300,1.483) (7.400,1.439) (7.500,1.387) (7.600,1.325) (7.700,1.257) (7.800,1.183) (7.900,1.104) (8.000,1.024) (8.100,0.943) (8.200,0.863) (8.300,0.786) (8.400,0.713) (8.500,0.644) (8.600,0.580) (8.700,0.521) (8.800,0.468) (8.900,0.420) (9.000,0.377) (9.100,0.339) (9.200,0.305) (9.300,0.274) (9.400,0.247) (9.500,0.223) (9.600,0.201) (9.700,0.181) (9.800,0.163) (9.900,0.147) (10.000,0.132) (10.100,0.119) (10.200,0.107) (10.300,0.096) (10.400,0.086) (10.500,0.077) (10.600,0.068) (10.700,0.061) (10.800,0.054) (10.900,0.048) (11.000,0.042)};
\draw[blue!60!black,thick] plot coordinates {(0.000,0.054) (0.017,0.055) (0.033,0.056) (0.050,0.057) (0.067,0.058) (0.083,0.059) (0.100,0.061) (0.117,0.062) (0.133,0.063) (0.150,0.064) (0.167,0.066) (0.183,0.067) (0.200,0.068) (0.217,0.069) (0.233,0.071) (0.250,0.072) (0.267,0.074) (0.283,0.075) (0.300,0.076) (0.317,0.078) (0.333,0.079) (0.350,0.081) (0.367,0.082) (0.383,0.084) (0.400,0.085) (0.417,0.087) (0.433,0.089) (0.450,0.090) (0.467,0.092) (0.483,0.094) (0.500,0.095) (0.517,0.097) (0.533,0.099) (0.550,0.101) (0.567,0.103) (0.583,0.104) (0.600,0.106) (0.617,0.108) (0.633,0.110) (0.650,0.112) (0.667,0.114) (0.683,0.116) (0.700,0.118) (0.717,0.120) (0.733,0.122) (0.750,0.124) (0.767,0.127) (0.783,0.129) (0.800,0.131) (0.817,0.133) (0.833,0.136) (0.850,0.138) (0.867,0.140) (0.883,0.143) (0.900,0.145) (0.917,0.147) (0.933,0.150) (0.950,0.152) (0.967,0.155) (0.983,0.157) (1.000,0.160) (1.017,0.163) (1.033,0.165) (1.050,0.168) (1.067,0.171) (1.083,0.173) (1.100,0.176) (1.117,0.179) (1.133,0.182) (1.150,0.185) (1.167,0.188) (1.183,0.191) (1.200,0.194) (1.217,0.197) (1.233,0.200) (1.250,0.203) (1.267,0.206) (1.283,0.209) (1.300,0.212) (1.317,0.216) (1.333,0.219) (1.350,0.222) (1.367,0.226) (1.383,0.229) (1.400,0.232) (1.417,0.236) (1.433,0.239) (1.450,0.243) (1.467,0.246) (1.483,0.250) (1.500,0.254) (1.517,0.257) (1.533,0.261) (1.550,0.265) (1.567,0.269) (1.583,0.273) (1.600,0.276) (1.617,0.214) (1.633,0.149) (1.650,0.082) (1.667,0.014) (1.683,0.056) (1.700,0.129) (1.717,0.203) (1.733,0.279) (1.750,0.313) (1.767,0.317) (1.783,0.321) (1.800,0.326) (1.817,0.330) (1.833,0.335) (1.850,0.339) (1.867,0.343) (1.883,0.348) (1.900,0.353) (1.917,0.357) (1.933,0.362) (1.950,0.366) (1.967,0.371) (1.983,0.376) (2.000,0.381) (2.017,0.385) (2.033,0.390) (2.050,0.395) (2.067,0.400) (2.083,0.386) (2.100,0.293) (2.117,0.198) (2.133,0.100) (2.150,0.000) (2.167,0.102) (2.183,0.207) (2.200,0.315) (2.217,0.425) (2.233,0.451) (2.250,0.457) (2.267,0.462) (2.283,0.467) (2.300,0.473) (2.317,0.478) (2.333,0.484) (2.350,0.489) (2.367,0.495) (2.383,0.500) (2.400,0.506) (2.417,0.511) (2.433,0.517) (2.450,0.523) (2.467,0.529) (2.483,0.534) (2.500,0.540) (2.517,0.546) (2.533,0.552) (2.550,0.558) (2.567,0.564) (2.583,0.569) (2.600,0.575) (2.617,0.581) (2.633,0.587) (2.650,0.593) (2.667,0.599) (2.683,0.606) (2.700,0.612) (2.717,0.618) (2.733,0.624) (2.750,0.630) (2.767,0.636) (2.783,0.643) (2.800,0.649) (2.817,0.655) (2.833,0.661) (2.850,0.668) (2.867,0.674) (2.883,0.680) (2.900,0.687) (2.917,0.693) (2.933,0.699) (2.950,0.706) (2.967,0.712) (2.983,0.719) (3.000,0.725) (3.017,0.731) (3.033,0.738) (3.050,0.744) (3.067,0.751) (3.083,0.757) (3.100,0.764) (3.117,0.770) (3.133,0.777) (3.150,0.783) (3.167,0.790) (3.183,0.797) (3.200,0.803) (3.217,0.810) (3.233,0.816) (3.250,0.823) (3.267,0.829) (3.283,0.836) (3.300,0.842) (3.317,0.849) (3.333,0.855) (3.350,0.862) (3.367,0.869) (3.383,0.875) (3.400,0.882) (3.417,0.888) (3.433,0.895) (3.450,0.901) (3.467,0.908) (3.483,0.914) (3.500,0.921) (3.517,0.927) (3.533,0.934) (3.550,0.940) (3.567,0.946) (3.583,0.953) (3.600,0.959) (3.617,0.966) (3.633,0.972) (3.650,0.978) (3.667,0.985) (3.683,0.991) (3.700,0.997) (3.717,0.765) (3.733,0.529) (3.750,0.290) (3.767,0.049) (3.783,0.196) (3.800,0.443) (3.817,0.694) (3.833,0.947) (3.850,1.053) (3.867,1.059) (3.883,1.065) (3.900,1.071) (3.917,1.077) (3.933,1.083) (3.950,1.089) (3.967,1.095) (3.983,1.100) (4.000,1.106) (4.017,1.112) (4.033,1.118) (4.050,1.123) (4.067,1.129) (4.083,1.134) (4.100,1.140) (4.117,1.145) (4.133,1.151) (4.150,1.156) (4.167,1.161) (4.183,1.111) (4.200,0.837) (4.217,0.560) (4.233,0.281) (4.250,0.000) (4.267,0.284) (4.283,0.570) (4.300,0.859) (4.317,1.149) (4.333,1.212) (4.350,1.216) (4.367,1.221) (4.383,1.226) (4.400,1.230) (4.417,1.235) (4.433,1.239) (4.450,1.244) (4.467,1.248) (4.483,1.252) (4.500,1.257) (4.517,1.261) (4.533,1.265) (4.550,1.269) (4.567,1.273) (4.583,1.277) (4.600,1.281) (4.617,1.284) (4.633,1.288) (4.650,1.292) (4.667,1.295) (4.683,1.299) (4.700,1.302) (4.717,1.305) (4.733,1.309) (4.750,1.312) (4.767,1.315) (4.783,1.318) (4.800,1.321) (4.817,1.324) (4.833,1.327) (4.850,1.330) (4.867,1.332) (4.883,1.335) (4.900,1.337) (4.917,1.340) (4.933,1.342) (4.950,1.344) (4.967,1.347) (4.983,1.349) (5.000,1.351) (5.017,1.353) (5.033,1.355) (5.050,1.356) (5.067,1.358) (5.083,1.360) (5.100,1.361) (5.117,1.363) (5.133,1.364) (5.150,1.365) (5.167,1.367) (5.183,1.368) (5.200,1.369) (5.217,1.370) (5.233,1.371) (5.250,1.372) (5.267,1.372) (5.283,1.373) (5.300,1.373) (5.317,1.374) (5.333,1.374) (5.350,1.375) (5.367,1.375) (5.383,1.375) (5.400,1.375) (5.417,1.375) (5.433,1.375) (5.450,1.375) (5.467,1.374) (5.483,1.374) (5.500,1.373) (5.517,1.373) (5.533,1.372) (5.550,1.372) (5.567,1.371) (5.583,1.370) (5.600,1.369) (5.617,1.368) (5.633,1.367) (5.650,1.365) (5.667,1.364) (5.683,1.363) (5.700,1.361) (5.717,1.360) (5.733,1.358) (5.750,1.356) (5.767,1.355) (5.783,1.353) (5.800,1.351) (5.817,1.349) (5.833,1.347) (5.850,1.344) (5.867,1.342) (5.883,1.340) (5.900,1.337) (5.917,1.335) (5.933,1.332) (5.950,1.330) (5.967,1.327) (5.983,1.324) (6.000,1.321) (6.017,1.318) (6.033,1.315) (6.050,1.312) (6.067,1.309) (6.083,1.305) (6.100,1.302) (6.117,1.299) (6.133,1.295) (6.150,1.292) (6.167,1.288) (6.183,1.284) (6.200,1.281) (6.217,1.277) (6.233,1.273) (6.250,1.269) (6.267,1.265) (6.283,1.261) (6.300,1.257) (6.317,1.252) (6.333,1.248) (6.350,1.244) (6.367,1.239) (6.383,1.235) (6.400,1.230) (6.417,1.226) (6.433,1.221) (6.450,1.216) (6.467,1.212) (6.483,1.207) (6.500,1.202) (6.517,1.197) (6.533,1.192) (6.550,1.187) (6.567,0.901) (6.583,0.616) (6.600,0.335) (6.617,0.056) (6.633,0.221) (6.650,0.495) (6.667,0.767) (6.683,1.036) (6.700,1.159) (6.717,1.419) (6.733,1.657) (6.750,1.872) (6.767,2.063) (6.783,2.229) (6.800,2.367) (6.817,2.477) (6.833,2.557) (6.850,2.605) (6.867,2.621) (6.883,2.604) (6.900,2.551) (6.917,2.462) (6.933,2.336) (6.950,2.172) (6.967,1.970) (6.983,1.728) (7.000,1.445) (7.017,1.122) (7.033,0.974) (7.050,0.726) (7.067,0.481) (7.083,0.239) (7.100,0.000) (7.117,0.236) (7.133,0.469) (7.150,0.699) (7.167,0.926) (7.183,0.966) (7.200,0.959) (7.217,0.953) (7.233,0.946) (7.250,0.940) (7.267,0.934) (7.283,0.927) (7.300,0.921) (7.317,0.914) (7.333,0.908) (7.350,0.901) (7.367,0.895) (7.383,0.888) (7.400,0.882) (7.417,0.875) (7.433,0.869) (7.450,0.862) (7.467,0.855) (7.483,0.849) (7.500,0.842) (7.517,0.836) (7.533,0.829) (7.550,0.823) (7.567,0.816) (7.583,0.810) (7.600,0.803) (7.617,0.797) (7.633,0.790) (7.650,0.783) (7.667,0.777) (7.683,0.770) (7.700,0.764) (7.717,0.757) (7.733,0.751) (7.750,0.744) (7.767,0.562) (7.783,0.383) (7.800,0.207) (7.817,0.034) (7.833,0.136) (7.850,0.302) (7.867,0.466) (7.883,0.627) (7.900,0.706) (7.917,0.956) (7.933,1.168) (7.950,1.346) (7.967,1.490) (7.983,1.602) (8.000,1.683) (8.017,1.735) (8.033,1.759) (8.050,1.757) (8.067,1.729) (8.083,1.679) (8.100,1.607) (8.117,1.515) (8.133,1.404) (8.150,1.275) (8.167,1.131) (8.183,0.973) (8.200,0.802) (8.217,0.619) (8.233,0.537) (8.250,0.398) (8.267,0.263) (8.283,0.130) (8.300,0.000) (8.317,0.127) (8.333,0.252) (8.350,0.373) (8.367,0.493) (8.383,0.511) (8.400,0.506) (8.417,0.500) (8.433,0.495) (8.450,0.489) (8.467,0.484) (8.483,0.478) (8.500,0.473) (8.517,0.467) (8.533,0.462) (8.550,0.457) (8.567,0.451) (8.583,0.446) (8.600,0.441) (8.617,0.436) (8.633,0.430) (8.650,0.425) (8.667,0.420) (8.683,0.415) (8.700,0.410) (8.717,0.405) (8.733,0.400) (8.750,0.395) (8.767,0.390) (8.783,0.385) (8.800,0.381) (8.817,0.376) (8.833,0.371) (8.850,0.366) (8.867,0.362) (8.883,0.357) (8.900,0.353) (8.917,0.348) (8.933,0.343) (8.950,0.339) (8.967,0.335) (8.983,0.330) (9.000,0.326) (9.017,0.321) (9.033,0.317) (9.050,0.313) (9.067,0.309) (9.083,0.305) (9.100,0.300) (9.117,0.296) (9.133,0.292) (9.150,0.288) (9.167,0.284) (9.183,0.280) (9.200,0.276) (9.217,0.273) (9.233,0.269) (9.250,0.265) (9.267,0.261) (9.283,0.257) (9.300,0.254) (9.317,0.250) (9.333,0.246) (9.350,0.243) (9.367,0.239) (9.383,0.236) (9.400,0.232) (9.417,0.229) (9.433,0.226) (9.450,0.222) (9.467,0.219) (9.483,0.216) (9.500,0.212) (9.517,0.159) (9.533,0.108) (9.550,0.058) (9.567,0.010) (9.583,0.037) (9.600,0.083) (9.617,0.127) (9.633,0.170) (9.650,0.185) (9.667,0.186) (9.683,0.185) (9.700,0.184) (9.717,0.183) (9.733,0.181) (9.750,0.179) (9.767,0.176) (9.783,0.173) (9.800,0.170) (9.817,0.167) (9.833,0.164) (9.850,0.160) (9.867,0.157) (9.883,0.153) (9.900,0.150) (9.917,0.146) (9.933,0.143) (9.950,0.139) (9.967,0.136) (9.983,0.127) (10.000,0.094) (10.017,0.061) (10.033,0.030) (10.050,0.000) (10.067,0.029) (10.083,0.057) (10.100,0.084) (10.117,0.111) (10.133,0.114) (10.150,0.112) (10.167,0.110) (10.183,0.108) (10.200,0.106) (10.217,0.104) (10.233,0.103) (10.250,0.101) (10.267,0.099) (10.283,0.097) (10.300,0.095) (10.317,0.094) (10.333,0.092) (10.350,0.090) (10.367,0.089) (10.383,0.087) (10.400,0.085) (10.417,0.084) (10.433,0.082) (10.450,0.081) (10.467,0.079) (10.483,0.078) (10.500,0.076) (10.517,0.075) (10.533,0.074) (10.550,0.072) (10.567,0.071) (10.583,0.069) (10.600,0.068) (10.617,0.067) (10.633,0.066) (10.650,0.064) (10.667,0.063) (10.683,0.062) (10.700,0.061) (10.717,0.059) (10.733,0.058) (10.750,0.057) (10.767,0.056) (10.783,0.055) (10.800,0.054) (10.817,0.053) (10.833,0.052) (10.850,0.051) (10.867,0.050) (10.883,0.049) (10.900,0.048) (10.917,0.047) (10.933,0.046) (10.950,0.045) (10.967,0.044) (10.983,0.043) (11.000,0.042)};
\node[gray,above] at (2.5,{0.590}) {\small $\rho_1$};
\draw[thin] (9.0,{0.377})--(9.6,1.75) node[right,inner sep=1pt]{\small $\rho_1+\rho_2$};
\draw[thin] (8.06,{1.938})--(9.0,2.55) node[right,inner sep=1pt]{\small $c_2\chi_2\rho_2$};
\draw[thin] (5.4,{1.375})--(5.0,2.55) node[left,inner sep=1pt]{\small $\chi_{\mathrm{out}}\rho_1$};
\end{scope}
\end{tikzpicture}
\caption{Two-dimensional representation of the random islands in the proof of \cref{lem:subadditivity_worse} (top) and the total density of the trial state along the dashed line (bottom).}\label{fig:islands}
\end{figure}

\medskip

\noindent\textbf{Step 1. (Random islands)} Let $\beta\in C^\infty_c([0,r_0))$ satisfy $0\le\beta\le1$ and $\beta\equiv1$ in a neighborhood of the origin, so that $\mathbf{x}\mapsto\beta(|\mathbf{x}|/r)$ is smooth. The constants $c_0$ and $r_0$ only depend on the tetrahedron $\tetra$ and are chosen so that
\begin{gather*}
0\le\beta(|\mathbf{x}|/r)\le(\iden_{(r-c_0\delta)\tetra}\ast\eta_\delta)(\mathbf{x})\le(\iden_{r\tetra}\ast\eta_\delta)(\mathbf{x})\le1,\\
\supp\big(\iden_{(r-c_0\delta)\tetra}\ast\eta_\delta\big)\subset\big\{\iden_{r\tetra}\ast\eta_\delta=1\big\},\qquad\supp\beta(|\cdot|/r)\subset\big\{\iden_{(r-c_0\delta)\tetra}\ast\eta_\delta=1\big\}.
\end{gather*}
We have
$$
\int_{\RR^3}\iden_{r\tetra}\ast\eta_\delta=|r\tetra|,\quad\int_{\RR^3}\iden_{(r-c_0\delta)\tetra}\ast\eta_\delta=\Big(1-\frac{c_0\delta}{r}\Big)^3|r\tetra|,\quad\int_{\RR^3}\beta(|\mathbf{x}|/r)\,\dd\mathbf{x}=\theta_2|r\tetra|
$$
for a universal constant $\theta_2\in(0,1)$. By Glaeser's inequality (\cref{thm:Glaeser}), $|\grad\sqrt{g}|^2\le\|\mathbf{D}^2g\|_{L^\infty}/2$ for each of the non-negative smooth functions $g=1-\iden_{r\tetra}\ast\eta_\delta$, $g=\iden_{(r-c_0\delta)\tetra}\ast\eta_\delta$ and $g=\beta(|\cdot|/r)$. This gives
\begin{multline}\label{eq:gradients_islands}
\int_{\RR^3}\big|\grad\sqrt{1-\iden_{r\tetra}\ast\eta_\delta}\big|^2+\int_{\RR^3}\big|\grad\sqrt{\iden_{(r-c_0\delta)\tetra}\ast\eta_\delta}\big|^2\le\frac{Cr^2}{\delta},\\
\int_{\RR^3}\big|\grad\sqrt{\beta(|\mathbf{x}|/r)}\big|^2\,\dd\mathbf{x}\le Cr.
\end{multline}
Let now $\{(\mathbf{x}_i,\mathbf{R}_i)\}_{i\in\NN}$ be a Poisson point process on $\RR^3\times \SO(3)$ with intensity $\mu\,\dd \mathbf{x}\,\dd \mathbf{R}$, where
$$
\mu:=\frac{v}{|r\tetra|},
$$
so that $v$ is the expected number of islands containing a given point. In other words the islands $\mathbf{x}_i+\mathbf{R}_ir\tetra$ are placed independently and uniformly in space, with random orientations, and cover a fraction of the order of $v$ of the space. They are allowed to overlap, see \cref{fig:islands}. We denote by $\EE$ the expectation with respect to this process. We define the three random functions
\begin{align*}
\chi_{\mathrm{out}}&:=\prod_i\big(1-\iden_{\mathbf{x}_i+\mathbf{R}_ir\tetra}\ast\eta_\delta\big),\\
\chi_1&:=1-\prod_i\big(1-\iden_{\mathbf{x}_i+\mathbf{R}_i(r-c_0\delta)\tetra}\ast\eta_\delta\big),\\
\chi_2(\mathbf{x})&:=1-\prod_i\big(1-\beta(|\mathbf{x}-\mathbf{x}_i|/r)\big).
\end{align*}
Almost surely, only finitely many islands meet a given bounded set, so that on such a set these products, as well as the sums over $i$ below, have only finitely many non-trivial factors and terms. In particular, on every ball, $\chi_{\mathrm{out}}$, $\chi_1$ and $\chi_2$ coincide with non-negative smooth functions with bounded second derivatives, obtained by keeping finitely many factors. By Glaeser's inequality (\cref{thm:Glaeser}), their square roots are therefore locally Lipschitz. They satisfy $0\le\chi_2\le\chi_1\le1-\chi_{\mathrm{out}}\le1$ as well as
\begin{equation}\label{eq:supports_islands}
\supp(\chi_1)\subset\{\chi_{\mathrm{out}}=0\},\qquad \supp(\chi_2)\subset\{\chi_1=1\}.
\end{equation}
For a Poisson process we have
$$
\EE\prod_i\Big(1-\phi\big(\mathbf{R}_i^{-1}(\mathbf{x}-\mathbf{x}_i)\big)\Big)=\exp\Big(-\mu\int_{\SO(3)}\dd\mathbf{R}\int_{\RR^3}\dd\mathbf{z}\,\phi\big(\mathbf{R}^{-1}(\mathbf{x}-\mathbf{z})\big)\Big)
$$
for every $0\le\phi\le1$. The expectations of our three functions are therefore the constants
$$
\EE\,\chi_{\mathrm{out}}=e^{-v},\qquad v_1:=\EE\,\chi_1=1-e^{-v(1-c_0\delta/r)^3},\qquad v_2:=\EE\,\chi_2=1-e^{-\theta_2v}.
$$
We define
$$
c_1:=\frac{1-e^{-v}}{v_1},\qquad c_2:=\frac1{v_2},\qquad\text{so that}\qquad 1\le c_1\le1+\frac{C\delta}{r},\qquad c_2\le\frac{C}{v}.
$$
Similarly, we find
\begin{equation}\label{eq:correlation_out}
\EE\big[\chi_{\mathrm{out}}(\mathbf{x})\chi_{\mathrm{out}}(\mathbf{y})\big]=e^{-2v}\exp\big(v\,\wt{h}_{r,\delta}(\mathbf{x}-\mathbf{y})\big)
\end{equation}
where
$$
\wt{h}_{r,\delta}(\mathbf{x}-\mathbf{y})=\frac1{|r\tetra|}\int_{\SO(3)}\dd\mathbf{R}\int_{\RR^3}\dd\mathbf{z}\,\big(\iden_{\mathbf{R}r\tetra+\mathbf{z}}\ast\eta_\delta\big)(\mathbf{x})\big(\iden_{\mathbf{R}r\tetra+\mathbf{z}}\ast\eta_\delta\big)(\mathbf{y})
$$
is the normalized overlap kernel used in the smeared Graf--Schenker inequality~\cite{graf1995molecular}, as given in~\cite[p.~55]{lewin2020local}. It satisfies $0\le\wt{h}_{r,\delta}\le1$ and it is supported in a ball of radius $Cr$. For $k=1,2$ we have in the same way
\begin{equation}\label{eq:cov_in}
0\le\EE\big[\chi_k(\mathbf{x})\chi_k(\mathbf{y})\big]-v_k^2\le \mu\int_{\SO(3)}\dd\mathbf{R}\int_{\RR^3}\dd\mathbf{z}\,g\big(\mathbf{R}^{-1}(\mathbf{x}-\mathbf{z})\big)g\big(\mathbf{R}^{-1}(\mathbf{y}-\mathbf{z})\big),
\end{equation}
where $g=\iden_{(r-c_0\delta)\tetra}\ast\eta_\delta$ for $k=1$ and $g=\beta(|\cdot|/r)$ for $k=2$ are the functions in the definition of $\chi_k$.
Finally, we estimate the gradients of the localization functions.

\begin{claim}
The random functions $\chi_{\mathrm{out}}$, $\chi_1$ and $\chi_2$ satisfy
\begin{equation}\label{eq:gradient_patterns}
\EE\,|\grad\sqrt{\chi_{\mathrm{out}}}|^2+\EE\,|\grad\sqrt{\chi_1}|^2\le\frac{Cv}{r\delta},\qquad \EE\,|\grad\sqrt{\chi_2}|^2\le\frac{Cv}{r^2}.
\end{equation}
\end{claim}

\begin{proof}
The function $\sqrt{\chi_{\mathrm{out}}}$ is the product of the functions $\sqrt{1-\iden_{\mathbf{x}_i+\mathbf{R}_ir\tetra}\ast\eta_\delta}$, which take their values in $[0,1]$. Hence $|\grad\sqrt{\chi_{\mathrm{out}}}|\le\sum_i|\grad\sqrt{1-\iden_{\mathbf{x}_i+\mathbf{R}_ir\tetra}\ast\eta_\delta}|$. For $k=1,2$ and $g$ as in~\cref{eq:cov_in}, we have in the same way $|\grad\chi_k|\le\sum_i|\grad g(\mathbf{R}_i^{-1}(\cdot-\mathbf{x}_i))|$. Since $|\grad\sqrt{\chi_k}|=|\grad\chi_k|/(2\sqrt{\chi_k})$ and $\chi_k\ge g\big(\mathbf{R}_i^{-1}(\cdot-\mathbf{x}_i)\big)$ for all $i$, this implies $|\grad\sqrt{\chi_k}|\le\sum_i\big|\grad\sqrt{g(\mathbf{R}_i^{-1}(\cdot-\mathbf{x}_i))}\big|$.

Let $K:=\supp(\iden_{r\tetra}\ast\eta_\delta)$ and let $N(\mathbf{x})$ be the number of islands such that $\mathbf{x}\in\mathbf{x}_i+\mathbf{R}_iK$. By the support properties of Step 1, only these islands contribute to the above sums, and the Cauchy--Schwarz inequality gives
$$
|\grad\sqrt{\chi_{\mathrm{out}}}|^2\le N\sum_if\big(\mathbf{R}_i^{-1}(\cdot-\mathbf{x}_i)\big),\qquad |\grad\sqrt{\chi_k}|^2\le N\sum_if_k\big(\mathbf{R}_i^{-1}(\cdot-\mathbf{x}_i)\big),
$$
where $f:=|\grad\sqrt{1-\iden_{r\tetra}\ast\eta_\delta}|^2$ and $f_k:=|\grad\sqrt{g}|^2$ are supported in $K$. Let $h\ge0$ be any function supported in $K$. Adding an island at $(\mathbf{z},\mathbf{R})$ increases $N(\mathbf{x})$ by one whenever $h(\mathbf{R}^{-1}(\mathbf{x}-\mathbf{z}))\neq0$. Hence the Mecke equation~\cite{last2017lectures} gives
\begin{align*}
\EE\Big[N(\mathbf{x})\sum_ih\big(\mathbf{R}_i^{-1}(\mathbf{x}-\mathbf{x}_i)\big)\Big]&=\mu\int_{\SO(3)}\dd\mathbf{R}\int_{\RR^3}\dd\mathbf{z}\,h\big(\mathbf{R}^{-1}(\mathbf{x}-\mathbf{z})\big)\big(\EE\,N(\mathbf{x})+1\big)\\
&=\mu\big(\EE\,N(\mathbf{x})+1\big)\int_{\RR^3}h.
\end{align*}
By Campbell's formula~\cite{last2017lectures}, $\EE\,N(\mathbf{x})=\mu|K|\le Cv$, since $|K|\le C|r\tetra|$ for $\delta\le r/C$. Applying this to $h=f$ and $h=f_k$ and using $\mu=v/|r\tetra|$, $v\le1/2$ and~\cref{eq:gradients_islands}, we obtain~\cref{eq:gradient_patterns}.
\end{proof}

\medskip

\noindent\textbf{Step 2. (The trial state)} Let $\Gamma_1$ be an optimal state for $\bm{\rho}_1$, which exists by \cref{thm:LL_attainment}, with total density $\rho_1$ and two-particle density $\rho_1^{(2)}(\mathbf{x},\mathbf{y})$. For every realization of the process we consider the spin-polarized density
$$
\bm{\rho}^{\mathrm{in}}:=c_1\chi_1\bm{\rho}_1+c_2\chi_2\bm{\rho}_2
$$
with total density $\rho^{\mathrm{in}}=c_1\chi_1\rho_1+c_2\chi_2\rho_2$, and the quasi-free state $\Gamma^{\mathrm{in}}:=\Gamma_{\bm{\gamma}^{\mathrm{in}}}$ of \cref{sec:a_priori}, where $\bm{\gamma}^{\mathrm{in}}$ is the trial state constructed for $\bm{\rho}^{\mathrm{in}}$ in the proof of \cref{thm:upper_bound_T}, with the same choice in~\cref{eq:cond_eta} for all realizations. It has the spin-polarized density $\bm{\rho}^{\mathrm{in}}$ and depends measurably on the realization of the process. Since the exchange term is non-positive, the proof of \cref{thm:upper_bound_T} and~\cref{eq:compare_trace_power} give
$$
\EC(\Gamma^{\mathrm{in}})\le\Tr(-\lapl)\bm{\gamma}^{\mathrm{in}}\le C\int_{\RR^3}(\rho^{\mathrm{in}})^{5/3}+C\int_{\RR^3}|\grad\sqrt{\bm{\rho}^{\mathrm{in}}}|^2.
$$
Let also $(\Gamma_1)_{|\sqrt{\chi_{\mathrm{out}}}}$ be the localization of $\Gamma_1$ outside of the islands. By~\cref{eq:supports_islands} these two states have disjoint supports and we can anti-symmetrize their tensor product. The so-obtained state $\Gamma_{\mathrm{rand}}$ has the spin-polarized density $\chi_{\mathrm{out}}\bm{\rho}_1+\bm{\rho}^{\mathrm{in}}$ and its energy is the sum of the energies of the two pieces. Our trial state is the average of $\Gamma_{\mathrm{rand}}$ over the realizations of the Poisson process,
$$
\Gamma:=\EE\,\Gamma_{\mathrm{rand}}.
$$
Since $\Gamma\mapsto\bm{\rho}_\Gamma$ is linear, the definition of $c_1$ and $c_2$ gives
$$
\bm{\rho}_\Gamma=\EE\big[\chi_{\mathrm{out}}+c_1\chi_1\big]\bm{\rho}_1+c_2\,\EE\big[\chi_2\big]\bm{\rho}_2=\bm{\rho}_1+\bm{\rho}_2.
$$
Let $\rho_{\mathrm{rand}}:=(\chi_{\mathrm{out}}+c_1\chi_1)\rho_1+c_2\chi_2\rho_2$ be the random density. The kinetic and Coulomb energies are linear in the state, the cross Coulomb terms of the two pieces cancel in $\EC$, and $\EE\,D(\rho_{\mathrm{rand}})-D(\rho_1+\rho_2)=\EE\,D(\rho_{\mathrm{rand}}-\rho_1-\rho_2)$ since $\EE\,\rho_{\mathrm{rand}}=\rho_1+\rho_2$. We obtain
\begin{equation}\label{eq:three_terms}
E(\bm{\rho}_1+\bm{\rho}_2)\le\EE\,\EC\big((\Gamma_1)_{|\sqrt{\chi_{\mathrm{out}}}}\big)+\EE\,\EC\big(\Gamma^{\mathrm{in}}\big)+\EE\,D\big(\rho_{\mathrm{rand}}-\rho_1-\rho_2\big).
\end{equation}
We estimate the three terms separately.

\medskip

\noindent\textbf{Step 3. (Energy in the islands)} Since $\chi_k^{5/3}\le\chi_k$, we have $\EE\int(\rho^{\mathrm{in}})^{5/3}\le Cv\int\rho_1^{5/3}+Cv^{-2/3}\int\rho_2^{5/3}$. For the gradient term we claim that, almost everywhere,
\begin{multline}\label{eq:gradient_inside}
|\grad\sqrt{\bm{\rho}^{\mathrm{in}}}|^2\le 3\chi_2\big|\grad\sqrt{c_1\bm{\rho}_1+c_2\bm{\rho}_2}\big|^2+3c_1\chi_1|\grad\sqrt{\bm{\rho}_1}|^2\\
+3c_2\rho_2|\grad\sqrt{\chi_2}|^2+2c_1\rho_1|\grad\sqrt{\chi_1}|^2.
\end{multline}
We prove~\cref{eq:gradient_inside} with its left side replaced by $\sum_j\QC(\bm{\rho}^{\mathrm{in}},\partial_{x_j}\bm{\rho}^{\mathrm{in}})$. Since the right side is locally integrable, \cref{lem:convexity}~(iii) then gives $\sqrt{\bm{\rho}^{\mathrm{in}}}\in H^1_{\mathrm{loc}}(\RR^3)$ almost surely, as well as~\cref{eq:gradient_inside} itself by~\cref{eq:gradient_Q}. The matrix $\bm{\rho}_2=(\bm{\rho}_1+\bm{\rho}_2)-\bm{\rho}_1$ is weakly differentiable, but we do not use that $\sqrt{\bm{\rho}_2}\in H^1(\RR^3)$. On the set $\{\chi_1<1\}$ we have $\chi_2=0$ and $\bm{\rho}^{\mathrm{in}}=c_1\chi_1\bm{\rho}_1$, so that~\cref{eq:gradient_inside} follows from $|\grad\sqrt{\chi_1\bm{\rho}_1}|^2\le2\chi_1|\grad\sqrt{\bm{\rho}_1}|^2+2\rho_1|\grad\sqrt{\chi_1}|^2$. On the set $\{\chi_1=1\}$ we have $\grad\chi_1=0$ almost everywhere and we write
$$
\partial_{x_j}\bm{\rho}^{\mathrm{in}}=\chi_2\,\partial_{x_j}(c_1\bm{\rho}_1+c_2\bm{\rho}_2)+(1-\chi_2)c_1\partial_{x_j}\bm{\rho}_1+c_2\bm{\rho}_2\,\partial_{x_j}\chi_2.
$$
Since $\mathbf{Y}\mapsto\QC(\mathbf{A},\mathbf{Y})$ is a quadratic form, we have $\QC(\mathbf{A},\mathbf{Y}_1+\mathbf{Y}_2+\mathbf{Y}_3)\le3\sum_k\QC(\mathbf{A},\mathbf{Y}_k)$. The matrix $\bm{\rho}^{\mathrm{in}}=c_1\bm{\rho}_1+c_2\chi_2\bm{\rho}_2$ is bounded from below by each of the three matrices $\chi_2(c_1\bm{\rho}_1+c_2\bm{\rho}_2)$, $c_1\bm{\rho}_1$ and $c_2\chi_2\bm{\rho}_2$. Using \cref{lem:convexity}~(ii) for each of the three terms, the homogeneity of $\QC$ and $\QC(\mathbf{A},\lambda \mathbf{A})=\lambda^2\Trs(\mathbf{A})/4$, we obtain~\cref{eq:gradient_inside}. Taking the expectation and using $c_2v_2=1$, $c_1\le2$, \cref{eq:gradient_patterns} and the homogeneity $v_2|\grad\sqrt{c_1\bm{\rho}_1+c_2\bm{\rho}_2}|^2=|\grad\sqrt{c_1v_2\bm{\rho}_1+\bm{\rho}_2}|^2$, we find
\begin{multline}\label{eq:inside}
\EE\,\EC\big(\Gamma^{\mathrm{in}}\big)\le Cv\int_{\RR^3}\rho_1^{5/3}+\frac{C}{v^{2/3}}\int_{\RR^3}\rho_2^{5/3}+C\int_{\RR^3}\big|\grad\sqrt{c_1v_2\bm{\rho}_1+\bm{\rho}_2}\big|^2\\
+Cv\int_{\RR^3}|\grad\sqrt{\bm{\rho}_1}|^2+\frac{Cv}{r\delta}\int_{\RR^3}\rho_1+\frac{C}{r^2}\int_{\RR^3}\rho_2.
\end{multline}

\medskip

\noindent\textbf{Step 4. (Energy outside of the islands)} We start with the kinetic energy. We consider the localization functions $\sqrt{\chi_{\mathrm{out}}}$, indexed by the realizations of the process, with the law of the process multiplied by $e^{v}$ as the measure. They are locally Lipschitz and satisfy the normalization condition of \cref{thm:IMS}, since $e^{v}\,\EE\big(\sqrt{\chi_{\mathrm{out}}}\big)^2=e^{v}\,\EE\,\chi_{\mathrm{out}}=1$, and $e^{v}\,\EE\,|\grad\sqrt{\chi_{\mathrm{out}}}|^2$ is bounded by~\cref{eq:gradient_patterns}. The IMS formula of \cref{thm:IMS} therefore reads
$$
e^{v}\,\EE\,\TC\big((\Gamma_1)_{|\sqrt{\chi_{\mathrm{out}}}}\big)=\TC(\Gamma_1)+\int_{\RR^3}e^{v}\,\EE\,|\grad\sqrt{\chi_{\mathrm{out}}}|^2\,\rho_1.
$$
The function $\EE\,|\grad\sqrt{\chi_{\mathrm{out}}}|^2$ is constant, since the law of the process is invariant under translations. Dividing by $e^{v}$ and using~\cref{eq:gradient_patterns} again, we obtain
$$
\EE\,\TC\big((\Gamma_1)_{|\sqrt{\chi_{\mathrm{out}}}}\big)=e^{-v}\TC(\Gamma_1)+\EE\,|\grad\sqrt{\chi_{\mathrm{out}}}|^2\int_{\RR^3}\rho_1\le e^{-v}\TC(\Gamma_1)+\frac{Cv}{r\delta}\int_{\RR^3}\rho_1.
$$
For a function $h$ we use the notation
$$
I_h:=\frac12\iint_{\RR^3\times\RR^3}\dd\mathbf{x}\,\dd\mathbf{y}\,h(\mathbf{x}-\mathbf{y})\frac{\rho_1^{(2)}(\mathbf{x},\mathbf{y})-\rho_1(\mathbf{x})\rho_1(\mathbf{y})}{|\mathbf{x}-\mathbf{y}|},
$$
so that $I_1=\Cc(\Gamma_1)-D(\rho_1)$. The two-particle density of the localized state is $\chi_{\mathrm{out}}(\mathbf{x})\chi_{\mathrm{out}}(\mathbf{y})\rho_1^{(2)}(\mathbf{x},\mathbf{y})$ and we obtain from~\cref{eq:correlation_out}
$$
\EE\Big[\Cc\big((\Gamma_1)_{|\sqrt{\chi_{\mathrm{out}}}}\big)-D(\chi_{\mathrm{out}}\rho_1)\Big]=e^{-2v}I_{\exp(v\wt{h}_{r,\delta})}.
$$
Since $0\le\wt{h}_{r,\delta}\le1$ we can write $\exp(v\wt{h}_{r,\delta})=1+v\wt{h}_{r,\delta}+g$ with $0\le g\le(e^v-1-v)\wt{h}_{r,\delta}$. Using $\rho_1^{(2)}\ge0$ and Young's inequality, we have
\begin{align*}
I_g&\le(e^v-1-v)\left(I_{\wt{h}_{r,\delta}}+\frac12\iint\dd\mathbf{x}\,\dd\mathbf{y}\,\wt{h}_{r,\delta}(\mathbf{x}-\mathbf{y})\frac{\rho_1(\mathbf{x})\rho_1(\mathbf{y})}{|\mathbf{x}-\mathbf{y}|}\right)\\
&\le (e^v-1-v)I_{\wt{h}_{r,\delta}}+Cv^2r^2\int_{\RR^3}\rho_1^2.
\end{align*}
On the other hand, the smeared Graf--Schenker inequality in the form of~\cite[Eq. (53)]{lewin2020local} states precisely that
$$
I_1\ge\left(1-\frac{C\delta}{r}\right)I_{\wt{h}_{r,\delta}}-\frac{C}{r}\int_{\RR^3}\rho_1-\frac{C\delta^3}{r}\int_{\RR^3}\rho_1^2.
$$
Collecting our estimates we find
\begin{equation*}
\EE\,\EC\big((\Gamma_1)_{|\sqrt{\chi_{\mathrm{out}}}}\big)\le e^{-v}\TC(\Gamma_1)+A\,I_1+\frac{Cv}{r}\Big(1+\frac1\delta\Big)\int_{\RR^3}\rho_1+Cv\Big(\frac{\delta^3}{r}+vr^2\Big)\int_{\RR^3}\rho_1^2
\end{equation*}
with
$$
A:=e^{-2v}\left(1+\frac{e^v-1}{1-C\delta/r}\right),\qquad\text{which satisfies}\qquad|A-1|\le Cv.
$$
We have $\TC(\Gamma_1)\ge0$ and $\TC(\Gamma_1)+I_1=E(\bm{\rho}_1)$, hence $I_1\le E(\bm{\rho}_1)$. On the other hand $I_1\ge-c_\LO\int\rho_1^{4/3}$ by the Lieb--Oxford inequality of \cref{thm:LO}. Therefore
\begin{multline*}
e^{-v}\TC(\Gamma_1)+A\,I_1\le \TC(\Gamma_1)+I_1+|A-1|\,|I_1|\\
\le E(\bm{\rho}_1)+Cv\left(|E(\bm{\rho}_1)|+\int_{\RR^3}\rho_1^{4/3}\right).
\end{multline*}
Using the bounds of~\cref{cor:upper_bound_E_rho,cor:lower_bound_E_rho} on $|E(\bm{\rho}_1)|$, we arrive at
\begin{multline}\label{eq:outside}
\EE\,\EC\big((\Gamma_1)_{|\sqrt{\chi_{\mathrm{out}}}}\big)\le E(\bm{\rho}_1)+Cv\int_{\RR^3}\big(\rho_1^{5/3}+\rho_1^{4/3}\big)+Cv\int_{\RR^3}|\grad\sqrt{\bm{\rho}_1}|^2\\
+\frac{Cv}{r}\Big(1+\frac1\delta\Big)\int_{\RR^3}\rho_1+Cv\Big(\frac{\delta^3}{r}+vr^2\Big)\int_{\RR^3}\rho_1^2.
\end{multline}

\medskip

\noindent\textbf{Step 5. (Fluctuations of the density)} We write $\rho_{\mathrm{rand}}-\rho_1-\rho_2=(F_1+F_2)\rho_1+F_3\rho_2$ with the three centered random functions
$$
F_1:=(c_1-1)(\chi_1-v_1),\qquad F_2:=\EE\,P-P,\qquad F_3:=c_2(\chi_2-v_2),
$$
where $P:=1-\chi_{\mathrm{out}}-\chi_1\ge0$. Since $D$ is a positive quadratic form, it suffices to estimate each term separately. For a function $K$ of the form $K(\mathbf{x}-\mathbf{y})=\mu\iint\dd\mathbf{z}\,\dd\mathbf{R}\,g(\mathbf{R}^{-1}(\mathbf{x}-\mathbf{z}))g(\mathbf{R}^{-1}(\mathbf{y}-\mathbf{z}))$, we have by the Hardy--Littlewood--Sobolev and H\"older inequalities
\begin{multline}\label{eq:HLS_islands}
\frac12\iint\dd\mathbf{x}\,\dd\mathbf{y}\,K(\mathbf{x}-\mathbf{y})\frac{\rho(\mathbf{x})\rho(\mathbf{y})}{|\mathbf{x}-\mathbf{y}|}=\mu\int_{\SO(3)}\dd\mathbf{R}\int_{\RR^3}\dd\mathbf{z}\,D\Big(g\big(\mathbf{R}^{-1}(\cdot-\mathbf{z})\big)\rho\Big)\\
\le C\mu\norm{g}_{L^{6/5}}^2\int_{\RR^3}\rho^2.
\end{multline}
From~\cref{eq:cov_in} and~\cref{eq:HLS_islands} we obtain
\begin{gather*}
\EE\,D(F_1\rho_1)\le C(c_1-1)^2\mu r^5\int_{\RR^3}\rho_1^2\le Cv\delta^2\int_{\RR^3}\rho_1^2,\\
\EE\,D(F_3\rho_2)\le Cc_2^2\mu r^5\int_{\RR^3}\rho_2^2\le \frac{Cr^2}{v}\int_{\RR^3}\rho_2^2.
\end{gather*}
The function $P$ lives in the thin layers at the boundary of the islands. We have $P\le\sum_is\big(\mathbf{R}_i^{-1}(\cdot-\mathbf{x}_i)\big)$ with $s:=\iden_{r\tetra}\ast\eta_\delta-\iden_{(r-c_0\delta)\tetra}\ast\eta_\delta\ge0$, which satisfies $\int s\le Cr^2\delta$ and $\norm{s}_{L^{6/5}}^2\le C(r^2\delta)^{5/3}$. The random variables $P(\mathbf{x})$ and $P(\mathbf{y})$ are independent when $|\mathbf{x}-\mathbf{y}|\ge Cr$. Otherwise we bound $P(\mathbf{x})P(\mathbf{y})$ by the double sum $\sum_{i,j}s\big(\mathbf{R}_i^{-1}(\mathbf{x}-\mathbf{x}_i)\big)s\big(\mathbf{R}_j^{-1}(\mathbf{y}-\mathbf{x}_j)\big)$. By the Mecke equation~\cite{last2017lectures} and Campbell's formula, the terms with $i=j$ give the first term below and those with $i\neq j$ give the second one:
$$
\EE\big[P(\mathbf{x})P(\mathbf{y})\big]\le\mu\int_{\SO(3)}\dd\mathbf{R}\int_{\RR^3}\dd\mathbf{z}\,s\big(\mathbf{R}^{-1}(\mathbf{x}-\mathbf{z})\big)s\big(\mathbf{R}^{-1}(\mathbf{y}-\mathbf{z})\big)+\Big(\mu\int_{\RR^3}s\Big)^2.
$$
Using again~\cref{eq:HLS_islands} and Young's inequality for the last term, this gives
$$
\EE\,D(F_2\rho_1)\le C\mu(r^2\delta)^{5/3}\int_{\RR^3}\rho_1^2+C\Big(\frac{v\delta}{r}\Big)^2r^2\int_{\RR^3}\rho_1^2\le Cv\big(r^{1/3}\delta^{5/3}+\delta^2\big)\int_{\RR^3}\rho_1^2.
$$
As a conclusion, we have proved that
\begin{equation}\label{eq:mixing_islands}
\EE\,D\big(\rho_{\mathrm{rand}}-\rho_1-\rho_2\big)\le Cv\big(r^{1/3}\delta^{5/3}+\delta^2\big)\int_{\RR^3}\rho_1^2+\frac{Cr^2}{v}\int_{\RR^3}\rho_2^2.
\end{equation}

\medskip

\noindent\textbf{Step 6. (Conclusion)} Inserting~\cref{eq:inside}, \cref{eq:outside} and~\cref{eq:mixing_islands} in~\cref{eq:three_terms}, we obtain
\begin{align}
E(\bm{\rho}_1+\bm{\rho}_2)\le& E(\bm{\rho}_1)+Cv\int_{\RR^3}\big(\rho_1^{5/3}+\rho_1^{4/3}\big)+Cv\int_{\RR^3}|\grad\sqrt{\bm{\rho}_1}|^2\nonumber\\
&+\frac{Cv}{r}\Big(1+\frac1\delta\Big)\int_{\RR^3}\rho_1+Cv\Big(\frac{\delta^3}{r}+vr^2+r^{1/3}\delta^{5/3}+\delta^2\Big)\int_{\RR^3}\rho_1^2\nonumber\\
&+\frac{C}{v^{2/3}}\int_{\RR^3}\rho_2^{5/3}\nonumber\\
&+C\int_{\RR^3}\big|\grad\sqrt{c_1v_2\bm{\rho}_1+\bm{\rho}_2}\big|^2+\frac{C}{r^2}\int_{\RR^3}\rho_2+\frac{Cr^2}{v}\int_{\RR^3}\rho_2^2
\label{eq:subadditivity_general}
\end{align}
for all $0<v\le1/2$, $r>0$ and $0<\delta\le r/C$. Let now $0<\epsilon\le1$ and $1\le r\le\epsilon^{-1/2}$. We choose
$$
v=\frac\epsilon2,\qquad\delta=\frac{r^{-1/5}}{C},
$$
with $C\ge1$ large enough so that the condition $\delta\le r/C$ is satisfied for all $r\ge1$. Then $r\delta\ge1/C$, $\delta\le1$, $r^{1/3}\delta^{5/3}\le1$ and $vr^2\le1/2$. All the terms involving $\rho_1$ are therefore bounded by $C\epsilon\int(\rho_1+\rho_1^2)$, whereas $r^2/v=2r^2/\epsilon$. It remains to deal with the gradient term. The number $\epsilon':=c_1v_2$ satisfies $\epsilon/C\le\epsilon'\le\epsilon$. With $\mathbf{A}=\epsilon \bm{\rho}_1+\bm{\rho}_2$ and $\mathbf{A}'=\epsilon'\bm{\rho}_1+\bm{\rho}_2$, we have $\mathbf{A}'\ge(\epsilon'/\epsilon)\mathbf{A}$, $\mathbf{A}'\ge\epsilon'\bm{\rho}_1$ and $\partial_{x_j}\mathbf{A}'=\partial_{x_j}\mathbf{A}-(\epsilon-\epsilon')\partial_{x_j}\bm{\rho}_1$. By \cref{lem:convexity}~(ii),
\begin{align*}
\QC(\mathbf{A}',\partial_{x_j}\mathbf{A}')&\le2\QC(\mathbf{A}',\partial_{x_j}\mathbf{A})+2(\epsilon-\epsilon')^2\QC(\mathbf{A}',\partial_{x_j}\bm{\rho}_1)\\
&\le C\QC(\mathbf{A},\partial_{x_j}\mathbf{A})+C\epsilon\,\QC(\bm{\rho}_1,\partial_{x_j}\bm{\rho}_1),
\end{align*}
that is, $|\grad\sqrt{c_1v_2\bm{\rho}_1+\bm{\rho}_2}|^2\le C|\grad\sqrt{\bm{\rho}_2+\epsilon \bm{\rho}_1}|^2+C\epsilon|\grad\sqrt{\bm{\rho}_1}|^2$. This concludes the proof of~\cref{eq:subadditivity_worse}.

Finally, assume that $\rho_2\le\lambda$ almost everywhere. Then $\epsilon^{-2/3}\rho_2^{5/3}\le(\lambda/\epsilon)^{2/3}\rho_2\le\big((\lambda/\epsilon)^{1/2}+\lambda/\epsilon\big)\rho_2$ and
$$
\frac{1}{r^2}\int_{\RR^3}\rho_2+\frac{r^2}{\epsilon}\int_{\RR^3}\rho_2^2\le\left(\frac{1}{r^2}+\frac{\lambda r^2}{\epsilon}\right)\int_{\RR^3}\rho_2.
$$
We take $r=1$ if $\lambda\ge\epsilon$, $r=(\epsilon/\lambda)^{1/4}$ if $\epsilon^3\le\lambda\le\epsilon$ and $r=\epsilon^{-1/2}$ if $\lambda\le\epsilon^3$. The parenthesis is then bounded by $2\lambda/\epsilon$, $2(\lambda/\epsilon)^{1/2}$ and $2\epsilon$, respectively, and~\cref{eq:subadditivity_lambda} follows from~\cref{eq:subadditivity_worse}.
\end{proof}

The inequality~\cref{eq:subadditivity_general} is more general than~\cref{eq:subadditivity_worse}. The size $r$ of the islands is limited from below by the term $(v/r)(1+\delta^{-1})\int\rho_1$, which is due to the localization of the state $\Gamma_1$ outside of the islands, and from above by the term $v^2r^2\int\rho_1^2$. In between, the term $r^{-2}\int\rho_2$, which is the kinetic energy necessary to confine the additional electrons in the islands, competes with the electrostatic energy $(r^2/v)\int\rho_2^2$ of the islands. We do not know if~\cref{eq:subadditivity_worse} holds without the last two terms, as would follow from the conjecture of~\cite{lewin2020local} that the term $D(\rho_2)/\epsilon$ can be dropped in~\cite[Lemma~19]{lewin2020local}.

\section{Thermodynamic limit for the SSW gas}\label{app:SSW_limit}
In this appendix we prove \cref{thm:spiral_limit} and \cref{prop:SSW_properties}. We first construct the SSW gas energy from direction-averaged tetrahedral energies and estimate their convergence rate. In \cref{sec:proof_spiral_limit} we identify this limit with the energy per unit volume along general domains.

\subsection{A convergence rate for tetrahedra}\label{sec:CV_rate_spiral}

For a bounded domain $\Omega\subset\RR^3$, a mollifier $\chi$ as in \cref{thm:spiral_limit} and $\delta>0$, we introduce the energy per unit volume of the spin-spiral wave cut off by $\iden_\Omega\ast\chi_\delta$ and, for tetrahedra, its average over the directions of $\mathbf{q}$,
\begin{equation}\label{eq:def_e_bar}
e_{\Omega,\chi_\delta}(\bm{\rho}_0,\mathbf{q}):=\frac{E\big(\bm{\rho}_{\mathbf{q}}(\iden_\Omega\ast\chi_\delta)\big)}{|\Omega|},\qquad \ol{e}_{\ell\tetra,\eta_\delta}(\bm{\rho}_0,\mathbf{q}):=\int_{\SO(3)}e_{\ell\tetra,\eta_\delta}(\bm{\rho}_0,\mathbf{R}\mathbf{q})\,\dd\mathbf{R}.
\end{equation}
The tetrahedral energy $e_{\ell\tetra,\eta_\delta}$ is a special case of the first definition, since $\eta_\delta$ is of the form $\chi_\delta$ by~\cref{eq:def_eta_delta}. The average depends only on $|\mathbf{q}|$.

Besides the covariance~\cref{eq:covariance_spiral} of the spin-spiral wave $\bm{\rho}_{\mathbf{q}}$ shown in \cref{sec:transformations}, we will use that
\begin{equation}\label{eq:bounds_spiral}
-c_\LO\rho_0^{4/3}\int_{\RR^3}(\iden_{\ell\tetra}\ast\eta_\delta)^{4/3}\le E\big(\bm{\rho}_{\mathbf{q}}(\iden_{\ell\tetra}\ast\eta_\delta)\big)\le E\big(\bm{\rho}_0(\iden_{\ell\tetra}\ast\eta_\delta)\big)+\frac{\rho_0|\mathbf{q}|^2}{4}|\ell\tetra|.
\end{equation}
The lower bound follows from~\cref{cor:lower_bound_E_rho} and the upper bound from~\cref{eq:spiral_upper} of \cref{thm:spirals}~(ii). Finally, $\mathbf{q}\mapsto E\big(\bm{\rho}_{\mathbf{q}}(\iden_{\ell\tetra}\ast\eta_\delta)\big)$ is the sum of a finite concave function and of a quadratic function by \cref{thm:spirals}~(i), hence it is continuous.

We have $0\le\iden_{\ell\tetra}\ast\eta_\delta\le1$, $\int_{\RR^3}|\grad\sqrt{\iden_{\ell\tetra}\ast\eta_\delta}|^2\le C\ell^2/\delta$ and $|\grad\sqrt{\bm{\rho}_0(\iden_{\ell\tetra}\ast\eta_\delta)}|^2=\rho_0|\grad\sqrt{\iden_{\ell\tetra}\ast\eta_\delta}|^2$. Hence~\cref{cor:upper_bound_E_rho} and~\cref{eq:bounds_spiral} give
\begin{equation}\label{eq:bounds_tetrahedron_spiral}
-c_\LO\rho_0^{4/3}\le e_{\ell\tetra,\eta_\delta}(\bm{\rho}_0,\mathbf{q})\le C\rho_0^{5/3}+\frac{C\rho_0}{\ell\delta}+\frac{\rho_0|\mathbf{q}|^2}{4}.
\end{equation}

\begin{proposition}[Convergence rate for tetrahedra]\label{prop:rate_SSW}
Let $\bm{\rho}_0\ge0$ be a constant Hermitian matrix and let $\mathbf{q}\in\RR^3$. For each fixed $\delta'>0$, the averaged energies $\ol e_{\ell'\tetra,\eta_{\delta'}}(\bm{\rho}_0,\mathbf{q})$ have a common limit $e_\SSW(\bm{\rho}_0,\mathbf{q})$ as $\ell'\to\infty$, independent of $\delta'$. For $0<\delta\le\ell/C$ we have the upper bound
\begin{equation}\label{eq:rate_SSW_upper}
\ol{e}_{\ell\tetra,\eta_\delta}(\bm{\rho}_0,\mathbf{q})\le e_\SSW(\bm{\rho}_0,\mathbf{q})+C\frac{\rho_0}{\ell}\Big(1+\delta^{-1}+\delta^3\rho_0+\delta\rho_0^{1/3}+\delta\rho_0^{2/3}+\delta|\mathbf{q}|^2\Big)
\end{equation}
and the averaged lower bound
\begin{multline}\label{eq:rate_SSW_lower}
\dashint_{1/2}^{3/2}\ol{e}_{t\ell\tetra,\eta_{t\delta}}(\bm{\rho}_0,\mathbf{q})\,\frac{\dd t}{t^4}\\
\ge e_\SSW(\bm{\rho}_0,\mathbf{q})-C\frac{\rho_0}{\ell}\Big(\delta^{-1}+\delta^3\rho_0+\delta\rho_0^{1/3}+\delta\rho_0^{2/3}+\delta|\mathbf{q}|^2\Big).
\end{multline}
The constant $C$ only depends on the chosen regularizing function $\eta_\star$.
\end{proposition}

At $\mathbf{q}=0$ these are finite-volume bounds for the uniform gas. For general $\mathbf{q}$ the proof of \cref{thm:LSSWA} uses these averages by choosing the comparison spin-spiral wave independently of the orientation of the tile.

Before proving the proposition, we derive a bound comparing the energy of a big domain to that of a small tetrahedron. This is the analog of the monotonicity estimate of~\cite[Lemma~5.11]{csirik2026non}. We state it for a general domain instead of a big tetrahedron, because it is used again in \cref{sec:proof_spiral_limit}.

\begin{lemma}[Monotonicity estimate]\label{lem:monotonicity_spiral}
Let $\bm{\rho}_0$ and $\mathbf{q}$ be as in \cref{prop:rate_SSW}, let $\Omega\subset\RR^3$ be a bounded domain and let $\chi$ be a mollifier as in \cref{thm:spiral_limit}. For any $\ell,\delta,\delta'>0$ such that $\delta/\ell\le1/C$, the bound
\begin{multline}\label{eq:two_scales_spiral}
 e_{\Omega,\chi_{\delta'}}(\bm{\rho}_0,\mathbf{q})
 \ge\left(1-\frac{C\delta}{\ell}\right)\ol e_{\ell\tetra,\eta_\delta}(\bm{\rho}_0,\mathbf{q})
 -\frac{C\rho_0}{\ell}\big(1+\delta^{-1}+\delta^3\rho_0\big)\\
 -C\,\frac{|\partial\Omega+B_{C(\ell+\delta+\delta')}|}{|\Omega|}\left(\rho_0^{4/3}+\rho_0^{5/3}+\rho_0|\mathbf{q}|^2+\frac{\rho_0}{\ell\delta}\right)
\end{multline}
holds true. Here $|\partial\Omega+B_{C(\ell+\delta+\delta')}|/|\Omega|$ is the volume fraction of a layer of width $C(\ell+\delta+\delta')$ around the boundary of $\Omega$. In particular, for any $\ell'\ge\ell+\delta+\delta'$,
\begin{multline}\label{eq:tetracomp_spiral}
 \ol e_{\ell'\tetra,\eta_{\delta'}}(\bm{\rho}_0,\mathbf{q})
 \ge\left(1-\frac{C\delta}{\ell}\right)\ol e_{\ell\tetra,\eta_\delta}(\bm{\rho}_0,\mathbf{q})
 -\frac{C\rho_0}{\ell}\big(1+\delta^{-1}+\delta^3\rho_0\big)\\
 -C\,\frac{\ell+\delta+\delta'}{\ell'}\left(\rho_0^{4/3}+\rho_0^{5/3}+\rho_0|\mathbf{q}|^2+\frac{\rho_0}{\ell\delta}\right).
\end{multline}
The constant $C$ only depends on $\eta_\star$ and $\chi$.
\end{lemma}

The left side of~\cref{eq:two_scales_spiral} involves the wavevector $\mathbf{q}$ itself, whereas the right side only depends on $|\mathbf{q}|$. The average over the directions of $\mathbf{q}$ comes from the rotations of the tiling.

\begin{proof}
We assume $\rho_0>0$, since all the energies vanish for $\rho_0=0$. Let $\delta/\ell\le1/C$ and write using \cref{prop:lower_bound}
\begin{align*}
 e_{\Omega,\chi_{\delta'}}(\bm{\rho}_0,\mathbf{q})
 &\ge\frac{1}{|\Omega|}\left(1-\frac{C\delta}{\ell}\right)\int_{\SO(3)}\dd\mathbf{R}\,\dashint_{C_\ell}\dd\bm{\tau}\\
 &\qquad\quad\times\sum_{\mathbf{z}\in\ZZ^3}\sum_{j=1}^{24}
 E\Big((\iden_{\ell\tetra_j}\ast\eta_\delta)(\mathbf{R}\,\cdot\,-\ell\mathbf{z}-\bm{\tau})(\iden_\Omega\ast\chi_{\delta'})\bm{\rho}_{\mathbf{q}}\Big)\\
 &\quad-\frac{C\rho_0}{\ell|\Omega|}\int_{\RR^3}\Big(\big(1+\delta^{-1}\big)(\iden_\Omega\ast\chi_{\delta'})+\delta^3\rho_0(\iden_\Omega\ast\chi_{\delta'})^2\Big)\\
 &=:\mathrm{(I)}+\mathrm{(II)}.
\end{align*}
Using $0\le\iden_\Omega\ast\chi_{\delta'}\le1$ and $\int_{\RR^3}\iden_\Omega\ast\chi_{\delta'}=|\Omega|$, the terms in (II) may be estimated as
\[
 \mathrm{(II)}\ge-\frac{C\rho_0}{\ell}\big(1+\delta^{-1}+\delta^3\rho_0\big).
\]

Next, we consider (I). For fixed $\mathbf{R}\in\SO(3)$ and $\bm{\tau}\in C_\ell$, define the index set
\[
 J:=\left\{(\mathbf{z},j)\in\ZZ^3\times\{1,\ldots,24\}\ :\
 \begin{aligned}
 &\supp\big((\iden_{\ell\tetra_j}\ast\eta_\delta)(\mathbf{R}\,\cdot\,-\ell\mathbf{z}-\bm{\tau})\big)\\
 &\quad\cap\supp(\iden_\Omega\ast\chi_{\delta'})\neq\emptyset
 \end{aligned}\right\},
\]
which collects all the small smeared tetrahedra that intersect the big smeared domain $\supp(\iden_\Omega\ast\chi_{\delta'})$. Furthermore, define
\[
 J_0:=\left\{(\mathbf{z},j)\in J\ :\
 \iden_\Omega\ast\chi_{\delta'}\equiv1\ \text{on}\
 \supp\big((\iden_{\ell\tetra_j}\ast\eta_\delta)(\mathbf{R}\,\cdot\,-\ell\mathbf{z}-\bm{\tau})\big)\right\},
\]
which contains the small smeared tetrahedra that are well inside $\Omega$. The two sets depend on the placement $(\mathbf{R},\bm{\tau})$ of the tiling, but the bounds on their cardinalities below do not. Only the pairs $(\mathbf{z},j)\in J$ contribute to (I). According to $J_0$ and $J\setminus J_0$, we may decompose the sum in (I) into (Ia) and (Ib), respectively.

To estimate (Ia), note that for $(\mathbf{z},j)\in J_0$ we simply have
\[
 (\iden_{\ell\tetra_j}\ast\eta_\delta)(\mathbf{R}\,\cdot\,-\ell\mathbf{z}-\bm{\tau})(\iden_\Omega\ast\chi_{\delta'})
 =(\iden_{\ell\tetra_j}\ast\eta_\delta)(\mathbf{R}\,\cdot\,-\ell\mathbf{z}-\bm{\tau}).
\]
Since $\ell\tetra_j=\ell\mathbf{R}_j\tetra-\ell\mathbf{z}_j$ and $\eta_\delta$ is radial, the right side is equal to
\[
 (\iden_{\ell\tetra}\ast\eta_\delta)\big(\mathbf{R}_j^{-1}\mathbf{R}\,\cdot\,-\mathbf{R}_j^{-1}(\ell\mathbf{z}-\ell\mathbf{z}_j+\bm{\tau})\big).
\]
By~\cref{eq:covariance_spiral}, the translation does not change the energy of the spin-spiral wave and the rotation $\mathbf{R}_j^{-1}\mathbf{R}$ of space rotates its wavevector. Using the definition~\cref{eq:def_e_bar}, we obtain
\[
 E\Big((\iden_{\ell\tetra_j}\ast\eta_\delta)(\mathbf{R}\,\cdot\,-\ell\mathbf{z}-\bm{\tau})\bm{\rho}_{\mathbf{q}}\Big)
 =|\ell\tetra|\,e_{\ell\tetra,\eta_\delta}(\bm{\rho}_0,\mathbf{R}_j^{-1}\mathbf{R}\mathbf{q})
\]
for all $(\mathbf{z},j)\in J_0$. Therefore,
\[
 \mathrm{(Ia)}=\left(1-\frac{C\delta}{\ell}\right)\int_{\SO(3)}\dd\mathbf{R}\,\dashint_{C_\ell}\dd\bm{\tau}\
 \frac1{24}\sum_{j=1}^{24}\frac{M^0_{\ell,j}}{|\Omega|}\,
 e_{\ell\tetra,\eta_\delta}(\bm{\rho}_0,\mathbf{R}_j^{-1}\mathbf{R}\mathbf{q}),
\]
where we have set $M^0_{\ell,j}=24|\ell\tetra|\,\#\{\mathbf{z}\in\ZZ^3\,:\,(\mathbf{z},j)\in J_0\}$. Since $24|\ell\tetra|=\ell^3$, this is the volume of the cubes of the tiling whose tetrahedron of type $j$ belongs to $J_0$.

Next, we estimate the contributions from the tetrahedra close to the boundary. For $(\mathbf{z},j)\in J\setminus J_0$, the Lieb--Oxford bound of \cref{cor:lower_bound_E_rho} and $0\le\iden_\Omega\ast\chi_{\delta'}\le1$ give
\begin{multline*}
 E\Big((\iden_{\ell\tetra_j}\ast\eta_\delta)(\mathbf{R}\,\cdot\,-\ell\mathbf{z}-\bm{\tau})(\iden_\Omega\ast\chi_{\delta'})\bm{\rho}_{\mathbf{q}}\Big)\\
 \ge-c_\LO\rho_0^{4/3}\int_{\RR^3}\iden_{\ell\tetra_j}\ast\eta_\delta
 =-c_\LO\rho_0^{4/3}|\ell\tetra|.
\end{multline*}
This can be further bounded as
\[
 \mathrm{(Ib)}\ge-\left(1-\frac{C\delta}{\ell}\right)c_\LO\rho_0^{4/3}
 \int_{\SO(3)}\dd\mathbf{R}\,\dashint_{C_\ell}\dd\bm{\tau}\ \frac{M^\partial_\ell}{|\Omega|},
\]
where we have set $M^\partial_\ell=|\ell\tetra|\,|J\setminus J_0|$.

Let us bound the volume ratios $M^0_{\ell,j}/|\Omega|$ and $M^\partial_\ell/|\Omega|$. Each cube $\mathbf{R}^{-1}(C_\ell+\ell\mathbf{z}+\bm{\tau})$ of the tiling contains one small tetrahedron of each type $j$, and the corresponding smeared tetrahedron is supported at a distance at most $b\delta$ from the cube. Let $c$ be the radius of the support of $\chi$. If the cube is at a distance larger than $b\delta+c\delta'$ from $\partial\Omega$, then either it lies inside $\Omega$ and its 24 tetrahedra belong to $J_0$, or it lies outside $\Omega$ and none of them belongs to $J$. All the other cubes are contained in $\partial\Omega+B_{C(\ell+\delta+\delta')}$. In particular, the index set $J\setminus J_0$ only collects small tetrahedra which are at a distance $O(\ell+\delta+\delta')$ from $\partial\Omega$. Therefore, uniformly in $\mathbf{R}$ and $\bm{\tau}$,
\begin{equation*}
 \max_{1\le j\le24}\left|\frac{M^0_{\ell,j}}{|\Omega|}-1\right|
 \le\frac{|\partial\Omega+B_{C(\ell+\delta+\delta')}|}{|\Omega|}
\end{equation*}
and
\begin{equation}\label{eq:volbdbound_spiral}
 \frac{M^\partial_\ell}{|\Omega|}\le\frac{|\partial\Omega+B_{C(\ell+\delta+\delta')}|}{|\Omega|}.
\end{equation}
With these at hand, we replace the volume ratios in (Ia) by one. In~\cite{csirik2026non} the energy of a small tetrahedron does not depend on its orientation and its sign can be used at this step. Here it depends on the orientation through the wavevector $\mathbf{R}_j^{-1}\mathbf{R}\mathbf{q}$, hence we use instead the two-sided bound
\[
 \big|e_{\ell\tetra,\eta_\delta}(\bm{\rho}_0,\mathbf{R}_j^{-1}\mathbf{R}\mathbf{q})\big|\le C\left(\rho_0^{4/3}+\rho_0^{5/3}+\rho_0|\mathbf{q}|^2+\frac{\rho_0}{\ell\delta}\right),
\]
which follows from~\cref{eq:bounds_tetrahedron_spiral}. By the invariance of the Haar measure and the definition~\cref{eq:def_e_bar}, we have
\[
 \frac1{24}\sum_{j=1}^{24}\int_{\SO(3)}
 e_{\ell\tetra,\eta_\delta}(\bm{\rho}_0,\mathbf{R}_j^{-1}\mathbf{R}\mathbf{q})\,\dd\mathbf{R}
 =\ol e_{\ell\tetra,\eta_\delta}(\bm{\rho}_0,\mathbf{q}).
\]
Since $0\le1-C\delta/\ell\le1$, we find
\begin{multline*}
 \mathrm{(Ia)}\ge\left(1-\frac{C\delta}{\ell}\right)\ol e_{\ell\tetra,\eta_\delta}(\bm{\rho}_0,\mathbf{q})\\
 -C\,\frac{|\partial\Omega+B_{C(\ell+\delta+\delta')}|}{|\Omega|}\left(\rho_0^{4/3}+\rho_0^{5/3}+\rho_0|\mathbf{q}|^2+\frac{\rho_0}{\ell\delta}\right).
\end{multline*}
Similarly, $\mathrm{(Ib)}\ge-C\rho_0^{4/3}|\partial\Omega+B_{C(\ell+\delta+\delta')}|/|\Omega|$ by~\cref{eq:volbdbound_spiral}. Adding the estimates on (Ia), (Ib) and (II) gives~\cref{eq:two_scales_spiral}.

Finally, let $\Omega=\ell'\tetra$ and $\chi_{\delta'}=\eta_{\delta'}$. Since the surface area of $\partial(\ell'\tetra)$ is $O(\ell'^2)$, the tetrahedra close to the boundary fill a volume $O(\ell'^2(\ell+\delta+\delta'))$, that is,
\[
 \frac{|\partial(\ell'\tetra)+B_{C(\ell+\delta+\delta')}|}{|\ell'\tetra|}\le C\,\frac{\ell+\delta+\delta'}{\ell'}\qquad\text{for $\ell'\ge\ell+\delta+\delta'$.}
\]
The right side of~\cref{eq:two_scales_spiral} only depends on $|\mathbf{q}|$. Replacing $\mathbf{q}$ by $\mathbf{S}\mathbf{q}$ and integrating over $\mathbf{S}\in\SO(3)$, we obtain~\cref{eq:tetracomp_spiral}.
\end{proof}

\begin{proof}[Proof of \cref{prop:rate_SSW}]
We follow the proof of~\cite[Theorem~5.10]{csirik2026non}.

\medskip

\noindent\textbf{Existence of the limit.} Fix $\delta,\delta'>0$. Taking the liminf as $\ell'\to\infty$ in~\cref{eq:tetracomp_spiral} of \cref{lem:monotonicity_spiral}, we find
\[
 \liminf_{\ell'\to\infty}\ol e_{\ell'\tetra,\eta_{\delta'}}(\bm{\rho}_0,\mathbf{q})
 \ge\left(1-\frac{C\delta}{\ell}\right)\ol e_{\ell\tetra,\eta_\delta}(\bm{\rho}_0,\mathbf{q})
 -\frac{C\rho_0}{\ell}\big(1+\delta^{-1}+\delta^3\rho_0\big)
\]
for all $\ell\ge C\delta$. By~\cref{eq:bounds_tetrahedron_spiral}, $\ol e_{\ell\tetra,\eta_\delta}(\bm{\rho}_0,\mathbf{q})$ stays bounded as $\ell\to\infty$. Taking the limsup as $\ell\to\infty$, we arrive at
\[
 \liminf_{\ell'\to\infty}\ol e_{\ell'\tetra,\eta_{\delta'}}(\bm{\rho}_0,\mathbf{q})\ge\limsup_{\ell\to\infty}\ol e_{\ell\tetra,\eta_\delta}(\bm{\rho}_0,\mathbf{q})
 \qquad\text{for all $\delta,\delta'>0$.}
\]
Taking $\delta=\delta'$ shows that $\ol e_{\ell\tetra,\eta_\delta}(\bm{\rho}_0,\mathbf{q})$ converges as $\ell\to\infty$ for each fixed smearing length $\delta$, and exchanging the roles of $\delta$ and $\delta'$ shows that the limit is independent of $\delta$. We denote it by $e_\SSW(\bm{\rho}_0,\mathbf{q})$.

\medskip

\noindent\textbf{Proof of the upper bound~\cref{eq:rate_SSW_upper}.} We split the left side of~\cref{eq:tetracomp_spiral} in \cref{lem:monotonicity_spiral} according to
\begin{align*}
 \ol e_{\ell'\tetra,\eta_{\delta'}}(\bm{\rho}_0,\mathbf{q})
 &=\left(1-\frac{C\delta}{\ell}\right)\ol e_{\ell'\tetra,\eta_{\delta'}}(\bm{\rho}_0,\mathbf{q})+\frac{C\delta}{\ell}\,\ol e_{\ell'\tetra,\eta_{\delta'}}(\bm{\rho}_0,\mathbf{q})\\
 &\le\left(1-\frac{C\delta}{\ell}\right)\ol e_{\ell'\tetra,\eta_{\delta'}}(\bm{\rho}_0,\mathbf{q})
 +\frac{C\delta}{\ell}\left(\rho_0^{5/3}+\rho_0|\mathbf{q}|^2+\frac{\rho_0}{\ell'\delta'}\right),
\end{align*}
where we used the upper bound in~\cref{eq:bounds_tetrahedron_spiral}. Rearranging the bound~\cref{eq:tetracomp_spiral} of \cref{lem:monotonicity_spiral}, we find
\begin{multline*}
 \left(1-\frac{C\delta}{\ell}\right)\ol e_{\ell\tetra,\eta_\delta}(\bm{\rho}_0,\mathbf{q})
 \le\left(1-\frac{C\delta}{\ell}\right)\ol e_{\ell'\tetra,\eta_{\delta'}}(\bm{\rho}_0,\mathbf{q})\\
 +\frac{C\rho_0}{\ell}\Big(1+\delta^{-1}+\delta^3\rho_0+\delta\rho_0^{2/3}+\delta|\mathbf{q}|^2\Big)\\
 +C\,\frac{\ell+\delta+\delta'}{\ell'}\left(\rho_0^{4/3}+\rho_0^{5/3}+\rho_0|\mathbf{q}|^2+\frac{\rho_0}{\ell\delta}\right)
 +\frac{C\delta}{\ell}\,\frac{\rho_0}{\ell'\delta'}.
\end{multline*}
Now taking the limit $\ell'\to\infty$ and dividing by $1-C\delta/\ell\ge1/2$, we arrive at
\[
 \ol e_{\ell\tetra,\eta_\delta}(\bm{\rho}_0,\mathbf{q})\le e_\SSW(\bm{\rho}_0,\mathbf{q})
 +\frac{C\rho_0}{\ell}\Big(1+\delta^{-1}+\delta^3\rho_0+\delta\rho_0^{2/3}+\delta|\mathbf{q}|^2\Big),
\]
which implies~\cref{eq:rate_SSW_upper}.

\medskip

\noindent\textbf{Proof of the averaged lower bound~\cref{eq:rate_SSW_lower}.} The proof is similar. As in \cref{lem:monotonicity_spiral} we work with a bounded domain $\Omega$ and a mollifier $\chi$, because the estimate~\cref{eq:two_scales_spiral_upper} below is used again in \cref{sec:proof_spiral_limit}. We use \cref{prop:upper_bound} with $\ell$ replaced by $\ell+\delta$ and let $\alpha\uparrow1/2$. With the notation $\beta=\ell/(\ell+\delta)$, where $1/2\le\beta<1$, we have
\begin{equation*}
\begin{aligned}
 e_{\Omega,\chi_{\delta'}}(\bm{\rho}_0,\mathbf{q})
 &\le\frac{1}{|\Omega|}\dashint_{1/2}^{3/2}\frac{\dd t}{t^4}\int_{\SO(3)}\dd\mathbf{R}\,\dashint_{C_{t(\ell+\delta)}}\dd\bm{\tau}
 \sum_{\mathbf{z}\in\ZZ^3}\sum_{j=1}^{24}E\big(\xi_{\mathbf{z},j}\,(\iden_\Omega\ast\chi_{\delta'})\bm{\rho}_{\mathbf{q}}\big)\\
 &\quad+\frac{C\delta}{\ell}\left(\rho_0^{5/3}+\rho_0|\mathbf{q}|^2
 +\frac{\rho_0}{|\Omega|}\int_{\RR^3}\big|\grad\sqrt{\iden_\Omega\ast\chi_{\delta'}}\big|^2\right)
 +\frac{C\rho_0}{\ell\delta}+\frac{C\delta^3\rho_0^2}{\ell}\\
 &=:\mathrm{(I)}+\mathrm{(II)},
\end{aligned}
\end{equation*}
where, for shortness, we denote by
\[
 \xi_{\mathbf{z},j}:=(\iden_{t(\ell+\delta)T_j(\beta\tetra)}\ast\eta_{t\delta})\big(\mathbf{R}\,\cdot\,-t(\ell+\delta)\mathbf{z}-\bm{\tau}\big)
\]
the small smeared tetrahedra. Notice that the scales are chosen so that the small tetrahedra are of size $t\ell$ and smeared at scale $t\delta$. To obtain (II) from the error terms of \cref{prop:upper_bound}, we used $0\le\iden_\Omega\ast\chi_{\delta'}\le1$, the product rule and $|\grad\sqrt{\bm{\rho}_{\mathbf{q}}}|^2\le\rho_0|\mathbf{q}|^2$.

Again, we split (I) according to $J_0$ and $J\setminus J_0$, which we define analogously to above for every fixed $t$, $\mathbf{R}$ and $\bm{\tau}$. First, we consider the summation over $J_0$ in (I), which we denote by (Ia). Using $t(\ell+\delta)T_j(\beta\tetra)=t\ell\mathbf{R}_j\tetra-t(\ell+\delta)\mathbf{z}_j$, we may write
\[
 \xi_{\mathbf{z},j}(\mathbf{x})
 =(\iden_{t\ell\tetra}\ast\eta_{t\delta})\Big(\mathbf{R}_j^{-1}\mathbf{R}\mathbf{x}-\mathbf{R}_j^{-1}\big(t(\ell+\delta)(\mathbf{z}-\mathbf{z}_j)+\bm{\tau}\big)\Big).
\]
By~\cref{eq:covariance_spiral}, we have
\[
 \mathrm{(Ia)}=\dashint_{1/2}^{3/2}\frac{\dd t}{t^4}\int_{\SO(3)}\dd\mathbf{R}\,\dashint_{C_{t(\ell+\delta)}}\dd\bm{\tau}\
 \frac1{24}\sum_{j=1}^{24}\frac{M^0_{t\ell,j}}{|\Omega|}\,
 e_{t\ell\tetra,\eta_{t\delta}}(\bm{\rho}_0,\mathbf{R}_j^{-1}\mathbf{R}\mathbf{q})
\]
with $M^0_{t\ell,j}=24|t\ell\tetra|\,\#\{\mathbf{z}\in\ZZ^3\,:\,(\mathbf{z},j)\in J_0\}$. The cubes of the tiling now have the volume $t^3(\ell+\delta)^3=24|t\ell\tetra|/\beta^3$. Here, the volume ratio may be bounded similarly as before,
\[
 \max_{1\le j\le24}\left|\frac{M^0_{t\ell,j}}{|\Omega|}-\beta^3\right|\le\frac{|\partial\Omega+B_{C(\ell+\delta+\delta')}|}{|\Omega|},
\]
uniformly in $t$, $\mathbf{R}$ and $\bm{\tau}$, after increasing the constant $C$ in the width of the layer if necessary. By~\cref{eq:bounds_tetrahedron_spiral} at the scales $t\ell$ and $t\delta$, the two-sided bound used in the proof of \cref{lem:monotonicity_spiral} holds for $e_{t\ell\tetra,\eta_{t\delta}}$, uniformly in $1/2\le t\le3/2$. As there, we obtain
\[
 \mathrm{(Ia)}\le\beta^3\dashint_{1/2}^{3/2}\ol e_{t\ell\tetra,\eta_{t\delta}}(\bm{\rho}_0,\mathbf{q})\,\frac{\dd t}{t^4}+C\,\frac{|\partial\Omega+B_{C(\ell+\delta+\delta')}|}{|\Omega|}\left(\rho_0^{4/3}+\rho_0^{5/3}+\rho_0|\mathbf{q}|^2+\frac{\rho_0}{\ell\delta}\right).
\]

Next, we estimate the contributions near the boundary using~\cref{eq:spiral_upper} of \cref{thm:spirals}~(ii) and \cref{cor:upper_bound_E_rho}. For $(\mathbf{z},j)\in J\setminus J_0$ we find
\begin{multline*}
 E\big(\xi_{\mathbf{z},j}\,(\iden_\Omega\ast\chi_{\delta'})\bm{\rho}_{\mathbf{q}}\big)
 \le C\big(\rho_0^{5/3}+\rho_0|\mathbf{q}|^2\big)\int_{\RR^3}\xi_{\mathbf{z},j}\\
 +C\rho_0\int_{\RR^3}\Big|\grad\sqrt{\xi_{\mathbf{z},j}\,(\iden_\Omega\ast\chi_{\delta'})}\Big|^2.
\end{multline*}
Here, the gradient term may be bounded using the Cauchy--Schwarz inequality by
\[
 C\rho_0\int_{\RR^3}\big|\grad\sqrt{\xi_{\mathbf{z},j}}\big|^2(\iden_\Omega\ast\chi_{\delta'})
 +C\rho_0\int_{\RR^3}\big|\grad\sqrt{\iden_\Omega\ast\chi_{\delta'}}\big|^2\,\xi_{\mathbf{z},j}.
\]
We have $\int_{\RR^3}\xi_{\mathbf{z},j}=|t\ell\tetra|$ and $\int_{\RR^3}|\grad\sqrt{\xi_{\mathbf{z},j}}|^2\le C\ell^2/\delta\le C|t\ell\tetra|/(\ell\delta)$. Moreover, the functions $\xi_{\mathbf{z},j}$ have disjoint supports and take their values in $[0,1]$. After summing over $J\setminus J_0$, we obtain
\begin{equation*}
 \mathrm{(Ib)}\le C\,\frac{\sup M^\partial_{t\ell}}{|\Omega|}
 \left(\rho_0^{5/3}+\rho_0|\mathbf{q}|^2+\frac{\rho_0}{\ell\delta}\right)
 +\frac{C\rho_0}{|\Omega|}\int_{\RR^3}\big|\grad\sqrt{\iden_\Omega\ast\chi_{\delta'}}\big|^2,
\end{equation*}
where $M^\partial_{t\ell}=|t\ell\tetra|\,|J\setminus J_0|$ and the supremum is taken over $t$, $\mathbf{R}$ and $\bm{\tau}$. As in~\cref{eq:volbdbound_spiral} we have $M^\partial_{t\ell}/|\Omega|\le|\partial\Omega+B_{C(\ell+\delta+\delta')}|/|\Omega|$. Collecting our estimates, we obtain
\begin{multline}\label{eq:two_scales_spiral_upper}
 e_{\Omega,\chi_{\delta'}}(\bm{\rho}_0,\mathbf{q})
 \le \beta^3\dashint_{1/2}^{3/2}\ol e_{t\ell\tetra,\eta_{t\delta}}(\bm{\rho}_0,\mathbf{q})\,\frac{\dd t}{t^4}
 +\frac{C\rho_0}{\ell}\big(\delta^{-1}+\delta^3\rho_0+\delta\rho_0^{2/3}+\delta|\mathbf{q}|^2\big)\\
 +C\,\frac{|\partial\Omega+B_{C(\ell+\delta+\delta')}|}{|\Omega|}\left(\rho_0^{4/3}+\rho_0^{5/3}+\rho_0|\mathbf{q}|^2+\frac{\rho_0}{\ell\delta}\right)\\
 +\frac{C\rho_0}{|\Omega|}\int_{\RR^3}\big|\grad\sqrt{\iden_\Omega\ast\chi_{\delta'}}\big|^2.
\end{multline}

Now we take $\Omega=\ell'\tetra$ and $\chi_{\delta'}=\eta_{\delta'}$ with a fixed $\delta'>0$. Then the volume fraction of the boundary layer is at most $C(\ell+\delta+\delta')/\ell'$, as in the proof of \cref{lem:monotonicity_spiral}, and
\[
 \frac{1}{|\ell'\tetra|}\int_{\RR^3}\big|\grad\sqrt{\iden_{\ell'\tetra}\ast\eta_{\delta'}}\big|^2\le\frac{C}{\ell'\delta'},
\]
so that the last two terms on the right side of~\cref{eq:two_scales_spiral_upper} vanish in the limit $\ell'\to\infty$. Averaging over the directions of $\mathbf{q}$ and taking the limit $\ell'\to\infty$, we obtain
\[
 e_\SSW(\bm{\rho}_0,\mathbf{q})\le\beta^3\dashint_{1/2}^{3/2}\ol e_{t\ell\tetra,\eta_{t\delta}}(\bm{\rho}_0,\mathbf{q})\,\frac{\dd t}{t^4}
 +\frac{C\rho_0}{\ell}\big(\delta^{-1}+\delta^3\rho_0+\delta\rho_0^{2/3}+\delta|\mathbf{q}|^2\big).
\]
Since $1-\beta^3\le3\delta/\ell$ and $\ol e_{t\ell\tetra,\eta_{t\delta}}(\bm{\rho}_0,\mathbf{q})\ge-c_\LO\rho_0^{4/3}$, replacing $\beta^3$ by one costs at most $C(\delta/\ell)\rho_0^{4/3}$. This proves~\cref{eq:rate_SSW_lower}.
\end{proof}

\subsection{Proof of \texorpdfstring{\cref{thm:spiral_limit} and \cref{prop:SSW_properties}}{the SSW gas limit and its properties}}\label{sec:proof_spiral_limit}

\begin{proof}[Proof of \cref{thm:spiral_limit}]
The limit constructed in \cref{prop:rate_SSW} uses tetrahedra and an average over the direction of $\mathbf{q}$. Fix $\mathbf{q}$ and apply~\cref{eq:two_scales_spiral} of \cref{lem:monotonicity_spiral} and~\cref{eq:two_scales_spiral_upper} with $\Omega=\Omega_N$ and $\delta'=\delta_N$. At fixed $\ell,\delta$, the boundary regularity in \cref{thm:spiral_limit} gives
\[
 \frac{|\partial\Omega_N+B_{C(\ell+\delta+\delta_N)}|}{|\Omega_N|}
 \le C\frac{\ell+\delta+\delta_N}{|\Omega_N|^{1/3}}\longrightarrow0 .
\]
The gradient of $\sqrt{\iden_{\Omega_N}\ast\chi_{\delta_N}}$ is supported in $\partial\Omega_N+B_{C\delta_N}$. By Cauchy--Schwarz its squared norm is bounded by $\iden_{\Omega_N}\ast|\grad\sqrt{\chi_{\delta_N}}|^2\le C\delta_N^{-2}$. Hence
\[
 \frac{1}{|\Omega_N|}\int_{\RR^3}\big|\grad\sqrt{\iden_{\Omega_N}\ast\chi_{\delta_N}}\big|^2
 \le\frac{C}{\delta_N|\Omega_N|^{1/3}}\longrightarrow0 .
\]

Fix $\delta>0$ and let $N\to\infty$ first. The lower estimate~\cref{eq:two_scales_spiral} of \cref{lem:monotonicity_spiral} gives
\[
 \liminf_{N\to\infty}e_{\Omega_N,\chi_{\delta_N}}(\bm{\rho}_0,\mathbf{q})
 \ge \left(1-\frac{C\delta}{\ell}\right)\ol e_{\ell\tetra,\eta_\delta}(\bm{\rho}_0,\mathbf{q})-\frac{C\rho_0}{\ell}\big(1+\delta^{-1}+\delta^3\rho_0\big).
\]
As $\ell\to\infty$, the right side tends to $e_\SSW(\bm{\rho}_0,\mathbf{q})$ by \cref{prop:rate_SSW}. The upper estimate~\cref{eq:two_scales_spiral_upper} gives
\begin{multline*}
 \limsup_{N\to\infty}e_{\Omega_N,\chi_{\delta_N}}(\bm{\rho}_0,\mathbf{q})
 \le \left(\frac{\ell}{\ell+\delta}\right)^3
 \dashint_{1/2}^{3/2}\ol e_{t\ell\tetra,\eta_{t\delta}}(\bm{\rho}_0,\mathbf{q})\,\frac{\dd t}{t^4}\\
 +\frac{C\rho_0}{\ell}\big(\delta^{-1}+\delta^3\rho_0+\delta\rho_0^{2/3}+\delta|\mathbf{q}|^2\big).
\end{multline*}
For every $t\in[1/2,3/2]$, $\ol e_{t\ell\tetra,\eta_{t\delta}}(\bm{\rho}_0,\mathbf{q})$ converges to the same $e_\SSW(\bm{\rho}_0,\mathbf{q})$, and~\cref{eq:bounds_tetrahedron_spiral} bounds the integrand uniformly for $\ell\ge1$. Dominated convergence gives the matching upper limit. This proves~\cref{eq:spiral_limit} for every admissible sequence and mollifier. At $\mathbf{q}=0$ this limit defines $e_\UEG$.
\end{proof}

\begin{proof}[Proof of \cref{prop:SSW_properties}]
The average in~\cref{eq:def_e_bar} depends only on $|\mathbf{q}|$, so its limit is isotropic. Since $\int_{\RR^3}(\iden_{\Omega_N}\ast\chi_{\delta_N})=|\Omega_N|$, \cref{thm:spirals}~(i) gives concavity of
\[
 \mathbf{q}\longmapsto
 e_{\Omega_N,\chi_{\delta_N}}(\bm{\rho}_0,\mathbf{q})
 -\frac14\rho_0|\mathbf{q}|^2
\]
for every $N$. Its pointwise limit is therefore concave, proving (ii). For (iii), apply \cref{thm:spirals}~(ii) and~(iii) to $\bm{\rho}=\bm{\rho}_0(\iden_{\Omega_N}\ast\chi_{\delta_N})$, for which $\int m_3/\int\rho=\zeta\cos\vartheta$. At $\mathbf{q}=0$ the energy per unit volume of each of the two constant matrices in those bounds converges to its uniform gas energy. The same specialization of~\cref{cor:lower_bound_E_rho,cor:upper_bound_E_rho} gives
\begin{equation}\label{eq:estim_simple_UEG}
-c_\LO\rho_0^{4/3}\le e_\UEG(\bm{\rho}_0)\le c_\TF\Trs\bm{\rho}_0^{5/3}\le c_\TF\rho_0^{5/3}.
\end{equation}
Indeed, the integral of the squared gradient of the square root of the cutoff, divided by $|\Omega_N|$, tends to zero as shown above. Taking $N\to\infty$ at fixed $\epsilon$ and then $\epsilon\to0$ in the upper bound gives the stated constant.
\end{proof}

Two matrices with the same eigenvalues and angle $\vartheta$ are conjugate by a spin rotation about $\mathbf{e}_3$. This rotation commutes with $\mathbf{U}_{\mathbf{q}}$ and preserves the finite-volume energy by~\cref{eq:invariance}, so the two limits agree. By \cref{prop:SSW_properties}~(i) and~(ii), $e_\SSW(\bm{\rho}_0,\mathbf{q})-\frac14\rho_0|\mathbf{q}|^2$ is non-increasing in $|\mathbf{q}|$.

\section{Regularity of the SSW and uniform gas energies}\label{app:Lipschitz_SSW}

We first describe the removal of electrons according to their spin. We will use this construction in the rotating frame of a finite tetrahedron.

\begin{lemma}[Removing electrons depending on their spin]\label{lem:monotone}
Let $\bm{\rho}\ge0$ be a spin-polarized density on $\RR^3$ with $\rho\in L^1(\RR^3)$ and $\sqrt{\bm{\rho}}\in H^1(\RR^3,\CC^{2\times2})$, and let $\mathbf{B}$ be a constant complex $2\times2$ matrix such that $\mathbf{B}^*\mathbf{B}\le\iden$. Then
\begin{equation}\label{eq:monotone_F}
F_\LL(\mathbf{B}\bm{\rho}\,\mathbf{B}^*)\le F_\LL(\bm{\rho}).
\end{equation}
In particular, for all constant Hermitian matrices $0\le \bm{\rho}_1\le \bm{\rho}_2$ with traces $\rho_1,\rho_2$ and every function $f\ge0$ such that $\sqrt{f}\in H^1(\RR^3)$,
\begin{equation}\label{eq:monotone}
E(\bm{\rho}_1f)\le E(\bm{\rho}_2f)+\big(\rho_2^2-\rho_1^2\big)D(f).
\end{equation}
\end{lemma}

\begin{proof}
Let $\Gamma$ be a state on Fock space with $\bm{\rho}_\Gamma=\bm{\rho}$ and let $\Gamma_{|\mathbf{B}}$ be the $\mathbf{B}$-localized state, where $\mathbf{B}$ acts on the spin variable of the one-particle space $L^2(\RR^3,\CC^2)$. This is the state which has the $k$-particle reduced density matrices $\mathbf{B}^{\otimes k}\Gamma^{(k)}(\mathbf{B}^*)^{\otimes k}$. For $0\le\mathbf{B}\le\iden$ it is constructed in~\cite{lewin2011geometric}. In general we write $\mathbf{B}=\mathbf{W}|\mathbf{B}|$ with $\mathbf{W}$ unitary and $0\le|\mathbf{B}|=(\mathbf{B}^*\mathbf{B})^{1/2}\le\iden$, and we set $\Gamma_{|\mathbf{B}}:=\GC_{\mathbf{W}}\Gamma_{|\,|\mathbf{B}|}\GC_{\mathbf{W}}^*$ with $\GC_{\mathbf{W}}=\bigoplus_n\mathbf{W}^{\otimes n}$. When $\mathbf{B}=\diag(\sqrt{\beta_\spu},\sqrt{\beta_\spd})$, it is obtained by removing the electrons independently of each other, an electron of spin $\sigma$ being kept with probability $\beta_\sigma$. The one-particle density matrix of $\Gamma_{|\mathbf{B}}$ is $\mathbf{B}\bm{\gamma}_\Gamma \mathbf{B}^*$, hence its spin-polarized density is $\mathbf{B}\bm{\rho}\,\mathbf{B}^*$. Since $\mathbf{B}$ commutes with the Laplacian and with the Coulomb potential $w(\mathbf{x}-\mathbf{y})=|\mathbf{x}-\mathbf{y}|^{-1}$, we have
\begin{gather*}
\TC(\Gamma_{|\mathbf{B}})=\Tr\Big(\mathbf{B}^*\mathbf{B}\,(-\lapl)^{1/2}\bm{\gamma}_\Gamma(-\lapl)^{1/2}\Big)\le\TC(\Gamma),\\
\Cc(\Gamma_{|\mathbf{B}})=\frac12\Tr\Big((\mathbf{B}^*\mathbf{B})^{\otimes 2}\,w^{1/2}\Gamma^{(2)}w^{1/2}\Big)\le\Cc(\Gamma)
\end{gather*}
and~\cref{eq:monotone_F} follows. If $0\le \bm{\rho}_1\le \bm{\rho}_2$ are two constant matrices, we take $\mathbf{B}=\bm{\rho}_1^{1/2}\bm{\rho}_2^{-1/2}$, where $\bm{\rho}_2^{-1/2}$ denotes the pseudo-inverse of $\bm{\rho}_2^{1/2}$, that is, its inverse on $\ran\bm{\rho}_2$ extended by $0$ on $\ker\bm{\rho}_2$. Since $\bm{\rho}_1\le\bm{\rho}_2$ we have $\ker\bm{\rho}_2\subset\ker\bm{\rho}_1$, hence $\ran\bm{\rho}_1\subset\ran\bm{\rho}_2$. Then $\mathbf{B}^*\mathbf{B}=\bm{\rho}_2^{-1/2}\bm{\rho}_1\bm{\rho}_2^{-1/2}\le \iden$ and $\mathbf{B}\bm{\rho}_2\mathbf{B}^*=\bm{\rho}_1$. We obtain $F_\LL(\bm{\rho}_1f)\le F_\LL(\bm{\rho}_2f)$, which is~\cref{eq:monotone} after subtracting the direct terms.
\end{proof}

For a tetrahedron of size $\ell$, the direct energy $D(f)$ in~\cref{eq:monotone} is of order $\ell^5$. We shall choose $\ell$ according to the density removed before passing to the thermodynamic energy.

\begin{proposition}[Regularity of $e_\SSW$]\label{prop:Lipschitz_SSW}
The function $e_\SSW$ has the following properties, with the convention that $e_\SSW(0,\mathbf{q})=0$.
\begin{itemize}
\item[(i)] \emph{(Bounds and scaling)}. For every Hermitian matrix $\bm{\rho}_0\ge0$ with trace $\rho_0$, every $\mathbf{q}\in\RR^3$ and every $0\le\alpha\le1$,
\begin{equation}\label{eq:bounds_SSW}
-c_\LO\rho_0^{4/3}\le e_\SSW(\bm{\rho}_0,\mathbf{q})\le c_\TF\rho_0^{5/3}+\frac{\rho_0|\mathbf{q}|^2}{4}
\end{equation}
and
\begin{equation}\label{eq:scaling_SSW}
\alpha^{5/3}\big(e_\SSW(\bm{\rho}_0,\mathbf{q})+c_\LO\rho_0^{4/3}\big)-\alpha^{4/3}c_\LO\rho_0^{4/3}\le e_\SSW\big(\alpha \bm{\rho}_0,\alpha^{1/3}\mathbf{q}\big)\le \alpha^{4/3}e_\SSW(\bm{\rho}_0,\mathbf{q}).
\end{equation}
\item[(ii)] \emph{(Convexity)}. For every $\mathbf{q}\in\RR^3$ and every $\rho_0>0$, the function $e_\SSW(\cdot,\mathbf{q})$ is convex on the set of the Hermitian matrices $\bm{\rho}_0\ge0$ with trace $\rho_0$.
\item[(iii)] \emph{(Lipschitz regularity in the wavevector)}. There exists a universal constant $C$ such that
\begin{equation}\label{eq:Lipschitz_q}
\big|e_\SSW(\bm{\rho}_0,\mathbf{q})-e_\SSW(\bm{\rho}_0,\mathbf{q}')\big|\le C\Big(\rho_0^{7/6}+\rho_0^{4/3}+\rho_0\max(|\mathbf{q}|,|\mathbf{q}'|)\Big)|\mathbf{q}-\mathbf{q}'|
\end{equation}
for every Hermitian matrix $\bm{\rho}_0\ge0$ with trace $\rho_0$ and all $\mathbf{q},\mathbf{q}'\in\RR^3$.
\item[(iv)] \emph{(Lipschitz regularity in the density)}. There exists a universal constant $C$ such that
\begin{equation}\label{eq:Lipschitz_SSW}
\big|e_\SSW(\bm{\rho}_1,\mathbf{q})-e_\SSW(\bm{\rho}_2,\mathbf{q})\big|\le \frac{C}{\nu}\Big(\max(\rho_1,\rho_2)^{1/3}+\max(\rho_1,\rho_2)^{2/3}+|\mathbf{q}|^2\Big)|\bm{\rho}_1-\bm{\rho}_2|
\end{equation}
for all $\mathbf{q}\in\RR^3$, all $0<\nu\le1/2$ and all non-negative Hermitian matrices $\bm{\rho}_1,\bm{\rho}_2$ with traces $\rho_1,\rho_2$ such that $\bm{\rho}_k\ge\nu\rho_k$.
\item[(v)] \emph{(H\"older regularity up to full polarization)}. For all Hermitian matrices $\bm{\rho}_1,\bm{\rho}_2\ge0$ with traces $\rho_1,\rho_2$ such that $r=\max\{\rho_1,\rho_2\}>0$ and all $\mathbf{q}\in\RR^3$,
\begin{equation}\label{eq:Holder_SSW}
\big|e_\SSW(\bm{\rho}_1,\mathbf{q})-e_\SSW(\bm{\rho}_2,\mathbf{q})\big|
\le C\big(r^{4/3}+r^{5/3}+r|\mathbf{q}|^2\big)
\left(\frac{|\bm{\rho}_1-\bm{\rho}_2|}{r}\right)^{1/3}
\end{equation}
with a universal constant $C$.
\item[(vi)] \emph{(Lipschitz regularity in the root of the density)}. For every $p\ge3$, under the assumptions of (v),
\begin{equation}\label{eq:root_Lipschitz_SSW}
\big|e_\SSW(\bm{\rho}_1,\mathbf{q})-e_\SSW(\bm{\rho}_2,\mathbf{q})\big|
\le C_p\big(r^{4/3}+r^{5/3}+r|\mathbf{q}|^2\big)
\frac{\big|\bm{\rho}_1^{1/p}-\bm{\rho}_2^{1/p}\big|}{r^{1/p}}.
\end{equation}
\end{itemize}
In particular, $e_\SSW$ is jointly continuous in $(\bm{\rho}_0,\mathbf{q})$ for all $\bm{\rho}_0\ge0$.
\end{proposition}

The constant in (iv) depends on the lower bound $\nu$ on the minority spin fraction. The scaling~\cref{eq:scaling_SSW} changes the wavevector, hence (iii) is needed to obtain the regularity of $e_\SSW(\cdot,\mathbf{q})$ along rays. The regularity in $\mathbf{q}$ is a consequence of the concavity stated in \cref{prop:SSW_properties}.

\begin{proof}
\noindent\textbf{Part (i).} The upper bound in~\cref{eq:bounds_SSW} follows from \cref{prop:SSW_properties}~(iii) and~\cref{eq:estim_simple_UEG}, and the lower bound from~\cref{eq:bounds_tetrahedron_spiral}. For $0<\alpha\le1$ set $\bm{\rho}_\alpha(\mathbf{x})=\alpha\bm{\rho}(\alpha^{1/3}\mathbf{x})$. A dilation of the many-body state gives
\[
 E(\bm{\rho}_\alpha)=\inf_{\bm{\rho}_\Gamma=\bm{\rho}}
 \left\{\alpha^{2/3}\TC(\Gamma)+\alpha^{1/3}\big(\Cc(\Gamma)-D(\rho)\big)\right\}.
\]
Since $\TC(\Gamma)\ge0$, this is at most $\alpha^{1/3}E(\bm{\rho})$. By the Lieb--Oxford inequality it is at least $\alpha^{2/3}E(\bm{\rho})-(\alpha^{1/3}-\alpha^{2/3})c_\LO\int\rho^{4/3}$. Apply these bounds to a spin-spiral wave of density $\bm{\rho}_0$ and wavevector $\mathbf{q}$ in a smeared tetrahedron of size $\ell$ with fixed smearing width. The rescaled tetrahedron has volume $\alpha^{-1}|\ell\tetra|$, density $\alpha\bm{\rho}_0$ and wavevector $\alpha^{1/3}\mathbf{q}$. Dividing by the rescaled volume and letting $\ell\to\infty$ gives~\cref{eq:scaling_SSW}, since $\int\rho^{4/3}/|\ell\tetra|\to\rho_0^{4/3}$. The case $\alpha=0$ follows from the convention at zero density.

\medskip

\noindent\textbf{Part (ii).} Let $\bm{\rho}_a,\bm{\rho}_b$ have the same trace. For a common cutoff $f$ and wavevector $\mathbf{q}$, the two spin-spiral waves have the same total density and hence the same direct energy. Mixing trial states with weights $t,1-t$ gives a trial state for the spin-spiral wave of density $t\bm{\rho}_a+(1-t)\bm{\rho}_b$. The variational principle gives convexity in finite volume, which passes to the limit in \cref{thm:spiral_limit}.

\medskip

\noindent\textbf{Part (iii).} By \cref{prop:SSW_properties}~(i) and~(ii) we can write $e_\SSW(\bm{\rho}_0,\mathbf{q})=G(|\mathbf{q}|)+\frac14\rho_0|\mathbf{q}|^2$, where $G$ is concave and non-increasing on $[0,\infty)$. By~\cref{eq:bounds_SSW} we have $-c_\LO\rho_0^{4/3}-\frac14\rho_0\sigma^2\le G(\sigma)\le c_\TF\rho_0^{5/3}$ for all $\sigma\ge0$. Hence, for $\sigma\ge0$ and $h>0$, the right derivative of $G$ satisfies
$$
0\le -G'(\sigma)\le\frac{G(\sigma)-G(\sigma+h)}{h}\le\frac{c_\TF\rho_0^{5/3}+c_\LO\rho_0^{4/3}}{h}+\frac{\rho_0(\sigma+h)^2}{4h}.
$$
Taking $h=\sigma+\rho_0^{1/6}+\rho_0^{1/3}$ we find $|G'(\sigma)|\le C(\rho_0^{7/6}+\rho_0^{4/3})+\rho_0\sigma$, which implies~\cref{eq:Lipschitz_q} since $\big||\mathbf{q}|-|\mathbf{q}'|\big|\le|\mathbf{q}-\mathbf{q}'|$.

\medskip

\noindent\textbf{Part (iv).} We start with the regularity along rays. Using~\cref{eq:scaling_SSW} and~\cref{eq:bounds_SSW}, and then~\cref{eq:Lipschitz_q} for the density $\alpha\bm{\rho}_0$ in order to replace $\alpha^{1/3}\mathbf{q}$ by $\mathbf{q}$, we find
\begin{equation}\label{eq:Lipschitz_ray_SSW}
\big|e_\SSW(\alpha \bm{\rho}_0,\mathbf{q})-e_\SSW(\bm{\rho}_0,\mathbf{q})\big|\le C(1-\alpha)\Big(\rho_0^{4/3}+\rho_0^{5/3}+\rho_0|\mathbf{q}|^2\Big),\qquad 0\le\alpha\le1.
\end{equation}
Here we have used $\rho_0^{7/6}|\mathbf{q}|\le\rho_0^{4/3}+\rho_0|\mathbf{q}|^2$ and $\rho_0^{4/3}|\mathbf{q}|\le\rho_0^{5/3}+\rho_0|\mathbf{q}|^2$. Next, let $\bm{\rho}_1\neq\bm{\rho}_2$ be two densities of the same trace $\rho_0>0$ such that $\bm{\rho}_k\ge\nu\rho_0$, and let
$$
\mathbf{Z}:=\bm{\rho}_2+\nu\rho_0\frac{\bm{\rho}_2-\bm{\rho}_1}{|\bm{\rho}_2-\bm{\rho}_1|}.
$$
The operator norm of a traceless Hermitian $2\times2$ matrix $\mathbf{H}$ is $|\mathbf{H}|/\sqrt2$, hence $\mathbf{Z}\ge0$ and $\Trs\mathbf{Z}=\rho_0$. We have $\bm{\rho}_2=(1-\lambda)\bm{\rho}_1+\lambda\mathbf{Z}$ with $\lambda:=|\bm{\rho}_2-\bm{\rho}_1|/(|\bm{\rho}_2-\bm{\rho}_1|+\nu\rho_0)\le|\bm{\rho}_2-\bm{\rho}_1|/(\nu\rho_0)$. The convexity (ii) and then~\cref{eq:bounds_SSW} provide
\begin{align*}
e_\SSW(\bm{\rho}_2,\mathbf{q})-e_\SSW(\bm{\rho}_1,\mathbf{q})&\le\lambda\Big(e_\SSW(\mathbf{Z},\mathbf{q})-e_\SSW(\bm{\rho}_1,\mathbf{q})\Big)\\
&\le\lambda\Big(c_\TF\rho_0^{5/3}+c_\LO\rho_0^{4/3}+\frac{\rho_0|\mathbf{q}|^2}4\Big)\\
&\le\frac{c_\TF\rho_0^{2/3}+c_\LO\rho_0^{1/3}+\frac14|\mathbf{q}|^2}{\nu}\,|\bm{\rho}_1-\bm{\rho}_2|.
\end{align*}
Exchanging $\bm{\rho}_1$ and $\bm{\rho}_2$, we obtain~\cref{eq:Lipschitz_SSW} for two densities of the same trace. For $0<\rho_1\le\rho_2$ set $\alpha=\rho_1/\rho_2$. Compare $\bm{\rho}_1$ with $\alpha\bm{\rho}_2$ at their common trace and then use~\cref{eq:Lipschitz_ray_SSW} to compare $\alpha\bm{\rho}_2$ with $\bm{\rho}_2$. Both differences have the required bound since $(1-\alpha)\rho_2\le\sqrt2|\bm{\rho}_1-\bm{\rho}_2|$ and $|\bm{\rho}_1-\alpha\bm{\rho}_2|\le|\bm{\rho}_1-\bm{\rho}_2|+(1-\alpha)\rho_2$. If $\rho_1=0$, the ray bound at $\alpha=0$ suffices.

\medskip

\noindent\textbf{Part (v).} Fix $\mathbf{q}$ and write $H(r)=r^{4/3}+r^{5/3}+r|\mathbf{q}|^2$. First suppose $0\le\mathbf{A}_1\le\mathbf{A}_2$, with $r_i=\Trs\mathbf{A}_i$. We may take $r_1<r_2$ since ordered densities with the same trace coincide. We shall prove
$$
e_\SSW(\mathbf{A}_1,\mathbf{q})-e_\SSW(\mathbf{A}_2,\mathbf{q})\le C H(r_2)h^{1/3},
\qquad h=(r_2-r_1)/r_2.
$$
As in \cref{lem:monotone}, take the contraction $\mathbf{B}=\sqrt{\mathbf{A}_1}\,\mathbf{A}_2^{-1/2}$ with the inverse on the support of $\mathbf{A}_2$. The inclusion $\ran\mathbf{A}_1\subset\ran\mathbf{A}_2$ gives $\mathbf{B}\sqrt{\mathbf{A}_2}=\sqrt{\mathbf{A}_1}$. Since $\sqrt{\mathbf{A}_1}\le\sqrt{\mathbf{A}_2}$, we have $\Trs(\sqrt{\mathbf{A}_1}\sqrt{\mathbf{A}_2})\ge r_1$, hence
$$
\big|(\iden-\mathbf{B})\sqrt{\mathbf{A}_2}\big|^2
=\big|\sqrt{\mathbf{A}_2}-\sqrt{\mathbf{A}_1}\big|^2\le r_2-r_1.
$$
Let $1/2\le t\le3/2$ and $\delta\le\ell/C$. Write a minimizing state for the spin-spiral wave of density $\mathbf{A}_2(\iden_{t\ell\tetra}\ast\eta_{t\delta})$, which exists by \cref{thm:LL_attainment}, as the rotation by $\mathbf{U}_{\mathbf{q}}$ of a state $\Gamma$ of spin-polarized density $\mathbf{A}_2(\iden_{t\ell\tetra}\ast\eta_{t\delta})$. Rotate the $\mathbf{B}$-localized state $\Gamma_{|\mathbf{B}}$ by $\mathbf{U}_{\mathbf{q}}$ to obtain a trial state for the spin-spiral wave of density $\mathbf{A}_1(\iden_{t\ell\tetra}\ast\eta_{t\delta})$. Both rotations preserve the Coulomb interaction, and the localization decreases it by the proof of \cref{lem:monotone}. Subtracting the direct energies therefore costs at most $(r_2^2-r_1^2)D(\iden_{t\ell\tetra}\ast\eta_{t\delta})\le C|t\ell\tetra|\,r_2^2 h\ell^2$.

Let $\bm{\gamma}=\sum_kn_k|\bm{\Phi}_k\rangle\langle\bm{\Phi}_k|$ be the one-particle density matrix of $\Gamma$. By~\cref{eq:kinetic_spiral} the kinetic energy after rotation is $\TC(\Gamma)-\mathbf{q}\cdot\mathbf{j}_\Gamma+|\mathbf{q}|^2r_2|t\ell\tetra|/4$. Localization decreases the first and last terms. To estimate the change in the middle term, set $P_{\mathbf{q}}=\bm{\sigma}_3\,\mathbf{q}\cdot(-i\grad)$. For $\bm{\Phi}\in H^1(\RR^3,\CC^2)$ integration by parts gives
\begin{align*}
\big|\pscal{\bm{\Phi},P_{\mathbf{q}}\bm{\Phi}}-\pscal{\mathbf{B}\bm{\Phi},P_{\mathbf{q}}\mathbf{B}\bm{\Phi}}\big|
&\le\|(\iden-\mathbf{B})\bm{\Phi}\|_2\big(\|P_{\mathbf{q}}\bm{\Phi}\|_2+\|P_{\mathbf{q}}\mathbf{B}\bm{\Phi}\|_2\big)\\
&\le 2|\mathbf{q}|\,\|(\iden-\mathbf{B})\bm{\Phi}\|_2\,\|\grad\bm{\Phi}\|_2 .
\end{align*}
Since $\sum_kn_k\|(\iden-\mathbf{B})\bm{\Phi}_k\|_2^2=|t\ell\tetra|\,|(\iden-\mathbf{B})\sqrt{\mathbf{A}_2}|^2\le|t\ell\tetra|\,r_2h$, the change in the current term is at most $2|\mathbf{q}|\sqrt{|t\ell\tetra|\,r_2h\,\TC(\Gamma)}$. By the Lieb--Oxford inequality, the kinetic energy of the minimizing state, that is, before undoing the rotation, is at most $|t\ell\tetra|\big(e_{t\ell\tetra,\eta_{t\delta}}(\mathbf{A}_2,\mathbf{q})+c_\LO r_2^{4/3}\big)$. Undoing the rotation at most doubles it, up to $|\mathbf{q}|^2r_2|t\ell\tetra|/2$, hence
$$
\TC(\Gamma)\le2|t\ell\tetra|\big(e_{t\ell\tetra,\eta_{t\delta}}(\mathbf{A}_2,\mathbf{q})+c_\LO r_2^{4/3}\big)+\frac{|\mathbf{q}|^2r_2|t\ell\tetra|}{2}.
$$
Consequently~\cref{eq:bounds_tetrahedron_spiral} gives
$$
\TC(\Gamma)\le C|t\ell\tetra|\big(H(r_2)+r_2/(\ell\delta)\big).
$$
Take $\delta=r_2^{-1/3}$ and $\ell=L\delta h^{-1/3}$ with a fixed sufficiently large $L$. Then $r_2/(\ell\delta)\le r_2^{5/3}$, and the current error per unit volume is bounded by $C H(r_2)\sqrt h$. The direct energy error per unit volume is bounded by $Cr_2^{4/3}h^{1/3}$. These estimates are uniform in $t$ and in the direction of $\mathbf{q}$. Averaging the trial-state inequality over directions and then over $t$ with the normalized weight $t^{-4}$, we apply~\cref{eq:rate_SSW_lower} of \cref{prop:rate_SSW} to $\mathbf{A}_1$ and~\cref{eq:rate_SSW_upper} to $\mathbf{A}_2$. Their errors are bounded by $C H(r_2)h^{1/3}$ with this choice of $\ell,\delta$. This proves the ordered estimate.

For general densities $\mathbf{A}_1,\mathbf{A}_2\ge0$, set $r=\max(r_1,r_2)>0$ and $d=|\mathbf{A}_1-\mathbf{A}_2|$. When $d\ge r/4$ the result follows from~\cref{eq:bounds_SSW}. Otherwise $r_1,r_2\ge r-\sqrt2d\ge r/2$. The density $\mathbf{Z}=\mathbf{A}_2+d\iden$ dominates $\mathbf{A}_1$ and has trace $s=r_2+2d$. Since $0\le s-r_1\le(2+\sqrt2)d$ and $r/2\le s\le 3r/2$, the ordered estimate bounds $e_\SSW(\mathbf{A}_1,\mathbf{q})-e_\SSW(\mathbf{Z},\mathbf{q})$ by $C H(r)(d/r)^{1/3}$. Set $\alpha=r_2/s$. Then $\alpha\mathbf{Z}=\alpha\mathbf{A}_2+(1-\alpha)(r_2/2)\iden$. Convexity at trace $r_2$ and~\cref{eq:bounds_SSW} give
$$
e_\SSW(\alpha\mathbf{Z},\mathbf{q})-e_\SSW(\mathbf{A}_2,\mathbf{q})\le C(1-\alpha)H(r).
$$
By~\cref{eq:Lipschitz_ray_SSW}, passing from $\alpha\mathbf{Z}$ to $\mathbf{Z}$ costs at most $C(1-\alpha)H(r)$, where $1-\alpha=2d/s\le4d/r$. Adding the bounds and exchanging $\mathbf{A}_1$ and $\mathbf{A}_2$ proves~\cref{eq:Holder_SSW}. Joint continuity follows from~\cref{eq:Lipschitz_q,eq:Holder_SSW} at positive trace and from~\cref{eq:bounds_SSW} at zero.

\medskip

\noindent\textbf{Part (vi).} For a unit vector $\mathbf{n}$ put $\mathbf{M}_{\mathbf{n}}(t)=(1-t)\mathbf{P}_{\mathbf{n}}+t\mathbf{P}_{-\mathbf{n}}$. At a fixed trace $r>0$ and wavevector $\mathbf{q}$, the function $F_{\mathbf{n}}(t)=e_\SSW(r\mathbf{M}_{\mathbf{n}}(t),\mathbf{q})$ is convex on $[0,1]$ by (ii). Its $1/3$-H\"older seminorm is at most $C H(r)$ by~\cref{eq:Holder_SSW}. The secant slopes to $0$ and $1$ thus bound its one-sided derivatives by
$$
-C H(r)t^{-2/3}\le F_{\mathbf{n},-}'(t)\le F_{\mathbf{n},+}'(t)\le C H(r)(1-t)^{-2/3},\qquad 0<t<1.
$$
Integrating on $0<s<t\le1/2$ and using continuity at $0$ gives
$$
|F_{\mathbf{n}}(t)-F_{\mathbf{n}}(s)|\le C H(r)|t^{1/3}-s^{1/3}|
\le C_p H(r)|t^{1/p}-s^{1/p}|,\qquad 0\le s,t\le1/2.
$$
The last inequality follows because $u\mapsto u^{p/3}$ is Lipschitz on $[0,1]$.

We also need the cost of changing the spin direction. Let $\mathbf{A}$ have trace $r$ and let $\mathbf{S}\in\SU(2)$ be constant. In the smeared tetrahedron $\iden_{t\ell\tetra}\ast\eta_{t\delta}$ of (v) use the state $\Gamma$ in the rotating frame as in (v), now for the density $\mathbf{A}$. Rotating $\Gamma$ by $\mathbf{S}$ in that frame and then by $\mathbf{U}_{\mathbf{q}}$ gives a trial state for the spin-spiral wave of density $\mathbf{S}\mathbf{A}\mathbf{S}^*$. By~\cref{eq:kinetic_spiral}, only the current term changes. With the orbitals of (v), its change in absolute value is at most
$$
|\mathbf{q}|\,\|\mathbf{S}^*\bm{\sigma}_3\mathbf{S}-\bm{\sigma}_3\|
\sum_k n_k\|\bm{\Phi}_k\|_2\|\grad\bm{\Phi}_k\|_2
\le 2|\mathbf{q}|\,\|\mathbf{S}-\iden\|\sqrt{r|t\ell\tetra|\,\TC(\Gamma)}.
$$
Taking $\delta=1$, we have $\TC(\Gamma)\le C|t\ell\tetra|\,(H(r)+r/\ell)$ as proved in (v). Dividing by $|t\ell\tetra|$ and passing to the limit $\ell\to\infty$ using \cref{thm:spiral_limit}, then applying the same argument to $\mathbf{S}^*$, yields
$$
\big|e_\SSW(\mathbf{S}\mathbf{A}\mathbf{S}^*,\mathbf{q})-e_\SSW(\mathbf{A},\mathbf{q})\big|\le C H(r)\|\mathbf{S}-\iden\|,
$$
since $|\mathbf{q}|\sqrt{rH(r)}\le H(r)$.

First consider two densities $r\mathbf{M}_i$ of the same trace, with $\mathbf{M}_i=\mathbf{M}_{\mathbf{n}_i}(t_i)$ and $0\le t_i\le1/2$. Write $d=|\mathbf{M}_1^{1/p}-\mathbf{M}_2^{1/p}|$. The smaller eigenvalues of the roots satisfy $|t_1^{1/p}-t_2^{1/p}|\le d$. If both $t_i\ge1/4$, \cref{eq:Lipschitz_SSW} with $\nu=1/4$ suffices since $|\mathbf{M}_1-\mathbf{M}_2|\le p d$ by the matrix Lipschitz bound used in~\cref{eq:Lipschitz_power}. Otherwise let $g_i=(1-t_i)^{1/p}-t_i^{1/p}$. The traceless part of $\mathbf{M}_i^{1/p}$ is $g_i\,\mathbf{n}_i\cdot\bm{\sigma}/2$, so $|g_1\mathbf{n}_1-g_2\mathbf{n}_2|\le\sqrt2 d$. Since $\max(g_1,g_2)\ge (3/4)^{1/p}-(1/4)^{1/p}>0$, we obtain
$$
|\mathbf{n}_1-\mathbf{n}_2|\le \frac{2|g_1\mathbf{n}_1-g_2\mathbf{n}_2|}{\max(g_1,g_2)}\le C_p d.
$$
Take the rotation of smallest angle from $\mathbf{n}_1$ to $\mathbf{n}_2$. It has a lift $\mathbf{S}\in\SU(2)$ with $\|\mathbf{S}-\iden\|\le|\mathbf{n}_1-\mathbf{n}_2|$. Changing first the direction by $\mathbf{S}$ and then the smaller eigenvalue fraction by the bound on $F_{\mathbf{n}_2}$ proves~\cref{eq:root_Lipschitz_SSW} at equal trace.

Finally suppose $0<r_1\le r_2=r$, write $\mathbf{M}_i=\mathbf{A}_i/r_i$ and set $\alpha=r_1/r$ and $D=|\mathbf{A}_1^{1/p}-\mathbf{A}_2^{1/p}|/r^{1/p}$. The Schatten $p$-norm of each $\mathbf{M}_i^{1/p}$ is one, and this norm is bounded by the Hilbert--Schmidt norm. Its reverse triangle inequality gives $1-\alpha^{1/p}\le D$. Hence
$$
|\mathbf{M}_1^{1/p}-\mathbf{M}_2^{1/p}|
\le D+(1-\alpha^{1/p})|\mathbf{M}_1^{1/p}|\le (1+\sqrt2)D .
$$
Apply the equal-trace bound at trace $r_1$, and then~\cref{eq:Lipschitz_ray_SSW} to compare $r_1\mathbf{M}_2$ with $r\mathbf{M}_2$. Since $1-\alpha\le pD$ and $H(r_1)\le H(r)$, this proves~\cref{eq:root_Lipschitz_SSW}. When $r_1=0$, we have $D\ge1$ and~\cref{eq:bounds_SSW} suffices.
\end{proof}

For the application to slowly varying densities we express the cube-root estimate in (vi) in terms of the prescribed power $\theta$. If $\bm{\rho}_1,\bm{\rho}_2\ge0$ have traces $\rho_1,\rho_2$, $r=\max\{\rho_1,\rho_2\}>0$ and $0<\theta<1$, then
\begin{equation}\label{eq:theta_to_cube_root}
\frac{\big|\bm{\rho}_1^{1/3}-\bm{\rho}_2^{1/3}\big|}{r^{1/3}}\le C_\theta\left(\frac{\big|\bm{\rho}_1^{\theta}-\bm{\rho}_2^{\theta}\big|}{r^{\theta}}\right)^{\min\{1,1/(3\theta)\}} .
\end{equation}
Indeed the spectra of $\mathbf{X}_k=\bm{\rho}_k^\theta/r^\theta$ lie in $[0,1]$. For $\gamma=1/(3\theta)\ge1$, the function $t^\gamma$ is Lipschitz on this interval and~\cref{eq:HS_functions} gives the same bound in Hilbert--Schmidt norm. For $\gamma<1$ use $|t^\gamma-s^\gamma|\le|t-s|^\gamma$ in~\cref{eq:HS_functions} and Jensen's inequality with the weights $|\pscal{\mathbf{u}_i,\mathbf{v}_k}|^2/2$, which sum to one. This gives $|\mathbf{X}_1^\gamma-\mathbf{X}_2^\gamma|\le2^{(1-\gamma)/2}|\mathbf{X}_1-\mathbf{X}_2|^\gamma$.

At $\mathbf{q}=0$ the proposition gives the corresponding estimates for $e_\UEG$. At fixed trace $r>0$, convexity and invariance under spin rotations imply that the function
\begin{equation*}
 f_r(\zeta)=e_\UEG\left(\tfrac{1+\zeta}{2}r,\tfrac{1-\zeta}{2}r\right)
\end{equation*}
is convex and even on $[-1,1]$. In particular it is non-decreasing on $[0,1]$. Applying~\cref{eq:Holder_SSW} at $\mathbf{q}=0$ to $\diag(r,0)$ and $\diag(r-\rho',\rho')$ yields
\begin{equation*}
 0\le e_\UEG(r,0)-e_\UEG(r-\rho',\rho')
 \le C(r^{4/3}+r^{5/3})(\rho'/r)^{1/3},\qquad 0\le\rho'\le r/2 .
\end{equation*}


\begin{thebibliography}{28}

\bibitem[Ando(1979)]{ando1979concavity}
Tsuyoshi Ando.
\newblock Concavity of certain maps on positive definite matrices and
  applications to {H}adamard products.
\newblock \emph{Linear Algebra and its Applications}, 26:\penalty0 203--241,
  1979.

\bibitem[Bach et~al.(1994)Bach, Lieb, and Solovej]{bach1994generalized}
Volker Bach, Elliott~H Lieb, and Jan~Philip Solovej.
\newblock Generalized {H}artree-{F}ock theory and the {H}ubbard model.
\newblock \emph{Journal of statistical physics}, 76:\penalty0 3--89, 1994.

\bibitem[Bhatia(1997)]{bhatia1997matrix}
Rajendra Bhatia.
\newblock \emph{Matrix analysis}, volume 169 of \emph{Graduate Texts in
  Mathematics}.
\newblock Springer-Verlag, New York, 1997.

\bibitem[Csirik and Laestadius(2026)]{csirik2026non}
Mih{\'a}ly~A Csirik and Andre Laestadius.
\newblock The non-uniform electron gas.
\newblock \emph{arXiv preprint arXiv:2603.21200}, 2026.

\bibitem[Cycon et~al.(2009)Cycon, Froese, Kirsch, and
  Simon]{cycon2009schrodinger}
Hans~L Cycon, Richard~G Froese, Werner Kirsch, and Barry Simon.
\newblock \emph{Schr{\"o}dinger operators: With application to quantum
  mechanics and global geometry}.
\newblock Springer, 2009.

\bibitem[Eich and Gross(2013)]{eich2013transverse}
Florian~G Eich and Eberhard K~U Gross.
\newblock Transverse spin-gradient functional for noncollinear
  spin-density-functional theory.
\newblock \emph{Physical Review Letters}, 111:\penalty0 156401, 2013.

\bibitem[Engel and Dreizler(2011)]{engel2011density}
Eberhard Engel and Reiner~M Dreizler.
\newblock \emph{Density Functional Theory: An Advanced Course}.
\newblock Theoretical and Mathematical Physics. Springer, Berlin, Heidelberg,
  2011.

\bibitem[Gilbarg and Trudinger(2001)]{gilbarg2001elliptic}
David Gilbarg and Neil~S Trudinger.
\newblock \emph{Elliptic Partial Differential Equations of Second Order}.
\newblock Classics in Mathematics. Springer, Berlin, Heidelberg, 2001.

\bibitem[Glaeser(1963)]{glaeser1963racine}
Georges Glaeser.
\newblock Racine carr{\'e}e d'une fonction diff{\'e}rentiable.
\newblock \emph{Annales de l'Institut Fourier}, 13\penalty0 (2):\penalty0
  203--210, 1963.

\bibitem[Gontier(2013)]{gontier2013n}
David Gontier.
\newblock {$N$}-representability in noncollinear spin-polarized
  density-functional theory.
\newblock \emph{Physical Review Letters}, 111\penalty0 (15):\penalty0 153001,
  2013.

\bibitem[Gontier(2015)]{gontier2015existence}
David Gontier.
\newblock Existence of minimizers for {K}ohn--{S}ham within the local spin
  density approximation.
\newblock \emph{Nonlinearity}, 28:\penalty0 57--76, 2015.

\bibitem[Graf and Schenker(1995)]{graf1995molecular}
Gian~Michele Graf and Daniel Schenker.
\newblock On the molecular limit of {C}oulomb gases.
\newblock \emph{Communications in mathematical physics}, 174:\penalty0
  215--227, 1995.

\bibitem[Hoffmann-Ostenhof and
  Hoffmann-Ostenhof(1977)]{hoffmann1977schrodinger}
Maria Hoffmann-Ostenhof and Thomas Hoffmann-Ostenhof.
\newblock ``{S}chr{\"o}dinger inequalities'' and asymptotic behavior of the
  electron density of atoms and molecules.
\newblock \emph{Physical Review A}, 16\penalty0 (5):\penalty0 1782, 1977.

\bibitem[K{\"u}bler et~al.(1988)K{\"u}bler, H{\"o}ck, Sticht, and
  Williams]{kubler1988density}
J.~K{\"u}bler, K.-H. H{\"o}ck, J.~Sticht, and A.~R. Williams.
\newblock Density functional theory of non-collinear magnetism.
\newblock \emph{Journal of Physics F: Metal Physics}, 18:\penalty0 469--483,
  1988.

\bibitem[Kubo and Ando(1980)]{kubo1980means}
Fumio Kubo and Tsuyoshi Ando.
\newblock Means of positive linear operators.
\newblock \emph{Mathematische Annalen}, 246:\penalty0 205--224, 1980.

\bibitem[Last and Penrose(2017)]{last2017lectures}
G{\"u}nter Last and Mathew Penrose.
\newblock \emph{Lectures on the {P}oisson Process}, volume~7 of \emph{Institute
  of Mathematical Statistics Textbooks}.
\newblock Cambridge University Press, Cambridge, 2017.

\bibitem[Levy(1979)]{levy1979universal}
Mel Levy.
\newblock Universal variational functionals of electron densities, first-order
  density matrices, and natural spin-orbitals and solution of the
  $v$-representability problem.
\newblock \emph{Proceedings of the National Academy of Sciences}, 76\penalty0
  (12):\penalty0 6062--6065, 1979.

\bibitem[Lewin(2011)]{lewin2011geometric}
Mathieu Lewin.
\newblock Geometric methods for nonlinear many-body quantum systems.
\newblock \emph{Journal of Functional Analysis}, 260\penalty0 (12):\penalty0
  3535--3595, 2011.

\bibitem[Lewin et~al.(2018)Lewin, Lieb, and Seiringer]{lewin2018statistical}
Mathieu Lewin, Elliott~H Lieb, and Robert Seiringer.
\newblock Statistical mechanics of the uniform electron gas.
\newblock \emph{Journal de l'{\'E}cole polytechnique-Math{\'e}matiques},
  5:\penalty0 79--116, 2018.
\newblock doi:~\href{https://doi.org/10.5802/jep.64}{10.5802/jep.64}.

\bibitem[Lewin et~al.(2020)Lewin, Lieb, and Seiringer]{lewin2020local}
Mathieu Lewin, Elliott~H Lieb, and Robert Seiringer.
\newblock The local density approximation in density functional theory.
\newblock \emph{Pure and Applied Analysis}, 2\penalty0 (1):\penalty0 35--73,
  2020.
\newblock doi:~\href{https://doi.org/10.2140/paa.2020.2.35}{10.2140/paa.2020.2.35}.

\bibitem[Lewin et~al.(2022)Lewin, Lieb, and Seiringer]{lewin2022improved}
Mathieu Lewin, Elliott~H Lieb, and Robert Seiringer.
\newblock Improved {L}ieb--{O}xford bound on the indirect and exchange
  energies.
\newblock \emph{Letters in Mathematical Physics}, 112\penalty0 (5):\penalty0
  92, 2022.
\newblock doi:~\href{https://doi.org/10.1007/s11005-022-01584-5}{10.1007/s11005-022-01584-5}.

\bibitem[Lewin et~al.(2023)Lewin, Lieb, and Seiringer]{lewin2023universal}
Mathieu Lewin, Elliott~H Lieb, and Robert Seiringer.
\newblock Universal functionals in density functional theory.
\newblock In {\'E}ric Canc{\`e}s and Gero Friesecke, editors, \emph{Density
  Functional Theory}, pages 115--182. Springer, 2023.
\newblock doi:~\href{https://doi.org/10.1007/978-3-031-22340-2_3}{10.1007/978-3-031-22340-2\_3}.

\bibitem[Lieb(1979)]{lieb1979lower}
Elliott~H Lieb.
\newblock A lower bound for {C}oulomb energies.
\newblock \emph{Physics Letters A}, 70\penalty0 (5-6):\penalty0 444--446, 1979.

\bibitem[Lieb(1981)]{lieb1981variational}
Elliott~H Lieb.
\newblock Variational principle for many-fermion systems.
\newblock \emph{Physical Review Letters}, 46:\penalty0 457--459, 1981.

\bibitem[Lieb(1983)]{lieb1983density}
Elliott~H Lieb.
\newblock Density functionals for {C}oulomb systems.
\newblock \emph{International Journal of Quantum Chemistry}, 24\penalty0
  (3):\penalty0 243--277, 1983.

\bibitem[Lieb and Oxford(1981)]{lieb1981improved}
Elliott~H Lieb and Stephen Oxford.
\newblock Improved lower bound on the indirect {C}oulomb energy.
\newblock \emph{International Journal of Quantum Chemistry}, 19\penalty0
  (3):\penalty0 427--439, 1981.

\bibitem[Lieb and Seiringer(2010)]{lieb2010stability}
Elliott~H Lieb and Robert Seiringer.
\newblock \emph{The stability of matter in quantum mechanics}.
\newblock Cambridge University Press, 2010.

\bibitem[von Barth and Hedin(1972)]{von1972local}
Ulf von Barth and Lars Hedin.
\newblock A local exchange-correlation potential for the spin polarized case:
  {I}.
\newblock \emph{Journal of Physics C: Solid State Physics}, 5:\penalty0
  1629--1642, 1972.

\end{thebibliography}
\end{document}